\documentclass[11pt]{article}
\usepackage{graphicx, yk} 
\usepackage[margin=2.5cm]{geometry}
\usepackage{float}
\usepackage{amsmath, amsfonts, framed}
\usepackage{amssymb}
\usepackage{tikz}
\usepackage{multirow}
\usepackage{hyperref}       
\usepackage{textcmds}
\usepackage{longtable}
\usepackage{array}
\usepackage{dsfont}
\usepackage{algorithm}
\usepackage{adjustbox}
\usepackage{circuitikz}
\usepackage{booktabs}
\usepackage{cleveref}
\usepackage{makecell}
\usepackage[table]{xcolor}
\usepackage[normalem]{ulem}
\usepackage{algpseudocode}
\usepackage[bottom]{footmisc}
\hypersetup{colorlinks,linkcolor={red},citecolor={blue},urlcolor={red}}  
\tikzset{every picture/.style={line width=0.75pt}} 
\usepackage{natbib}
\setcitestyle{author year,open={(},close={)}}

\newcommand\blfootnote[1]{
  \begingroup
  \renewcommand\thefootnote{}\footnote{#1}
  \addtocounter{footnote}{-1}
  \endgroup
}

\providecommand{\keywords}[1]
{
  \small	
  \textbf{\textit{Keywords---}} #1
}

\newcommand{\indep}{\perp\!\!\!\!\perp}
\newcommand{\independent}{\perp\!\!\!\!\perp}

\title{\bf Debiased Causal Mediation Analysis in Ultra-High-Dimensional Settings in the Presence of Interaction Effects}
\author{Shi Bo, AmirEmad Ghassami$^\dagger$, and Debarghya Mukherjee$^\dagger$ \\
Department of Mathematics and Statistics, Boston University}
\date{First Version: December 11, 2024; Current Version:  August 11, 2026}

\begin{document}

\maketitle 

\begin{abstract}
Mediation analysis is a crucial tool for uncovering the mechanisms through which a treatment affects an outcome, providing deeper causal insights and guiding effective interventions. Despite substantial advances in mediation analysis with fixed- or low-dimensional mediators and covariates, estimation and inference for mediation functionals remain limited when both mediators and covariates are ultra-high-dimensional. In this paper, we propose an estimator for the mediation functional in a high-dimensional setting that accommodates the treatment--covariate interactions in the mediator model, as well as treatment--covariate and treatment--mediator interactions in the outcome model. As the parameter of interest involves high-dimensional components from different treatment arms and regression equations, existing debiasing approaches are not directly applicable, motivating our multi-step debiasing technique for handling such structurally complex functionals. We establish that the proposed estimator is $\sqrt{n}$-consistent and asymptotically normal, enabling valid inference for natural direct and indirect effects. We evaluate our proposed methodology through extensive simulation studies and apply it to the TCGA lung cancer dataset to estimate the effect of smoking, mediated by DNA methylation, on the survival time of lung cancer patients.
\end{abstract}

\blfootnote{$^\dagger$ Corresponding authors.}

\newpage

\begingroup
\hypersetup{linkcolor=black}
\tableofcontents
\endgroup

\newpage

\section{Introduction}

The most commonly targeted quantity in causal inference is the total causal effect, which captures the effect of a treatment, action, or policy on an outcome variable of interest. Causal mediation analysis goes one step further by identifying the mechanisms through which treatment influences the outcome variable. Hence, it plays a crucial role in deepening our understanding of the intricate causal relationships within the variables of the system under
study. Mediation analysis is widely used in many fields of science, including social science, behavioral science, economics, decision-making, epidemiology, and neuroscience. In the past three decades, the topic has also received much attention in the statistical literature (see, e.g., \citep{albert2008mediation, lindquist2012functional, baron1986moderator, richiardi2013mediation, ten2012review}).

The advancement of technology has enabled the collection of an extensive array of variables in many scientific domains, including brain imaging, genetics, epidemiology, and public health studies. Consequently, mediation analysis is increasingly carried out in high-dimensional settings, where the number of potential mediators can be comparable to or even larger than the sample size. This challenge is particularly pronounced in environmental epigenetics and other omics applications, where numerous molecular measurements may serve as mediating mechanisms.
Early approaches to high-dimensional mediation analysis often relied on heuristic screening, dimension reduction, or pathway selection, with limited theoretical guarantees. \textcolor{black}{Furthermore, the existing theory typically does not simultaneously address ultra-high-dimensional covariates, mediators, and treatment interactions in both the mediator and outcome models, which are important components in many real-world applications. 
Developing a methodology that provides valid inference for natural direct and indirect effects in this full, interaction-rich, ultra-high-dimensional setting, therefore, addresses an important gap in the literature and is the focus of the current paper.}

\medskip
\noindent
{\bf Our contribution: }
In this paper, we propose a new methodology for estimating mediation functionals in the presence of ultra-high-dimensional (UHD) covariates and mediators, where the ambient dimensions may be exponentially large compared to the sample size. We consider a structural linear model in which the response depends on both UHD covariates and UHD mediators, and the mediators themselves are generated from a linear model in the UHD covariates. Unlike the majority of the existing literature on high-dimensional mediation analysis, our setup allows for interaction with binary exposure for both data-generating processes. To the best of our knowledge, this is the first work to provide a methodology for estimating direct and indirect treatment effects in the presence of both UHD covariates and mediators, along with their interaction with a binary exposure variable. 

\textcolor{black}{Our method of estimating the mediation functional essentially combines two high-dimensional regressions with two carefully designed balancing steps. We first regress the response on both covariates and mediators using treatment observations. Rather than estimating the entire high-dimensional relationship between the mediators and covariates, which would require a large number of high-dimensional regressions, we use control observations to construct balancing weights that directly approximate only the component of this relationship needed for estimating the parameter of interest.
Because the mediation functional involves parameters from different treatment arms and different regression equations, the resulting target is not a standard linear functional of a single high-dimensional regression coefficient, nor a standard bilinear functional or inner product between two regression coefficients; it contains composite terms involving both UHD vectors and UHD matrices (where both the number of rows and columns may be exponentially large in sample size). Consequently, existing debiasing methods for linear functionals \cite{javanmard2014confidence,van2014asymptotically,zhang2014confidence} and bilinear or inner-product functionals \cite{guo2019optimal,cai2021optimal} are not directly applicable; see Remark~\ref{rem:existing_debiasing_functional} for more details.
To overcome this difficulty, we develop a multi-stage debiasing strategy; in particular, we introduce another carefully designed high-dimensional regression, coupled with the preceding balancing step, to remove the first-order bias introduced by regularized estimation and by the intermediate approximation used in constructing the mediation functional.
We show that, under fairly general sparsity, regularity, and restricted-eigenvalue-type conditions, the proposed estimator of the mediation functional (and consequently the natural direct and indirect effects) is $\sqrt{n}$-consistent and asymptotically normal. We also construct a consistent estimator of its asymptotic variance, which leads to asymptotically valid confidence intervals for the mediation functional, and for the natural direct and indirect effects.}

The rest of the paper is organized as follows. 
After introducing the problem setting in Section \ref{sec:setting}, we describe our debiased estimation procedure in Section \ref{sec:method}.
In Section \ref{section:3}, we provide the bias analysis of our estimator and its connection to the fixed-dimensional setting. In Section \ref{sec:theory}, we provide a theoretical justification for our approach under high-dimensional asymptotics. The proof of the main theorem, along with technical lemmas, can be found in the supplementary material. In Section \ref{section:5}, we conduct some simulation experiments and a real data experiment on the TCGA lung cancer dataset. A software implementation in R is available at \textcolor{blue}{UHDmedi}\footnote{https://github.com/shibo769/UHDmedi/tree/main}. We conclude in Section \ref{section:7} and discuss future work.

\subsection{Related Work}
Early approaches to high-dimensional mediation analysis often relied on screening, dimension reduction, or pathway selection.
For example, \cite{zhang2016estimating} and \cite{perera2022hima2} proposed to first screen mediators that are strongly associated with the outcome and then conduct standard mediation analysis on the selected low-dimensional set. \cite{huang2016hypothesis} and \cite{chen2018high} suggested dimension-reduction approaches that map high-dimensional mediators to low-dimensional representations, while \cite{zhang2018distance} developed a distance-based approach that aggregates individually weak signals. \cite{zhao2016pathway} proposed a pathwise Lasso method for selecting sparse mediation pathways in the presence of high-dimensional mediators. These methods provide useful tools for empirical analysis, but none of them provide full inferential theory for natural direct and indirect effects in ultra-high-dimensional settings. 
More recent work has developed theoretical guarantees for high-dimensional mediation inference. In particular, \cite{guo2022high,guo2023statistical,guo2024estimations} study estimation and testing for mediation models with high-dimensional mediators and fixed-dimensional exposures or observed covariates, including extensions to generalized mediation models. 
{\color{black} However, our work differs in several key aspects; we not only allow both UHD covariates and mediators in the response generation model, but also both treatment-covariate and treatment-mediator interactions in the outcome generation model and treatment-covariate interaction in the mediator generation model, which makes our model significantly more general and makes the analysis substantially more complicated. 
Moreover, our theoretical analysis avoids the strong beta-min/support-recovery-type requirements used in that line of work and is established under weaker high-dimensional assumptions; see  Section \ref{appendix:existing-high-dimensional-mediation-comparison} of the supplementary material for more details.}
\cite{lin2025} proposed a testing procedure for high-dimensional mediation effects that allows both the dimensions of mediators and exposures to grow with the sample size,  \textcolor{black}{but does not accommodate UHD covariates or interactions between treatment variable and covariates/mediators.}
\cite{rakshit2024statistical} developed high-dimensional Poisson regression inference with applications to mediation analysis for count outcomes, allowing high-dimensional mediators and treatment--mediator interactions. While their framework addresses an important nonlinear outcome setting, \textcolor{black}{our work studies a different source of difficulty under a linear link: the simultaneous presence of ultra-high-dimensional covariates and mediators, together with their interactions with the treatment variable, which leads to a substantially more complex parameter of interest.}
In addition, \cite{jones2025causal} studied causal mediation analysis \textcolor{black}{from a post-selection inference perspective with fixed-dimensional mediators, as opposed to the high-dimensional setup considered in this paper.} 
\cite{chakrabortty2018inference} and \cite{shi2022testing} investigated high-dimensional mediation problems under sparse linear structural equation or graphical models, \textcolor{black}{with a focus on causal discovery and individual or controlled mediation effects, as opposed to natural direct/indirect effects considered in this paper. Furthermore, they allow the dimension to grow polynomially with $n$, whereas our setup allows an exponentially large ambient dimension.} 
\cite{loh2022nonlinear} proposed an outcome-regression-based averaging approach for estimating interventional direct/indirect effects with high-dimensional mediators, \textcolor{black}{but their causal targets differ from the natural direct/indirect effects considered in our work. Moreover, their approach does not provide theoretical guarantees and relies on plug-in outcome-regression estimators (which can be potentially biased in high dimension), whereas our method develops a debiased estimator with valid asymptotic inference in the ultra-high-dimensional regime.}
Taken together, these studies make important contributions to high-dimensional mediation analysis, while their main focus is complementary to the one pursued in this work.

Our debiasing approach is primarily motivated by the large body of literature on debiasing techniques for estimating a contrast of high-dimensional parameters in a linear model (see, e.g., \citep{javanmard2014confidence,van2014asymptotically,zhang2014confidence}) or bilinear/quadraric functionals \citep{guo2019optimal,cai2021optimal}. 
{\color{black} However, these existing tools are not directly applicable to our setting, because the mediation functional depends on nuisance components learned across different treatment arms and different structural equations, and further involves both an unknown UHD matrix and an unknown UHD vector; see Remark \ref{rem:existing_debiasing_functional} and Section \ref{appendix:existing-debiasing-functional} of the supplementary material for more details.} 
The use of debiasing techniques in causal inference was also recently employed by \cite{athey2018approximate}, where the authors developed a methodology to estimate the total effect of a binary treatment on an outcome in the presence of high-dimensional covariates. 
Our work is a generalization of their work towards estimating both the natural direct and indirect causal effects separately in the presence of high-dimensional mediators and interaction with the treatment variable. 
{\color{black} Related balancing/Riesz-representer-based approaches, e.g., \citep{hirshberg2021augmented,chernozhukov2022automatic,ben2021balancing}, provide general tools for debiasing linear functionals; however, in our setting, these balancing steps are only intermediate components of a larger multi-stage debiasing procedure.}
Our proposed method can also be viewed as a high-dimensional counterpart of the influence function-based approach of \cite{tchetgen2012semiparametric}. In particular, we demonstrate that our estimator exhibits a second-order bias in product form, similar to the influence function-based approach in the fixed-dimensional setting, as established by \cite{liu2024bias}. However, our framework is different from this line of work, as the framework of \citep{tchetgen2012semiparametric,liu2024bias} is 
fundamentally fixed-dimensional and relies on nuisance objects such as the conditional mediator density, which are not directly estimable in our UHD regime without severely restrictive assumptions. Kindly see Section \ref{appendix:existing-high-dimensional-mediation-comparison} of the supplementary material for more details.

\section{Notations and Problem Setting}\label{sec:setting}
For a random variable/vector $X$, we denote $\mathbb{E}[X]$ and $\Sigma_X$ as the expectation and the variance/covariance matrix of $X$, respectively. 
For a matrix $\bX \in \reals^{m \times n}$, we use $ X_{i}$ (respectively $ X_{.,i}$) to denote its $i^{th}$ row (respectively $i^{th}$ column). We also use the notation $\|\bX\|_{p, q}$ to denote $(\sum_{i = 1}^m \|X_i\|_p^q)^{1/q}$, i.e. $\ell_q$ norm of the $\{\|X_i\|_p: 1 \le i \le m\}$. We also use the notation $\|\bX\|_{\rm op}$ and $\|\bX\|_F$ to denote its operator norm and the Frobenius norm, respectively. 
The notations $\mathbf{1}_q$ and $\mathbf{0}_q$ are used to denote the vectors consisting of all ones and all zeros, respectively, in $\mathbb{R}^q$. For a sequence of random variables $\{X_n\}_{n \in \bbN}$, we say that it admits an asymptotically deterministic equivalent sequence if there exists a deterministic sequence $\{x_n\}_{n \in \bbN}$ such that $X_n - x_n = o_p(1)$. 

We consider a setting with a binary treatment variable $A \in\{0,1\}$ (also called \textit{exposure} or \textit{action variable}), a set of pre-treatment covariates 
$ X = [X_1 \cdots X_p]^\top \in \mathbb{R}^p$, an outcome variable of interest, denoted by $Y$, and a set of post-treatment, pre-outcome variables $ M = [M_1 \cdots M_q]^\top \in \mathbb{R}^q$.  
We consider the challenging setting where $p \wedge q \gg n$. 
Using the potential outcome notations \citep{rubin1974estimating}, let $Y^{(a,m)}$ denote the potential outcome of  $Y$, had the treatment and mediator variables been set to values $A = a$ and $M = m$. Similarly, we define $M^{(a)}$ as the potential outcome of $M$ had the treatment variables been set to the value $A = a$. Based on variables $Y^{(a,m)}$ and $M^{(a)}$, we define $Y^{(a)} = Y^{(a,M^{(a)})}$. 
An i.i.d. sample $(X_i, A_i, M_i, Y_i)_{i=1}^n$ of size $n$ is given, where we assume
\begin{itemize}
    \item $Y = Y^{(a,m)}$ if $A = a$, $M = m$, and
    \item $M = M^{(a)}$ if $A = a$,
\end{itemize}
which is referred to as the \emph{consistency assumption}, requiring that the realized outcome is equal to the potential outcome corresponding to the observed treatment and mediator, and the realized mediator is equal to the potential mediator corresponding to the observed treatment. We also assume that the probability of receiving treatment and control and any value of the mediator are bounded away from zero, i.e., for some $ e > 0$, we have 
\begin{equation*}
    \textstyle
    0 < e < p(a \mid x) \ \ \forall (a, x), \quad \text{and} \quad  0 < e < p(m \mid a, x) \ \ \forall (a, x, m). 
\end{equation*}
which is referred to as the \emph{positivity assumption}.
The average treatment effect (ATE) is defined as $\mathbb{E}[Y^{(a=1)} - Y^{(a=0)}]$. This parameter captures the difference in the potential outcome mean if all units received treatment $A = 1$ and if all units received treatment $A = 0$ (e.g., placebo treatment).  In mediation analysis, the focus is on the fact that this causal effect is partly mediated through the variable $M$, and it is of interest to quantify the direct and indirect portions of the causal effect by partitioning the ATE as follows \citep{robins1992identifiability, pearl2022direct}. 
\begin{equation}\label{eq:NDENIE}
    \begin{aligned}
        \mathbb{E}\left[Y^{(1)}-Y^{(0)}\right]&=\mathbb{E}\left[Y^{\left(1, M^{(1)}\right)}-Y^{\left(0, M^{(0)}\right)}\right]\\
   & =\mathbb{E}\left[Y^{\left(1, M^{(1)}\right)}-Y^{\left(1, M^{(0)}\right)}\right]+\mathbb{E}\left[Y^{\left(1, M^{(0)}\right)}-Y^{\left(0, M^{(0)}\right)}\right].
    \end{aligned}
\end{equation}
The first and the second terms in the last expression are called the total indirect effect and the pure direct effect, respectively by \cite{robins1992identifiability}, and are called the natural indirect effect (NIE) and the natural direct effect (NDE) of the treatment on the outcome, respectively, by \cite{pearl2022direct}.
We will use the latter terminology in this work. NIE captures the change in the expectation of the response variable in a hypothetical scenario where the value of the treatment variable is fixed at $A = 1$, while the mediator behaves as if the treatment had been changed from value 0 to 1. NDE captures the change in the expectation of the outcome in a hypothetical scenario where the value of the treatment variable is changed from value 0 to 1, while the mediator behaves as if the treatment is fixed at 0. 
Note that if one can identify and estimate the parameter $\mathbb{E}[Y^{(a^\prime, M^{(a)})}]$ for $a,a^\prime \in \{0,1\}$, then based on Equation \eqref{eq:NDENIE}, NIE, NDE and ATE can all be estimated. Henceforth, we focus on the parameter $\mathbb{E}[Y^{(1,M^{(0)})}]$ in this paper.

We operate under the assumption of sequential exchangeability, which implies there are no unobserved confounders for the treatment-outcome, mediator-outcome, and treatment-mediator relationships \citep{imai2010identification}, formalized as follows. 
\begin{assumption}[Sequential exchangeability]\label{assumption:2.1}
Let $X_1 \independent X_2 \mid X_3$ indicate that the random variables $X_1$ and $X_2$ are conditionally independent given the random variable $X_3$.
\begin{enumerate}
    \item $Y^{(a,m)} \independent \{A,M\} \mid X$, for all $a,m$, 
    \item $M^{(a)} \independent A \mid X$, for all $a$, 
    \item $Y^{(a,m)} \independent M^{(a')}\mid X$, for all $a,a',m$. 
\end{enumerate}
\end{assumption}
As established in \citep{imai2010identification}, under Assumption \ref{assumption:2.1}, the parameter $\mathbb{E}[Y^{(1,M^{(0)})}]$ can be identified as 
\begin{equation} 
\textstyle
\label{theta:functional}
    \mathbb{E}[Y^{(1,M^{(0)})}]  =\iint \mathbb{E}[Y \mid X=x, M=m, A=1] f(m \mid X=x, A=0) f(x) d m d x,
\end{equation}
where we use $f$ to represent densities over variables. The right hand side is known as the \emph{mediation functional} in the literature, and will be our \emph{parameter of interest} in this work. 

The estimation of the mediation effect is a well-studied problem in statistics in both parametric and non-parametric setups. Yet, in the presence of UHD variables with both $p$ and $q$ larger than the sample size, the preference often leans toward simple linear models instead of intricate non-linear models due to the developed theory as well as the interpretability of the model. 
Hence, we focus on the following data generating mechanisms for the outcome and mediator variables:
\begin{equation}
    \begin{aligned}
Y &=  (1-A)\left(X^{\top} \beta_{0}+M^{\top} \gamma_{0} + \epsilon'\right)+A\left(X^{\top} \beta_{1}+M^{\top} \gamma_{1} + \epsilon\right), \\
M& =(1-A)\left(\mathbf B_{0} X + U\right)+A\left(\mathbf B_{1} X + U'\right), \label{eq:dgp}
\end{aligned}
\end{equation}
where $(\epsilon, \epsilon')$ and $(U, U')$ are centered random shocks and the matrices $\mathbf B_1, \mathbf B_0 \in \mathbb R^{q \times p}$, and vectors $\beta_1, \beta_0 \in \mathbb R^{p}$, $\gamma_1, \gamma_0 \in \mathbb R^{q}$ contain unknown regression coefficients that encode the relationships among the exposures, mediators and outcome. 
Note that this data generating process (DGP) is flexible in the sense that 
\allowdisplaybreaks
\begin{enumerate}
    \item it allows for UHD mediators $M$ and UHD pre-treatment covariates $X$,
    \item it allows for $A-X$ and $A-M$ interactions (although it requires no $X-M$ interaction), and 

    \item it does not require any particular modeling assumptions or restrictions on the propensity score, the conditional distribution of $A$ given $X$. 
\end{enumerate}
This DGP implies the following.
\begin{align}
   & \mu_1(X,M) := \mathbb{E}[Y \mid X, M, A = 1] = X^\top \beta_1 + M^\top \gamma_1,\label{eq:mu1}\\
    & \mu_{10}(X) := \mathbb{E}[\mu_1(X,M)\mid X, A = 0] = X^\top \beta_1 + X^\top \mathbf B_0^\top \gamma_1, \label{eq:mu10}
\end{align}
for all $X \in \mathbb{R}^p, M \in \mathbb{R}^q$, where we absorbed the intercepts in the means. Therefore, the parameter of interest in Equation \eqref{theta:functional} reduces to
\begin{equation}
\textstyle
\label{eq:pop_mean_def}
\mathbb E [\mu_{10}(X)] = \mu_X^\top( \beta_1 + \mathbf B_0^\top \gamma_1), \quad \mu_X = \bbE[X].
\end{equation}
Henceforth, we denote the number of treated units by \(n_t\), and the number of control units by \(n_c\). We use the notations \( \bX_c\), \( \bX_t\), and $\mathbf M_c$, $\mathbf M_t$ to refer to the feature matrices corresponding to control and treated units, respectively, with rows corresponding to observations from different individuals. 
Similar to \cite{athey2018approximate}, we primarily focus on the sample version of the parameter of interest, which for the mediation functional, will be
\begin{equation}
\label{eq:2.3}
\textstyle
    \theta_0 \equiv \theta_{10} =\frac{1}{n} \sum_{i=1}^n \mu_{10}(X_i)= \bar X^\top (\beta_1 + \mathbf B_0^\top \gamma_1),
\end{equation}
where $\bar{X} = \frac{1}{n} \sum_{i=1}^n X_i$. 
{\color{black} Note that, under Assumption \ref{assumption:2.1}, the expected values of $Y^{(1,M(1))} \equiv Y^{(1)}$ and $Y^{(0,M(0))} \equiv Y^{(0)}$ can be expressed similarly as
\begin{equation}
   \label{eq:theta11-00} 
\textstyle
\begin{aligned}
\theta_{11} & := \frac1n \sum_{i = 1}^n \mathbb E[\mathbb E[Y \mid X_i, M, A = 1] \mid X_i, A = 1] = \bar X^\top (\beta_1 + \mathbf B_1^\top \gamma_1), \\
\theta_{00} & := \frac1n \sum_{i = 1}^n \mathbb E[\mathbb E[Y \mid X_i, M, A = 0] \mid X_i, A = 0] = \bar X^\top (\beta_0 + \mathbf B_0^\top \gamma_0).
\end{aligned}
\end{equation}
Using these quantities, the (sample average) natural indirect effect (NIE) can be written as $\theta_{\rm NIE} = \theta_{11} - \theta_{10}$ whereas the natural direct effect (NDE) can be written as $\theta_{\rm NDE} = \theta_{10} - \theta_{00}$. We use the superscript ``pop" to denote corresponding population versions, i.e. $\theta_{10}^{\rm pop} = \mu_X^\top (\beta_1 + \bB_0^\top \gamma_1)$, $\theta_{11}^{\rm pop} = \mu_X^\top (\beta_1 + \bB_1^\top \gamma_1)$ and $\theta_{00}^{\rm pop} = \mu_X^\top(\beta_0 + \bB_0^\top \gamma_0)$. As a consequence, we have $\theta_{\rm NDE}^{\rm pop} = \theta_{10}^{\rm pop} - \theta_{00}^{\rm pop}$ and $\theta_{\rm NIE}^{\rm pop} = \theta_{11}^{\rm pop} - \theta_{10}^{\rm pop}$. 
The estimation of $\theta_{10}$ (with the same estimator also applicable to $\theta_{10}^{\rm pop}$), is substantially more involved than that of $\theta_{11}$ and $\theta_{00}$. This is because $\theta_{10}$ involves a cross-arm interaction between parameters learned from different treatment groups (i.e., $\bB_0$ from the control arm and $\gamma_1$ from the treatment arm), whereas $\theta_{aa}$ depends only on parameters associated with arm $a$, for $a \in \{0,1\}$. Accordingly, in the next section, we first present our proposed methodology for estimating $\theta_{10}$ (which we denote by $\theta_0$ for notational simplicity). We then describe how this estimator can be combined with estimators of $\{\theta_{aa}:a\in\{0,1\}\}$ to construct estimators of $\theta_{\rm NDE}$ and $\theta_{\rm NIE}$.
}

\section{Debiased Estimator for the Mediation Functional}
\label{sec:method}
In this section, we present our methodology for estimating the mediation functional
$\theta_{10} =\bar X^\top\left(\beta_1+\mathbf B_0^\top\gamma_1\right)$.
A naive approach for estimating this parameter is a simple plug-in-based approach:
in the first step, we estimate $(\beta_1,\gamma_1)$ by regressing $Y$ on $(X,M)$
using the treatment observations, and then estimate $\mathbf B_0$ by regressing
$M$ on $X$ coordinate-wise using the control observations. The resulting estimator
is $\hat \theta^{\rm plug-in} = \bar X^\top(\hat \beta_1 + \hat \bB_0^\top \hat \gamma_1)$. 
As the setting under consideration is high-dimensional, penalized regression
methods, such as the Lasso, are needed to estimate the regression parameters.
However, directly plugging the resulting regularized estimators into the mediation
functional generally introduces a substantial regularization bias. Hence, our goal
is to devise a debiasing procedure for estimating $\theta_0$ such that the refined
estimator is $\sqrt n$-consistent and asymptotically normal.

To motivate our debiasing procedure, we first consider a simpler scenario in which $\bB_0$ is known. In that case, we can debias $(\hat\beta_1,\hat\gamma_1)$ in the direction $(\bar X, \ \ \bB_0\bar X)$ to obtain a debiased estimator. Using tools from the literature on standard high-dimensional debiased Lasso estimators, one can show that the resulting estimator is $\sqrt n$-consistent and asymptotically normal. However, this debiasing procedure is not directly applicable when $\mathbf B_0$ is unknown. Importantly, estimation of the full UHD matrix $\bB_0$ is not necessary for estimating $\theta_0$; only the vector $\bB_0 \bar X$ enters the direction along which $(\hat\beta_1,\hat\gamma_1)$ needs to be debiased. Motivated by this observation, we directly approximate $\bB_0 \bar X$ with respect to the $\ell_\infty$ norm through a balancing procedure using the control observations. Note that $\bB_0\bar X$ can be completely dense under our general setup (see Remark \ref{rem:existing_debiasing_functional} for more details), in which case consistent estimation of the entire vector with respect to the $\ell_p$ for $p \in [1, \infty)$ norm is not possible in the UHD regime. 
However, as we establish later, it suffices to estimate $\bB_0\bar X$ in the $\ell_\infty$ norm, which can be achieved under much weaker assumptions. 
\begin{figure}[htbp]
    \centering
    \includegraphics[width=1.0\linewidth]{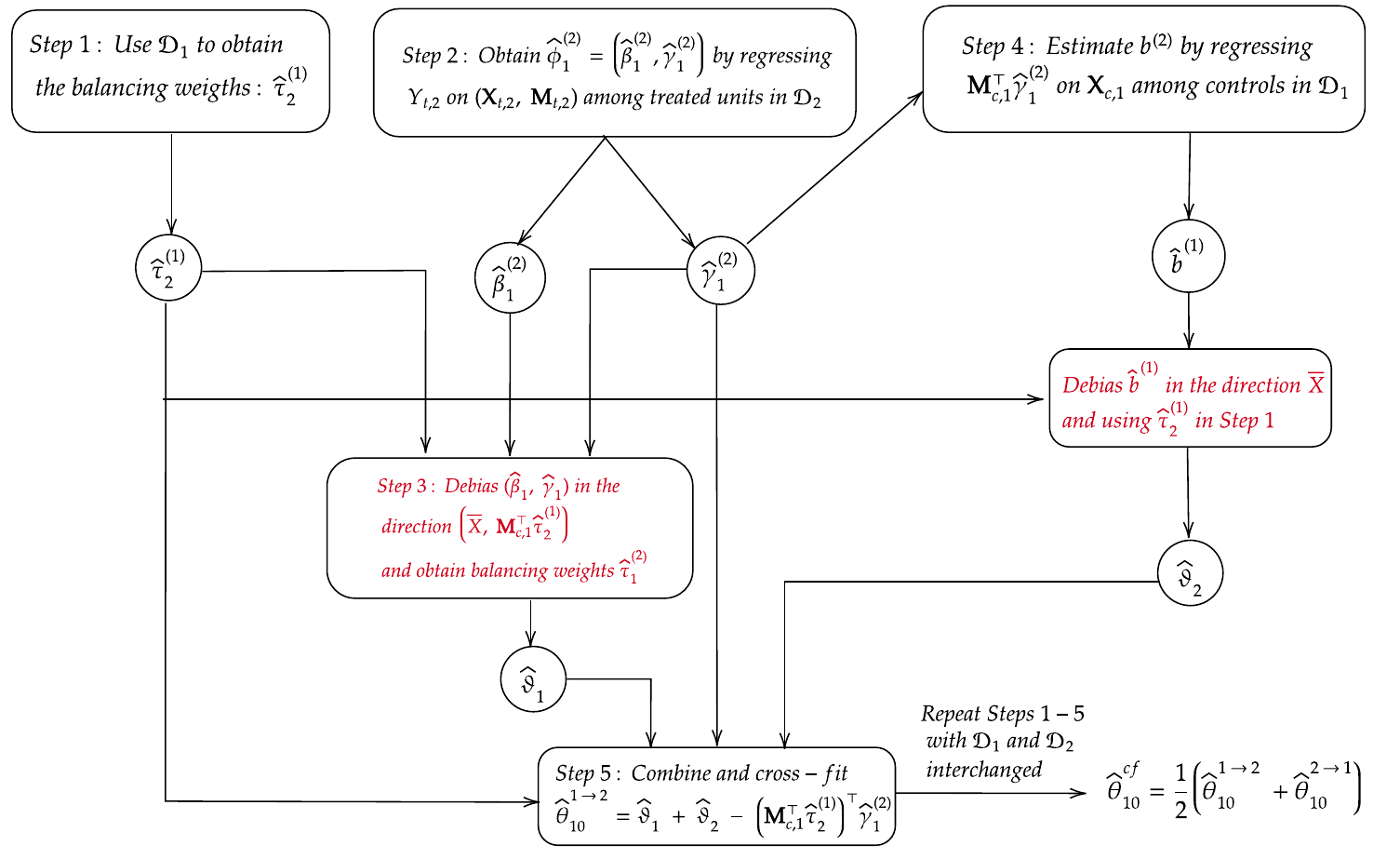}
    \caption{{\color{black}Schematic illustration of Algorithm~\ref{algomain_modified}.}}
    \label{fig:est}
\end{figure}
We now present the details of our methodology, which is summarized in Algorithm~\ref{algomain_modified} and Figure~\ref{fig:est}. Our method essentially comprises six key steps. 
We first divide the dataset into two subsets $\mathcal D_1$ and $\mathcal D_2$. Then in Step 1, using the control observations in $\mathcal D_1$, we construct a balancing weight $\hat\tau_2^{(1)}$ such that $\mathbf X_{c,1}^\top\hat\tau_2^{(1)}$ approximates $\bar X$ by solving optimization problem \eqref{eq:balancing_opt_1}, and then {\color{black} set $\hat m_{c,1} = \mathbf M_{c,1}^\top\hat\tau_2^{(1)}$. As per our DGP (Equation \eqref{eq:dgp}), $M_i = \bB_0X_i + U_i$ on the control arm (i.e., $\bM_{c, 1} = \bX_{c, 1} \bB_0^\top + \bU_{c, 1}$ in matrix form), it is immediate that the vector $\hat m_{c,1}$ approximates $\mathbf B_0\bar X$}.
In Step 2, we use the treated observations in $\cD_2$ to estimate $\phi_1=(\beta_1^\top\ \gamma_1^\top)^\top$ by regressing $Y$ on $(X,M)$ along with an $\ell_1$ penalty (call this estimator $\hat \phi_1^{(2)} = (\hat \beta_1^{(2)}, \hat \gamma_1^{(2)})$). 

\begin{algorithm}[!ht]
\caption{{\color{black}Estimation Procedure for Mediation Functional under Cross-fitting}}
\begin{algorithmic}
\small
\State \textbf{Step 0:}  Randomly split the sample into two parts $(\cD_1, \cD_2)$ with $n/2$ samples in  each. 
\vspace{1em}
\State \textbf{Step 1:} Use $\cD_1$ to obtain the following balancing weights: 
\begin{equation}
\label{eq:balancing_opt_1}
\hat \tau^{(1)}_2 = \argmin_{\tau}\left\{\frac12 \|\tau\|_2^2: \ \|\bar X - \bX_{c, 1}^\top \tau\|_\infty \le K_2 \sqrt{\frac{\log{p}}{n_{c,1}}}, \ \|\tau\|_\infty \le n_{c,1}^{-\alpha}\right\} \,.
\end{equation}

\State \textbf{Step 2:} Use $\mathcal D_2$ to estimate ${\phi}_1 = ({\beta}_1^\top \quad {\gamma}_1^\top)^\top$ by regressing $\by_{t, 2}$ on $\mathbf W_{t, 2}=(\mathbf X_{t, 2} \quad \mathbf M_{t, 2})$ along with a $\ell_1$ penalty
\[
\hat{\phi}^{(2)}_1 = ((\hat \beta^{(2)}_1)^{\top} \quad  (\hat \gamma^{(2)}_1)^{\top})^\top 
= 
\argmin_{\phi} 
\left\{ \frac{1}{2n_{t,2}}\|\by_{t, 2} - \bW_{t, 2}\phi\|_2^2 
+ \lambda_{n,1} \|\phi\|_1 
\right\}.
\]

\State \textbf{Step 3:} Compute bias correction weights with contrast $a^{(1)} = \big( \bar X^\top \quad \hat m_{c,1}^\top \big)^\top$, where $\hat m_{c,1} = \bM_{c, 1}^\top\hat \tau^{(1)}_2$
and tuning parameter $K_1$ using $\mathcal D_2$,
\begin{equation}
\label{eq:balancing_opt_2}
        \hat \tau^{(2)}_1 =  \operatorname{argmin}_{\tau}\left\{ \frac12\|\tau\|_2^2: \ \ 
        \| a^{(1)} - \mathbf{W}_{t,2}^\top \tau\|_{\infty} \leqslant K_1   \sqrt{\frac{\log(p+q)}{n_{t,2}}},\ \ 
        \|\tau\|_\infty \leq n_{t, 2}^{-\alpha} \right\}, 
\end{equation}
and set $\hat \vartheta_1 = (a^{(1)})^\top \hat \phi^{(2)}_1 + (\hat \tau^{(2)}_1)^\top (\by_{t, 2} - \bW_{t, 2}\hat \phi^{(2)}_1)$. 
\vspace{1em}
\State \textbf{Step 4:} On $\cD_1$, approximate $b^{(2)} = \bB_0^\top \hat \gamma_1^{(2)}$ by regressing $\bM_{c, 1} \hat \gamma^{(2)}_1$ on $\bX_{c,1}$ along with a $\ell_1$ penalty:
$$
\hat b^{(1)} = \argmin_{b} \left\{\frac{1}{2n_{c,1}}\|\bM_{c, 1} \hat \gamma^{(2)}_1 - \bX_{c, 1} b\|_2^2 +  \lambda_{n, 2} \|b\|_1\right\} \,.
$$
Set $\hat \vartheta_2 = \bar X^\top \hat b^{(1)} + (\hat \tau_2^{(1)})^\top (\bM_{c, 1}\hat \gamma^{(2)}_1 - \bX_{c, 1}\hat b^{(1)})$. 
\vspace{1em}

\State \textbf{Step 5:} Compute the estimator as: $\hat \theta_{10}^{1\to2} = \hat \vartheta_1 + \hat \vartheta_2 - \hat m_{c, 1}^\top \hat \gamma^{(2)}_1 \,.$
\vspace{1em}
\State \textbf{Step 6:} Interchange the role of $\cD_1$ and $\cD_2$ and perform Step 1 -- 5 to obtain analogous estimator $\hat \theta_{10}^{2\to1}$. 

\vspace{1em}
\State {\bf Return:}  The final cross-fitted estimator $\hat \theta^{\rm cf}_{10} = (\hat \theta_{10}^{1\to2} + \hat \theta_{10}^{2\to1})/2$. 
\end{algorithmic}
\label{algomain_modified}
\end{algorithm}

In Step 3, we debias this estimator in the direction {\color{black} $a^{(1)} = (\bar X^\top \ \ \hat m_{c, 1}^\top)^\top$ to obtain: 
\begin{equation*}
    \hat \vartheta_1 = (a^{(1)})^\top \hat \phi^{(2)}_1 + (\hat \tau^{(2)}_1)^\top (\by_{t, 2} - \bW_{t, 2}\hat \phi^{(2)}_1)
\end{equation*}
Here $\hat \tau_1^{(2)}$ is the debiasing/correction weights obtained by solving Equation \eqref{eq:balancing_opt_2}. 
This step can be viewed as a natural extension of the standard debiasing procedure that would be used if $\bB_0$ were known, as in that case we would debias $\hat\phi_1$ in the direction $(\bar X^\top \ \ (\bB_0 \bar X)^\top)^\top$. {\color{black} However, since $\bB_0\bar X$ is unknown, we replace it by its approximation $\hat m_{c, 1}$ and then debias $\hat \phi_1$ in the approximated direction $(\bar X^\top \ \ \hat m_{c, 1}^\top)^\top$.} However, this approximated debiasing direction introduces an additional layer of bias, because $\hat m_{c,1}$ is generally a biased estimator of $\bB_0\bar X$. Therefore, although the correction in Step 3 reduces the regularization bias associated with $\hat\phi_1^{(2)}$, additional corrections are needed to remove the bias introduced by replacing $\bB_0\bar X$ by $\hat m_{c,1}$.}

Towards that end, in Step 4, we regress $M^\top \hat \gamma^{(2)}$ on $X$ along with an $\ell_1$ penalty using the control observations in $\cD_1$. Since
\begin{equation*}
\textstyle
\bbE[M^\top\hat\gamma_1^{(2)} \mid X,A=0] = X^\top\bB_0^\top\hat\gamma_1^{(2)},
\end{equation*}
this regression estimates $b^{(2)} = \bB_0^\top\hat\gamma_1^{(2)}$. As we used the control observations in $\cD_1$ to estimate $b^{(2)}$, we call this estimator $\hat b^{(1)}$. 
We then debias the resulting estimator $\hat b^{(1)}$ in the direction $\bar X$ using the same control-side balancing weight $\hat\tau_2^{(1)}$ that was used to construct $\hat m_{c,1}$ to obtain 
\begin{equation*}
\hat \vartheta_2 = \bar X^\top \hat b^{(1)} + (\hat \tau_2^{(1)})^\top (\bM_{c, 1}\hat \gamma^{(2)}_1 - \bX_{c, 1}\hat b^{(1)})
\end{equation*}
{\color{black} This coupling (using the same $\hat \tau^{(1)}_2$ in Step 1 and Step 4) is crucial for removing the first-order bias introduced by the estimated direction and leads to a remaining bias of product form.} 

In Step 5, we combine the two intermediate estimators to obtain a fold-specific estimator of $\theta_{10}$.
Thus, the estimator consists of a regression-based initial estimator together with two correction terms constructed using the treatment and control observations, respectively. As will be established in Proposition~\ref{prop:bias_structure}, the use of the same balancing weight $\hat\tau_2^{(1)}$ in the construction of $\hat m_{c,1}$ and in the second correction term ensures that the remaining bias has a product form and is
asymptotically negligible.
Finally, in Step 6, we interchange the roles of $\mathcal D_1$ and $\mathcal D_2$ and average the two fold-specific estimators to obtain the final cross-fitted estimator $\hat \theta_{10}^{\rm cf}$. 
\\\\
\noindent
\textcolor{black}{\bf Estimation of NDE and NIE: }
So far, we have described our procedure for estimating $\theta_{10}$, as defined in Equation \eqref{eq:2.3}. As mentioned previously, estimating $\theta_{00}$ and $\theta_{11}$ is simpler than estimating $\theta_{10}$, since these parameters do not involve quantities from different treatment arms. For any $a \in \{0,1\}$, recall that $\theta_{aa} = \bar X^\top (\beta_a + \bB_a^\top \gamma_a)$ (Equation \eqref{eq:theta11-00}). Moreover, it follows directly from the DGP (Equation \eqref{eq:dgp}) that $\bbE[Y \mid X, A = a] = X^\top (\beta_a + \bB_a^\top \gamma_a)$. 

\begin{algorithm}[!ht]
\caption{\textcolor{black}{Estimation Procedure for NDE}}
\begin{algorithmic}
\small

\State \textbf{Step 0: } Start with same split $\cD_1$ and $\cD_2$ as in Algorithm \ref{algomain_modified}. 
\vspace{1em}
\State \textbf{Step 1: } Regress $\by_{c, 1}$ on $\bX_{c, 1}$ using $\cD_1$: 
\begin{equation*}
\textstyle
\hat \rho_c^{(1)} = \argmin_{\rho}\ \left\{\frac{1}{n_{c,1}} \|\by_{c, 1} - \bX_{c, 1} \rho\|_2^2 + \lambda_{n, 3} \|\rho\|_1\right\}
\end{equation*}

\State \textbf{Step 2: }Use $\hat \tau_2^{(1)}$ obtained from the Step 1 of Algorithm \ref{algomain_modified} to obtain: 
\begin{equation*}
\textstyle
\hat \theta_{00}^{(1)} = \bar X^\top \hat \rho_c^{(1)} + (\hat \tau_2^{(1)})^\top(\by_{c, 1} - \bX_{c, 1} \hat \rho_c^{(1)}) \,.
\end{equation*}

\State \textbf{Step 3: }Swap the role of the dataset; use the control observations from $\cD_2$ to obtain $\hat \rho_c^{(2)}$ using Step 1 with $(\by_{c, 2}, \bX_{c, 2})$ and use $\hat \tau_2^{(2)}$ obtained in Step 6 of Algorithm \ref{algomain_modified} to obtain $\hat \theta_{00}^{(2)}$. 
\vspace{1em}
\State \textbf{Return: }The estimator $\hat \theta_{\rm NDE}^{\rm cf} = \hat \theta_{10}^{\rm cf} - (\hat \theta_{00}^{(1)} + \hat \theta_{00}^{(2)})/2$, where $\hat \theta_{10}^{\rm cf}$ is obtained from Algorithm \ref{algomain_modified}. 

\end{algorithmic}
\label{algo_NDE}
\end{algorithm}

\begin{algorithm}[!ht]
\caption{\textcolor{black}{Estimation Procedure for NIE}}
\begin{algorithmic}
\small

\State \textbf{Step 0: } Start with same split $\cD_1$ and $\cD_2$ as in Algorithm \ref{algomain_modified}.
\vspace{1em}

\State \textbf{Step 1: }On $\cD_1$ use the treatment observation; regress $\by_{t,1}$ on $\bX_{t, 1}$ to obtain $\hat \rho_t^{(1)}$: 
\begin{equation*}
\textstyle
\hat \rho_t^{(1)} = \argmin_\rho \left\{\frac{1}{n_{t, 1}}\|\by_{t, 1} - \bX_{t,1}\rho\|_2^2 + \lambda_{n, 4} \|\rho\|_1\right\}
\end{equation*}

\State \textbf{Step 2: }Construct the balancing weight for debiasing: 
\begin{equation}
\textstyle
\label{eq:balancing_opt_3}
\hat \omega^{(1)} = \argmin\left\{\frac12\|\omega\|_2^2, \ \ \|\bar X - \bX_{t, 1}^\top \omega\|_\infty \le K_3 \sqrt{\frac{\log{p}}{n_{t, 1}}}, \ \ \|\omega\|_\infty \le n_{t, 1}^{-\alpha}\right\}
\end{equation}
and set $\hat \theta_{11}^{(1)}$ as: 
\begin{equation*}
\textstyle
\hat \theta^{(1)}_{11} = \bar X^\top \hat \rho_t^{(1)} + (\hat \omega^{(1)})^\top(\by_{t, 1} - \bX_{t, 1} \hat\rho_t^{(1)}) \,.
\end{equation*}

\State \textbf{Step 3: }Apply Step 1 and Step 2 to the treatment observations in $\cD_2$ to obtain $\hat \theta_{11}^{(2)}$ analogously. 

\vspace{1em}
\State \textbf{Return: }The estimator $\hat \theta_{\rm NIE}^{\rm cf} = (\hat \theta_{11}^{(1)} + \hat \theta_{11}^{(2)})/2 - \hat \theta_{10}^{\rm cf}$, where $\hat \theta_{10}^{\rm cf}$ is obtained from Algorithm \ref{algomain_modified}.

\end{algorithmic}
\label{algo_NIE}
\end{algorithm}
This representation naturally suggests a simple Lasso-based debiasing procedure for estimating $\theta_{aa}$. Specifically, using the observations from treatment arm $a$, we regress $Y$ on $X$ to estimate the reduced-form coefficient $\beta_a+\bB_a^\top\gamma_a$, and then debias the resulting estimator in the direction of $\bar X$. To ensure compatibility with our estimator of $\theta_{10}$, we use an analogous data-splitting and cross-fitting procedure to construct $\hat\theta_{aa}^{\rm cf}$ for each $a\in\{0,1\}$. We then set our estimators for NDE and NIE as $\hat \theta_{\rm NDE}^{\rm cf} = \hat \theta_{10}^{\rm cf} - \hat \theta_{00}^{\rm cf}$ and $\hat \theta_{\rm NIE}^{\rm cf} = \hat \theta_{11}^{\rm cf} - \hat \theta_{10}^{\rm cf}$. 
The complete estimation procedures for the NDE and NIE are summarized in Algorithms~\ref{algo_NDE} and~\ref{algo_NIE}, respectively. A detailed theoretical analysis of these estimators is provided in Section \ref{sec:proof_NDE} and Section \ref{sec:proof_NIE} of the supplementary document, where we establish that $\hat\theta_{\rm NDE}^{\rm cf}$ and $\hat\theta_{\rm NIE}^{\rm cf}$ are $\sqrt{n}$-CAN estimators of $\theta_{\rm NDE}$ and $\theta_{\rm NIE}$ (as well as their population counterparts), respectively. 
{\color{black}
\begin{remark}
\label{rem:choice_alpha}
{\color{black}
The choice of $\alpha$ in Equation \eqref{eq:balancing_opt_1} and \eqref{eq:balancing_opt_2} of Algorithm \ref{algomain_modified} and in Equation \eqref{eq:balancing_opt_3} of Algorithm \ref{algo_NIE} is quite flexible; one can choose any $\alpha \in (1/2, 1)$ as long as $n^{1-\alpha} \gg \sqrt{\log{(n(p+q))}}$. The condition $\alpha < 1$ is sufficient to guarantee feasibility of the optimization problems with high probability, while the lower bound $\alpha>1/2$ is needed in the asymptotic normality argument (see the proof of Lemma \ref{lem:feasibility} and Theorem \ref{thm:main_revised} for more details).}
\end{remark}}

{\color{black}
\begin{remark}[Relation to debiased inference for high-dimensional functionals]
\label{rem:existing_debiasing_functional}
{\color{black} Our debiasing technique, as presented in Algorithm~\ref{algomain_modified}, is structurally distinct from existing debiasing approaches for high-dimensional linear and bilinear functionals. In particular, \citep{cai2021optimal,javanmard2014confidence,van2014asymptotically,zhang2014confidence} study inference for linear functionals of the form $x^\top\beta$, while \cite{guo2019optimal} develops inference for quadratic and bilinear functionals such as $\beta^\top\gamma$ in high-dimensional linear models. In contrast, our parameter of interest contains the term $\bar X^\top\bB_0^\top\gamma_1=(\bB_0\bar X)^\top\gamma_1$, which involves a UHD matrix $\bB_0$ and a UHD vector $\gamma_1$ that are associated with the control and treatment arms, respectively. Moreover, although $\bB_0$ is assumed to be row-sparse, the vector $\bB_0\bar X$ need not be sparse and can be completely dense under our general setup. Consequently, without imposing additional structural assumptions, $\bB_0\bar X$ cannot generally be consistently estimated as a full UHD vector with respect to the $\ell_1/\ell_2$ norm, and consequently the method of \cite{guo2019optimal} cannot be directly applied. Our procedure simply obtains an approximation of $\bB_0 \bar X$ with respect to the $\ell_\infty$ norm along with careful debiasing of the bias introduced by this approximation.
Therefore, our problem cannot be reduced to estimating an inner product between two separately $\ell_2$-consistently estimable UHD vectors. This, together with the fact that the two components are associated with different treatment arms and different regression equations, prevents the techniques developed in \cite{guo2019optimal} from being directly applicable. A detailed explanation is provided in Section~\ref{appendix:existing-debiasing-functional} of the supplementary document.}
\end{remark}}

\section{Bias Analysis and Connection to Fixed-Dimensional Setting}
\label{section:3}

\subsection{Bias Analysis}
In this section, we present the bias structure of the debiased estimator $\hat\theta_{10}^{\rm cf}$ and show that the bias is asymptotically negligible, i.e., it decays at a faster rate than $n^{-1/2}$. We would like to point out that \emph{debiasing} a high-dimensional parameter does not make it \emph{unbiased}, i.e., it does not remove the bias completely, but rather reduces the bias sufficiently to achieve $\sqrt n$-consistency (e.g., see \citep{zhang2014confidence,van2014asymptotically,javanmard2014confidence}).
Recall the definition of $b = \bB_0^\top \hat \gamma_1$, $\hat \phi_1 = (\hat \beta_1, \hat \gamma_1)$ and $a = (\bar X^\top \ \ \hat m_c^\top)^\top$. 
For notational simplicity, throughout this subsection, we suppress the superscript/subscript $j$ used in the algorithms to index the data fold, with all quantities understood to correspond to a generic fold. The following proposition establishes the exact bias structure of $\hat\theta_{10}^{\mathrm{cf}}$.

\begin{proposition}
\label{prop:bias_structure}
The estimation error of $\hat \theta^{\rm cf}_{10}$, obtained via Algorithm \ref{algomain_modified} can be decomposed as  $\hat\theta_{10}^{\mathrm{cf}}-\theta_{10} = \Delta_n^{\mathrm{cf}}+V_n^{\mathrm{cf}}$
where
\begin{equation*}
    \textstyle
    V_n^{\rm cf} = \frac12\sum_{j = 1}^2\left[\sum_{\substack{i \in \cD_j, A_i = 1}} \hat \tau_{1, i}^{(j)} \eps_i + \sum_{\substack{i \in \cD_j, A_i = 0}} \hat \tau_{2, i}^{(j)}U_i^\top \gamma_1\right],
\end{equation*}
and $\Delta_n^{\rm cf} = (\Delta_n^{(1)} + \Delta_n^{(2)})/2$ with 
\begin{equation}
\textstyle
\label{eq:exact_bias_new}
\begin{aligned}
\Delta_n^{(j)} & = (a^{(j)} - \bW_{t, j'}^\top \hat \tau_1^{(j')})^\top (\hat \phi^{(j')}- \phi_1) + (\bar X - \bX_{c, j}^\top \hat \tau_2^{(j)})^\top(\hat b^{(j)} - b^{(j')}) \\
& \qquad +  (\bar X - \bX_{c, j}^\top \hat \tau_2^{(j)})^\top \bB_0^\top(\hat \gamma_1^{(j')} - \gamma_1) 
\end{aligned}
\end{equation}
for $j \in \{1, 2\}$ and $j' = 3 - j$. Consequently, we have, for $j \in \{1,2\}$:   
\begin{equation}
\textstyle
\label{eq:deltanbound_cf}
{\color{black}\begin{aligned}
|\Delta_n^{(j)}| &\le
\|\bB_0(\bar X - \bX_{c, j}^\top \hat \tau_2^{(j)})\|_\infty
\|\hat\gamma^{(j')}_1-\gamma_1\|_1  +
\|a^{(j)}-\mathbf W_{t, j'}^\top \hat\tau^{(j')}_1\|_\infty
\|\hat\phi_1^{(j')}-\phi_1\|_1 \\
& \qquad +
\|\bar X-\mathbf X_{c, j}^\top \hat\tau^{(j)}_2\|_\infty
\|\hat b^{(j)}-b^{(j')}\|_1 \,.
\end{aligned}}
\end{equation}
\end{proposition}
The proof can be found in Section \ref{sec:proof_prop_bias_structure} of the supplementary material. Observe that each of the three terms in the upper bound of $|\Delta_n^{(j)}|$ has a \emph{product form}, i.e., it is the product of an estimation error and an approximation error. More specifically, the first summand is the product of the estimation error of $\gamma_1$ and the error in approximating $\bar X$ by $\bX_c^\top\tau_2$. 
The second summand is the product of the estimation error of $\phi_1$ and the approximation error of $a$ by $\bW_t^\top\tau_1$, 
while the last summand is the product of the estimation error of $b$ and the approximation error of $\bar X$ by $\bX_c^\top\tau_2$. Although each individual estimation or approximation error can be significantly larger than $n^{-1/2}$ in order, the product form makes each summand asymptotically negligible. In Section~\ref{sec:theory}, we provide sufficient conditions on the sparsity of the parameters and the distributions of the covariates and error terms under which $|\Delta_n^{\rm cf}| = o_p(n^{-1/2})$ and $\sqrt{n}V_n^{\rm cf}$ converges to a normal distribution. This establishes that $\hat\theta_0^{\mathrm{cf}}$ is a $\sqrt{n}$-CAN estimator of $\theta_0$.

\subsection{Connection to Fixed-Dimensional Setting}
In a fixed-dimensional setting, i.e., when the number of covariates and potential mediators is fixed, \cite{tchetgen2012semiparametric} provided a debiased estimator of the mediation functional constructed based on the influence function of the parameter of interest. Our proposed estimator can be viewed as a counterpart of their approach in the UHD setting. The debiased estimator proposed by \cite{tchetgen2012semiparametric} can be
written as
\begin{equation}
    \begin{aligned}
\textstyle
\label{eq:IF}
    \hat{\theta}^{\rm IF} &=\mathbb{E}_n \bigg[\hat \mu_{10}(X) + \frac{A}{\hat p(A = 1 \mid X)} \frac{\hat p(M \mid A=0, X)}{\hat p(M \mid A=1, X)} \Big(Y-\hat \mu_1(X, M)\Big)\\
    &~~~~~~~~~~~~~~~~~~~~~~~~~~+\frac{1-A}{1-\hat p(A = 1 \mid X)}\Big(\hat \mu_1(X, M)- \hat \mu_{10}(X)\Big)\bigg] \,,
\end{aligned}
\end{equation}
where $\hat\mu_{10}(X) = \int \hat \mu_1(X, M)\hat p(M \mid A = 0, X)dM$. We now show that our estimator $\hat\theta_{10}^{\mathrm{cf}}$, obtained via Algorithm~\ref{algomain_modified}, is structurally similar to $\hat\theta^{\rm IF}$ when the underlying data-generating process is linear. Consider the nuisance-function estimators $\hat\mu_1(X,M) = X^\top\hat\beta_1+M^\top\hat\gamma_1$ and $\hat\mu_{10}(X) = X^\top(\hat\beta_1+\hat b)$, where $\hat b$ estimates $b=\mathbf B_0^\top\hat\gamma_1$. Then the estimator $\hat\theta^{\rm IF}$ takes the form

\allowdisplaybreaks
\begin{align}
\label{eq:IFbased}
\hat \theta^{\rm IF}&= \mathbb{E}_n \bigg[ X^{\top}\left(\hat\beta_{1}+ \hat b\right) + \frac{A}{\hat p(A = 1 \mid X)} \frac{\hat p(M \mid A=0, X)}{\hat p(M \mid A=1, X)}\left(Y-X^{\top} \hat\beta_{1}-M^{\top} \hat\gamma_{1}\right) \notag \\
& \quad +\frac{1-A}{1-\hat p(A = 1 \mid X)}\left(X^{\top} \hat\beta_{1}+M^{\top} \hat\gamma_{1}-X^{\top}\left(\hat\beta_{1}+ \hat b\right)\right)\bigg ] \notag\\
&=\mathbb{E}_n \bigg[ X^{\top}\left(\hat\beta_{1}+ \hat b\right) + \omega_{1}A\left(Y-X^{\top} \hat\beta_{1}-M^{\top} \hat\gamma_{1}\right) +\omega_{2}(1-A)\left(M^{\top} \hat\gamma_{1}-X^{\top} \hat b\right) \bigg]  \notag \\
& = \bar X^\top \left(\hat \beta_1 + \hat b\right) + \frac1n \sum_{\{i: A_i = 1\}}\omega_{1, i}\left(Y_i - X_i^\top \hat \beta_1 - M_i^\top \hat \gamma_1\right) \notag \\
& \hspace{15em}+ \frac1n \sum_{\{i: A_i = 0\}} \omega_{2, i}\left(M_i^\top \hat \gamma_1 - X_i^\top \hat b\right) \,,
\end{align}
where
\begin{equation*}
\textstyle
\begin{aligned}
    \omega_{1,i} = \frac{1}{\hat p(A_i = 1 \mid X_i)}\frac{\hat p(M_i \mid A_i = 0, X_i)}{\hat p(M_i \mid A_i = 1, X_i)}, \quad \omega_{2, i} = \frac{1}{1-\hat p(A_i = 1 \mid X_i)}.
\end{aligned}
\end{equation*}
As can be seen, the estimator $\hat\theta^{\rm IF}$ consists of a regression-based initial estimator together with two weighted terms for bias correction. This is the same structure as our proposed estimator. For simplicity, consider the fold-specific estimator $\hat\theta_{10}^{1 \to 2}$, suppressing the fold indices. Using $\hat m_c=\mathbf M_c^\top\tau_2$, this estimator in Algorithm~\ref{algomain_modified} can be decomposed as
\begin{equation}
\textstyle
\label{eq:our_est}
\begin{aligned}
\hat\theta_{10}^{1\to2}  &= \bar X^\top (\hat\beta_1+ \hat b) + \sum_{\{i:A_i=1\}} \hat \tau_{1,i} (Y_i-X_i^\top\hat\beta_1-M_i^\top\hat\gamma_1)\\
&\hspace{15em} + \sum_{\{i:A_i=0\}} \hat \tau_{2,i}(M_i^\top\hat\gamma_1 - X_i^\top \hat b).
\end{aligned}
\end{equation}
It is evident from Equations \eqref{eq:IFbased} and \eqref{eq:our_est} that our estimator has the same general form as $\hat\theta^{\rm IF}$, with $\hat \tau_1$ and $\hat \tau_2$ playing the roles of the weights $\omega_1/n$ and $\omega_2/n$, respectively.
Consider the nuisance functions 
\begin{equation*}
\textstyle
\pi_1(X) =
1/p(A=0\mid X), \qquad \pi_2(M,X) = p(M\mid A=0,X)/
p(M,A=1\mid X),
\end{equation*}
and recall that $\mu_1(X,M) = \bbE[Y\mid X,M,A=1]$ and $\mu_{10}(X) = \bbE[\mu_1(X,M)\mid X,A=0]$. 
In a recent work, \cite{liu2024bias} showed that, writing the estimator in Equation~\eqref{eq:IF} as
\begin{equation*}
\textstyle
\begin{aligned}
\hat{\theta}^{mr} & = \mathbb{E}_{n}\left[\hat{\mu}_{10}(X)+(1-A)\hat{\pi}_{1}(X)\left\{\hat{\mu}_{1}(M, X)-\hat{\mu}_{10}(X)\right\} \right. \\
& \hspace{15em} \left. +A \hat{\pi}_{2}(M, X)\left\{Y-\hat{\mu}_{1}(M, X)\right\}\right],
\end{aligned}
\end{equation*}
the bias has the following form:
\begin{equation}
\textstyle
\begin{aligned}
\label{eq:2.12}
\mathbb{E}\left[\hat{\theta}^{m r}\right]-\theta_0= & \mathbb{E}[\left\{(1-A)\hat{\pi}_{1}(X)-1\right\}\left\{\mu_{10}(X)-\hat{\mu}_{10}(X)\right\} \\
& +\left\{A \hat{\pi}_{2}(M, X)-(1-A)\hat{\pi}_{1}(X)\right\}\left\{\mu_{1}(M, X)-\hat{\mu}_{1}(M, X)\right\}] .
\end{aligned}
\end{equation}
In the fixed-dimensional setting, by H\"older's inequality and plugging in $\hat\mu_1(X,M) = X^\top\hat\beta_1+M^\top\hat\gamma_1$ and $\hat\mu_{10}(X) = X^\top (\hat\beta_1+\hat b)$, Equation~\eqref{eq:2.12} can be upper bounded by
\begin{align*}
\left|
\mathbb E[\hat\theta^{mr}]-\theta_0
\right| &\le
\mathbb E\left[
\left\|
X^\top
\left\{
A\hat\pi_2(M,X)-1
\right\}
\right\|_\infty
\left\|
\beta_1-\hat\beta_1
\right\|_1
\right]\\
&+
\mathbb E\left[
\left\{
\left\|
M^\top \{A\hat\pi_2(M,X)
-
(1-A)\hat\pi_1(X)\}
\right\|_\infty
\right. \right. \\
& \hspace{7em} \left. \left. + \left\|
(\mathbf B_0X)^\top
\left\{
(1-A)\hat\pi_1(X)-1
\right\}
\right\|_\infty
\right\}
\left\|
\gamma_1-\hat\gamma_1
\right\|_1
\right]\\
&+
\mathbb E\left[
\left\|
X^\top
\left\{
(1-A)\hat\pi_1(X)-1
\right\}
\right\|_\infty
\left\|
\mathbf B_0^\top\hat\gamma_1-\hat b
\right\|_1
\right].
\end{align*}
Coming back to our estimator $\hat\theta_{10}^{1\to2}$, the remaining bias $\Delta_{n, 1}$, as shown in Proposition~\ref{prop:bias_structure}, is bounded by
\allowdisplaybreaks
\begin{align*}
|\Delta_n|
&\le
\left\|
a-\mathbf W_t^\top\tau_1
\right\|_\infty
\left\|
\hat\phi_1-\phi_1
\right\|_1
+
\left\|
\mathbf B_0
\left(
\bar X-\mathbf X_c^\top\tau_2
\right)
\right\|_\infty
\left\|
\hat\gamma_1-\gamma_1
\right\|_1\\
&\qquad
+
\left\|
\bar X-\mathbf X_c^\top\tau_2
\right\|_\infty
\left\|
\hat b-\mathbf B_0^\top\hat\gamma_1
\right\|_1\\
&\le
\left\|
a-\mathbf W_t^\top\tau_1
\right\|_\infty
\left\|
\hat\beta_1-\beta_1
\right\|_1\\
&\qquad
+
\left\{
\left\|
a-\mathbf W_t^\top\tau_1
\right\|_\infty
+
\|\mathbf B_0\|_{1,\infty}
\left\|
\bar X-\mathbf X_c^\top\tau_2
\right\|_\infty
\right\}
\left\|
\hat\gamma_1-\gamma_1
\right\|_1\\
&\qquad
+
\left\|
\bar X-\mathbf X_c^\top\tau_2
\right\|_\infty
\left\|
\hat b-\mathbf B_0^\top\hat\gamma_1
\right\|_1.
\end{align*}
It is apparent that the bias structures of $\hat\theta^{mr}$ and $\hat\theta_{10}^{1 \to 2}$—and consequently that of $\hat\theta_{10}^{\mathrm{cf}}$—are similar. Both biases are the sum of three terms, each of which has the form of a product of errors.

\noindent
{\bf Relation to Balancing: }
We also draw a connection between our debiasing technique and balancing
techniques in the causal inference literature. In the influence-function-based
estimator for the ATE, the correction term contains the propensity score as a
nuisance function. \cite{imai2014covariate} showed that the propensity score
satisfies a certain balancing property and proposed a balancing estimator for
the propensity score based on this observation.
\cite{zubizarreta2015stable} proposed an optimization-based approach for
achieving the same balance.

In the high-dimensional setting, \cite{athey2018approximate} illustrated that
standard debiasing techniques for penalized regression-based estimators
(e.g., the methods proposed in
\citep{javanmard2014confidence,van2014asymptotically,
zhang2014confidence}) are inherently related to the balancing approach of
\cite{zubizarreta2015stable}. However, these approaches are not directly
applicable in the presence of mediators for estimating the mediation
functional. Recently, \cite{liu2024bias} presented counterparts of the
balancing property and balancing estimators of \cite{imai2014covariate} for
the mediation functional in a fixed-dimensional setting.

The counterparts of the balancing errors appearing in \cite{liu2024bias} are $\|\bar X - \bX_c^\top\tau_2\|_\infty$ and $\|a-\mathbf W_t^\top\tau_1\|_\infty$ in our setting. In addition, the first balancing weight $\tau_2$ is used to construct the direct approximation $\hat m_c=\mathbf M_c^\top\tau_2$ of $\mathbf B_0\bar X$, and the same weight is reused in the second bias-correction term. Hence, the two balancing optimizations in our proposed estimation strategy can be viewed as UHD adaptations of the balancing construction in \cite{liu2024bias}, together with an additional coupling that is needed to handle the cross-arm structure of the mediation functional.

\section{Theoretical Analysis}
\label{sec:theory}

In this section, we provide a theoretical justification for our proposed methodology. We first state the assumptions used throughout our analysis and then establish that the proposed estimator of $\theta_{10}$ is $\sqrt{n}$-consistent and asymptotically normal. We subsequently extend this result to conditionally heteroskedastic outcome errors, the corresponding population-average mediation functional, and the estimators of the NDE and NIE introduced in Algorithms~\ref{algo_NDE} and \ref{algo_NIE}. Throughout this section, we write $d=p+q$ and allow the dimensions, sparsity levels, model parameters, and underlying distributions to vary with $n$, although this dependence is suppressed for notational simplicity. All constants appearing below are independent of $(n,p,q)$ unless stated otherwise.
\subsection{Assumptions}
\label{sec:assumptions}

We first recall the definitions of the sub-Gaussian and sub-exponential norms
used in our analysis.

\begin{definition}[Sub-Gaussian and sub-exponential norms]
For a random variable $Z$, its sub-Gaussian and sub-exponential norms are
defined, respectively, as
\begin{equation*}
\textstyle
\|Z\|_{\psi_2} = \sup_{r\ge1}r^{-1/2}\left(\bbE|Z|^r\right)^{1/r}, \qquad \|Z\|_{\psi_1} = \sup_{r\ge1}r^{-1}\left(\bbE|Z|^r\right)^{1/r}.
\end{equation*}
For a random vector $Z\in\reals^m$, we define $\|Z\|_{\psi_2} = \sup 
\{\|v^\top Z\|_{\psi_2}: \|v\|_2 = 1\}$. 
\end{definition}
We will also use the standard restricted eigenvalue condition for the
high-dimensional regressions involved in our estimation procedure.
\begin{definition}[Restricted eigenvalue and compatibility conditions]
For an index set $S\subseteq\{1,\ldots,m\}$ and a constant $L\ge1$, define $\cC(S,L) = \left\{\delta\in\reals^m:\|\delta_{S^c}\|_1 \le L\|\delta_S\|_1\right\}$. 
A positive semi-definite matrix $\Sigma\in\reals^{m\times m}$ is said to
satisfy the restricted eigenvalue condition at sparsity level $r$ with
constants $(L,\kappa_{\rm RE})$ if $\delta^\top\Sigma\delta \ge \kappa_{\rm RE}\|\delta\|_2^2$ for every $S\subseteq\{1,\ldots,m\}$ with $|S|\le r$ and every
$\delta\in\cC(S,L)$. 
The corresponding compatibility constant is defined by
\begin{equation*}
\textstyle
\phi^2(\Sigma;r,L) = \inf_{\substack{|S|\le r,\ \delta\in\cC(S,L)\\ \delta\neq0}}\frac{r  \delta^\top\Sigma\delta}{\|\delta_S\|_1^2}.
\end{equation*}
\end{definition}
\noindent
We now state the two main assumptions used in our theoretical analysis.

\begin{assumption}[\textcolor{black}{Sampling, covariates, and errors}]
\label{assm:cov_err}
The following conditions hold.

\begin{enumerate}
\item
The observations $\{(Y_i,M_i,X_i,A_i)\}_{i=1}^n$ are independent and
identically distributed, and the random sample split is independent of the
observations. The treatment probability satisfies $0< \pi \le \bbP(A=1) \le 1-\pi <1$ for some constant $\pi>0$. Consequently, each arm-specific sample size within each data fold is proportional to $n$ with probability tending to one. 

\item For $a\in\{0,1\}$, let $\mu_{X,a} = \bbE[X\mid A=a]$, $\Sigma_{X,a} = \var(X\mid A=a)$ and $\mu_X = \bbE[X]$. 
The conditional distribution of $X-\mu_{X,a}$ given $A=a$ is uniformly sub-Gaussian, and $\|\mu_{X,a}\|_\infty\le C_\mu, \|\mu_X\|_\infty\ge\kappa_X>0$. Furthermore, there exist constants $0<c_X<C_X<\infty$ such that
\begin{equation*}
\textstyle
c_X \le \lambda_{\min}(\Sigma_{X,a}) \le \lambda_{\max}(\Sigma_{X,a}) \le C_X, \qquad a\in\{0,1\}.
\end{equation*}
We also assume that $\mu_X^\top\Sigma_{X,a}^{-1}\mu_X \le C_{\mu,\Sigma}$ for $a \in \{0, 1\}$. 

\item Let $W = (X^\top \ \ M^\top)^\top$ and $\Omega_{W,a} = \bbE[WW^\top\mid A=a]$. Let $a_{10}^*= (\mu_X^\top \ \ (\bB_0 \mu_X)^\top)^\top$ and the target-specific population comparison vector $\eta_{10}^* = \Omega_{W,1}^{-1}a_{10}^*$. Assume $\|\eta_{10}^*\|_1 \le C_\eta$ for some constant $C_\eta<\infty$.

\item Let $\epsilon_a$ denote the outcome-regression error on arm $a$. We assume $\bbE[\epsilon_a\mid X,M,A=a]=0$ and $\var(\epsilon_a\mid X,M,A=a) = \nu_a(X,M)$ where, for some constants $0<c_\epsilon<C_\epsilon<\infty$, $c_\epsilon \le \nu_a(X,M) \le C_\epsilon$
almost surely. Furthermore, the errors $\epsilon_a$ are assumed to be uniformly conditionally sub-Gaussian given $(X,M,A=a)$. The homoskedastic setup corresponds to the special case $\nu_a(X,M)=\sigma_a^2$. Let $U_a$ denote the mediator-regression error on arm $a$. We assume $\bbE[U_a\mid X,A=a]=0$ and $\var(U_a\mid X,A=a)=\Sigma_{U,a}$ where
\begin{equation*}
\textstyle
0<c_U \le \lambda_{\min}(\Sigma_{U,a}) \le \lambda_{\max}(\Sigma_{U,a}) \le C_U <\infty.
\end{equation*}
Moreover, $U_a$ is uniformly conditionally sub-Gaussian given $(X,A=a)$.
\end{enumerate}
\end{assumption}

\begin{assumption}[{\color{black} Parameters and sparsity}]
\label{assm:params}
The following conditions hold.
\begin{enumerate}
\item
There exist constants $C_\beta,C_\gamma,C_{\bB}<\infty$ such that, for
$a\in\{0,1\}$, $\|\beta_a\|_2\le C_\beta, \|\gamma_a\|_2\le C_\gamma$ and $\|\bB_a\|_{1,\infty} \vee \|\bB_a\|_{\rm op} \le C_{\bB}$.

\item
There exist sparsity levels $k_\beta,k_\gamma$ such that $\|\beta_a\|_0\le k_\beta, \|\gamma_a\|_0\le k_\gamma$ for $a \in \{0, 1\}$. Let $k = k_\beta + k_\gamma$ denote the total sparsity. 
The mediator-regression matrices are row-wise sparse, i.e., $\max_{1 \le j \le q} \|\bB_{a,j\cdot}\|_0\le s$ for $a \in \{0, 1\}$. 
This implies that the reduced-form coefficients $\rho_a = \beta_a+\bB_a^\top\gamma_a$ satisfies $\|\rho_a\|_0 \le k_\beta+sk_\gamma \le (1+s)k$ and $\|\bB_a^\top \gamma_{a'}\|_0 \le sk_\gamma \le sk$. 

\item
The sparsity levels and ambient dimensions satisfy $(1+s)(1+k)\log(p+q) \ll \sqrt{n}$. 
The Lasso tuning parameters are chosen at the usual high-dimensional orders $\lambda_{n, 1} \asymp \sqrt{\log{(p+q)}/n_t}$ and $\lambda_{n, \ell} \asymp \sqrt{\log{p}/n_a}$ for $\ell \in \{2, 3, 4\}$, where $n_t$ and $n_a$ denote the relevant arm- and fold-specific sample
sizes.
\end{enumerate}
\end{assumption}
Assumptions of this type are commonly used in the analysis of high-dimensional statistical models (e.g., see \cite{buhlmann2011statistics} and \cite{lin2025}).
In particular, the sub-Gaussian conditions impose restrictions on the tail behavior of the design vectors and regression errors, while the eigenvalue conditions prevent the arm-specific covariance matrices from becoming degenerate or excessively ill-conditioned. 
These assumptions facilitate the use of standard concentration inequalities for empirical means, covariance matrices, and high-dimensional regression scores. 
The condition $\|\eta_{10}^*\|_1\le  C_\eta$ is analogous to the bounded-$\ell_1$ conditions on the coefficient vectors of Riesz representers commonly imposed in high-dimensional functional inference. Indeed, since $\Omega_{W,1}\eta_{10}^*=a_{10}^*$, the linear function $W^\top\eta_{10}^*$ satisfies $\bbE[W(W^\top\eta_{10}^*)\mid A=1]=a_{10}^*$, and therefore serves as the population balancing function associated with the target direction $a_{10}^*$. Conditions of this form are commonly used in the high-dimensional literature for debiasing and inference (e.g., see Condition SC(a) of \cite{chernozhukov2022debiased}, Assumption 6.3 of \cite{escanciano2023automatic}, and the conditions in Corollary D.2 of \cite{chernozhukov2026adversarial}).
For expositional clarity, the sub-Gaussian norms, covariance eigenvalues, parameter norms, and rowwise matrix norms in Assumptions~\ref{assm:cov_err} and \ref{assm:params} are uniformly bounded by constants independent of $(n,p,q)$. 
It is possible to relax these conditions by allowing the corresponding bounds to grow with $n$. Such an extension would, however, require careful tracking of these sequences throughout the proofs, and would consequently modify the required convergence rates. Since this generalization would not contribute substantially to the main methodological ideas of the paper, we do not pursue it here.

We next discuss the sparsity-rate condition in
Assumption~\ref{assm:params}. Conditions of the general form $r\log{(d)}/\sqrt{n} \to 0$, 
where $r$ denotes the sparsity level and $d$ is the ambient dimension, commonly arise in the debiased Lasso literature when establishing $\sqrt{n}$-consistency and asymptotic normality for linear contrasts of sparse high-dimensional parameters; see, for example, \citep{javanmard2014confidence,van2014asymptotically,zhang2014confidence}. Moreover, without imposing additional structural assumptions, the necessity of such a condition for constructing a $\sqrt{n}$-consistent estimator of a linear contrast was demonstrated by \cite{cai2017confidence}, who established a corresponding minimax lower bound for the estimation error.
In our setting, estimation of $\theta_{aa}$ for $a\in\{0,1\}$ requires estimating a linear contrast (in the direction of $\bar X$ or $\mu_X$), of the reduced-form parameter $\beta_a+\bB_a^\top\gamma_a$. Under Assumption~\ref{assm:params}, $\|\beta_a+\bB_a^\top\gamma_a\|_0 \le k_\beta+s k_\gamma \le (1+s)k$. 
Similarly $\theta_{10}$ involves a linear contrast (again in the direction of $\bar X$ or $\mu_X$) of $\beta_1 + \bB_0^\top \gamma_1$, and by the same argument $\|\beta_1 + \bB_0^\top \gamma_1\|_0 \le (1+s)k$ under Assumption \ref{assm:params}. 
Consequently, the standard sparsity-rate requirement discussed above translates in our setting to $(1+s)k\log{p}/\sqrt{n} \to 0$. 
This condition is therefore naturally required to obtain a $\sqrt{n}$-consistent and asymptotically normal estimator of $\theta_{aa}$ for each $a\in\{0,1\}$. 
As a consequence, this condition involving product sparsity $sk$ cannot be relaxed without further structural assumptions.
The consolidated condition $(1+s)(1+k)\log(p+q)/\sqrt{n} \to 0$ in Assumption~\ref{assm:params} is a convenient sufficient condition that simultaneously controls this remainder, the treated outcome-regression remainder, and the analogous reduced-form remainders required for the NDE and NIE analyses. It is stated in a slightly stronger but simpler worst-case form to avoid introducing separate rate conditions for every
high-dimensional regression.
\begin{remark}
    In the proof of Theorem \ref{thm:main_revised}, we require $\|\mu_X\|_\infty \ge \kappa > 0$. This is required to rule out the fact that if the contrasts are relatively small, then we would end up with $\hat \tau_1, \hat \tau_2 \approx 0$, resulting in a super-efficient estimator as noted in \cite{athey2018approximate}. 
\end{remark}
{\color{black}
\begin{remark}
\label{rem:sub-G_W}
{\color{black}Under Assumptions~\ref{assm:cov_err} and \ref{assm:params}, the arm-specific
design vector $W$ is also uniformly sub-Gaussian. Indeed, on the arm $A=a$,
we can write
\begin{equation*}
\textstyle
W =
\begin{pmatrix}
X \\ M
\end{pmatrix}
=
\begin{pmatrix}
X \\ \bB_aX+U_a
\end{pmatrix}
=
\begin{pmatrix}
\bI_p & \mathbf 0\\
\bB_a & \bI_q
\end{pmatrix}
\begin{pmatrix}
X \\ U_a
\end{pmatrix}.
\end{equation*}
The uniform sub-Gaussianity of $X-\mu_{X,a}$ and $U_a$, together with the
uniform bound on $\|\bB_a\|_{\rm op}$, implies that
$W-\mu_{W,a}$ is uniformly sub-Gaussian conditional on $A=a$. Moreover,
since $\cov(X, U_a \mid A = a) = 0$, the covariance matrix of $W$ can be written as
\begin{equation*}
\textstyle
\Sigma_{W,a} =
\begin{pmatrix}
\bI_p & \mathbf 0\\
\bB_a & \bI_q
\end{pmatrix}
\begin{pmatrix}
\Sigma_{X,a} & \mathbf 0\\
\mathbf 0 & \Sigma_{U,a}
\end{pmatrix}
\begin{pmatrix}
\bI_p & \bB_a^\top\\
\mathbf 0 & \bI_q
\end{pmatrix}.
\end{equation*}
Therefore, the eigenvalues of $\Sigma_{W,a}$ are uniformly bounded above
and away from zero.}
\end{remark}
}
\begin{remark}[Sample restricted eigenvalue conditions]
\label{rem:sample_RE}
The restricted eigenvalue and compatibility conditions required for the high-dimensional regressions below need not be imposed as separate assumptions. Under Assumptions~\ref{assm:cov_err} and
\ref{assm:params}, the centered arm-specific design vectors $X$ and $W$ are uniformly sub-Gaussian, and their covariance eigenvalues are uniformly bounded away from zero and infinity. Therefore, the sample covariance/second moment matrices satisfy restricted eigenvalue conditions with probability
tending to one, as soon as $n \gg s\log{(ep/s)}$ where $p$ is the ambient dimension and $s$ is the sparsity level (e.g., see \citep{raskutti2010restricted,rudelson2012reconstruction,vershynin2018high}). 
A formal statement and proof are provided in
Section \ref{sec:additional_lemmas} of the supplementary material.
\end{remark}

\subsection{Consistency and Asymptotic Normality of \texorpdfstring{$\hat{\theta}_{0}^{\mathrm{cf}}$}{Consistency and Asymptotic Normality}}
\label{sec:4.2}
In this subsection, we present our key theoretical finding, i.e., our estimators $\hat \theta_{10}^{\rm cf}, \hat \theta_{\rm NDE}^{\rm cf}, \hat \theta_{\rm NIE}^{\rm cf}$ are $\sqrt{n}$-consistent and asymptotically normal for both the sample and the population version of the parameters of interest. 
It is immediate from the definition of $\hat \theta_{10}^{\rm cf}$ in Algorithm \ref{algomain_modified} that its construction relies on $\hat \tau_1$ and $\hat \tau_2$, obtained by solving optimization problems Equation \eqref{eq:balancing_opt_2} and \eqref{eq:balancing_opt_1} in Algorithm \ref{algomain_modified} respectively. 
Thus, it is imperative to establish that these optimization problems are feasible (with probability going to $1$), a result we formalize in the following lemma: 

\begin{lemma}
    \label{lem:feasibility}
 Under Assumptions \ref{assm:cov_err} and \ref{assm:params}, there exist sufficiently large fixed constants $(K_1, K_2)$, such that the optimization problems Equation \eqref{eq:balancing_opt_1} and \eqref{eq:balancing_opt_2} are feasible with probability going to $1$. Moreover, the minimizers $\{\hat \tau_1^{(j)}, \hat \tau_2^{(j)}\}$ satisfy for $j \in \{1, 2\}$, 
    \begin{equation*}
    \textstyle
    n\left(\|\hat \tau_1^{(j)}\|_2^2 + \|\hat \tau_2^{(j)}\|_2^2\right) = O_p(1) \,.
    \end{equation*}
\end{lemma}
\vspace{-0.4em}
The proof of this lemma is deferred to Section \ref{sec:proof_lem_feasibility}   of the supplementary material. Lemma~\ref{lem:feasibility} shows that the optimization problems defining the debiasing weights are feasible with probability tending to one, and hence the proposed estimator is well-defined with high probability. Note that for obtaining $\hat \theta^{\rm cf}_{\rm NIE}$ via Algorithm \ref{algo_NIE}, we need to perform another optimization (Equation \eqref{eq:balancing_opt_3}). However, this optimization problem is very closely related to that in Equation \eqref{eq:balancing_opt_1}, and the proof for the existence of $\hat \omega$ is almost verbatim to that for the existence of $\hat \tau_2$, changing the notion of the treatment arms and data folds appropriately (recall that $\hat \tau_2$ involves control covariates and $\hat \omega$ involves treatment covariates). Therefore, we skip the proof for brevity. 

We next establish the asymptotic normality of the proposed estimators. We begin with $\hat\theta_{10}^{\rm cf}$ and consider inference for the sample-average mediation functional $\theta_{10}$ under homoskedastic outcome-regression errors, as the proof of this result contains the main technical ingredients of our analysis. We then extend the result to conditionally heteroskedastic outcome errors and to inference for the population-average mediation functional $\theta_{10}^{\rm pop}$. Finally, we present the corresponding results for $\hat\theta_{\rm NDE}^{\rm cf}$ and $\hat\theta_{\rm NIE}^{\rm cf}$. The following theorem establishes that, under homoskedasticity, $\hat\theta_{10}^{\rm cf}$ is a $\sqrt{n}$-consistent and asymptotically normal estimator of $\theta_{10}$: 

\begin{theorem}
\label{thm:main_revised}
Let $\hat\theta_{10}^{\rm cf}$ be the estimator of $\theta_{10}$ obtained from Algorithm \ref{algomain_modified}. Under Assumptions~\ref{assm:cov_err} and \ref{assm:params} along with homoskedasticity, i.e. $\nu_a(X, M) \equiv \sigma_a^2$ for $a \in \{0, 1\}$, we have: 
\begin{equation*}
\frac{\sqrt{n}(\hat\theta_{10}^{\rm cf}-\theta_{10})}{\sigma_n^{\rm cf}} \xrightarrow{d} \cN(0,1) \,, \text{ with }
\end{equation*}
\begin{equation*}
\textstyle
{\color{black}(\sigma_n^{\rm cf})^2 = \frac n4 \sum_{j = 1}^2 (\sigma_1^2 \lVert \hat \tau_1^{(j)} \rVert_2^2+ \gamma_1^\top\Sigma_{U,0}\gamma_1 \lVert \hat \tau_2^{(j)}\rVert_2^2) \,.}
\end{equation*}
\end{theorem}

\begin{remark}
    When $\gamma_1 = 0$, i.e., the response is unaffected by the mediators. In that case, Theorem \ref{thm:main_revised} implies $\textstyle{(\hat \theta_{10}^{\mathrm{cf}} - \theta_{10})/((\sigma_1/2) \sqrt{(\| \hat \tau_1^{(1)}\|_2^2 + \| \hat \tau_1^{(2)}\|_2^2)})}$ is asymptotically normal, which is Theorem 3 of \cite{athey2018approximate}. Therefore, our result can be viewed as a generalized version of the results in \citep{athey2018approximate} under cross-fitting in the presence of high-dimensional mediators along with the high-dimensional covariates. 
\end{remark}
Theorem~\ref{thm:main_revised} establishes that $\hat \theta_{10}^{\rm cf}$ is a $\sqrt{n}$-CAN estimator of the sample-average mediation functional $\theta_{10}$ under fairly general conditions. We next extend this result to the population-average mediation functional $\theta_{10}^{\rm pop}$. The sample-average target is often preferred in high-dimensional debiasing and approximate residual-balancing analyses because it effectively yields inference conditional on the observed covariates (e.g., see, \cite{athey2018approximate}). Nevertheless, the population-average functional is also a natural target in a superpopulation framework. The following corollary shows that $\hat\theta_{10}^{\rm cf}$ remains a $\sqrt{n}$-CAN estimator of $\theta_{10}^{\rm pop}$, with an additional variance component accounting for the fluctuation of the empirical covariate average around its population mean.
{\color{black}
\begin{corollary}
    \label{cor:pop_extension}
  {\color{black}  Consider the same setup as in Theorem \ref{thm:main_revised}. Define $\Lambda = \beta_1 + \bB_0^\top \gamma_1$ and set $\sigma^2_X = \Lambda^\top \Sigma_X \Lambda$ where $\Sigma_X = \var(X)$. If $\sigma_X^2 \le C$ for some constant $C$ and $(\sigma_n^{\rm cf})^2/((\sigma_n^{\rm cf})^2 + \sigma_X^2)$ admits an asymptotically deterministic equivalent sequence, then }
    \begin{equation*}
    {\color{black}\frac{\sqrt{n}(\hat \theta_{10}^{\rm cf} - \theta^{\rm pop}_{10})}{\sqrt{(\sigma_n^{\rm cf})^2 + \sigma_X^2}} \xrightarrow{d} \cN(0, 1) \,.}
    \end{equation*}
\end{corollary}
}
Compared with Theorem~\ref{thm:main_revised}, the variance now contains the additional term $\sigma_X^2=\Lambda^\top\Sigma_X\Lambda$, accounting for the fluctuation of $\bar X^\top\Lambda$ around its population counterpart $\mu_X^\top\Lambda$. The main technical difficulty remains the sample-average result established in Theorem~\ref{thm:main_revised}. For the population-average target, we additionally combine this result with the central limit theorem for $\sqrt{n}(\bar X-\mu_X)^\top\Lambda$. Since both components depend on the realized covariates, their asymptotic joint behavior cannot be obtained by treating them as independent and directly adding their variances. We therefore use a conditional central limit theorem argument and impose the additional condition that the relative contribution of the sample-target variance to the total variance admits an asymptotically deterministic equivalent.

Our next corollary now further extends these results to the heteroskedastic outcome regression error setting of Assumption \ref{assm:cov_err}: 
{\color{black}
\begin{corollary}
\label{cor:heteroskedastic_extension}
    {\color{black}Consider the same setup as in Theorem \ref{thm:main_revised}, but instead of assuming outcome regression errors are homoskedastic, we assume there exist functions $\nu_a(x, m) \in [c_-, c_+]$ with $c_- > 0$ and $c_+ < \infty$ such that $\var(\eps \mid X = x, M = m, A = a) = \nu_a(x, m)$. Then $\hat \theta_{10}^{\rm cf}$ satisfies: 
    \begin{equation*}
    {\color{black}\frac{\sqrt{n}(\hat \theta_{10}^{\rm cf} - \theta_{10})}{\sigma_{n}^{\rm cf, het}} \xrightarrow{d} \cN(0, 1),} 
    \end{equation*}
    \begin{equation*}
    \textstyle
    {\color{black}(\sigma_n^{\rm cf, het})^2 = \frac{n}{4}\sum_{j = 1}^2 \left[\sum_{i \in \cD_j, A_i = 1} \left\{(\hat \tau_{1, i}^{(j)})^2\nu_1(X_i, M_i)\right\} + \gamma_1^\top \Sigma_{U, 0} \gamma_1 \|\hat \tau_2^{(j)}\|_2^2\right].}
    \end{equation*}
    Furthermore, under the additional assumption as in Corollary \ref{cor:pop_extension} ($\sigma^2_X \le C$ and $(\sigma_n^{\rm cf, het})^2/((\sigma_n^{\rm cf, het})^2 + \sigma_X^2)$ admits an asymptotically deterministic equivalent sequence) we have
    \begin{equation*}
    {\color{black}\frac{\sqrt{n}(\hat \theta_{10}^{\rm cf} - \theta^{\rm pop}_{10})}{\sqrt{(\sigma_n^{\rm cf, het})^2 + \sigma_X^2}} \xrightarrow{d} \cN(0, 1) \,.}
    \end{equation*}}
\end{corollary}
}
The key difference between Corollary~\ref{cor:heteroskedastic_extension} and Theorem~\ref{thm:main_revised}, together with its population extension in Corollary~\ref{cor:pop_extension}, lies in the variance normalization. Under conditional heteroskedasticity, the homoskedastic scale $\sigma_n^{\rm cf}$ is replaced by $\sigma_n^{\rm cf,het}$, which explicitly incorporates the conditional outcome-error variance $\nu_1(X_i,M_i)$. In particular, setting $\nu_1(x,m)\equiv\sigma_1^2$ recovers the homoskedastic variance and, consequently, the conclusions of the preceding results. In practice, the heteroskedastic variance can be estimated using the corresponding squared regression residuals, i.e. replacing $\nu_1(X_i,M_i)$ with $(Y_i-X_i^\top\hat\beta_1-M_i^\top\hat\gamma_1)^2$, 
while the control-arm mediator-error contribution can be estimated by  replacing $\gamma_1^\top \Sigma_{U, 0}\gamma_1$ by  $(M_i^\top\hat\gamma_1-X_i^\top\hat b)^2$. Under our assumptions, this estimator is consistent and can therefore be used to construct asymptotically valid confidence intervals; see Section~\ref{section:5} and Section~\ref{Appendix:more_simu_sensi} of the supplementary material for numerical evidence.

Our next theorem presents an analogous result for $\hat \theta_{\rm NDE}^{\rm cf} \text{ and } \hat \theta_{\rm NIE}^{\rm cf}$. In analogy with Corollary \ref{cor:pop_extension}, define $\sigma^2_{X, \rm NDE} = \Lambda_{\rm NDE}^\top \Sigma_X \Lambda_{\rm NDE}$ with $\Lambda_{\rm NDE}= (\beta_1 - \beta_0) + \bB_0^\top(\gamma_1 - \gamma_0)$ and $\sigma^2_{X, \rm NIE} = \Lambda_{\rm NIE}^\top \Sigma_X \Lambda_{\rm NIE}$ with $\Lambda_{\rm NIE} = (\bB_1 - \bB_0)^\top \gamma_1$: 
{\color{black}
\begin{theorem}[NDE]
\label{thm:NDE}
{\color{black}Under Assumptions \ref{assm:cov_err} and \ref{assm:params}, the estimator $\hat \theta_{\rm NDE}^{\rm cf}$, as obtained from Algorithm \ref{algo_NDE} satisfies: }
\begin{equation*}
{\color{black}\frac{\sqrt{n}(\hat \theta_{\rm NDE}^{\rm cf}  - \theta_{\rm NDE})}{\sigma_{n, \rm NDE}^{\rm cf, het}} \xrightarrow{d} \cN(0, 1) \,,}
\end{equation*}
{\color{black}where, with} {\color{black}$\textstyle{\omega_0(X) = (\gamma_1 - \gamma_0)^\top \Sigma_{U, 0}(\gamma_1 - \gamma_0) + \bbE[\nu_0(X, M) \mid X, A = 0]}$}, 
\begin{equation*}
\textstyle
{\color{black}(\sigma_{n, \rm NDE}^{\rm cf, het})^2 = \frac n4\sum_{j=1}^2\left[\sum_{i \in \cD_j, A_i = 1}(\hat \tau_{1,i}^{(j)})^2\nu_1(X_i, M_i) +  \sum_{i\in \cD_j, A_i = 0}(\hat \tau_{2,i}^{(j)})^2\omega_0(X_i)\right].}
\end{equation*}
{\color{black}Furthermore, if $\omega^{\rm het}_{n, \rm NDE} = (\sigma_{n, \rm NDE}^{\rm cf, het})^2/((\sigma_{n, \rm NDE}^{\rm cf, het})^2 + \sigma^2_{X, \rm NDE})$ admits an asymptotically deterministic equivalent sequence and $\sigma^2_{X, \rm NDE} \le C_X$ for some $C_X > 0$, then }
\begin{equation*}
{\color{black}\frac{\sqrt{n}\left(\hat\theta_{\rm NDE}^{\rm cf} - \theta_{\rm NDE}^{\rm pop}\right)}{\sqrt{
(\sigma_{n,\rm NDE}^{\rm cf,het})^2 + \sigma_{X,\rm NDE}^2}} \xrightarrow{d} \cN(0,1).
\,}
\end{equation*}
\end{theorem}
}
{\color{black}
\begin{theorem}[NIE]
\label{thm:NIE}
    {\color{black}Under Assumptions \ref{assm:cov_err} and \ref{assm:params} and $\|\gamma_1\|_2 \ge c_\gamma > 0$ the estimator $\hat \theta_{\rm NIE}^{\rm cf}$, as obtained via Algorithm \ref{algo_NIE} satisfies}
    \begin{equation*}
    {\color{black}\frac{\sqrt{n}(\hat \theta_{\rm NIE}^{\rm cf} - \theta_{\rm NIE})}{\sigma_{n, \rm NIE}^{\rm cf, het}} \xrightarrow{d} \cN(0,1) \,,}
    \end{equation*}
{\color{black}    where, with $v^2_{U, a, 1} =\gamma_1^\top \Sigma_{U, a} \gamma_1$, }
    \begin{equation*}
    \textstyle
    {\color{black}(\sigma_{n, \rm NIE}^{\rm cf, het})^2 = \frac n4\sum_{j = 1}^2 \left[\sum_{\substack{i \in \cD_j \\ A_i = 1}}(\hat\omega_i^{(j)}-\hat\tau_{1,i}^{(j)})^2\nu_1(X_i, M_i) + v_{U,1,1}^2  \|\hat \omega^{(j)}\|_2^2 + v_{U,0,1}^2 \|\hat \tau_2^{(j)}\|_2^2 
    \right].}
    \end{equation*}
   {\color{black} Furthermore is the sequence $\textstyle{\omega_{n,\rm NIE}^{\rm het} = (\sigma_{n, \rm NIE}^{\rm cf, het})^2/((\sigma_{n, \rm NIE}^{\rm cf, het})^2 + \sigma^2_{X, \rm NIE})}$ admits an asymptotically deterministic equivalent sequence and \(\sigma_{X,\mathrm{NIE}}^2\le C_X < \infty\), then}
    \begin{equation*}
    {\color{black}\frac{\sqrt{n}(\hat \theta_{\rm NIE}^{\rm cf} - \theta_{\rm NIE}^{\rm pop})}{\sqrt{(\sigma_{n, \rm NIE}^{\rm cf, het})^2 + \sigma_{X, \rm NIE}^2}} \xrightarrow{d} \cN(0, 1) \,.}
    \end{equation*}
\end{theorem}
}
Taken together, our theorems and corollaries establish the $\sqrt{n}$-CAN property of our proposed estimators for the (population and sample average) mediation functional, NDE, and NIE under conditional heteroskedasticity in the presence of UHD covariates and mediators and their interaction with the treatment assignment. These theoretical guarantees provide the basis for constructing asymptotically valid confidence intervals for the relevant causal effects.

\noindent
\textcolor{black}{\bf Consistent estimation of asymptotic variance: }In practice, the asymptotic variance can be estimated directly using the corresponding squared regression residuals. Specifically, define
\begin{equation*}
\textstyle
\hat\epsilon_{1,i}^{(j)} = Y_i-X_i^\top\hat\beta_1^{(j)} -M_i^\top\hat\gamma_1^{(j)}, \qquad 
\hat r_{0,i}^{(j)} = M_i^\top\hat\gamma_1^{(j')} -X_i^\top\hat b^{(j)},
\end{equation*}
for the treated and control observations in \(\mathcal D_j\), respectively, and set
\begin{equation*}
\textstyle
(\hat\sigma_{n}^{\mathrm{cf,het}})^2 = \frac{n}{4}\sum_{j=1}^2 \left[\sum_{\substack{i\in\mathcal D_j\\ A_i=1}}\bigl(\hat\tau_{1,i}^{(j)}\bigr)^2\bigl(\hat\epsilon_{1,i}^{(j)}\bigr)^2 + \sum_{\substack{i\in\mathcal D_j\\ A_i=0}} \bigl(\hat\tau_{2,i}^{(j)}\bigr)^2 \bigl(\hat r_{0,i}^{(j)}\bigr)^2
\right].
\end{equation*}
Under the assumptions of Corollary \ref{cor:heteroskedastic_extension}, this estimator is consistent for $(\sigma_{n}^{\mathrm{cf,het}})^2$. Intuitively speaking, the Lasso estimation rates ensure that the fitted residuals are asymptotically close to the corresponding true regression errors, while \(n\|\hat\tau_k^{(j)}\|_2^2=O_p(1)\) and the maximum-weight constraint \(\|\hat\tau_k^{(j)}\|_\infty\leq n^{-\alpha}\) ensure that no individual observation dominates the weighted sums. Consequently, the weighted squared residuals consistently estimate the conditional-variance contributions appearing in \((\sigma_{n}^{\mathrm{cf,het}})^2\), so that $(\hat\sigma_{n}^{\mathrm{cf,het}})^2 - (\sigma_{n}^{\mathrm{cf,het}})^2 =o_p(1)$. 
Analogous residual-based estimators can be constructed for the asymptotic variances of the NDE and NIE estimators, which we skip for brevity.

\section{Simulation and Real Data Analysis} 
\label{section:5}
\subsection{Simulation Designs}
\label{sec:simulation}
In this section, we present simulation studies to assess the performance of our proposed methodology. 
Specifically, we compare our debiased estimator ($\hat \theta_{10}^{\mathrm{cf}}$ in Algorithm \ref{algomain_modified}) to a naive (non-debiased) estimator $\hat \theta^{\rm plug-in}$ (defined at the beginning of Section \ref{sec:method}) in terms of their root mean squared error (RMSE), standard deviation (SD), {\color{black}confidence interval length and coverage.} To understand the behavior of the proposed estimator across different regimes, we consider {\color{black} four data-generating processes that share the same generative mechanism (described below) but differ in terms of the sparsity, the signal strength of the nonzero coefficients, and heterogeneity.}

\begin{enumerate}
\item \textcolor{black}{We generate covariate $X \sim \mathcal N(\mu_X, \Sigma_X)$ with a dense but dimension-rescaled mean, $\mu_X = \frac{c}{\sqrt p}\mathbf 1$ with $c = 0.5\sqrt{50}$. The covariance matrix is generated as $\Sigma_X = R \Lambda R^\top$, where $R$ is an orthonormal matrix and $1 \leq \lambda_{\min}(\Sigma_X) \leq \lambda_{\max}(\Sigma_X) \leq 2$.}
    \item We generate treatment $A \sim\text{Ber}\{\text{logit}(X^\top \alpha)\}$, where $\alpha \in \mathbb{R}^p$ is a sparse vector with $\| \alpha\|_0 = 3$ with nonzero elements randomly sampled from $\mathcal U(0,2)$.
    \item Upon generating $(A, X)$, we generate the mediators from the model: 
    \begin{equation*}
        M = (1-A)(\mathbf B_0 X + U) + A(\mathbf B_1 X + U')  \,,
    \end{equation*}
    where $\mathbf B_1, \mathbf B_0 \in \mathbb R^{q \times p}$ are row-sparse matrices, where each row of both $\mathbf B_0$ and $\mathbf B_1$ contains $s$ (positions are randomly chosen) nonzero elements sampled from $\mathcal U(0.5,1.5)$. \textcolor{black}{The errors $U, U' {\sim} \mathcal N(0, \Sigma_U), \Sigma_U = R_U \Lambda_U R_U^\top$, where $R_U \in \mathbb{R}^{q \times q}$ is orthonormal and $\Lambda_U=\mathrm{diag}(\lambda_1,\ldots,\lambda_q)$ with $1 \le \lambda_{\min}(\Sigma_U) \le \lambda_{\max}(\Sigma_U) \le 2$.}
    \item Finally, we generate the response variable from the model:
    \begin{equation*}
        Y = (1-A)(X^\top \beta_0 + M^\top \gamma_0 + \epsilon')  + A(X^\top \beta_1 + M^\top \gamma_1 + \epsilon)\,,
    \end{equation*}
    where $\beta_0, \beta_1 \in \mathbb{R}^p$ are sparse vectors with $k_\beta$ non-zero elements. The non-zero indices are chosen randomly, and the values corresponding to those indices are sampled from $\mathcal U(0.5, 1.5)$. Similarly, $\gamma_0, \gamma_1 \in \mathbb{R}^q$ are also sparse vectors with $k_\gamma$ non-zero elements. As before, the non-zero indices are sampled randomly, and values corresponding to those indices are sampled from $\mathcal U(0.5, 1.5)$, and $\epsilon, \epsilon' \overset{i.i.d.}{\sim} \mathcal N(0, \sigma^2)$.
\end{enumerate}

\textcolor{black}{With the above-mentioned data-generating process, we study four regimes:}

\begin{itemize}
    \item {\color{black}\textbf{Base.}} The sparsity levels are fixed at $s = k_\beta = k_\gamma = 3$ across all dimensions, every nonzero coefficient is a moderate signal sampled from $\mathcal U(0.5, 1.5)$, and the errors are homoskedastic, $\epsilon, \epsilon' \sim \mathcal N(0, \sigma^2)$.
    \item {\color{black}\textbf{Increasing sparsity.}} The number of nonzero coefficients grows with the
ambient dimension, with $s = k_\beta = k_\gamma \in \{5, 10, 15, 20\}$ corresponding to
$p+q \in \{100, 800, 1500, 2500\}$; nonzero values are again moderate, sampled
from $\mathcal U(0.5, 1.5)$. To keep the mediator model well-scaled as its
support grows, the rows of the mediator coefficient matrices $\mathbf B_0,
\mathbf B_1 \in \mathbb R^{q \times p}$ are normalized to unit $\ell_1$ norm.
    \item {\color{black}\textbf{Vanishing minimum non-zero signal.}} The sparsity is now fixed at $s = k_\beta = k_\gamma = 3$, but approximately half of the nonzero coefficients of the regression coefficients are now generated from near the detection boundary, with magnitudes of order $n^{-1/2}_{\mathrm{max}}$ ($n_{\max} = 1250$, the maximum number of samples used in our simulation, and these coordinates are sampled from $\mathcal U(0.8, 1.2)/\sqrt{n_{\max}}$), while the remaining coefficients are moderate as in the base case. This regime assesses the performance of the estimator when part of the signal is only weakly identifiable.
    \item {\color{black}\textbf{Heteroskedastic noise.}} We retain the sparsity ($s = k_\beta = k_\gamma = 3$) and moderate signals as in the base case, but replace the homoskedastic errors in item (4) by conditionally heteroskedastic ones whose variance is now a function of both the covariates and the mediators. Specifically, we generate errors as  $\epsilon_i \mid (X_i, M_i) \sim \mathcal N\!\big(0,\, \sigma^2\, v(X_i, M_i)\big)$ (analogously for $\epsilon'_i$), where
    \begin{equation}\label{eq:simu_hetero}
        v(X_i, M_i) \;=\; \frac{1 + \tfrac{1}{2}\big[(X_i^\top a_\epsilon)^2 + (M_i^\top b_\epsilon)^2\big]}{\;\mathbb{E}\big[\,1 + \tfrac{1}{2}\big((X^\top a_\epsilon)^2 + (M^\top b_\epsilon)^2\big)\big]\;}\,,
    \end{equation}
    with sparse $a_\epsilon \in \mathbb{R}^p,\ b_\epsilon \in \mathbb{R}^q$ satisfying $\|a_\epsilon\|_0 = \|b_\epsilon\|_0 = 3$ and nonzero entries of $(a_\epsilon, b_\epsilon)$ drawn from $\mathcal U(0,1)$. The normalization ensures $\mathbb{E}[v(X, M)] = 1$, so the marginal noise level matches the homoskedastic case. 
\end{itemize}
In Assumption \ref{assm:cov_err}, we claim that $0<c_\epsilon<C_\epsilon<\infty$ and $c_\epsilon \le \nu_a(X,M) \le C_\epsilon$ almost surely, whereas in Equation \eqref{eq:simu_hetero}, $v(X_i, M_i) $ can potentially be unbounded. We include this setting as a stress test, allowing the conditional variance to be unbounded in order to assess the robustness of our estimator beyond the theoretical conditions. Our Assumption \ref{assm:cov_err} could also be extended to accommodate such settings via an appropriate truncation that controls extreme conditional variances.

The number of samples $n$ varies over $\{250, 500, 750, 1000, 1250\}$. For each value of $n$ in this set, we vary $p + q$ over the set $\{100, 800, 1500, 2500\}$. Furthermore, for the homoskedastic case, the noise variance $\sigma^2 \in \{0.1, 1\}$. Recall that our parameter of interest is $\theta_{10} = \bar X^\top (\beta_1 + \mathbf B_0^\top \gamma_1)$ where $\bar X$ is the sample average of the covariates. We keep $X$ and other parameters (e.g., $(\beta_a, \gamma_a, \alpha, \mathbf B_a)$ for $a \in \{0, 1\}$) fixed in the simulation study to keep the parameter of interest unchanged.  
\begin{table}[ht!]
\centering
\small
\renewcommand{\arraystretch}{1.4}
\resizebox{1\textwidth}{!}{
\begin{tabular}{cc|cccc|ccc|cccc} \hline
\multicolumn{2}{c|}{$s=3,\ k_1=k_2=3$} & \multicolumn{11}{c}{$\sigma^2 = 1$}\\ \hline
\multicolumn{2}{c|}{Estimators} & \multicolumn{7}{c|}{$\hat\theta_{10}^{\mathrm{cf}}$ (Debiasing)} & \multicolumn{4}{c}{$\hat \theta^{\rm plug-in}$}\\ \hline
 & & \multicolumn{4}{c|}{Monte Carlo CI} & \multicolumn{3}{c|}{Analytical CI} & \multicolumn{4}{c}{Monte Carlo CI}\\
$p+q$ & \multicolumn{1}{c|}{$n$} & RMSE & SD & CI Len & Cov & SE & CI Len & Cov & RMSE & SD & CI Len & Cov\\ \hline
\multirow{5}{*}{$100$} & \multicolumn{1}{c|}{$250$}  & 0.691 & 0.334 & 1.308 & 0.892& 0.304 & 1.194 & 0.902& 0.951 & 0.244 & 0.958 & 0.034
\\
 & \multicolumn{1}{c|}{$500$}  & 0.197 & 0.179 & 0.700 & 0.912 & 0.187 & 0.734 & 0.926 & 0.678 & 0.152 & 0.596 & 0.014
\\
 & \multicolumn{1}{c|}{$750$}  & 0.213 & 0.142 & 0.556 & 0.902& 0.150 & 0.588 & 0.926& 0.595 & 0.124 & 0.486 & 0.008
\\
 & \multicolumn{1}{c|}{$1000$} & 0.128 & 0.121 & 0.473 & 0.932 & 0.130 & 0.509 & 0.930& 0.529 & 0.103 & 0.405 & 0.000
\\
 & \multicolumn{1}{c|}{$1250$} & 0.132 & 0.125 & 0.490 & 0.936 & 0.121 & 0.474 & 0.948 & 0.483 & 0.103 & 0.403 & 0.008
\\ \hline
\multirow{5}{*}{$800$} & \multicolumn{1}{c|}{$250$}  & 0.238 & 0.225 & 0.883 & 0.948 & 0.384 & 1.505 & 0.996 & 0.484 & 0.215 & 0.841 & 0.450
\\
 & \multicolumn{1}{c|}{$500$}  & 0.169 & 0.146 & 0.574 & 0.906 & 0.204 & 0.801 & 0.982 & 0.432 & 0.056 & 0.221 & 0.000
\\
 & \multicolumn{1}{c|}{$750$}  & 0.159 & 0.131 & 0.512 & 0.892 & 0.152 & 0.597 & 0.954 & 0.430 & 0.056 & 0.219 & 0.000
\\
 & \multicolumn{1}{c|}{$1000$} & 0.118 & 0.118 & 0.461 & 0.948 & 0.132 & 0.519 & 0.966 & 0.349 & 0.041 & 0.160 & 0.000
\\
 & \multicolumn{1}{c|}{$1250$} & 0.098 & 0.097 & 0.382 & 0.952 & 0.114 & 0.446 & 0.976 & 0.322 & 0.039 & 0.153 & 0.000\\ \hline
\multirow{5}{*}{$1500$} & \multicolumn{1}{c|}{$250$}  & 0.445 & 0.254 & 0.995 & 0.682 & 0.375 & 1.471 & 0.908& 0.245 & 0.173 & 0.679 & 0.817
\\
 & \multicolumn{1}{c|}{$500$}  & 0.175 & 0.150 & 0.588 & 0.920 & 0.199 & 0.781 & 0.980 & 0.372 & 0.044 & 0.173 & 0.000
\\
 & \multicolumn{1}{c|}{$750$}  & 0.168 & 0.124 & 0.488 & 0.854 & 0.148 & 0.580 & 0.950& 0.321 & 0.036 & 0.141 & 0.000
\\
 & \multicolumn{1}{c|}{$1000$} & 0.097 & 0.097 & 0.381 & 0.946 & 0.122 & 0.477 & 0.984 & 0.280 & 0.029 & 0.115 & 0.000
\\
 & \multicolumn{1}{c|}{$1250$} & 0.104 & 0.096 & 0.376 & 0.926 & 0.108 & 0.424 & 0.952 & 0.248 & 0.026 & 0.101 & 0.000
\\ \hline
\multirow{5}{*}{$2500$} & \multicolumn{1}{c|}{$250$}  & 0.721 & 0.313 & 1.227 & 0.464 & 0.466 & 1.825 & 0.798 & 0.454 & 0.183 & 0.718 & 0.358
\\
 & \multicolumn{1}{c|}{$500$}  & 0.168 & 0.157 & 0.617 & 0.944 & 0.226 & 0.885 & 0.894& 0.270 & 0.030 & 0.117 & 0.000
\\
 & \multicolumn{1}{c|}{$750$}  & 0.208 & 0.132 & 0.516 & 0.774 & 0.170 & 0.666 & 0.966& 0.326 & 0.033 & 0.130 & 0.000
\\
 & \multicolumn{1}{c|}{$1000$} & 0.118 & 0.114 & 0.448 & 0.948& 0.139 & 0.543 & 0.962& 0.255 & 0.025 & 0.098 & 0.000
\\
 & \multicolumn{1}{c|}{$1250$} & 0.111 & 0.108 & 0.424 & 0.950& 0.122 & 0.480 & 0.984 & 0.171 & 0.018 & 0.071 & 0.000\\ \hline
\end{tabular}%
}
\caption{{\color{black}Simulation results for the debiased ($\hat\theta_{10}^{\mathrm{cf}}$) and naive ($\hat \theta^{\rm plug-in}$) estimators under the base DGP with $\sigma^2=1$. Numbers are averaged over 500 replications with $K_1=K_2=2.75$.}}
\label{tab:simu_combined_base_sigma1}
\end{table}

\textcolor{black}{For brevity, the main text reports only the base regime with $\sigma^2 = 1$ (Table \ref{tab:simu_combined_base_sigma1}). The remaining configurations, namely the base regime with $\sigma^2 = 0.1$, the increasing-sparsity, vanishing minimal non-zero signal, and the heteroskedastic-noise regimes, together with the sensitivity analyses, can be found in Section \ref{Appendix:more_simu_sensi} of  the supplementary material (Tables \ref{tab:simu_combined_base_sigma0p1}-- \ref{tab:simu_hetero_sigma0p5} and \ref{tab:sens_diag_n750}--\ref{tab:sens_grid}).}

We use Algorithm \ref{algomain_modified} to compute our estimator. The Lasso estimates in Step 2 and Step 4 are obtained using the R package \texttt{glmnet}, where the tuning parameter is selected by cross-validation using the \texttt{lambda.1se} rule from R function \texttt{cv.glmnet} (\cite{friedman2010regularization}). The debiasing weights in Step 1 and Step 3 are solved by solving the corresponding optimization problem via the R package \texttt{pogs} with $K_1 = K_2 = 2.75$. This particular choice of $(K_1, K_2)$ is motivated by our sensitivity analysis, as elaborated in Table \ref{tab:sens_diag_n750} - \ref{tab:sens_grid} in Section \ref{Appendix:more_simu_sensi} of the supplementary material. This choice remains consistent with previous studies (e.g., \cite{javanmard2014confidence}). We also record the computational cost under a representative simulation setting. \textcolor{black}{Specifically, when $n=1000$ and $p+q=1500$, running the $500$ Monte Carlo replications across $30$ CPU cores takes approximately $12$ minutes of wall-clock time, which amortizes to about $1.4$ seconds per replication ($12\times 60/500 \approx 1.4$).}

\begin{figure}
    \centering
    \includegraphics[width=1\linewidth]{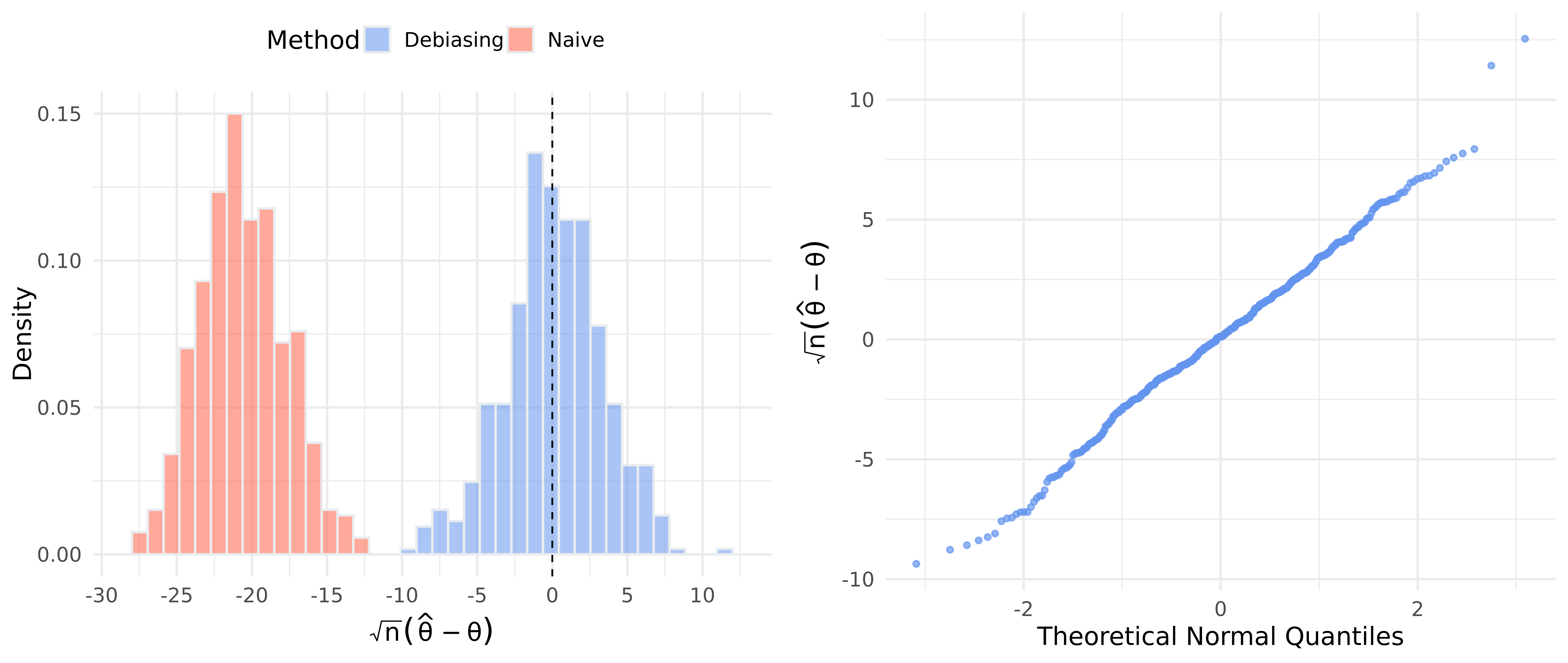}
    \caption{Histogram and QQ plot of $\sqrt{n}(\hat \theta - \theta)$ under $n = 750, p + q = 1500, \sigma_{\epsilon} = 0.1$.}
    \label{fig:histqq}
\end{figure}
The simulation results are summarized in Table \ref{tab:simu_combined_base_sigma1} and in Section  \ref{Appendix:more_simu_sensi} of supplementary materials. Each table reports the root-mean-squared error (RMSE) and standard deviation (SD) of both $\hat \theta^{\rm plug-in}$ and $\hat \theta_{10}^{\rm cf}$, averaged over 500 Monte Carlo replications, \textcolor{black}{along with the estimated confidence-interval length and empirical coverage under both a Monte Carlo standard deviation and our estimated analytical standard error.} 
The results indicate that the debiased estimator $\hat \theta_{10}^{\rm cf}$ consistently achieves a lower RMSE than the naive estimator, with the performance gap becoming increasingly prominent as the sample size $n$ grows. Furthermore, the standard deviations of the two estimators are comparable, suggesting that the improved RMSE of our debiased estimator is attributable to the successful bias correction. \textcolor{black}{This pattern is consistent across all four regimes: in the base, minimal-signal, and heteroskedastic-noise regimes, $\hat \theta_{10}^{\rm cf}$ achieves close-to-nominal coverage as $n$ increases, and the estimated analytical standard error $\hat \sigma_n^{\rm cf}$ closely matches the Monte Carlo standard deviation. In the increasing-sparsity regime, the same conclusions hold in moderate dimensions, while the most challenging high-dimensional, small-sample cells ($p+q = 2500$, $n = 250$) are, as expected, the hardest, with $\hat \theta_{10}^{\rm cf}$ nonetheless remaining more accurate and better calibrated than $\hat \theta^{\rm plug-in}$.}
To further illustrate the performance of the estimators, we display histograms of the naive (red) and debiased (blue) estimators in the left column of Figure \ref{fig:histqq}, based on 500 Monte Carlo replications. The estimators are centered around the true parameter and scaled by $\sqrt{n}$, where $n$ is the number of samples. 
The histogram clearly reveals that the naive estimator is substantially biased, while the debiased estimator is centered around zero as a result of successful bias correction. Additionally, the Q-Q plot of the debiased estimator, shown in the right column of Figure \ref{fig:histqq}, aligns closely with the standard normal distribution, providing empirical evidence for its asymptotic normality and supporting our theoretical results.

{\color{black}
\begin{remark}[Comparison with \cite{guo2022high}]
    \label{rem:comparison_guo}
 {\color{black} We also conducted additional numerical experiments to compare the performance of our proposed estimator with the estimator of \cite{guo2022high} in a simplified setting where both methods can be applied.
    A direct comparison is not possible, since the framework of \cite{guo2022high,guo2023statistical,guo2024estimations} does not accommodate treatment-covariate interactions in the mediator model or treatment--covariate/treatment--mediator interactions in the outcome model. Therefore, in the supplementary material, we consider a simplified common setting: we remove the treatment interactions from our model and include UHD covariates with a natural modification of their procedure. The details can be found in Section \ref{appendix:comparison-guo-trilogy} of the supplementary document. It is evident from the results that, especially when the sparsity level is high or the minimum active signal strength is small, our debiased estimator outperforms the modified estimator in \cite{guo2022high} in both RMSE and asymptotic coverage.}
\end{remark}
}

\subsection{Real Data Experiments} 
DNA methylation, an epigenetic modification characterized by the addition of a methyl group to the DNA molecule, is essential for regulating gene expression and is linked to various biological functions, and is often regarded as a potential mediator linking exposures to health outcomes \citep{fujii2022dna}. Recent studies suggest that smoking leads to widespread DNA methylation changes, which may contribute to cancer development and progression \citep{tan2013epigenomic, zhang2016estimating}. Meanwhile, certain smoking-associated methylation markers have been linked to increased mortality risk among lung cancer patients \citep{zhang2016estimating, mehta2015epigenetics}. 
As an application of our developed methodology, we apply our method to the data from the lung cancer cohort of The Cancer Genome Atlas (TCGA) project to study the direct and indirect effects of smoking on the survival time of patients diagnosed with lung squamous cell carcinoma and lung adenocarcinoma. The data is publicly available at (\url{https://xenabrowser.net/datapages/}).
The dataset includes DNA methylation data from 907 individuals, feature data for 1,299 individuals, and survival data for 1,145 individuals, among which $833$ individuals were common. 
\begin{itemize}
    \item Our response variable, $Y$, is the logarithm of the survival time, where the survival time is measured as the number of days from diagnosis to death or the last follow-up date. 
    We used the inter-quartile range to detect and remove the outliers in the response variable. 

    \item We selected $20$ clinically relevant features, $X$ (e.g., age, sex, etc.; see full list in Section \ref{appendix:pheno} of the supplementary material) for our study by excluding all those features which has more than $100$ missing values. See the Supplementary Materials for the complete list of the chosen features. We then imputed the remaining missing values; for numerical features (11), we used \texttt{softImpute} \citep{mazumder2010spectral} and for categorical variables (9), we used \texttt{mice} \citep{van2011mice} imputation. 
    Furthermore, we considered up to third-order polynomial interactions for better expressiveness. These operations ultimately yielded a $ 42$-dimensional covariate.

    \item  The DNA methylation data comprises $350,019$ DNA methylation marker values ($M$), represented as continuous ratios ranging from 0 to 1, reflecting the intensity of methylation, where higher values indicate a greater degree of methylation. For more details, see (\url{https://www.cancer.gov/ccg/research/genome-sequencing/tcga}). We selected only those mediators with no missing values. This reduces the number of mediators to {\color{black}$264,797$}. 

\end{itemize}
After the data pre-processing, we are left with $n = 738$ units, each with $42$ features $(X)$ and $264,797$ mediators ($M$). The treatment indicator ($A$) is the smoking status of the units ($A = 1$ indicates a smoker). 
Among these samples, there are $42.6\% (315)$ females and $57.3\% (423)$ males. 
The patients' ages range from 33 to 90 years, with a median age of 67 years. The median survival time is 603 days (1.65 years). 
\\\\
To estimate the direct and indirect effects (NDE and NIE) of smoking on the survival time of lung cancer patients, we fit the following high-dimensional linear model to the pre-processed data:
\begin{align}
\label{eq:dgp_real_data}
    M_i &= (1 - A_i)(\mathbf{B}_0 X_i + U_i)  + A_i (\mathbf{B}_1 X_i + U'_i), \hspace{.2in} i = 1, \ldots, 738, \notag \\
    Y_i & = (1 - A_i)( X_i^\top \beta_0 + M_i^\top \gamma_0 + \epsilon_i') + A_i ( X_i^\top \beta_1 + M_i^\top \gamma_1 + \epsilon_i) \,.
\end{align}
\textcolor{black}{We then apply Algorithm~\ref{algo_NDE} and~\ref{algo_NIE} to estimate the NDE, NIE, and total treatment effect. For our real data application, one full run of the proposed algorithm (one bootstrap replicate) takes approximately $48$ seconds using $10$ CPU cores.}

Table~\ref{tab:results1} summarizes the estimates of the NDE, NIE, and total effect of smoking on log survival time. The estimate of the total effect implies that, on average, smokers have \textbf{92.14\%} ($e^{-0.0819} \times 100\%$) of the survival time of non-smokers, i.e., smoking reduces approximately \textbf{7.86\%} of the survival time of lung cancer patients. Of this total effect, our results demonstrate that approximately \textbf{5.59\%} $= 100 \times (1 - e^{-0.0575})\%$ corresponds to the indirect effect (NIE), i.e., the impact of smoking on survival time mediated via DNA methylation. In comparison, approximately \textbf{2.41\%} $= 100 \times (1 - e^{-0.0244})\%$ corresponds to the direct effect (NDE), which can be attributed to pathways beyond the methylation mechanism considered in this analysis.

Since the confidence interval for the indirect effect estimate excludes zero, while the confidence interval for the direct effect estimate contains zero, our findings suggest that the estimated effect of smoking on survival time is primarily mediated through the DNA methylation mechanism captured in this analysis. This indicates that DNA methylation plays an important mediating role in the current dataset, although additional methylation mechanisms and other biological pathways may remain to be investigated.

\begin{table}[t]
\centering
\small
\begin{tabular}{cccc}
\hline
& Estimate & Bootstrap Mean & Quantile CI (95\%) \\ 
\hline
Indirect Effect 
& $-0.0575$ & $-0.0506$ & $(-0.1536,\ -0.0395)$\\

Direct Effect 
& $-0.0244$ & $-0.0288$ & $(-0.1507,\ 0.0773)$ \\

Total Effect 
& $-0.0819$ & $-0.0778$ & $(-0.2044,\ -0.0369)$\\
\hline
\end{tabular}
\caption{{\color{black}Summary of the estimated effects, bootstrap means, and 95\% quantile confidence intervals based on 300 bootstrap samples.}}
\label{tab:results1}
\end{table}

We compared the results with a fixed-dimensional OLS-based approach. Note that if both $X$ and $M$ are low-dimensional in Equation \eqref{eq:dgp_real_data}, we can easily estimate the NDE, NIE (and consequently the total effect) by computing ordinary least squared (OLS) estimators via regressing i) $Y$ on $(X, M)$ (on treatment and control observations separately), and ii) $M$ on $X$ (on treatment and control observations separately). To make the problem low-dimensional, we selected CpG sites (cg19757631, cg05147638, cg24720672, cg08108679, cg05575921, cg24859433) which were identified as significant mediators in previous studies \citep{cui2021high, zhang2021review, zhang2021mediation}, as our $M$, and $20$ clinically relevant features as our $X$ (i.e., we do not consider their higher-order terms and interactions).

In Table \ref{tab:results2}, we present the results in a fixed-dimensional setting. The results reveal that the OLS-based approach produces conclusions about the indirect and total effects that are contrary to intuition. The estimate of the indirect and total effect imply that smokers have $128.80\%$ ($e^{0.2531} \times 100\%$) (mediated through DNA methylation), and $126.52\%$ ($e^{0.2353} \times 100\%$) of survival time of non-smokers, i.e., smoking increases $\approx 29\%$ $=100 \times (1-e^{0.2531})\%$ (corresponds to the NIE) and $\approx 27\%$ $ = 100 \times (1-e^{0.2353})\%$ of the survival time of lung cancer patients. Furthermore, the confidence interval for the direct effect and total effect includes zero, suggesting a lack of statistical significance. These findings further emphasize the limitations of low-dimensional mediator selection and conventional modeling approaches, reinforcing the importance of incorporating high-dimensional methodologies for accurate mediation analysis in complex biological systems.

\begin{table}[t]
\centering
\small 
\begin{tabular}{cccc}
\hline
\multicolumn{1}{c}{}                & \multicolumn{1}{c}{Estimate} & Bootstrap Mean&Quantile CI (95\%)\\ \hline
\multicolumn{1}{c}{Indirect Effect} & \multicolumn{1}{c}{0.2600
}     & 0.2531
&(0.1279, 0.3752) 
\\ \hline
\multicolumn{1}{c}{Direct Effect} & \multicolumn{1}{c}{-0.0223
}     & -0.0178
&(-0.2496, 0.1688) 
\\ \hline
\multicolumn{1}{c}{Total Effect} & \multicolumn{1}{c}{0.2378}     & 0.2353&(-0.0079, 0.4346) \\ \hline
\end{tabular}
\caption{Summary of the effects in OLS (CpG sites selected from prior studies) with mean and confidence interval (CI) under 300-sample bootstrap.} 
\label{tab:results2}
\end{table}

\section{Conclusion} \label{section:7}
In this paper, we introduced a novel methodology for estimating the mediation functional and, consequently, of the NDE, NIE in the presence of ultra-high-dimensional covariates and mediators. Our model is sufficiently flexible to accommodate interactions between the treatment assignment and both covariates and mediators in the conditional mean of the response, as well as interactions between the treatment assignment and covariates in the conditional expectation of the mediators. The primary contribution is to develop a novel debiasing framework to obtain a $\sqrt{n}$-CAN estimate of the mediation functional and, consequently, of the NDE, NIE in the presence of these challenges, which is potentially applicable to other problems sharing a similar structural equation model. To the best of our knowledge, this is the first study to address the challenge of estimating NDE and NIE when dealing with high-dimensional covariates, mediators, and their interactions with the treatment variable simultaneously. While this paper focuses on a linear model, extending the methodology to generalized linear models is a natural next step, which we leave as an intriguing direction for future research.

\bibliographystyle{apalike}
\bibliography{references}

\newpage
\appendix 


\section{\texorpdfstring{\color{black}Proof of Proposition \ref{prop:bias_structure}}{Proof of Proposition \ref{prop:bias_structure}}}
\label{sec:proof_prop_bias_structure}

\begin{proof}
We first consider $\hat \theta_{10}^{1 \to 2}$. Recall that $\theta_{10}
=
\bar X^\top(\beta_1+\bB_0^\top\gamma_1),
\
\phi_1^\top=(\beta_1^\top\ \ \gamma_1^\top).$
We also recall the definitions of $\hat\vartheta_1$ and
$\hat\vartheta_2$ in Algorithm~\ref{algomain_modified}. Adding and
subtracting $(a^{(1)})^\top\phi_1$ in $\hat\vartheta_1$, we have
\begin{align*}
       \hat\vartheta_1
&=
\hat\vartheta_1-(a^{(1)})^\top\phi_1
+(a^{(1)})^\top\phi_1 \\
&=
\left(
a^{(1)}-\bW_{t,2}^\top\hat\tau_1^{(2)}
\right)^\top
\left(
\hat\phi^{(2)}-\phi_1
\right)
+
(a^{(1)})^\top\phi_1
+
(\hat\tau_1^{(2)})^\top\beps_t^{(2)}.
\end{align*}
Similarly, adding and subtracting $\bar X^\top b^{(2)}$ in
$\hat\vartheta_2$ gives
\begin{align*}
\hat\vartheta_2
&=
\hat\vartheta_2-\bar X^\top b^{(2)}
+\bar X^\top b^{(2)} \\
&=
\left(
\bar X-\bX_{c,1}^\top\hat\tau_2^{(1)}
\right)^\top
\left(
\hat b^{(1)}-b^{(2)}
\right)
+
\bar X^\top b^{(2)}
+
(\hat\tau_2^{(1)})^\top
\bU_c^{(1)}\hat\gamma_1^{(2)}.
\end{align*}
Therefore,
\begin{align}
\label{eq:bias_exp_1}
\hat\theta_{10}^{1\to2}
&=
\hat\vartheta_1+\hat\vartheta_2
-\hat m_{c,1}^\top\hat\gamma_1^{(2)}
\notag\\
&=
\left(
a^{(1)}-\bW_{t,2}^\top\hat\tau_1^{(2)}
\right)^\top
\left(
\hat\phi^{(2)}-\phi_1
\right)
+
\left(
\bar X-\bX_{c,1}^\top\hat\tau_2^{(1)}
\right)^\top
\left(
\hat b^{(1)}-b^{(2)}
\right)
\notag\\
&\quad
+
(a^{(1)})^\top\phi_1
+
\bar X^\top b^{(2)}
+
(\hat\tau_2^{(1)})^\top
\bU_c^{(1)}
\left(
\hat\gamma_1^{(2)}-\gamma_1
\right)
-
\hat m_{c,1}^\top\hat\gamma_1^{(2)}
\notag\\
&\quad
+
(\hat\tau_1^{(2)})^\top\beps_t^{(2)}
+
(\hat\tau_2^{(1)})^\top\bU_c^{(1)}\gamma_1.
\end{align}
\noindent
Next, note that $\hat m_{c,1}
=
\bM_{c,1}^\top\hat\tau_2^{(1)}
=
\bB_0\bX_{c,1}^\top\hat\tau_2^{(1)}
+
(\bU_c^{(1)})^\top\hat\tau_2^{(1)}.$
Moreover, $(a^{(1)})^\top\phi_1
=
\bar X^\top\beta_1+\hat m_{c,1}^\top\gamma_1,
 \
\bar X^\top b^{(2)}
=
\bar X^\top\bB_0^\top\hat\gamma_1^{(2)}.$
The second line in \eqref{eq:bias_exp_1} follows that
\begin{align*}
&
(a^{(1)})^\top\phi_1
+
\bar X^\top b^{(2)}
+
(\hat\tau_2^{(1)})^\top\bU_c^{(1)}
\left(
\hat\gamma_1^{(2)}-\gamma_1
\right)
-
\hat m_{c,1}^\top\hat\gamma_1^{(2)}
\\
&=
\bar X^\top\beta_1
+
\hat m_{c,1}^\top\gamma_1
+
\bar X^\top\bB_0^\top\hat\gamma_1^{(2)}
+
(\hat\tau_2^{(1)})^\top\bU_c^{(1)}
\left(
\hat\gamma_1^{(2)}-\gamma_1
\right)
-
\hat m_{c,1}^\top\hat\gamma_1^{(2)}
\\
&=
\bar X^\top\beta_1
+
\bar X^\top\bB_0^\top\hat\gamma_1^{(2)}
+
\left[
\hat m_{c,1}
-
(\bU_c^{(1)})^\top\hat\tau_2^{(1)}
\right]^\top
\left(
\gamma_1-\hat\gamma_1^{(2)}
\right)
\\
&=
\bar X^\top\beta_1
+
\bar X^\top\bB_0^\top\hat\gamma_1^{(2)}
+
\left(
\bB_0\bX_{c,1}^\top\hat\tau_2^{(1)}
\right)^\top
\left(
\gamma_1-\hat\gamma_1^{(2)}
\right)
\\
&=
\bar X^\top
\left(
\beta_1+\bB_0^\top\gamma_1
\right)
+
\left(
\bar X-\bX_{c,1}^\top\hat\tau_2^{(1)}
\right)^\top
\bB_0^\top
\left(
\hat\gamma_1^{(2)}-\gamma_1
\right)
\\
&=
\theta_{10}
+
\left(
\bar X-\bX_{c,1}^\top\hat\tau_2^{(1)}
\right)^\top
\bB_0^\top
\left(
\hat\gamma_1^{(2)}-\gamma_1
\right).
\end{align*}
Substituting this identity into \eqref{eq:bias_exp_1}, we obtain
\begin{align*}
\hat\theta_{10}^{1\to2}-\theta_{10}
&=
\left(
a^{(1)}-\bW_{t,2}^\top\hat\tau_1^{(2)}
\right)^\top
\left(
\hat\phi^{(2)}-\phi_1
\right)
\\
&\quad
+
\left(
\bar X-\bX_{c,1}^\top\hat\tau_2^{(1)}
\right)^\top
\left(
\hat b^{(1)}-b^{(2)}
\right)
\\
&\quad
+
\left(
\bar X-\bX_{c,1}^\top\hat\tau_2^{(1)}
\right)^\top
\bB_0^\top
\left(
\hat\gamma_1^{(2)}-\gamma_1
\right)
\\
&\quad
+
(\hat\tau_1^{(2)})^\top\beps_t^{(2)}
+
(\hat\tau_2^{(1)})^\top\bU_c^{(1)}\gamma_1
\\
&=: \Delta_n^{(1)}+V_n^{(1)}.
\end{align*}
\noindent
By the same argument with the roles of $\cD_1$ and $\cD_2$ reversed, we have
\begin{align*}
\hat\theta_{10}^{2\to1}-\theta_{10}
&=
\left(
a^{(2)}-\bW_{t,1}^\top\hat\tau_1^{(1)}
\right)^\top
\left(
\hat\phi^{(1)}-\phi_1
\right)
\\
&\quad
+
\left(
\bar X-\bX_{c,2}^\top\hat\tau_2^{(2)}
\right)^\top
\left(
\hat b^{(2)}-b^{(1)}
\right)
\\
&\quad
+
\left(
\bar X-\bX_{c,2}^\top\hat\tau_2^{(2)}
\right)^\top
\bB_0^\top
\left(
\hat\gamma_1^{(1)}-\gamma_1
\right)
\\
&\quad
+
(\hat\tau_1^{(1)})^\top\beps_t^{(1)}
+
(\hat\tau_2^{(2)})^\top\bU_c^{(2)}\gamma_1
\\
&=: \Delta_n^{(2)}+V_n^{(2)}.
\end{align*}
\noindent
Letting $j'=3-j$, for $j\in\{1,2\}$,
\begin{align*}
\Delta_n^{(j)}
&=
\left(
a^{(j)}-\bW_{t,j'}^\top\hat\tau_1^{(j')}
\right)^\top
\left(
\hat\phi^{(j')}-\phi_1
\right)
\\
&\quad
+
\left(
\bar X-\bX_{c,j}^\top\hat\tau_2^{(j)}
\right)^\top
\left(
\hat b^{(j)}-b^{(j')}
\right)
\\
&\quad
+
\left(
\bar X-\bX_{c,j}^\top\hat\tau_2^{(j)}
\right)^\top
\bB_0^\top
\left(
\hat\gamma_1^{(j')}-\gamma_1
\right),
\end{align*}
which gives \eqref{eq:exact_bias_new}. Since $\hat\theta_{10}^{\rm cf}
=
\frac{1}{2}
\left(
\hat\theta_{10}^{1\to2}
+
\hat\theta_{10}^{2\to1}
\right),$
we have
\begin{align*}
\hat\theta_{10}^{\rm cf}-\theta_{10}
&=
\frac{1}{2}
\left[
\left(
\hat\theta_{10}^{1\to2}-\theta_{10}
\right)
+
\left(
\hat\theta_{10}^{2\to1}-\theta_{10}
\right)
\right]
\\
&=
\frac{\Delta_n^{(1)}+\Delta_n^{(2)}}{2}
+
\frac{V_n^{(1)}+V_n^{(2)}}{2}
\\
&=
\Delta_n^{\rm cf}+V_n^{\rm cf}.
\end{align*}
\noindent
Moreover, by the definitions of the fold-specific error, $(\hat\tau_1^{(j)})^\top\beps_t^{(j)}
=
\sum_{\substack{i\in\cD_j\\A_i=1}}
\hat\tau_{1,i}^{(j)}\eps_i,
\
(\hat\tau_2^{(j)})^\top\bU_c^{(j)}\gamma_1
=
\sum_{\substack{i\in\cD_j\\A_i=0}}
\hat\tau_{2,i}^{(j)}U_i^\top\gamma_1.$
we obtain,
\begin{equation*}
V_n^{\rm cf}
=
\frac12
\sum_{j=1}^2
\left[
\sum_{\substack{i\in\cD_j\\A_i=1}}
\hat\tau_{1,i}^{(j)}\eps_i
+
\sum_{\substack{i\in\cD_j\\A_i=0}}
\hat\tau_{2,i}^{(j)}U_i^\top\gamma_1
\right].
\end{equation*}
\noindent
The bound in \eqref{eq:deltanbound_cf} can be proved directly via applying H\"older's inequality to each of the three terms, yielding
\begin{align*}
|\Delta_n^{(j)}|
&\le
\left\|
\bB_0
\left(
\bar X-\bX_{c,j}^\top\hat\tau_2^{(j)}
\right)
\right\|_\infty
\left\|
\hat\gamma_1^{(j')}-\gamma_1
\right\|_1
\\
&\quad
+
\left\|
a^{(j)}-\bW_{t,j'}^\top\hat\tau_1^{(j')}
\right\|_\infty
\left\|
\hat\phi^{(j')}-\phi_1
\right\|_1
\\
&\quad
+
\left\|
\bar X-\bX_{c,j}^\top\hat\tau_2^{(j)}
\right\|_\infty
\left\|
\hat b^{(j)}-b^{(j')}
\right\|_1,
\end{align*}
which proves \eqref{eq:deltanbound_cf}. This completes the proof.
\end{proof}

\section{\texorpdfstring{\color{black}Proof of Results from Section \ref{sec:theory}}{Proof of Results from Section \ref{sec:theory}}}
\label{sec:proofs_theory}
\subsection{{\color{black}Proof of Lemma \ref{lem:feasibility}}
\label{sec:proof_lem_feasibility}}
\begin{proof}
    Let us consider the optimization problem Equation \eqref{eq:balancing_opt_1} on $\cD_1$, and the proof for $\cD_2$ is the same. To establish the feasibility, we will construct a vector $\tau_2^0$ which satisfies the constraints in Equation \eqref{eq:balancing_opt_1}. Towards that end, define: 
    \begin{equation*}
    \textstyle
    \tau^0_{2, i} = \frac{1}{n_{c,1}} \mu_X^\top \Sigma_{X, 0}^{-1}(X_i - \mu_{X, 0}), \qquad i \in \cD_1 \text{ with } A_i = 0 \,.
    \end{equation*}
    Here $\Sigma_{X, 0} = \var(X \mid A = 0)$ and $\mu_{X, 0} = \bbE[X \mid A = 0]$. From Assumption \ref{assm:cov_err} we have $\mu_X^\top \Sigma_{X, 0}^{-1} \mu_X < \infty$, which, along with the fact that $\lambda_{\min}(\Sigma_{X, 0}) \ge c_X > 0$ implies $\mu_X^\top \Sigma_{X, 0}^{-2}\mu_X \le C$ for some constant $C$. Furthermore, by Assumption \ref{assm:cov_err}, $\zeta_i = \mu_X^\top \Sigma_{X, 0}^{-1}(X_i - \mu_{X, 0})$ is a centered sub-Gaussian random variable with sub-Gaussian constant $\mu_X^\top \Sigma_{X, 0}^{-2}\mu_X$. Therefore, from the standard sub-Gaussian inequality, we have with probability going to 1: 
    \begin{equation*}
    \textstyle
    \max_i |\zeta_i| \le C_1 \sqrt{\log{n_{c, 1}}} \implies \|\tau_2^0\|_\infty \le C \frac{\sqrt{\log{n_{c, 1}}}}{n_{c, 1}} \le n_{c, 1}^{-\alpha}
    \end{equation*}
    for any $\alpha \in (0, 1)$ and for all large $n$. For the other constraint:
    \begin{align*}
    \|\bar X - \bX_{c, 1}^\top \tau_2^0\|_\infty  & = \left\|\bar X - \frac{1}{n_{c, 1}}\bX_{c, 1}^\top(\bX_{c, 1} - \mathbf{1}_{n_{c, 1}}\mu_{X, 0}^\top)\Sigma_{X, 0}^{-1} \mu_X\right\|_\infty \\
    & \le \|\bar X - \mu_X\|_\infty + \left\|\mu_X - \frac{1}{n_{c, 1}}\bX_{c, 1}^\top(\bX_{c, 1} - \mathbf{1}_{n_{c, 1}}\mu_{X, 0}^\top)\Sigma_{X, 0}^{-1} \mu_X\right\|_\infty
    \end{align*}
The first term on the right-hand side is again $\le C\sqrt{\log{p}/n}$ with probability going to $1$ via simple sub-Gaussian concentration. Now we establish a bound on the second term on the right-hand-side of the above equation. 
For notational simplicity set $v=\Sigma_{X,0}^{-1}\mu_X$ and $Z_i = X_i - \mu_{X, 0}$. Then
    \begin{equation*}
    \textstyle
\frac{1}{n_{c, 1}}\bX_c^\top\left(\bX_c-\mathbf 1_{n_{c,1}}\mu_{X,0}^\top\right)\Sigma_{X,0}^{-1}\mu_X=\frac{1}{n_{c, 1}}\sum_{i=1}^{n_{c,1}} X_iZ_i^\top v.
    \end{equation*}
Furthermore, we have: 
    \begin{equation*}
    \textstyle
\bbE[X_iZ_i^\top\mid A_i=0]=\bbE[(\mu_{X,0}+Z_i)Z_i^\top\mid A_i=0]=\Sigma_{X,0}
    \end{equation*}
and consequently $\bbE[X_i Z_i^\top v \mid A = 0] = \mu_X$. Therefore, defining $\tilde Z_i = X_iZ_i^\top v-\mu_X$, we have: 
    \begin{equation*}
    \textstyle
\frac{1}{n_{c, 1}}\bX_c^\top\left(\bX_c-\mathbf 1_{n_{c, 1}}\mu_{X,0}^\top\right)\Sigma_{X,0}^{-1}\mu_X-\mu_X=\frac{1}{n_{c, 1}}\sum_{i=1}^{n_{c, 1}} \tilde Z_i,
    \end{equation*}
It is immediate that $\tilde Z_1, \cdots, \tilde Z_{n_{c, 1}}$ are $p$-dimensional independent centered random vectors conditional on the treatment indicators. Next, we argue that $\tilde Z_i$ is a sub-exponential random vector. Towards that end, define the whitened covariate vector $P_i:=\Sigma_{X,0}^{-1/2}(X_i-\mu_{X,0})$ for $1 \le i \le n_{c,1}$ and also define $z=\Sigma_{X,0}^{-1/2}\mu_X$ and $u_j=\Sigma_{X,0}^{1/2}e_j$ for $1 \le j \le p$. Then, we have $Z_i= X_i-\mu_{X,0}=\Sigma_{X,0}^{1/2}P_i$ which further implies $Z_i^\top v=P_i^\top z$ and $Z_{ij}=P_i^\top u_j$ for $1 \le j \le p$. As a consequence,
    \begin{equation*}
    \textstyle
X_{ij}Z_i^\top v=\left(\mu_{X,0,j}+P_i^\top u_j\right)(P_i^\top z).
    \end{equation*}
Moreover, since $\bbE[P_iP_i^\top\mid A_i=0]=I_p$, we also have: 
    \begin{equation*}
    \textstyle
\bbE[(P_i^\top u_j)(P_i^\top z)\mid A_i=0]=u_j^\top z=e_j^\top\mu_X=\mu_{X,j}.
    \end{equation*}
Therefore,
\begin{equation}
\textstyle
\label{eq:expansion_z_tilde}
\tilde Z_{ij}=\mu_{X,0,j}P_i^\top z+ (P_i^\top u_j)(P_i^\top z)-\bbE[(P_i^\top u_j)(P_i^\top z)\mid A_i=0].
\end{equation}
Let $K_X<\infty$ denote the uniform conditional sub-Gaussian norm bound for $X-\mu_{X,a}$ in Assumption \ref{assm:cov_err}. Since $\lambda_{\min}(\Sigma_{X,0})\geq c_X$, the vectors $P_i$ are uniformly conditionally sub-Gaussian with
\begin{equation*}
\textstyle
\|P_i\|_{\psi_2\mid A_i=0} \leq K_X/\sqrt{c_X} =:K_P.
\end{equation*}
Furthermore, Assumption \ref{assm:cov_err} implies $\|u_j\|_2^2 \le e_j^\top\Sigma_{X,0}e_j \le C_X$, and $\|z\|_2^2 = \mu_X^\top\Sigma_{X,0}^{-1}\mu_X \le C_{\mu,\Sigma}$. Therefore,
\begin{equation*}
\textstyle
\|P_i^\top u_j\|_{\psi_2\mid A_i=0} \le K_P\sqrt{C_X}, \qquad \|P_i^\top z\|_{\psi_2\mid A_i=0} \le K_P\sqrt{C_{\mu,\Sigma}}.
\end{equation*}
Since the product of two sub-Gaussian random variables is sub-exponential, we have: 
\begin{equation*}
\textstyle
\|(P_i^\top u_j)(P_i^\top z)\|_{\psi_1\mid A_i=0} \leq C K_P^2\sqrt{C_XC_{\mu,\Sigma}}.
\end{equation*}
Also, since $|\mu_{X,0,j}|\leq\|\mu_{X,0}\|_\infty\leq C_\mu$, we have 
\begin{equation*}
\textstyle
\|\mu_{X,0,j}P_i^\top z\|_{\psi_1\mid A_i=0} \le C C_\mu K_P\sqrt{C_{\mu,\Sigma}}.
\end{equation*}
Using the standard centering inequality $\|Y-\mathbb E[Y]\|_{\psi_1}\leq C\|Y\|_{\psi_1}$, we conclude that
\begin{equation*}
\textstyle
\max_{1\leq j\leq p}
\|\widetilde Z_{ij}\|_{\psi_1\mid A_i=0} \le C\left\{C_\mu K_P\sqrt{C_{\mu,\Sigma}} + K_P^2\sqrt{C_XC_{\mu,\Sigma}}\right\} =: L_X.
\end{equation*}
As $L_X$ does not depend on $j$, we have $\max_{1 \le j \le p} \|\tilde Z_{ij}\|_{\psi_1} \le L_X$, and consequently an application of Bernstein's inequality yields, for $t > 0$,   
    \begin{equation*}
    \textstyle
\bbP\left(\left|\frac{1}{n_{c, 1}}\sum_{i=1}^{n_{c, 1}} \tilde Z_{ij}\right|>t\right)\le 2\exp\left[-c n_{c, 1}\min\left\{\frac{t^2}{L_X^2},\frac{t}{L_X}\right\}\right].
    \end{equation*}
Taking a union bound over $j=1,\ldots,p$ yields
    \begin{equation*}
    \textstyle
\bbP\left(\left\|\frac{1}{n_{c, 1}}\sum_{i=1}^{n_{c, 1}} \tilde Z_i\right\|_\infty >t\right)\le 2\exp\left[\log{p} -c n_{c, 1}\min\left\{\frac{t^2}{L_X^2},\frac{t}{L_X}\right\}\right]
\end{equation*}
Finally, if we choose $t = AL_X\sqrt{\log{p}/n_{c,1}}$ for some sufficiently large constant $A > 0$, we have: 
    \begin{equation*}
    \textstyle
\bbP\left(\left\|\frac{1}{n_{c, 1}}\sum_{i=1}^{n_{c, 1}} \tilde Z_i\right\|_\infty \le AL\sqrt{\frac{\log{p}}{n_{c, 1}}}\right) \longrightarrow 1, \quad \text{as } n \uparrow \infty \,.
\end{equation*}
This immediately implies, with probability going to $1$: 
    \begin{equation*}
    \textstyle
\left\|\mu_X - \frac{1}{n_{c, 1}}\bX_{c, 1}^\top(\bX_{c, 1} - \mathbf{1}_{n_{c, 1}}\mu_{X, 0}^\top)\Sigma_{X, 0}^{-1} \mu_X\right\|_\infty \le AL_X\sqrt{\frac{\log{p}}{n_{c, 1}}} \,.
\end{equation*}
As a consequence, $\tau_2^0$ is a feasible solution and consequently the optimization problem in Equation \eqref{eq:balancing_opt_1} is feasible. Furthermore, from the optimality of $\hat \tau_2^{(1)}$, we have $n\|\hat \tau_2^{(1)}\|_2^2 \le n\|\tau_2^0\|_2^2$. On the other hand, from the definition, 
\begin{align*}
    n\|\tau_2^0\|_2^2 & = \frac{n}{n_{c, 1}} \frac{1}{n_{c, 1}} \sum_{\substack{i \in \cD_1 \\ A_i = 0}} v^\top (X_i - \mu_{X, 0})(X_i - \mu_{X, 0})^\top v \\
    & = \frac{n n_1}{n_{c, 1}^2} \frac{1}{n_1} \sum_{\substack{i \in \cD_1}} v^\top (X_i - \mu_{X, 0})(X_i - \mu_{X, 0})^\top v \mathds{1}_{A_i = 0}
    = \frac{n}{n_{c, 1}} v^\top \hat \Sigma^{(1)}_{X, 0} v 
\end{align*}
Now, 
\begin{align*}
& \bbE\left[ \frac{1}{n_1} \sum_{\substack{i \in \cD_1}} (X_i - \mu_{X, 0})(X_i - \mu_{X, 0})^\top \mathds{1}_{A_i = 0}\right] \\
& = \bbE\left[(X - \mu_{X, 0})(X - \mu_{X, 0})^\top \mathds{1}_{A = 0}\right]  = \Sigma_{X, 0} \bbP(A = 0) 
\end{align*}
As $\Sigma_{X, 0}$ has bounded maximum eigenvalue and $\|v\|_2^2$ is bounded, 
    \begin{equation*}
    \textstyle
\bbE\left[ \frac{1}{n_1} \sum_{\substack{i \in \cD_1}} v^\top (X_i - \mu_{X, 0})(X_i - \mu_{X, 0})^\top v \mathds{1}_{A_i = 0}\right] \le C \,. 
\end{equation*}
Therefore, an application of Markov inequality yields: 
    \begin{equation*}
    \textstyle
\frac{1}{n_1} \sum_{\substack{i \in \cD_1}} v^\top (X_i - \mu_{X, 0})(X_i - \mu_{X, 0})^\top v \mathds{1}_{A_i = 0}
    = \frac{n}{n_{c, 1}} v^\top \hat \Sigma^{(1)}_{X, 0} v  = O_p(1)
\end{equation*}
This, along with the fact that $n \sim n_1 \sim n_{c, 1}$ (as $\bbP(A = 1)$ is bounded away from $0$ and $1$ and each split has half of the samples), yields the conclusion. 

Next, we will establish the feasibility of the optimization problem Equation \eqref{eq:balancing_opt_2}. Towards that end recall the definition of $\eta^*$ and $a^*$ from part (3) of Assumption \ref{assm:cov_err}; $a^* = (\mu_X^\top \ \ (\bB_0 \mu_X)^\top)^\top$ and $\eta^* = \Omega_{W, 1}^{-1} a^*$, where $\Omega_{W, 1} = \bbE[WW^\top \mid A = 1]$. Define a vector $\tau_1^0$ as: 
    \begin{equation*}
    \textstyle
\tau^0_{1, i} = \frac{1}{n_{t, 2}}W_i^\top \eta^*, \qquad i \in \cD_2, \text{ with } A_i = 1 \,. 
\end{equation*}
Therefore, we have $\tau_1^0 = (\bW_{t, 2} \eta^*)/n_{t, 2}$. Hence, we have: 
    \begin{equation*}
    \textstyle
a  - \bW_{t, 2}^\top \tau_1^0 = a - \frac{1}{n_{t, 2}} \bW_{t, 2}^\top \bW_{t, 2} \eta^* = a - \hat \Omega_{n, W, t}\eta^* \,.
\end{equation*}
As a consequence, 
\allowdisplaybreaks
\begin{align*}
& \left\|a  - \bW_{t, 2}^\top \tau_1^0\right\|_\infty \\
& = \left\|a - a^* + a^* - a  - \bW_{t, 2}^\top \tau_1^0\right\|_\infty \\
& = \left\|a - a^* + (\Omega_{W, t} - \hat \Omega_{n, W, t}) \eta^*\right\|_\infty \\
& \le \|a - a^*\|_\infty + \|\Omega_{W, t} - \hat \Omega_{n, W, t}\|_{\infty, \infty}\|\eta^*\|_1 \\
& \le \|\bar X - \mu_X\|_\infty + \|\hat m_{c, 1} - \bB_0 \mu_X\|_\infty + C_\eta \|\Omega_{W, t} - \hat \Omega_{n, W, t}\|_{\infty, \infty} \\
& \le (1 + \|\bB_0\|_{1, \infty})\|\bar X - \mu_X\|_\infty + \|\hat m_{c, 1} - \bB_0 \bar X\|_\infty +  C_\eta \|\Omega_{W, t} - \hat \Omega_{n, W, t}\|_{\infty, \infty} \\
& \le C\|\bar X - \mu_X\|_\infty + \|\bB_0(\bX_{c, 1}^\top \hat \tau^{(1)}_2 - \bar X)\|_\infty + \|(\bU_c^{(1)})^\top \hat \tau_2^{(1)}\|_\infty \\
& \hspace{20em} +  C_\eta \|\Omega_{W, t} - \hat \Omega_{n, W, t}\|_{\infty, \infty} \\
& \le C \left(\|\bar X - \mu_X\|_\infty + \|(\bX_{c, 1}^\top \hat \tau^{(1)}_2 - \bar X)\|_\infty +  \|\Omega_{W, t} - \hat \Omega_{n, W, t}\|_{\infty, \infty}\right) + \|(\bU_c^{(1)})^\top \hat \tau_2^{(1)}\|_\infty
\end{align*}
In the previous inequalities, we have used the fact that $\|\eta^*\|_1 \le C_\eta$ (Assumption \ref{assm:cov_err}) and $\|\bB_0\|_{1, \infty} \le C_{\bB}$ (Assumption \ref{assm:params}). 
The first and second summands on the right-hand side are $\le C \sqrt{\log{p}/n}$ for a large enough constant $C$ with probability going to $1$ by standard sub-Gaussian concentration for $\bar X$ and the definition of $\hat \tau_2^{(1)}$ (Equation \eqref{eq:balancing_opt_1}). For the fourth term, we use the fact that conditionally on $\bX_c$, $U_{c, i}$'s independent sub-Gaussian vector with sub-Gaussian constants uniformly bounded by $\sigma_U$. Therefore, conditionally on $\bX$, 
    \begin{equation*}
    \textstyle
\|(\bU_c^{(1)})^\top \hat \tau_2^{(1)}\|_\infty = O_p\left(\|\hat \tau_2^{(1)}\|_2 \ \sqrt{\log{q}}\right) = O_p\left(\sqrt{\frac{\log{q}}{n}}\right)
\end{equation*}
where the last conclusion follows from the fact that $n\|\hat \tau_2^{(1)}\|_2^2 = O_p(1)$ as established previously. Last but not least $ \|\Omega_{W, t} - \hat \Omega_{n, W, t}\|_{\infty, \infty} = O_p(\sqrt{\log{(p+q)}/n})$ which follows from standard sub-exponential Bernstein inequality, along with a union bound on the co-ordinates of $\hat \Omega_{n, W, t}$. Therefore, we conclude: 
    \begin{equation*}
    \textstyle
\left\|a^{(1)} - \bW_{t, 2}^\top \tau_1^0\right\|_\infty \le K_1 \sqrt{\frac{\log{(p+q)}}{n_{t, 2}}}
\end{equation*}
with probability going to $1$ for large enough constant $K_1$. We next argue that $\|\tau_1^0\|_\infty \le n_{t, 2}^{-\alpha}$. Towards that end: 
\begin{align*}
    \max_{1 \le i \le n_{t, 2}} |\tau_{1, i}^0| & = \frac{1}{n_{t,2}} \ \max_{1 \le i \le n_{t, 2}} \ |W_i^\top \eta^*| \\
    & 
    \le \frac{|\mu_{W, 1}^\top \eta^*|}{n_t} + \sqrt{\lambda_{\max}(\Sigma_{W,1})} \  \frac{1}{n_{t,2}} \ \max_{1 \le i \le n_{t, 2}} \ |\Sigma_{W, 1}^{-1/2}(W_i - \mu_{W, 1})^\top \eta^*| \\
    & \le O_p\left(\frac{1}{n_{t,2}}\right) + C O_p\left(\frac{\sqrt{\log{n_{t, 2}}}}{n_{t,2}}\right) \le n_{t, 2}^{-\alpha}
\end{align*}
for any $\alpha \in (0, 1)$. Here we have used the fact that $|\mu_{W, 1}^\top \eta^*| \le C$ which follows from the fact that $\|\mu_{W, 1}\|_\infty \le C_\mu$,  $\|\eta^*\|_1 \le C_\eta$, $\lambda_{\max}(\Sigma_{W,1}) \le C$ and $\Sigma_{W, 1}^{-1/2}(W_i - \mu_{W, 1})$ is sub-Gaussian (Assumption \ref{assm:cov_err} and Remark \ref{rem:sub-G_W}). Finally, we will show that $n\|\hat\tau_1^{(2)}\|_2^2 = O_p(1)$ by showing that $n\|\tau_1^0\|_2^2 = O_p(1)$ and then using the fact that $n\|\hat \tau_1^{(2)}\|_2^2 \le n\|\tau_1^0\|_2^2$ as $\hat \tau_1^{(2)}$ is the optimal solution of Equation \eqref{eq:balancing_opt_2}, whereas $\tau_1^0$ is only a feasible solution. Now for $\tau_1^0$, we have: 
    \begin{equation*}
    \textstyle
n\|\tau_1^0\|_2^2 = \frac{n n_2}{n_{t, 2}^2} \ \frac{1}{n_2}\sum_{i \in \cD_2}  (W_i^\top \eta^*)^2 \mathds{1}_{A_i = 1} \le (\eta^*)^\top \hat \Omega \hat \eta^* \le \|\hat \Omega\|_\infty \|\eta^*\|_1^2 \,.
\end{equation*}
Now $\|\hat \Omega\|_\infty = O_p(1)$ as $\|\bbE[WW^\top \mid A = 1]\|_\infty \le \|\Sigma_{W, 1}\|_\infty + \|\mu_{W, 1}\|_\infty^2$. (Assumption \ref{assm:cov_err}). Furthemore, $\|\eta^*\|_1 \le C_\eta$ by Assumption \ref{assm:cov_err}. Hence, $n\|\tau_1^0\|_2^2 = O_p(1)$. This completes the proof. 

\end{proof}
\subsection{{\color{black}Proof of Theorem \ref{thm:main_revised}}}
\label{sec:proof_thm_main}
\begin{proof}
From Proposition \ref{prop:bias_structure}, we have: 
\begin{equation}
\label{eq:decomp_first_10}
\frac{\sqrt{n}(\hat\theta_{10}^{\rm cf}-\theta_{10})}{\sigma_n^{\rm cf}} = \frac{\sqrt{n}(\Delta_n^{\rm cf} + V_n^{\rm cf})}{\sigma_n^{\rm cf}} \,.
\end{equation}
Therefore, we divide the rest of the proof into three parts: 
\begin{enumerate}
    \item[a)] We show that $\sqrt{n}\Delta_n^{\rm cf} = o_p(1)$, 
    \item[b)] We show that $\sigma_n^{\cf}$ is bounded away from 0 with probability going to $1$. 
    \item[c)] We show that $\sqrt{n}V_n^{\rm cf}/\sigma_n^{\rm cf}$ converges to $\cN(0, 1)$. 
\end{enumerate}
\noindent
{\bf Asymptotic negligibility of $\sqrt{n}\Delta_n^{\rm cf}$.} We show that $\sqrt{n}|\Delta_{n, 1}| = o_p(1)$. The proof for $\Delta_{n, 2}$ is same and hence skipped. For our proof we will use upper on $|\Delta_{n, 1}|$ as established in Proposition \ref{prop:bias_structure}. 
By standard analysis to establish the convergence rate of the Lasso estimator with respect to the $\ell_1$-norm under the sparsity constraint (e.g., \cite[Corollary 2]{negahban2012unified}), we have the following (we omit the notation $j$ from the next paragraph for notational simplicity, as the argument holds for each $j \in \{1, 2\}$): 
\begin{equation}
\label{eq:lassobound}
\textstyle
\|\hat \phi_1  - \phi_1 \|_1 = \mathcal{O}_p\left(k \sqrt{\frac{\log(p+q)}{n_t}}\right), \ \ \  \|\hat b - b\|_1  = \mathcal{O}_p \left( sk_\gamma\sqrt{\frac{\log (p)}{n_c}} + k \sqrt{\frac{\log{(p+q)}}{n_t}}\right) \,. 
\end{equation}
Note that in order to apply Corollary 2 of \cite{negahban2012unified}, we need the design matrices to satisfy the RE condition, and the columns should be approximately standardized. Regarding the RE condition, we need $(\bW_t^\top \bW_t)/n_t$ to satisfy the RE condition with high probability for establishing the bound on $\|\hat \phi_1 - \phi_1\|_1$ and $(\bX_c^\top \bX_c)/n_c$ to satisfy the RE condition with high probability to establish the bound on $\|\hat b - b\|_1$. For the former, note that, by Assumption \ref{assm:cov_err}, the population second moment matrix $\bbE[WW^\top \mid A = 1] = \Sigma_{W, t} + \mu_{W, t}\mu_{W, t}^\top \succeq \Sigma_{W, t}$ has minimum eigenvalue bounded away from $0$ and $\bW_t$ has i.i.d. sub-Gaussian rows (Assumption \ref{assm:cov_err} and Remark \ref{rem:sub-G_W}). Therefore, by \cite[Theorem 6]{rudelson2012reconstruction}, we can conclude that $(\bW_t^\top \bW_t)/n_t$ also satisfies RE condition with probability going to $1$. Similar logic also applies for $(\bX_c^\top \bX_c)/n_c$ and its population second moment matrix $\bbE[XX^\top \mid A = 0]$ as minimum eigenvalues bounded away from $0$ and $\bX_c$ has i.i.d. sub-Gaussian rows (Assumption \ref{assm:cov_err}). Now, as for column normalization, we need $\|\bW_{t, *j}\|_2/\sqrt{n_t} \le C$ and $\|\bX_{c, *j}\|_2/\sqrt{n_c} \le C$ for some constant $C > 0$ with probability going to $1$. This follows from standard sub-Gaussian concentration; For $\bW_t$, fix $1 \le j \le p+q$ and set 
\vspace{-1em}
\begin{equation*}
\textstyle
\mu_{W, t, j}^{(2)} = (\Sigma_{W, t} + \mu_{W, t}\mu_{W, t}^\top)_{jj} = \bbE[W_{t, ij}^2] \,.
\vspace{-1em}
\end{equation*}
Therefore, we have:
\allowdisplaybreaks
\begin{align}
\label{eq:W_square_conc}
\max_{1 \le j \le p+q} \frac{1}{n_t} \sum_{i \in \cD_2: A_i = 1}  W_{t, ij}^2 & \le \max_{1 \le j \le p+q}  \left|\frac{1}{n_t} \sum_{i \in \cD_2: A_i = 1}  (W_{t, ij}^2 - \mu^{(2)}_{W, t, j})\right| + \max_{1 \le j \le p+q} \mu^{(2)}_{W, t, j} \notag \\
& \le \max_{1 \le j \le p+q} \mu^{(2)}_{W, t, j} + C\sqrt{\frac{\log{(p+q)}}{n}},
\end{align}
with probability going to $1$. The last bound follows from a standard sub-Gaussian concentration argument. We next verify that \(\max_{1\leq j\leq p+q}\mu_{W,t,j}^{(2)}\) is uniformly bounded. Recall that $\mu_{W,t,j}^{(2)} = (\Sigma_{W,t})_{jj}+\mu_{W,t,j}^2$. Therefore, by Remark \ref{rem:sub-G_W} and Assumptions \ref{assm:cov_err} and \ref{assm:params}, $\lambda_{\max}(\Sigma_{W,t}) \le C_W := (1+C_B)^2(C_X\vee C_U)$, where \(C_X\), \(C_U\), and \(C_B\) are the constants as defined in
Assumptions~2 and 3. Moreover, since $\mu_{W,t} =  (\mu_{X,1}^\top, (B_1\mu_{X,1})^\top)^\top$, we have
\begin{equation*}
\textstyle
\|\mu_{W,t}\|_\infty \le \max\left\{\|\mu_{X,1}\|_\infty, \  \|B_1\|_{1,\infty}\|\mu_{X,1}\|_\infty\right\} \le (1\vee C_B)C_\mu =:C_{\mu,W}.
\end{equation*}
As a consequence, $\max_{1\leq j\leq p+q}\mu_{W,t,j}^{(2)} \le C_W+C_{\mu,W}^2 := C_{W,2}<\infty$. Therefore, Equation \eqref{eq:W_square_conc} implies that, with probability going to one,
\begin{equation*}
\textstyle
\max_{1\leq j\leq p+q}\frac{1}{n_t}\sum_{\substack{i\in\mathcal D_2\\A_i=1}}W_{t,ij}^2 \le 2C_{W,2} \implies \max_{1\leq j\leq p+q}\frac{\|\mathbf W_{t,*j}\|_2}{\sqrt{n_t}} \le \sqrt{2C_{W,2}} \,,
\end{equation*}
with probability going to one. Similarly, for the control-arm covariate design, $\bbE[X_j^2\mid A=0] = (\Sigma_{X,0})_{jj}+\mu_{X,0,j}^2 \le C_X+C_\mu^2 := C_{X,2}$. The same concentration argument therefore gives $\max_{1\leq j\leq p}\|\mathbf X_{c,*j}\|_2/\sqrt{n_c} \le \sqrt{2C_{X,2}}$ with probability going to one.
Therefore, conditions required for Corollary 2.1 of \cite{negahban2012unified} are satisfied by the design matrices $\bW_t$ and $\bX_c$, and consequently Equation \eqref{eq:lassobound} holds. 
The rate for $\|\hat b - b\|_1$ needs some additional calculation as $b = \bB_0^\top \hat \gamma_1$ may not be exactly sparse and is deferred to Lemma \ref{lem:rate_b_hat}.  
Now, going back to $\Delta_{n, 1}$, we have, from \eqref{eq:deltanbound_cf}: 
\begin{align*}
\sqrt{n}|\Delta_{n, 1}| &\le
\sqrt{n}\|\bB_0(\bar X - \bX_{c, j}^\top \hat \tau_2^{(1)})\|_\infty
\|\hat\gamma^{(2)}_1-\gamma_1\|_1 \\
&\qquad
+
\sqrt{n}\|a^{(1)}-\mathbf W_{t, 2}^\top \hat\tau^{(2)}_1\|_\infty
\|\hat\phi_1^{(2)}-\phi_1\|_1 \\
& \qquad +
\sqrt{n}\|\bar X-\mathbf X_{c, 1}^\top \hat\tau^{(1)}_2\|_\infty
\|\hat b^{(1)}-b^{(2)}\|_1 := T_1 + T_2 + T_3 \,.
\end{align*}
Now, by definition of $\hat \tau_2^{(1)}$ (Step 1 of Algorithm \ref{algomain_modified}), and the fact that $\|\bB_0\|_{1, \infty} \le C_\bB$ (Assumption \ref{assm:params}), we have:  
\begin{align*}
T_1 & \le \sqrt{n}\|\bB_0\|_{1, \infty} \|\bar X - \bX_{c, j}^\top \hat \tau_2^{(j)}\|_\infty\|\hat\gamma^{(2)}_1-\gamma_1\|_1 \\
& \le C_\bB K_2 \sqrt{n}\sqrt{\frac{\log{p}}{n_{c, 1}}} \|\hat \gamma^{(2)}_1 - \gamma_1\|_1 \\
& = O_p\left(\sqrt{n} \times \sqrt{\frac{\log{p}}{n_{c, 1}}} \times k \sqrt{\frac{\log(p+q)}{n_t}}\right) = o_p(1)
\end{align*}
The last conclusion follows from Assumption \ref{assm:params}. Now, from Equation \eqref{eq:balancing_opt_2}, we have $\|a^{(1)} - \bW_{t, 2}^\top \hat \tau_1^{(2)}\|_\infty \le K_1 \sqrt{\log{(p+q)}/n_{t,2}}$. Therefore, this, along with Equation \eqref{eq:lassobound} implies: 
\begin{align*}
T_2 & = \sqrt{n}\|a^{(1)}-\mathbf W_{t, 2}^\top \hat\tau^{(2)}_1\|_\infty
\|\hat\phi_1^{(2)}-\phi_1\|_1 \\
& = O_p\left(\sqrt{n}  \times \sqrt{\frac{\log{(p+q)}}{n}} \times k \sqrt{\frac{\log{(p+q)}}{n}} \right) = O_p\left(\frac{k\log{(p+q)}}{\sqrt{n}}\right) = o_p(1),
\end{align*}
where the last conclusion follows from Assumption \ref{assm:params}. Similarly for $T_3$, we have $\|\bar X-\mathbf X_{c, 1}^\top \hat\tau^{(1)}_2\|_\infty \le K_2 \sqrt{\log{p}/n_{c,1}}$ from Equation \eqref{eq:balancing_opt_1}. Therefore, this and Equation \eqref{eq:lassobound} implies: 
\begin{align*}
    T_3  & = \sqrt{n}\|\bar X-\mathbf X_c^\top\tau_2\|_\infty \|\hat b-b\|_1 \\
    & = O_p\left(\sqrt{n}  \times \sqrt{\frac{\log{(p)}}{n}} \times \left(sk_\gamma \sqrt{\frac{\log{p}}{n}} + k \sqrt{\frac{\log{(p+q)}}{n}}\right) \right) \\
    & = O_p\left(\frac{sk_\gamma\log{p}}{\sqrt{n}} + \frac{k\log{(p+q)}}{\sqrt{n}}\right) = o_p(1),
\end{align*}
where, as before, the last conclusion follows from Assumption \ref{assm:params}. This completes the proof that $\sqrt{n}|\Delta_{n, 1}| = o_p(1)$. 

\noindent
{\bf Lower bound on $\sigma_n^{\rm cf}$: }Our next goal is to show that $\sigma_n^{\rm cf}$ remains bounded away from zero with probability tending to one. Recall that
\begin{equation}
\begin{aligned}
\label{eq:sigma_n_bound}
(\sigma_n^{\rm cf})^2 = \frac n4 \sum_{j = 1}^2 (\sigma_1^2 \lVert \hat \tau_1^{(j)} \rVert_2^2+ \gamma_1^\top\Sigma_{U,0}\gamma_1 \lVert \hat \tau_2^{(j)}\rVert_2^2)
 \ge \frac{n\sigma_1^2 \lVert \hat \tau_1^{(1)} \rVert_2^2}{4}  \,.
\end{aligned}
\end{equation}
Therefore, we next show that $n\| \hat \tau_1^{(1)}\|_2^2 \ge c$ for some constant $c > 0$ with probability going to $1$. This, along with the fact that $\sigma_1^2 > 0$ (Assumption \ref{assm:cov_err}) will yield the conclusion. In fact, we will establish both $n\|\hat \tau_1^{(2)}\|_2^2$ and $n\|\hat \tau_2^{(1)}\|_2^2$ are bounded away from $0$ with probability going to $1$, which will be used in the later part of the proof. The conclusion for the balancing weights using another fold of the data (i.e., $(\hat \tau_1^{(1)}, \hat \tau_2^{(2)})$) is similar and hence skipped. We start with $n\|\hat \tau^{(2)}_1\|_2^2$. From the definition of $a^{(1)}$, we have: 
\begin{align*}
    \|a^{(1)}\|_\infty  \ge \|\bar X\|_\infty \ge \|\mu_X\|_\infty - \|\bar X - \mu_X\|_\infty & \ge \kappa - \|\bar X - \mu_X\|_\infty \ge \kappa/2
\end{align*}
where the last inequality follows with probability approaching to $1$ as $\|\bar X - \mu_X\|_\infty = O_p(\sqrt{\log{p}/n})$ from standard sub-Gaussian concentration and $\kappa > 0$. Here we have also used the fact that $\|\mu_X\|_\infty \ge \kappa$ (Assumption \ref{assm:cov_err}). Now on this event, there exists $1 \le j \le p+q$ such that $|a_j| \ge \kappa/2$ and $1 \le j \le p$ such that $|\bar X_j| \ge \kappa/2$. 
For any feasible solution $\tau_1$ of Equation \eqref{eq:balancing_opt_2} and $\tau_2$ of Equation \eqref{eq:balancing_opt_1} satisfy for all large $n$: 
\begin{align*}
|(\mathbf W_{t,2}^\top \tau_1)_j|
&\ge
|a^{(1)}_j|-|a^{(1)}_j-(\mathbf W_{t,2}^\top\tau_1)_j|
\ge
\kappa/2-K_1\sqrt{\frac{\log(p+q)}{n_t}}
\ge
\kappa/4,\\
|(\mathbf X_{c,1}^\top \tau_2)_j|
&\ge
|\bar X_j|-|\bar X_j-(\mathbf X_{c,1}^\top\tau_2)_j|
\ge
\kappa/2-K_2\sqrt{\frac{\log p}{n_c}}
\ge
\kappa/4,
\end{align*}
with probability tending to one, since $\log(p+q)/(n_t\wedge n_c)\to 0$. Applying Cauchy--Schwarz yields
\begin{align*}
\kappa_1^2
\le
(\mathbf W_{t,2}^\top\tau_1)_j^2
\le
\|\tau_1\|_2^2
\sum_{\substack{i \in \cD_2 \\ A_i=1}} W_{t,ij}^2,
\qquad
\kappa_2^2
\le
(\mathbf X_c^\top\tau_2)_j^2
\le
\|\tau_2\|_2^2
\sum_{\substack{i \in \cD_1 \\ A_i=0}} X_{c,ij}^2.
\end{align*}
Therefore,
\begin{align*}
n\|\tau_1\|_2^2
\ge
\frac{\kappa^2/16}{
\frac{n_{t,2}}{n}\cdot \frac{1}{n_t}\sum_{\substack{i \in \cD_2 \\ A_i=1}}W_{t,ij}^2
},
\qquad
n\|\tau_2\|_2^2
\ge
\frac{\kappa^2/16}{
\frac{n_{c,1}}{n}\cdot \frac{1}{n_c}\sum_{\substack{i \in \cD_1 \\ A_i=0}}X_{c,ij}^2
}.
\end{align*}
For the denominator, $n_{t,2}/n \le 1$ and $(\sum_{i: A_i = 1} W^2_{t, ij})/n_{t,2} \le 2 \mu^{(2)}_{W, t, j}$, which follows from the same calculation as in Equation \eqref{eq:W_square_conc}. Therefore, we conclude that with probability going to $1$,
\begin{equation}
    \label{eq:lower_bound_tau}
    n\|\hat \tau^{(2)}_1\|_2^2 \ \wedge \ n\|\hat \tau^{(1)}_2\|_2^2 \ge \frac{\kappa^2}{C} \ \ \text{with probability going to }1. 
\end{equation}
This completes the proof. 

\noindent
{\bf Asymptotic normality: }Finally we will establish the asymptotic normality of $\sqrt{n}V_n^{\rm cf}/\sigma_n^{\rm cf}$. Recall that, from Proposition \ref{prop:bias_structure}, we have: 
$$
\frac{\sqrt{n}V_n^{\rm cf}}{\sigma_n^{\rm cf}} = \frac{\sqrt{n}}{2\sigma_n^{\rm cf}} \left[\sum_{\substack{i \in \cD_2 \\ A_i = 1}} \hat \tau^{(2)}_{1, i} \eps_i + \sum_{\substack{i \in \cD_1 \\ A_i = 0}} \hat \tau^{(1)}_{2, i} U_i^\top \gamma_1 + \sum_{\substack{i \in \cD_1 \\ A_i = 1}} \hat \tau^{(1)}_{1, i} \eps_i + \sum_{\substack{i \in \cD_2 \\ A_i = 0}} \hat \tau^{(2)}_{2, i} U_i^\top \gamma_1\right] := T
$$
where we introduce the notation $T$ for notational simplicity. For the rest of the analysis define $\tau_1^{(j),0}$ as \emph{an oracle version of} $\hat \tau^{(j)}_1$ where $\tau_1^{(j),0}$ is obtained by solving optimization problem Equation \eqref{eq:balancing_opt_2} with replacing $a^{(j')}$ by $a^{(j'),0} \equiv a^0$ where $a^0 = (\bar X^\top \ \ (\bB_0 \bar X)^\top)^\top$. In other words, we obtain $a^0$ by replacing the $\hat m_{c, j}$ part by its oracle counterpart $\bB_0 \bar X$. Note that now $\{\hat \tau_1^{(j), 0}\}$ for $j \in \{1, 2
\}$ only depends on $\cH_n := \sigma(A_{1:n}, X_{1:n}, \{M_i: A_i = 1\})$. Furthermore, by definition, $\{\hat \tau_2^{(j)}\}$ also is measurable function of  $\sigma(A_{1:n}, X_{1:n})$, and consequently a measurable function of $\cH_n$.  Our following Lemma shows that $\hat \tau_1^{(j), 0}$ is asymptotically close to $\hat \tau_1^{(j)}$: 

\begin{lemma}
    \label{lem:tau_concentration_revised}
    Under Assumption \ref{assm:cov_err} and \ref{assm:params}, we have: 
    \begin{equation*}
    \textstyle
    \sqrt{n}\|\hat \tau_1^{(j)} - \hat \tau_1^{(j), 0}\|_2 = o_p(1), \quad j \in \{1, 2\}. 
    \end{equation*}
\end{lemma}
Now, using this orcale $\hat \tau_1^{(j), 0}$, we define: 
\allowdisplaybreaks
\begin{align*}
T^0 & = \frac{\sqrt{n}}{2\sigma_n^{\rm cf, 0}} \left[\sum_{\substack{i \in \cD_2 \\ A_i = 1}} \hat \tau^{(2),0}_{1, i} \eps_i + \sum_{\substack{i \in \cD_1 \\ A_i = 0}} \hat \tau^{(1)}_{2, i} U_i^\top \gamma_1 + \sum_{\substack{i \in \cD_1 \\ A_i = 1}} \hat \tau^{(1),0}_{1, i} \eps_i + \sum_{\substack{i \in \cD_2 \\ A_i = 0}} \hat \tau^{(2)}_{2, i} U_i^\top \gamma_1\right] \\
& := \sum_{i = 1}^n \zeta_{n, i} 
\end{align*}
where 
\begin{align}
\label{eq:oracle_asym_var}
\zeta_{n, i} &= \frac{\sqrt{n}}{2\sigma_n^{\rm cf, 0}} \left\{\eps_i\left(\hat \tau^{(1),0}_{1, i}\mathds{1}_{i \in \cD_1, A_i = 1} + \hat \tau^{(2),0}_{1, i}\mathds{1}_{i \in \cD_2, A_i = 1}\right) \right. \notag \\
& \qquad \qquad \left. + U_i^\top \gamma_1 \left(\hat \tau^{(1)}_{2, i}\mathds{1}_{i \in \cD_1, A_i = 0} + \hat \tau^{(2)}_{2, i}\mathds{1}_{i \in \cD_2, A_i = 0}\right)\right\} \notag \\
(\sigma_n^{\rm cf, 0})^2 & = \frac n4 \sum_{j = 1}^2 (\sigma_1^2 \lVert \hat \tau_1^{(j), 0} \rVert_2^2+ \gamma_1^\top\Sigma_{U,0}\gamma_1 \lVert \hat \tau_2^{(j)}\rVert_2^2)
\end{align}
We next show that $T^0$ is asymptotically normal and then we will show that $T - T^0 = o_p(1)$. This would complete the proof. To establish that $T^0$ is asymptotically normal, first observe that conditional on $\cH_n$ the weights $\{(\hat \tau_1^{(j),0}, \hat \tau_2^{(j)}): j \in \{1, 2\}\}$ are fixed and $(U_i, \eps_i)$ are independent across $i$. From $\bbE[\eps_i \mid X_i, M_i, A_i = 1] = 0$ and $\bbE[U_i \mid X_i, A_i = 0] = 0$ (Assumption \ref{assm:cov_err}), we also have $\bbE[\zeta_{n, i} \mid \cH_n] = 0$ and 
\begin{align*}
\var(T^0 \mid \cH_n) & = \sum_i \var(\zeta_{n, i} \mid \cH_n) \\
& = \frac{n}{4(\sigma_n^{\rm cf, 0})^2} \sum_{j = 1}^2 (\sigma_1^2 \lVert \hat \tau_1^{(j), 0} \rVert_2^2+ \gamma_1^\top\Sigma_{U,0}\gamma_1 \lVert \hat \tau_2^{(j)}\rVert_2^2) = 1\,.
\end{align*}
Here we have used the fact that $\var(\eps_i \mid X_i, M_i, A_i = 1) = \sigma_1^2$ and $\var(U_i \mid X_i, A_i = 0) = \Sigma_{U, 0}$ (Assumption \ref{assm:cov_err}). 
Now, we will verify Lyapunov's condition for the third moment, i.e., we show that $\sum_i \bbE[|\zeta_{n,i}|^3 \mid \cH_n] \to 0$ in probability. Towards that end, we will use the fact that $\eps_i$ and $U_i^\top \gamma_1/\|\gamma_1\|_2$ are sub-Gaussian (uniformly over $(X, M, A)$) and consequently they have finite third moment (Assumption \ref{assm:cov_err}). For notational simplicity, we will use $C$ to denote these bounds, i.e. $\bbE[|\eps_i|^3 \mid \cH_n] \le C$ and $\bbE[|U_i^\top \gamma_1|^3 \mid \cH_n] \le C \|\gamma_1\|_2^3$. Using this we have: 
\begin{align*}
    & \sum_i \bbE[|\zeta_{n, i}|^3 \mid \cH_n] \\ 
    & \le C\frac{n^{3/2}}{8(\sigma_n^{\rm cf, 0})^3} \sum_{i = 1}^n \left[|\hat \tau_{1, i}^{(1), 0}|^3 \mathds{1}_{i \in D_1, A = 1} + |\hat \tau_{1, i}^{(2), 0}|^3 \mathds{1}_{i \in D_2, Ai = 1} \right. \\
    & \qquad \qquad \qquad \qquad \qquad \qquad \left. + |\hat \tau_{2, i}^{(1)}|^3 \mathds{1}_{i \in \cD_1, A_i = 0} + |\hat \tau_{2, i}^{(2)}|^3 \mathds{1}_{i \in \cD_2, A_i = 0}\right] \\
    & \le C\frac{n^{3/2}}{8(\sigma_n^{\rm cf, 0})^3} \ \sum_{j = 1}^2 \left(\|\hat \tau_{1, i}^{(j), 0}\|_2^2 \|\hat \tau_{1, i}^{(j), 0}\|_\infty + \|\hat \tau_{2}^{(j)}\|_2^2 \|\hat \tau_{2, i}^{(j)}\|_\infty\right) \,.
\end{align*}
In Lemma \ref{lem:feasibility} we have established $n\|\hat \tau_{1, i}^{(j)}\|_2^2 = O_p(1)$ and $n\|\hat \tau_{2, i}^{(j)}\|_2^2 = O_p(1)$. This, along with Lemma \ref{lem:tau_concentration_revised}, implies $n\|\hat \tau_1^{(j), 0}\|_2^2 = O_p(1)$. 
On the other hand, by definition of the optimization problem (Equation \eqref{eq:balancing_opt_1} and \eqref{eq:balancing_opt_2}), we have $\|\hat \tau_1^{(j), 0}\|_\infty \vee \|\hat \tau_{2}^{(j)}\| \le Cn^{-\alpha}$ where we have used the fact that $n_{c, j} \sim n$ and $n_{t, j} \sim n$ with probability going to $1$ as $\bbP(A = 1)$ is bounded away from $0$ and $1$, as well as each split has $n/2$ observations.
(Recall that $\hat \tau_1^{(j), 0}$ is obtained by solving Equation \eqref{eq:balancing_opt_2} with $a^{(j')}$ replaced by $a^0$, so by feasibility it satisfies $\|\hat \tau_1^{(j), 0}\|_\infty \le n_{t, 2}^{-\alpha} \le Cn^{-\alpha}$).  Therefore, we can conclude: 
\begin{equation*}
\textstyle
 n^{3/2} \sum_{j = 1}^2 \left(\|\hat \tau_{1, i}^{(j), 0}\|_2^2 \|\hat \tau_{1, i}^{(j), 0}\|_\infty + \|\hat \tau_{2}^{(j)}\|_2^2 \|\hat \tau_{2, i}^{(j)}\|_\infty\right) = O_p(n^{-1-\alpha + 3/2}) = o_p(1)
\end{equation*}
where the last conclusion follows from the fact that $\alpha > 1/2$. 
On the other hand, we have established in the previous part of the proof that $\sigma_n^{\rm cf}$ is bounded away from $0$ with probability going to $1$. We now argue that  $\sigma_n^{\rm cf, 0}$ is also bounded away from 0 with high probability. 
\begin{align}
\label{eq:diff_oracle_actual_sigma}
    \left|(\sigma_n^{\rm cf})^2 - (\sigma_n^{\rm cf, 0})^2\right| &  = \frac{n\sigma_1^2}{4} \sum_{j = 1}^2 |\lVert \hat \tau_1^{(j), 0} \rVert_2^2 - \lVert \hat \tau_1^{(j)} \rVert_2^2| \notag\\
    & \le \frac{n\sigma_1^2}{4} \sum_{j = 1}^2 \|\hat \tau_1^{(j), 0} + \hat \tau_1^{(j)}\|_2 \|\hat \tau_1^{(j), 0} - \hat \tau_1^{(j)}\|_2 \notag\\
    & \le \frac{\sigma_1^2}{4}\sum_{j = 1}^2 \left(\sqrt{n}(\|\hat \tau_1^{(j), 0}\|_2 + \|\hat \tau_1^{(j)}\|_2)\right)\left(\sqrt{n}\|\hat \tau_1^{(j), 0} - \hat \tau_1^{(j)}\|_2\right) \notag \\
    & = o_p(1) \,.
\end{align}
Here, the last conclusion follows from the fact that $\sqrt{n}\|\hat \tau_1^{(j), 0} - \hat \tau_1^{(j)}\|_2 = o_p(1)$ (Lemma \ref{lem:tau_concentration_revised}) and $\sqrt{n}(\|\hat \tau_1^{(j), 0}\|_2 + \|\hat \tau_1^{(j)}\|_2) = O_p(1)$ (Lemma \ref{lem:feasibility} and Lemma \ref{lem:tau_concentration_revised}). As a consequence, $\sigma_n^{\rm cf,0}$ will be bounded away from $0$ with probability going to $1$. Hence, $\sum_i \bbE[|\zeta_{n, i}|^3 \mid \cH_n] = o_p(1)$. Therefore, by an application of conditional central limit theorem \cite{bulinski2017conditional}, we conclude $T^0 \xrightarrow{d} \cN(0, 1)$.

Next we argue that $T - T^0 = o_p(1)$ which would establish that $T \xrightarrow{d} \cN(0, 1)$. Towards that end, write $T = L/\sigma_n^{\rm cf}$ and $T^0 = L^0/\sigma_n^{\rm cf, 0}$. Therefore, we can split the difference $T - T^0$ into two parts: 
\begin{equation}
\label{eq:t_t0_diff}
T - T^0 = \frac{L}{\sigma_n^{\rm cf}} - \frac{L^0}{\sigma_n^{\rm cf, 0}} = \frac{L - L^0}{\sigma_n^{\rm cf}} + \frac{L^0}{\sigma_n^{\rm cf, 0}}\left(\frac{\sigma_n^{\rm cf, 0}}{\sigma_n^{\rm cf}} - 1\right) := T_1 + T_2 \,.
\end{equation}
Our goal is to shwo that both $T_{1}$ and $T_{2}$ are $o_p(1)$. For $T_{1}$, it is enough to show that $L - L^0 = o_p(1)$  as we have already established that $\sigma_n^{\rm cf}$ is bounded away from $0$. Towards that end, define $ \cG_n = \sigma(A_{1:n}, X_{1:n}, M_{1:n})$. We have: 
\begin{align*}
    & \bbE[(L - L^0)^2 \mid \cG_n] \\
    & = \frac{n}{4}\bbE\left[\left(\sum_{i \in \cD_2, A_i = 1} (\hat \tau^{(2)}_{1, i} - \hat \tau^{(2), 0}_{1, i})\eps_i + \sum_{i \in \cD_1, A_i = 1}(\hat \tau^{(1)}_{1, i} - \hat \tau^{(1), 0}_{1, i})\eps_i\right)^2 \mid \cG_n\right] \\
    & = \frac{n\sigma_1^2}{4} \sum_{j = 1}^2 \|\hat \tau^{(j)}_1 - \hat \tau^{(j),0}_1\|_2^2  = o_p(1) \,. 
\end{align*}
where the last conclusion again follows from Lemma \ref{lem:tau_concentration_revised}. 
Therefore $|L - L^0| = o_p(1)$ and consequently $T_{1} = o_p(1)$. For $T_{2}$, we have already shown $|(\sigma_n^{\rm cf})^2 - (\sigma_n^{\rm cf, 0})^2| = o_p(1)$. This, along with the fact that $\sigma_n^{\rm cf}$ and $\sigma_n^{\rm cf. 0}$ are bounded away from $0$, implies $|\sigma_n^{\rm cf} - \sigma_n^{\rm cf, 0}| = o_p(1)$. Now, this, along with the fact that $\sigma_n^{\rm cf}$ is bounded away from $0$ immplies $(\sigma_n^{\rm cf. 0}/\sigma_n^{\rm cf} - 1) = o_p(1)$. As we have already established $L^0/\sigma_n^{\rm cf. 0}$ is asymptotically normal (hence $O_p(1)$), we therefore have $T_{2} = o_p(1)$. This completes the proof that $T \xrightarrow{d} \cN(0, 1)$ and hence concludes Theorem \ref{thm:main_revised}. 
\end{proof}

\subsection{Proof of Corollary \ref{cor:pop_extension}}
\label{sec:proof_cor_pop_ext}
\begin{proof}
First, we claim, as a corollary of Theorem \ref{thm:main_revised} that our proof essentially yields the following conditional version of the central limit theorem: 
\begin{lemma}[Conditional asymptotic normality]
\label{lem:conditional-clt}
Let $\cX_n=\sigma(X_{1:n})$ be the $\sigma$-field generated by $\{X_1,\ldots,X_n\}$. Under the same assumptions as Theorem~\ref{thm:main_revised}, we have, for every $t\in\reals$,
$$
\bbP\left(\frac{\sqrt{n}(\hat\theta^{\rm cf}_{10}-\theta_{10})}{\sigma_n^{\rm cf}} \le t \,\middle|\, \cX_n \right) \longrightarrow \Phi(t), \qquad\text{in probability}.
$$
Here $\Phi(t)$ denotes the CDF of the standard normal distribution.
\end{lemma}
The proof of this Lemma is deferred to Section \ref{sec:additional_lemmas}. 
Now, we present the proof of Corollary~\ref{cor:pop_extension} using this lemma. Towards that end, let us define $A_n := \sqrt{n}(\hat\theta_{10}^{\rm cf}-\theta_{10})$ and $B_n := \sqrt{n}(
\theta_{10}-\theta_{10}^{\rm pop})$. Recall that, by definition, $\theta_{10} = \bar X^\top\Lambda$, $\theta_{10}^{\rm pop} = \mu_X^\top\Lambda$ with $\Lambda = \beta_1+\bB_0^\top\gamma_1$. 
Therefore,
\begin{equation*}
\textstyle
B_n = \sqrt{n}(\bar X-\mu_X)^\top\Lambda = \frac{1}{\sqrt{n}}\sum_{i=1}^n (X_i-\mu_X)^\top\Lambda,
\end{equation*}
and consequently $B_n$ is measurable with respect to $\cX_n$. By Lemma~\ref{lem:conditional-clt}, for every fixed $u\in\reals$,
\begin{equation*}
\textstyle
\bbE\left[\exp\left\{iu(A_n/\sigma_n^{\rm cf})\right\}\,\middle|\, \cX_n\right] \xrightarrow{p} e^{-u^2/2}.
\end{equation*}
Since the absolute difference between the two characteristic functions is bounded by $2$, convergence in probability implies
convergence in $L_1$. Consequently,
\begin{equation}
\textstyle
\label{eq:conditional_cf_L1_population}
\bbE\left|\bbE\left[\exp\left\{iu(A_n/\sigma_n^{\rm cf})\right\}\,\middle|\, \cX_n\right]  - e^{-u^2/2} \right| \longrightarrow 0.
\end{equation}
We next establish the asymptotic joint behavior of
$(A_n/\sigma_n^{\rm cf},B_n)$. For any fixed
$u,v\in\reals$, since $B_n$ is measurable with respect to $\cX_n$,
\begin{equation*}
\textstyle
\bbE\left[\exp\left\{iu(A_n/\sigma_n^{\rm cf}) + ivB_n\right\}\right] = \bbE\left[e^{ivB_n}\bbE\left[\exp\left\{iu(A_n/\sigma_n^{\rm cf})\right\}\,\middle|\, \cX_n\right]\right].
\end{equation*}
As a consequence,
\begin{align}
\label{eq:diff_char_fun_zero_population}
& \left|\bbE\left[\exp\left\{iu(A_n/\sigma_n^{\rm cf}) + ivB_n\right\}\right] - e^{-u^2/2}\bbE[e^{ivB_n}] \right| \notag \\
& \le
\bbE\left|\bbE\left[\exp\left\{iu(A_n/\sigma_n^{\rm cf})\right\}\,\middle|\, \cX_n\right] - e^{-u^2/2}\right| \longrightarrow 0,
\end{align}
where the last conclusion follows from
Equation~\eqref{eq:conditional_cf_L1_population}.
For the rest of the analysis define $\omega_n = (\sigma_n^{\rm cf})^2/(\sigma^2_X + (\sigma_n^{\rm cf})^2)$. By our assumption, there exists a bounded deterministic sequence $\bar \omega_n$ such that $\omega_n - \bar \omega_n \to 0$ in probability. 
As $\sigma_X^2\le C$, given any sequence, there exists a further subsequence such that on that subsequence (we use the same notation $n$ for the sub-sequence for notational simplicity)
\begin{equation*}
\textstyle
\sigma_X^2\longrightarrow\zeta^2, \qquad \bar\omega_n \longrightarrow \omega,
\end{equation*}
for some $\zeta^2\in[0,C]$ and $\omega\in[0,1]$ and consequently $\omega_n \xrightarrow{p} \omega$. By the sub-Gaussian design condition, the triangular array $\left\{(X_i-\mu_X)^\top\Lambda: 1\le i\le n\right\}$ satisfies the Lindeberg condition, and consequently, along this sub-sequence, by Lindeberg-Feller CLT, we have: 
\begin{equation*}
\textstyle
B_n \xrightarrow{d} Z_2 \sim \cN(0, \zeta^2) \,.
\end{equation*}
Combining this conclusion with Equation~\eqref{eq:diff_char_fun_zero_population}, we have, for any fixed $u,v\in\reals$,
\begin{equation*}
\textstyle
\bbE\left[\exp\left\{iu(A_n/\sigma_n^{\rm cf}) + ivB_n\right\}\right] \longrightarrow e^{-u^2/2}e^{-v^2\zeta^2/2}.
\end{equation*}
This immediately implies,
\begin{equation}
\label{eq:joint_clt_population}
\left(
\frac{A_n}{\sigma_n^{\rm cf}}, B_n\right) \xrightarrow{d} (Z_1,Z_2),
\end{equation}
where $Z_1\sim\cN(0,1), Z_2\sim\cN(0,\zeta^2)$ and $Z_1\indep Z_2$. 
Now define $D_n^2 := (\sigma_n^{\rm cf})^2+\sigma_X^2$. Then, from our definition of $\omega_n$, we have $\omega_n = (\sigma_n^{\rm cf})^2/D_n^2$, and consequently 
\begin{equation}
\textstyle
\label{eq:population_total_decomposition}
\textstyle
(A_n + B_n)/D_n = \sqrt{\omega_n}\ (A_n/\sigma_n^{\rm cf}) + (B_n/D_n).
\end{equation}
Furthermore, we also have $D_n = \sigma_X/\sqrt{1 - \omega_n}$. First consider the case $\zeta > 0$. Then $\omega_n \to \omega$ in probability and $\sigma_X \to \zeta > 0$. Therefore, by Slutsky's theorem, along with Equation \eqref{eq:joint_clt_population} we have: 
\begin{align*}
\frac{A_n+B_n}{D_n} &\xrightarrow{d} \sqrt{\omega}\,Z_1 + \frac{\sqrt{1-\omega}}{\zeta} Z_2 \sim \cN(0, 1) \,. 
\end{align*}
Suppose next that $\zeta^2=0$. In this case, $\bbE[B_n^2] = \sigma_X^2 \to 0$, and consequently $B_n = o_p(1)$. Furthermore, Theorem~\ref{thm:main_revised} establishes that, for
some constant $c_0>0$, $\bbP\left((\sigma_n^{\rm cf})^2\ge c_0 \right) \to 1$ and consequently $D_n$ is also bounded away from 0 with probability going to $1$. Therefore, $\omega_n \to 1$ in probability and $B_n/D_n= o_p(1)$. 
Then it follows from Equation~\eqref{eq:population_total_decomposition} that
\begin{equation*}
\textstyle
\frac{A_n+B_n}{D_n} = \frac{A_n}{\sigma_n^{\rm cf}} + o_p(1) \xrightarrow{d} \cN(0,1),
\end{equation*}
where the last conclusion follows from Theorem~\ref{thm:main_revised}. Hence, we have proved that given any sequence, there exists a further subsequence along which
\begin{equation*}
\textstyle
\frac{A_n+B_n}{\sqrt{(\sigma_n^{\rm cf})^2+\sigma_X^2}} \xrightarrow{d} \cN(0,1).
\end{equation*}
As the limiting distribution is the same for all converging subsequences, we conclude: 
\begin{equation*}
\textstyle
\frac{A_n+B_n}{\sqrt{(\sigma_n^{\rm cf})^2+\sigma_X^2}} \xrightarrow{d} \cN(0,1).
\end{equation*}
This completes the proof recalling $A_n+B_n = \sqrt{n}(\hat\theta_{10}^{\rm cf} - \theta_{10}^{\rm pop})$.  
\end{proof}

\subsection{{\color{black}Proof of Corollary \ref{cor:heteroskedastic_extension}}}
\label{sec:proof_cor_hetero}
\begin{proof}
    For the first part, the proof is very closely related to the proof of Theorem \ref{thm:main_revised}, so we will highlight the key differences here. The first part of the proof of Theorem \ref{thm:main_revised}, i.e., $\sqrt{n}\Delta_n^{\rm cf} = o_p(1)$, remains identical, as the lasso rates and the conclusion of Lemma \ref{lem:feasibility} do not depend directly on the conditional heteroskedasticity as long as errors are uniformly sub-Gaussian. The second part, i.e., lower bounding $(\sigma_n^{\rm cf, het})^2$ also remains almost identical upon observing that $\var(\eps \mid X, M, A = 1) \ge c_-$. The only difference comes in the third part. Recall the oracle balancing weights $\tau_1^{(j), 0}$ as introduced in the proof of Theorem \ref{thm:main_revised}, which is obtained by replacing $a$ with $a^0 = (\bar X^\top \ \ (\bB_0 \bar X)^\top)^\top$ in Step 3 of Algorithm \ref{algomain_modified}. Also recall the definition of $T$ and $T^0$ introduced in the proof of the asymptotic normality part of Theorem \ref{thm:main_revised} with $\sigma_n^{\rm cf}$ replaced by $\sigma_n^{\rm cf, het}$ as defined in the statement of Corollary \ref{cor:heteroskedastic_extension}. As before, we first establish that $T^0$ converges to $\cN(0, 1)$, where recall that $T^0 = \sum_{i = 1}^n \zeta_{n, i, \rm het}$ with:  
\allowdisplaybreaks
\begin{align*}
    \zeta_{n, i, \rm het} &= \frac{\sqrt{n}}{2\sigma_n^{\rm cf, het, 0}} \left\{\eps_i\left(\hat \tau^{(1),0}_{1, i}\mathds{1}_{i \in \cD_1, A_i = 1} + \hat \tau^{(2),0}_{1, i}\mathds{1}_{i \in \cD_2, A_i = 1}\right) \right. \\
    & \qquad \qquad \qquad \qquad \qquad \left. + U_i^\top \gamma_1 \left(\hat \tau^{(1)}_{2, i}\mathds{1}_{i \in \cD_1, A_i = 0} + \hat \tau^{(2)}_{2, i}\mathds{1}_{i \in \cD_2, A_i = 0}\right)\right\} \notag \\
    (\sigma_n^{\rm cf, het, 0})^2 & = \frac n4 \sum_{j = 1}^2 \left(\sum_{i \in \cD_j, A_i = 1} \{(\hat \tau_{1, i}^{(j),0})^2\nu_1(X_i, M_i)\}+ \gamma_1^\top\Sigma_{U,0}\gamma_1 \lVert \hat \tau_2^{(j)}\rVert_2^2\right) \,.
\end{align*}
Define $\cH_n = \sigma(A_{1:n}, X_{1:n}, \{M_i: A_i = 1\})$. 
It is immediate that conditional on $\cH_n$, $\{\zeta_{n, i, \rm het}\}$ are independent across $i$. 
Same calculation as in the proof of Theorem \ref{thm:main_revised} yields $\bbE[\zeta_{n, i, \rm het} \mid \cH_n] = 0$ and $\var(\sum_i \zeta_{n, i, \rm het} \mid \cH_n) = 1$. 
We next show that the difference between $(\sigma_n^{\rm cf, het})^2$ and $(\sigma_n^{\rm cf, het, 0})^2$ is asymptotically negligible (similar to Equation \eqref{eq:diff_oracle_actual_sigma}): 
\begin{align}
\label{eq:sigma_hetero_diff_bound}
    & \left|(\sigma_n^{\rm cf, het})^2 - (\sigma_n^{\rm cf, het, 0})^2\right| \notag \\
    &  = 
    \frac{n}{4}\sum_{j = 1}^2 \left[\sum_{i \in \cD_j, A_i = 1} \left\{\left((\hat \tau_{1, i}^{(j)})^2 - (\hat \tau_{1, i}^{(j),0})^2\right)\nu_1(X_i, M_i)\right\}\right] \notag \\
    & \le \frac{nc_+}{4}\sum_{j = 1}^2  \sum_{i \in \cD_j, A_i = 1} \left|(\hat \tau_{1, i}^{(j)})^2 - (\hat \tau_{1, i}^{(j),0})^2\right| \notag \\
    & \le \frac{nc_+}{4}\sum_{j = 1}^2  \sum_{i \in \cD_j, A_i = 1} \left|\hat \tau_{1, i}^{(j)} + \hat \tau_{1, i}^{(j),0}\right|\left|\hat \tau_{1, i}^{(j)} - \hat \tau_{1, i}^{(j),0}\right| \notag \\
    & \le \frac{nc_+}{4}\sum_{j = 1}^2 \sqrt{  \sum_{i \in \cD_j, A_i = 1} \left|\hat \tau_{1, i}^{(j)} + \hat \tau_{1, i}^{(j),0}\right|^2}\sqrt{ \sum_{i \in \cD_j, A_i = 1}\left|\hat \tau_{1, i}^{(j)} - \hat \tau_{1, i}^{(j),0}\right|^2} \notag \\
    & \le  \frac{nc_+}{2\sqrt{2}} \sum_{j = 1}^2 (\|\hat \tau_{1}^{(j)}\|_2 + \| \hat \tau_{1}^{(j),0}\|_2)\|\hat \tau_{1}^{(j)} - \hat \tau_{1}^{(j),0}\|_2 \notag \\
    & = \frac{c_+}{2\sqrt{2}} \sum_{j = 1}^2 (\sqrt{n}\|\hat \tau_{1}^{(j)}\|_2 + \sqrt{n}\| \hat \tau_{1}^{(j),0}\|_2)\sqrt{n}\|\hat \tau_{1}^{(j)} - \hat \tau_{1}^{(j),0}\|_2  = o_p(1) \,.
\end{align}
The last conclusion follows from Lemma \ref{lem:feasibility} and Lemma \ref{lem:tau_concentration_revised}. Therefore, 
$\sigma_n^{\rm cf, het, 0}$ is also bounded away from 0 as probability tend to $1$. 
As a consequence, $\sum_i \bbE[|\zeta_{n, i, \rm het}|^3 \mid \cH_n] \to 0$ by the exact same calculation as in the proof of Theorem \ref{thm:main_revised}; the only difference is $(\sigma_n^{\rm cf, 0})^3$ is replaced by $(\sigma_n^{\rm cf, het, 0})^3$ but as we have argued, $\sigma_n^{\rm cf, het, 0}$ is bounded away from $0$ with probability going to $1$ as $\nu_1(x, m) \ge c_- > 0$. Therefore, by conditional CLT, $T^0 \xrightarrow{d} \cN(0, 1)$. Next, we need to show $T - T^0 = o_p(1)$. Following same expansion as in Equation \eqref{eq:t_t0_diff} we need to show $L- L^0 = o_p(1)$ and $|\sigma_n^{\rm cf, het} - \sigma_n^{\rm cf, het, 0}| = o_p(1)$. 
The latter is already established. For the former, we have, with $\cG_n = \sigma(A_{1:n}, X_{1:n}, M_{1:n})$: 
\begin{align*}
    & \bbE[(L - L^0)^2 \mid \cG_n] \\
    & = \frac{n}{4}\bbE\left[\left(\sum_{i \in \cD_2, A_i = 1} (\hat \tau^{(2)}_{1, i} - \hat \tau^{(2), 0}_{1, i})\eps_i + \sum_{i \in \cD_1, A_i = 1}(\hat \tau^{(1)}_{1, i} - \hat \tau^{(1), 0}_{1, i})\eps_i\right)^2 \mid \cG_n\right] \\
    & \le \frac{nc_+}{4} \sum_{j = 1}^2 \|\hat \tau^{(j)}_1 - \hat \tau^{(j),0}_1\|_2^2  = o_p(1) \,. 
\end{align*}
where the last conclusion follows from Lemma \ref{lem:tau_concentration_revised}. Consequently, by conditional Markov's inequality, \(L-L^0=o_p(1)\). This completes the proof of asymptotic normality of $\hat \theta_{10}^{\rm cf}$ around $\theta_{10}$. 

Next, we will establish the asymptotic normality of $\hat \theta_{10}^{\rm cf}$ around $\theta^{\rm pop}_{10}$. The proof here is closely related to that of Corollary \ref{cor:pop_extension}. We use the same notation and definition for $V_n^{\rm cf}, V_n^{\rm cf, 0}, T_n^{\rm cf}, A_n, B_n$ and $L_n^{\rm cf, 0}$ as used/introduced in the proof of Corollary \ref{cor:pop_extension}. Our first goal is to prove an analogous result of Lemma \ref{lem:conditional-clt} with $\sigma_n^{\rm cf}$ replaced by $\sigma_n^{\rm cf, het}$. The proof is almost exactly the same, the only thing we need to use is $(\sigma_n^{\rm cf, het, 0}/\sigma_n^{\rm cf, het}) - 1 = o_p(1)$ along with the fact that $\sigma_n^{\rm cf, het}$ and $\sigma_n^{\rm cf, het, 0}$ are bounded away from $0$ with probabibility $1$ as we have already argued in the previous part. Therefore, we also have: 
$$
\bbP\left(\frac{\sqrt{n}(\hat \theta_{10}^{\rm cf} - \theta_{10})}{\sigma_n^{\rm cf, het}} \le t \mid \cX_n\right) \longrightarrow \Phi(t), \quad \text{in probability with respect to }\cX_n \,.
$$Then the exact same calculation as in Corollary \ref{cor:pop_extension} yields 
\begin{equation}
\label{eq:an_bn_indep}
\left|\bbE\left[\exp\left\{iu(A_n/\sigma_n^{\rm cf, het}) + ivB_n\right\}\right] - e^{-u^2/2}\bbE[e^{ivB_n}] \right| \longrightarrow 0,
\end{equation}
The last part is also a similar sub-sequence argument. Define $\omega_n^{\rm het} = (\sigma_n^{\rm cf, het})^2/((\sigma_n^{\rm cf, het})^2 + \sigma^2_X)$, where, as in Corollary \ref{cor:pop_extension} $\sigma^2_X = \Lambda^\top \var(X) \Lambda$ with $\Lambda = (\beta_1 + \bB_0^\top \gamma_1)$. As per our assumption, there exists a deterministic sequence $\bar \omega_n^{\rm het} \in [0, 1]$ such that $\omega_n^{\rm het} - \bar \omega_n^{\rm het} = o_p(1)$. As $\sigma_X^2 \le C$ (bounded), given any sequence, there exists a further subsequence such that along that subsequence $\bar \omega_n^{\rm het} \to \omega \in [0, 1]$ and $\sigma_X^2 \to \zeta^2 \ge 0$. Assume $\zeta > 0$. 
We use the same index $n$ for this subsequence for notational simplicity. 
Now, by sub-Gaussianity on the design, $\{(X_i - \mu_X)^\top \Lambda: 1 \le i \le n\}$ satisfies the Lindeberg condition, and consequently we have along this subsequence: 
\begin{equation*}
\textstyle
B_n = \sqrt{n}(\bar X - \mu_X)^\top \Lambda \xrightarrow{d} \cN(0, \zeta^2) \,.
\end{equation*}
This, along with Equation \eqref{eq:an_bn_indep} implies:
\begin{equation*}
\textstyle
\left(\frac{A_n}{\sigma_n^{\rm cf, het}}, \ B_n\right) \xrightarrow{d} (Z_1, Z_2), \ \ Z_1 \sim \cN(0, 1), Z_2 \sim \cN(0, \zeta^2), \ Z_1 \indep Z_2 \,.
\end{equation*}
Define $D^2_n = (\sigma_n^{\rm cf, het})^2 + \sigma^2_X$. Then $D_n = \sigma_X/\sqrt{1 - \omega_n^{\rm het}}$. Then: 
\begin{equation*}
\textstyle
(A_n + B_n)/D_n = \sqrt{\omega_n^{\rm het}}(A_n/\sigma_n^{\rm cf, het}) + (B_n/D_n) 
\end{equation*}
Therefore, along the subsequence, by Slutsky's theorem, we have: 
\begin{equation*}
\textstyle
\frac{A_n + B_n}{D_n} \xrightarrow{d} \sqrt{\omega} Z_1 + \frac{\sqrt{1 - \omega}}{\zeta} Z_2 \sim \cN(0, 1) \,.
\end{equation*}
The case $\zeta = 0$ is handled exactly in the same way as in the proof of Corollary \ref{cor:pop_extension} and hence is skipped. Therefore, we have concluded that given any sequence, there exists a further sub-sequence such that along that sub-sequence $(A_n + B_n)/D_n \xrightarrow{d} \cN(0, 1)$. As the limiting distribution is the same along all such subsequences, we conclude that the entire sequence $(A_n + B_n)/D_n$ converges to $\cN(0, 1)$. This completes the proof. 
\end{proof}

\subsection{{\color{black}Proof of Theorem \ref{thm:NDE}}}
\label{sec:proof_NDE}
\begin{proof}
    Recall from the definition of NDE, we have $\hat \theta_{\rm NDE}^{\rm cf} = \hat \theta_{10}^{\rm cf} - \hat \theta_{00}^{\rm cf}$, where 
    \begin{equation*}
    \textstyle
   \hat \theta_{00}^{\rm cf} = \frac{\hat \theta_{00}^{(1)} + \hat \theta_{00}^{(2)}}{2} = \frac12 \sum_{j = 1}^2 \left(\bar X^\top \hat \rho_c^{(j)} + (\hat \tau_2^{(j)})^\top(\by_{c, j} - \bX_{c, j}\hat \rho_c^{(j)})\right)
    \end{equation*}
From the reduced form regression we have: 
\begin{align*}
Y & = A(X^\top \beta_1 + M^\top \gamma_1 + \eps) + (1-A)(X^\top \beta_0 + M^\top \gamma_0 + \eps')  \\
& = A(X^\top \beta_1 + X^\top \bB_1^\top \gamma_1 + \gamma_1^\top U' + \eps) + (1-A)(X^\top \beta_0 + X^\top \bB_0^\top \gamma_0 + \gamma_0^\top U + \eps')
\end{align*}
Therefore, on the control arm, $Y = X^\top \rho_c + \xi_0$, where $\rho_c = \beta_0 + \bB_0^\top \gamma_0$ and $\xi_0 = U^\top \gamma_0 + \eps'$. Using this we can decompose the estimation error of $\hat \theta_{00}^{\rm cf}$ as: 
\begin{align*}
   \hat \theta_{00}^{\rm cf} - \theta_{00} & =  \frac12 \sum_{j = 1}^2 \left(\bar X^\top (\hat \rho_c^{(j)} - \rho_c) + (\hat \tau_2^{(j)})^\top(\by_{c, j} - \bX_{c, j}\hat \rho_c^{(j)})\right) \\
   & = \frac12 \sum_{j = 1}^2 (\bar X - \bX_{c, j}^\top \hat \tau_2^{(j)})^\top (\hat \rho_c^{(j)} - \rho_c) + \frac12 \sum_{j = 1}^2 (\hat \tau_2^{(j)})^\top \bxi_0 := \Delta_{n, 00}^{\rm cf} + V_{n, 00}^{\rm cf}
\end{align*}
This, along with the decomposition of $\hat \theta_{\rm 10}^{\rm cf}$ (Equation \eqref{eq:decomp_first_10}) yields: 
\begin{equation*}
\frac{\sqrt{n}(\hat \theta_{\rm NDE}^{\rm cf}  - \theta_{\rm NDE})}{\sigma_{n, \rm NDE}^{\rm cf, het}} = \frac{\sqrt{n}(\Delta_{n}^{\rm cf} - \Delta_{n, 00}^{\rm cf}+ V_n^{\rm cf} - V_{n, 00}^{\rm cf})}{\sigma_{n, \rm NDE}^{\rm cf, het}}
\end{equation*}
The proof proceeds similarly to that of Theorem \ref{thm:main_revised}. We first show $\sqrt{n}(\Delta_{n}^{\rm cf} - \Delta_{n, 00}^{\rm cf}) = o_p(1)$, then we show $\sigma_{n, \rm NDE}^{\rm cf, het}$ is bounded away from $0$ with probability going to $1$, and finally we establish that $\sqrt{n}(V_n^{\rm cf} - V_{n, 00}^{\rm cf})/\sigma_{n, \rm NDE}^{\rm cf, het}$ converges to $\cN(0, 1)$.  
We have already established in the proof of Theorem \ref{thm:main_revised} that $\sqrt{n}\Delta_n^{\rm cf}= o_p(1)$. Here, we will first show that $\sqrt{n}\Delta_{n, 00}^{\rm cf} = o_p(1)$. Given that $\|\rho_c\|_0 = \|\beta_0 + \bB_0^\top \gamma_0\|_0 \le k_\beta + sk_\gamma \le (1+s)k$ and the error $\xi_0$ satisfies $\bbE[\xi_0 \mid X, A = 0] = 0$ and sub-Gaussian (uniformly over $X$,  Assumption \ref{assm:cov_err}) and the design matrix satisfies RE condition with probability going to $1$, the lasso estimator $\hat \rho_c^{(j)}$ satisfies by same argument in Theorem \ref{thm:main_revised} 
\begin{equation*}
\textstyle
\|\hat \rho_c^{(j)} - \rho_c\|_1 = O_p\left(sk\sqrt{\frac{\log{p}}{n}}\right) \,, \ \|\hat \rho_c^{(j)} - \rho_c\|_2 = O_p\left(\sqrt{\frac{sk\log{p}}{n}}\right) \,.
\end{equation*}
As a consequence: 
\begin{align*}
\sqrt{n}|\Delta_{n, 00}^{\rm cf}| & \le \frac{1}{2}\sum_{j= 1}^2 \|\bar X - \bX_{c, j}^\top \hat \tau_2^{(j)}\|_\infty \|\hat \rho_c^{(j)} - \rho_c\|_1 \\
& = O_p\left(\sqrt{n} \times \sqrt{\frac{\log{p}}{n}} \times sk\sqrt{\frac{\log{p}}{n}}\right) = O_p\left(\frac{sk\log{p}}{\sqrt{n}}\right) = o_p(1)
\end{align*}
where the last conclusion follows from Assumption \ref{assm:params}. Here we have also used the fact that $\|\bar X - \bX_c^\top \hat \tau^{(j)}_2\|_\infty \le K_2 \sqrt{\log{p}/n}$ which is by definition of the optimization problem \eqref{eq:balancing_opt_1}. 
This completes the proof of $\sqrt{n}(\Delta_{n}^{\rm cf} - \Delta_{n, 00}^{\rm cf}) = o_p(1)$. 

We next argue that $\sigma_{n, \rm NDE}^{\rm cf, het}$ is bounded away from $0$ with probability going to $1$. The proof is very similar to that of the second step of Theorem \ref{thm:main_revised}; from the definition, we have: 
\begin{align*}
     (\sigma_{n, \rm NDE}^{\rm cf, het})^2 & = \frac n4\sum_{j=1}^2\left[\sum_{i \in \cD_j, A_i = 1}(\hat \tau_{1,i}^{(j)})^2\nu_1(X_i, M_i) +  \sum_{i\in \cD_j, A_i = 0}(\hat \tau_{2,i}^{(j)})^2\omega_0(X_i)\right] \\
     & \ge \frac{c_-}{4} \ n\|\hat \tau^{(1)}_1\|_2^2
\end{align*}
Here we have used the fact that $\nu_1(x, m) \ge c_-$ (Assumption \ref{assm:cov_err}). 
Now, we have already established in the proof of Theorem \ref{thm:main_revised} $n\|\hat \tau^{(1)}_1\|_2^2$ is bounded away from $0$ with probability going to $1$. This immediately implies $\sigma_{n, \rm NDE}^{\rm cf, het}$ is bounded away from $0$ with probability going to $1$. 

The final part is to establish the asymptotic normality. Similar to the proof of Theorem \ref{thm:main_revised}. From the definition of $V_n^{\rm cf}$ and $V_{n, 00}^{\rm cf}$, we have: 
$$
\frac{\sqrt{n}(V_n^{\rm cf} - V_{n, 00}^{\rm cf})}{\sigma_{n, \rm NDE}^{\rm cf, het}} = \frac{\sqrt{n}}{2\sigma_{n, \rm NDE}^{\rm cf, het}} \sum_{j = 1}^2 \left\{(\hat \tau_1^{(j)})^\top \beps^{(j)} + (\hat \tau_2^{(j)})^\top (\bU^{(j)} (\gamma_1 - \gamma_0) - \beps^{'{(j)}})\right\}
$$
Recall that definition of $\hat \tau_1^{(j), 0}$ as introduced in the proof of Theorem \ref{thm:main_revised}. We have established in Lemma \ref{lem:tau_concentration_revised} that $\sqrt{n}\|\hat \tau_1^{(j)} - \hat \tau_1^{(j), 0}\|_2 = o_p(1)$. Define $T^0$ as: 
$$
T^0 = \frac{\sqrt{n}}{2\sigma_{n, \rm NDE}^{\rm cf, het, 0}} \sum_{j = 1}^2 \left[\sum_{\substack{i \in \cD_j \\ A_i = 1}} \hat \tau^{(j),0}_{1, i} \eps_i + \sum_{\substack{i \in \cD_j \\ A_i = 0}} \hat \tau^{(j)}_{2, i} (U_i^\top (\gamma_1 - \gamma_0) - \eps'_i)\right] := \sum_{i = 1}^n \zeta^{\rm het}_{n, i} 
$$
with 
\begin{align*}
\zeta^{\rm het}_{n, i} & = \frac{\sqrt{n}}{2\sigma_{n, \rm NDE}^{\rm cf, het, 0}} \left\{\eps_i\left(\hat \tau^{(1),0}_{1, i}\mathds{1}_{i \in \cD_1} + \hat \tau^{(2),0}_{1, i}\mathds{1}_{i \in \cD_2}\right)\mathds{1}_{A_i = 1} \right. \\
& \qquad \qquad \qquad \left. + (U_i^\top (\gamma_1 - \gamma_0) - \eps'_i) \left(\hat \tau^{(1)}_{2, i}\mathds{1}_{i \in \cD_1} + \hat \tau^{(2)}_{2, i}\mathds{1}_{i \in \cD_2}\right)\mathds{1}_{A_i = 0}\right\}\\
(\sigma_{n, \rm NDE}^{\rm cf, het, 0})^2 & = \frac n4\sum_{j=1}^2\left[\sum_{i \in \cD_j, A_i = 1}(\hat \tau_{1,i}^{(j), 0})^2\nu_1(X_i, M_i) +  \sum_{i\in \cD_j, A_i = 0}(\hat \tau_{2,i}^{(j)})^2\omega_0(X_i)\right] \,.
\end{align*}
Define $\cH_n = \sigma(A_{1:n}, X_{1:n}, \{M_i: A_i = 1\})$. Then it is immediate that conditional on $\cH_n$, $\zeta^{\rm het}_i$ are independent across $i$ with $\bbE[\zeta^{\rm het}_i \mid \cH_n] = 0$ and 
\begin{align*}
& \var(T^0 \mid \cH_n) \\
& = \sum_i \var(\zeta^{\rm het}_i \mid \cH_n) \\
& = \frac{n}{4(\sigma_{n, \rm NDE}^{\rm cf, het, 0})^2}\sum_{j=1}^2\left[\sum_{i \in \cD_j, A_i = 1}(\hat \tau_{1,i}^{(j), 0})^2\nu_1(X_i, M_i) +  \sum_{i\in \cD_j, A_i = 0}(\hat \tau_{2,i}^{(j)})^2\omega_0(X_i)\right] = 1 \,.
\end{align*}
The proof that $\sum_i \bbE[|\zeta^{\rm het}_{n, i}|^3 \mid \cH_n] = o_p(1)$ is same as that of $\sum_i \bbE[|\zeta_{n, i}|^3 \mid \cH_n] = o_p(1)$ in the proof of Theorem \ref{thm:main_revised} upon using
\begin{align*}
    \nu_1(x, m) \vee \omega_0(x) \le C,  \qquad & (\text{Assumption } \ref{assm:cov_err})  \\
    (\gamma_1- \gamma_0)^\top \Sigma_{U, 0} (\gamma_1 - \gamma_0) \le \lambda_{U, 0, +} \|\gamma_1 - \gamma_0\|_2^2 \le 2C_\gamma^2\lambda_{U, 0, +}\,. \qquad & (\text{Assumption } \ref{assm:cov_err} \ \ \& \ \ \ref{assm:params})
\end{align*}
Furthermore, as we have already argued $\sigma_{n, \rm NDE}^{\rm cf, het}$ is bounded away from $0$ with probability going to $1$, the same calculation as of Equation \eqref{eq:sigma_hetero_diff_bound} yields that $\sigma_{n, \rm NDE}^{\rm cf, het, 0}$ is also  bounded away from $0$ with probability going to $1$ (observe that $(\sigma_{n, \rm NDE}^{\rm cf, het})^2 - (\sigma_{n, \rm NDE}^{\rm cf, het, 0})^2$ is same as $(\sigma_{n}^{\rm cf, het})^2 - (\sigma_{n}^{\rm cf, het})^2$). 
Therefore, $T^0 \xrightarrow{d} \cN(0, 1)$. Finally defining $T = \sqrt{n}(V_n^{\rm cf} - V_{n, 00}^{\rm cf})/\sigma_{n, \rm NDE}^{\rm cf, het}$, all we need to argue is $T - T^0$.
This is also very similar to that of Theorem \ref{thm:main_revised} and Corollary \ref{cor:heteroskedastic_extension}. 
To keep the notation consistent with the previous proofs, let us define $L^0 = T^0 \sigma_{n, \rm NDE}^{\rm cf, het, 0}$ and $L = T \sigma_{n, \rm NDE}^{\rm cf, het}$. 
Similar decomposition as in Equation \eqref{eq:t_t0_diff} yields: 
\begin{equation*}
T - T^0 = \frac{L}{\sigma_{n, \rm NDE}^{\rm cf, het}} - \frac{L^0}{\sigma_{n, \rm NDE}^{\rm cf, het, 0}} = \frac{L - L^0}{\sigma_{n, \rm NDE}^{\rm cf, het}} + \frac{L^0}{\sigma_{n, \rm NDE}^{\rm cf, het, 0}}\left(\frac{\sigma_n^{\rm cf, het, 0}}{\sigma_{n, \rm NDE}^{\rm cf, het}} - 1\right) := T_1 + T_2 \,.
\end{equation*}
Therefore, to establish asymptotic normality of $T$ all we need to show is $L - L^0 = o_p(1)$ and $|\sigma_{n, \rm NDE}^{\rm cf, het, 0} - \sigma_{n, \rm NDE}^{\rm cf, het}| = o_p(1)$ (as this along with the fact that $\sigma_{n, \rm NDE}^{\rm cf, het}$ is bounded away from $0$ with probability going to $1$ implies $|(\sigma_{n, \rm NDE}^{\rm cf, het, 0}/\sigma_{n, \rm NDE}^{\rm cf, het}) - 1| = o_p(1)$. 
The second part can be proved immediately by the same calculation as in Equation \eqref{eq:sigma_hetero_diff_bound}. 
For the first part, define $\cG_n = \sigma(A_{1:n}, X_{1:n}, M_{1:n})$. 
Then we have: 
\begin{equation*}
\textstyle
L - L^0 = \frac{\sqrt{n}}{2}\sum_{j = 1}^2 \sum_{\substack{i \in \cD_j \\ A_i = 1}}(\hat \tau_{1, i}^{(j)} - \hat \tau_{1, i}^{(j), 0})\eps_i \,. 
\end{equation*}
As a consequence, 
\begin{equation*}
\textstyle
\bbE\left[(L- L^0)^2 \mid \cG_n\right] \le \frac{nc_+}{4}\sum_{j = 1}^2 \|\hat \tau^{(j)}_1 - \hat \tau_1^{(j), 0}\|_2^2 = o_p(1)
\end{equation*}
where the conclusion follows from Lemma \ref{lem:tau_concentration_revised}. Therefore, an application of the Markov inequality yields that $L - L^0 = o_p(1)$. This completes the proof. 

The proof for asymptotic normality of $\hat \theta_{\rm NDE}^{\rm cf}$ around $\theta_{\rm NDE}^{\rm pop}$ follows the same argument as that of
Corollary~\ref{cor:pop_extension}; we only highlight the required
changes. Define
\begin{equation*}
\textstyle
A_n = \sqrt n\left(
\hat\theta_{\rm NDE}^{\rm cf} - \theta_{\rm NDE}
\right), \qquad B_n = \sqrt{n}\left(\theta_{\rm NDE} - \theta_{\rm NDE}^{\rm pop}\right).
\end{equation*}
Since $\theta_{\rm NDE} = \bar X^\top\Lambda_{\rm NDE}$ and $\theta_{\rm NDE}^{\rm pop} = \mu_X^\top\Lambda_{\rm NDE}$, we have
\begin{equation*}
\textstyle
B_n = \sqrt{n}(\bar X-\mu_X)^\top\Lambda_{\rm NDE} = \frac1{\sqrt{n}}\sum_{i=1}^n(X_i-\mu_X)^\top\Lambda_{\rm NDE}.
\end{equation*}
Consequently, $B_n$ is measurable with respect to $\sigma(X_{1:n}) = \cX_n$. Similar argument as in Lemma~\ref{lem:conditional-clt} yields,
for every fixed $u\in\reals$,
\begin{equation*}
\textstyle
\bbE\left[\exp\left\{iu\frac{A_n}{\sigma_{n,\rm NDE}^{\rm cf,het}}\right\} \,\middle|\, \cX_n\right] \xrightarrow{p} e^{-u^2/2}.
\end{equation*}
Since $B_n$ is $\cX_n$-measurable, the same conditional characteristic-function calculation as in the proof of Corollary~\ref{cor:pop_extension} gives, for any $u,v\in\reals$,
\begin{align}
\label{eq:joint_cf_NDE_population}
&
\left|
\bbE\left[\exp\left\{iu\frac{A_n}{\sigma_{n,\rm NDE}^{\rm cf,het}} + ivB_n\right\}\right] - e^{-u^2/2}\bbE[e^{ivB_n}]\right| \longrightarrow 0.
\end{align}
Since $\sigma_{X,\rm NDE}^2\le C$ and
$\bar\omega_{n,\rm NDE}^{\rm het}\in[0,1]$, given any subsequence, there exists a further
subsequence, which we continue to index by $n$, such that $\sigma_{X,\rm NDE}^2\to\zeta^2$ and $\bar\omega_{n,\rm NDE}^{\rm het}\to\omega$ simultaneously along that subsequence for some $0 \le \zeta^2 \le C_X$ and $0 \le \omega \le 1$. 
By the assumed asymptotic determinism, $\omega_{n,\rm NDE}^{\rm het} \to \omega$ in probability along this subsequence. 
First assume $\zeta>0$. By the Lindeberg condition and Equation~\eqref{eq:joint_cf_NDE_population},
\begin{equation*}
\textstyle
\left(\frac{A_n}{\sigma_{n,\rm NDE}^{\rm cf,het}},\frac{B_n}{\sigma_{X,\rm NDE}} \right)\xrightarrow{d} (Z_1,Z_2), \ \ Z_1,Z_2 \xrightarrow{\iid} \cN(0, 1)  \,.
\end{equation*}
Define $D_{n,\rm NDE}^2 = (\sigma_{n,\rm NDE}^{\rm cf,het})^2 + \sigma_{X,\rm NDE}^2$. Then, by Slutsky's theorem, 
\begin{equation*}
\textstyle
\frac{A_n+B_n}{D_{n,\rm NDE}} = \sqrt{\omega_{n,\rm NDE}^{\rm het}}\frac{A_n}{\sigma_{n,\rm NDE}^{\rm cf,het}} + \sqrt{1-\omega_{n,\rm NDE}^{\rm het}}\frac{B_n}{\sigma_{X,\rm NDE}} \xrightarrow{d} \sqrt{\omega}\,Z_1 + \sqrt{1-\omega}\,Z_2 \sim\cN(0,1).
\end{equation*}
On the other hand, if $\zeta=0$, then $\bbE[B_n^2] = \sigma_{X,\rm NDE}^2 \to 0$ and consequently $B_n=o_p(1)$. Moreover, since
$(\sigma_{n,\rm NDE}^{\rm cf,het})^2$ is bounded away from zero with probability going to one, $D_{n,\rm NDE}/\sigma_{n,\rm NDE}^{\rm cf,het} \to 1$ in probability. Thus,
\begin{equation*}
\textstyle
\frac{A_n+B_n}{D_{n,\rm NDE}} = \frac{A_n}{\sigma_{n,\rm NDE}^{\rm cf,het}} + o_p(1) \xrightarrow{d} \cN(0,1).
\end{equation*}
We have therefore established that given any subsequence, there exists a further
subsequence along which 
\begin{equation*}
\textstyle
\frac{A_n+B_n}{\sqrt{(\sigma_{n,\rm NDE}^{\rm cf,het})^2 + \sigma_{X,\rm NDE}^2}} \xrightarrow{d} \cN(0,1).
\end{equation*}
As the limit is the same, the convergence holds for the entire sequence. This completes the proof.
\end{proof}

\subsection{{\color{black}Proof of Theorem \ref{thm:NIE}}}
\label{sec:proof_NIE}
\begin{proof}
    On the treatment arm, the reduced form of the regression model satisfies $Y_i = X_i^\top \rho_t + \xi_{1, i}$ where $\rho_t = \beta_1 + \bB_1^\top \gamma_1$ and $\xi_{1, i} = \gamma_1^\top U'_i + \eps_i$. Recall the definition of $\hat \theta_{11}^{\rm cf}$ (Algorithm \ref{algo_NIE}): 
    \begin{align*}
    \hat \theta_{11}^{\rm cf} = \frac{\hat \theta_{11}^{(1)} + \hat \theta_{11}^{(2)}}{2} & = \frac{1}{2}\sum_{j = 1}^2\left(\bar X^\top \hat \rho_t^{(j)} + (\hat \omega^{(j)})^\top(\by_{t, j} - \bX_{t, j} \hat \rho_t^{(j)})\right) \\
    & = \theta_{11} +  \frac{1}{2}\sum_{j = 1}^2\left((\bar X - \bX_{t, j}^\top \hat \omega^{(j)})^\top(\hat \rho_t^{(j)} - \rho_t)\right) + \frac{1}{2}\sum_{j = 1}^2 (\hat \omega^{(j)})^\top \bxi^{(j)}_1 \\
    & := \theta_{11} + \Delta_{n, 11}^{\rm cf} + V_{n, 11}^{\rm cf}. 
    \end{align*}
As a consequence, we have: 
\allowdisplaybreaks
\begin{align*}
     \frac{\sqrt{n}(\hat \theta_{\rm NIE}^{\rm cf} - \theta_{\rm NIE})}{\sigma_{n, \rm NIE}^{\rm cf, het}} & =  \frac{\sqrt{n}((\hat \theta_{11}^{\rm cf} - \hat \theta_{10}^{\rm cf}) - (\theta_{11} - \theta_{10}))}{\sigma_{n, \rm NIE}^{\rm cf, het}} \\
     & = \frac{\sqrt{n}(\Delta_{n, 11}^{\rm cf} - \Delta_n^{\rm cf})}{\sigma_{n, \rm NIE}^{\rm cf, het}} + \frac{\sqrt{n}(V_{n, 11}^{\rm cf} - V_n^{\rm cf})}{\sigma_{n, \rm NIE}^{\rm cf, het}}
\end{align*}
where $(\Delta_n^{\rm cf}, V_n^{\rm cf})$ are same as defined in Equation \eqref{eq:decomp_first_10}. The proof structure is closely related to that of Theorem \ref{thm:main_revised} and Theorem \ref{thm:NDE}. First, we show that the bias term $\sqrt{n}(\Delta_{n, 11}^{\rm cf} - \Delta_n^{\rm cf}) = o_p(1)$. Then we argue $\sigma_{n, \rm NIE}^{\rm cf, het}$ is bounded away from $0$ with probability going to $1$, and finally we show that the second term on the right-hand side of the above equation converges to $\cN(0, 1)$. For the first part, all we need to show is $\sqrt{n}\Delta_{n, 11}^{\rm cf} = o_p(1)$ as we have already established in the proof of Theorem \ref{thm:main_revised} that $\sqrt{n}\Delta_n^{\rm cf} = o_p(1)$. Towards that end, first observe that, by Assumption \ref{assm:cov_err} and \ref{assm:params} $\xi_{1, i} = \gamma_1^\top U_i' + \eps_i$ is uniformly sub-Gaussian conditional on $\sigma(X_{1:n})$. Also, by Assumption \ref{assm:params}, $\|\rho_t\|_0 = \|\beta_1 + \bB_1^\top \gamma_1\|_0 \le k_\beta + sk_\gamma \le (1 + s)k$. Furthermore, by Assumption \ref{assm:cov_err} and Remark \ref{rem:sample_RE}, the design matrix $\bX_{t, j}$ satisfies RE condition with probability going to $1$. 
Therefore, by standard Lasso concentration, we have: 
\begin{equation*}
\textstyle
\|\hat \rho_t^{(j)} - \rho_t\|_1 = O_p\left(sk\sqrt{\frac{\log{p}}{n}}\right), \quad \|\hat \rho_t^{(j)} - \rho_t\|_2 = O_p\left(\sqrt{\frac{sk\log{p}}{n}}\right) \,.
\end{equation*}
As a consequence, we have: 
\begin{align*}
    \sqrt{n}|\Delta_{n, 11}^{\rm cf}| & \le \sqrt{n}\sum_{j = 1}^2 \left|(\bar X - \bX_{t, j}^\top \hat \omega^{(j)})^\top(\hat \rho_t^{(j)} - \rho_t)\right|  \\
    & \le \sqrt{n}\sum_{j = 1}^2 \|\bar X - \bX_{t, j}^\top \hat \omega^{(j)}\|_\infty \|\hat \rho_t^{(j)} - \rho_t\|_1 \\
    & = O_p\left(\sqrt{n} \times \sqrt{\frac{\log{p}}{n}} \times sk\sqrt{\frac{\log{p}}{n}}\right) = o_p(1) \,.
\end{align*}
where, in the last conclusion we have used $(sk\log{p})/\sqrt{n}= o(1)$ (Assumption \ref{assm:params}). Furthermore, $\|\bar X - \bX_{t, j}^\top \hat \omega^{(j)}\|_\infty = O_p(\sqrt{\log{p}/n})$ follows from the definition of the optimization problem \eqref{eq:balancing_opt_3}. This establishes the first part. Now, to argue $\sigma_{n, \rm NIE}^{\rm cf, het}$ is bounded away from $0$ with probability going to $1$, it is immediate from the definition of $\sigma_{n, \rm NIE}^{\rm cf, het}$ that
\begin{equation*}
\textstyle
(\sigma_{n, \rm NIE}^{\rm cf, het})^2 \ge \frac{n}{4}v^2_{U, 0, 1} \|\hat \tau^{(1)}_2\|_2^2 \ge \frac{\lambda_{\min}(\Sigma_{U, 0})c_\gamma^2}{4} n \|\hat \tau^{(1)}_2\|_2^2
\end{equation*}
We have already argued in the proof of Theorem \ref{thm:main_revised} that $n\|\hat \tau^{(1)}_2\|_2^2$ bounded away from $0$ with probability going to $1$ (see Equation \eqref{eq:lower_bound_tau}). Hence $\sigma_{n, \rm NIE}^{\rm cf, het}$ is bounded away from $0$ with probability going to $1$. Finally, we will establish the asymptotic normality. Towards that end, define an oracle balancing weight $\tau_1^{(j),*}$ as: 
\begin{equation*}
\textstyle
\tau_{1, i}^{(j), *} = \frac{1}{n_{t,j}}W_i^\top \eta^*, \quad \eta^* = \Omega_{W, 1}^{-1}a^*_{10} = \Omega_{W, 1}^{-1}\begin{pmatrix}
    \mu_X \\ \bB_0 \mu_X
\end{pmatrix}, \quad 1 \le i \le n_{t, j} \,.
\end{equation*}
First we argue that: 
\begin{equation}
\textstyle
    \label{eq:closeness_tau_tau_star}
    \sqrt{n}\|\hat \tau_1^{(j)} - \tau_1^{(j), *}\|_2 = o_p(1) \,.
\end{equation}
The proof is similar to the proof of Lemma \ref{lem:tau_concentration_revised}. Fix $j = 2$ (we choose $j= 2$ because it is explicitly mentioned in Algorithm \ref{algomain_modified}; the proof for $j = 1$ is verbatim). Recall that the main difference between $\hat \tau_1^{(2)}$ is obtained by solving the optimization problem \eqref{eq:balancing_opt_2} with $a^{(1)} = (\bar X^\top, \hat m_{c, 1}^\top)^\top$. Let $\hat \eta_n := \hat \eta_n(a^{(1)})$ be the rescaled dual of $\hat \tau_1^{(2)}$. Then from Equation \eqref{eq:KKT_dual_opt_prob} inside the proof of Lemma \ref{lem:tau_concentration_revised}, we have: 
\begin{align*}
\hat \eta_n & = \argmin_\eta \left\{\frac{1}{n_t}
\sum_{i=1}^{n_t}
H_{n_tb_n}\left((\bW_t\eta)_i\right) - \eta^\top a^{(1)} + \check \lambda_{n_t} \|\eta\|_1\right\} \\
& = \argmin_\eta \ \left\{\cL_{n, a^{(1)}}(\eta) + \check \lambda_{n_t} \|\eta\|_1\right\} \,.
\end{align*}
where $n_t$ is a shorthand for $n_{t,2}$ and $\check \lambda_{n_t} = K_1 \sqrt{\log{(p+q)}/n_t}$ for some appropriately chosen large constant $K_1$. Now observe that: 
\begin{equation*}
\textstyle
\max_{1 \le i \le n_t} |W_{t, i}^\top \eta^*| \le \|\eta^*\|_1 \max_i \|W_{t, i}\|_\infty = O_p(\sqrt{\log{n} + \log{d}}) = o(n^{1-\alpha}) \,.
\end{equation*}
Therefore, all large $n$, with probability going to $1$, $H_{n_t b_n}(W_{t, i}^\top \eta^*) = (W_{t, i}^\top \eta^*)^2/2$ for all $1 \le i \le n_t$. Now from optimality of $\hat \eta_n$, we have: 
\begin{align*}
    \cL_{n, a^{(1)}}(\hat \eta_n) + \check \lambda_{n_t} \|\hat \eta_n\|_1  \le \cL_{n, a^{(1)}}(\eta^*) + \check \lambda_{n_t} \|\eta^*\|_1 
\end{align*}
Then, by the same calculation as used to obtain Equation \eqref{eq:bound_eta_hat_l1} using convexity of $\cL_{n, a^{(1)}}(\cdot)$, we conclude that $\|\hat \eta_n\|_1 \le 3\|\eta^*\|_1 \le 3C_\eta$ with probability going $1$. We next claim that $\|\bW_t(\hat \eta_n - \eta^*)\|_2^2/n_t = o_p(1)$. Now, as we have argued $\|\hat \eta_n\|_1 \le 3C_\eta$ with probability going to $1$, hence on this event, 
\begin{equation*}
\textstyle
\max_{1 \le i \le n_t} |W_{t, i}^\top \hat \eta_n| \le \|\hat \eta_n\|_1 \max_i \|W_{t, i}\|_\infty = O_p(\sqrt{\log{n} + \log{(p+q)}}) = o(n^{1-\alpha}) \,.
\end{equation*}
Hence, we also have on this event, $H_{n_t b_n}(W_{t, i}^\top \hat \eta_n) = (W_{t, i}^\top \hat \eta_n)^2/2$. Therefore, we have, again from the optimality of $\hat \eta_n$: 
\begin{equation*}
\textstyle
    \frac{1}{2n_t}\|\bW_t \hat \eta_n\|_2^2 - \hat \eta_n^\top a^{(1)} + \check \lambda_{n_t} \|\hat \eta_n\|_1 \le \frac{1}{2n_t}\|\bW_t \eta^*\|_2^2 -(\eta^*)^\top a^{(1)} + \check \lambda_{n_t} \|\eta^*\|_1
\end{equation*}
Hence, 
\begin{align}
\label{eq:del_eta_close}
& \frac{1}{2n_t}\|\bW_t(\hat \eta_n - \eta^*)\|_2^2 + \check \lambda_{n_t} \|\hat \eta_n\|_1 \notag \\
& \le (\eta^*)^\top \hat \Omega_{W, t} (\eta^* - \hat \eta_n) + (\hat \eta_n - \eta^*)^\top a^{(1)} + \check \lambda_{n_t} \|\eta^*\|_1 \notag \\
& = ( \eta^* - \hat \eta_n)^\top(\hat \Omega_{W, t} \eta^* - a^{(1)}) + \check \lambda_{n_t} \|\eta^*\|_1 \notag \\
& =  ( \eta^* - \hat \eta_n)^\top(\hat \Omega_{W, t} \eta^* - a_{10}^* + a^*_{10} - a^{(1)}) + \check \lambda_{n_t} \|\eta^*\|_1 \notag \\
& = ( \eta^* - \hat \eta_n)^\top((\hat \Omega_{W, t} - \Omega_{W, 1})\eta^* + a^*_{10} - a^{(1)}) + \check \lambda_{n_t} \|\eta^*\|_1
\end{align}
Now, 
\begin{equation*}
\textstyle
\|(\hat \Omega_{W, t} - \Omega_{W, 1})\eta^*\|_\infty \le \|\eta^*\|_1 \|\hat \Omega_{W, t} - \Omega_{W, 1}\|_{\infty, \infty} \le C_\eta C \sqrt{\frac{\log{p+q}}{n}}
\end{equation*}
with probability going to $1$, where the last inequality follows from the fact that the elements of $\hat \Omega_{W, t}$ are sub-exponential and consequently an application of Bernstein's inequality, along with a union bound, yields the conclusion. Furthermore, by same calculation as in Equation \eqref{eq:ahat_astar_close}, we have, with probability going to $1$: 
\begin{equation*}
\textstyle
\|a^*_{10} - a^{(1)}\|_\infty \le C'\sqrt{\frac{\log{(p+q)}}{n}} \,.
\end{equation*}
Hence, choosing $K_1$ large enough, we conclude with probability going to $1$: 
\begin{equation*}
\textstyle
\|(\hat \Omega_{W, t} - \Omega_{W, 1})\eta^*\|_\infty + \|a^*_{10} - a^{(1)}\|_\infty \le \frac{\check \lambda_{n_t}}{2} \,.
\end{equation*}
Therefore, going back to Equation \eqref{eq:del_eta_close}, we have: 
\begin{equation*}
\textstyle
\frac{1}{2n_t}\|\bW_t(\hat \eta_n - \eta^*)\|_2^2 + \check \lambda_{n_t} \|\hat \eta_n\|_1 \le \|\hat \eta_n - \eta^*\|_1 \frac{\check \lambda_{n_t}}{2} + \check \lambda_{n_t} \|\eta^*\|_1 
\end{equation*}
Therfore, from the fact $\|\hat \eta_n\|_1 \le 3C_\eta$, $\|\eta^*\|_1 \le C_\eta$ and $\check \lambda_{n_t} = O(\sqrt{\log{(p+q)/n}})$, we have: 
\begin{equation*}
\textstyle
\frac{1}{2n_t}\|\bW_t(\hat \eta_n - \eta^*)\|_2^2 = O_p\left(\sqrt{\frac{\log{(p+q)}}{n}}\right) = o_p(1) \,.
\end{equation*}
Now, going back to KKT condition of $\hat \tau_1^{(2)}$ (see Equation \eqref{eq:KKT_cond_tau} in the proof of Lemma \ref{lem:tau_concentration_revised} for more details): 
\begin{equation*}
\textstyle
\hat \tau_1^{(2)} = \bW_{t, 2} \hat \eta^{\rm KKT}(a^{(1)}) - \hat \zeta^{\rm KKT}(a^{(1)}) = \frac{1}{n_t} \bW_t \hat \eta_n - \hat \zeta^{\rm KKT}(a^{(1)})
\end{equation*}
where we used the fact $\hat \eta_n = n_t \hat \eta^{\rm KKT}$. We next argue that $\hat \zeta^{\rm KKT}(a^{(1)}) = 0$. Recall that from the definition of $\hat \zeta^{\rm KKT}(a^{(1)})$, it is $0$ if the condition $\|\hat \tau^{(1)}\|_\infty \le n^{-\alpha}$ is strict, i.e. $\|\hat \tau^{(1)}\|_\infty < n^{-\alpha}$. Towards that end, recall the rescaled dual optimization problem (Equation \eqref{eq:rescaled_dual_KKT} in the proof of Lemma \ref{lem:tau_concentration_revised}): 
\begin{equation*}
\textstyle
(\hat \eta_n, \hat \zeta_n) = \argmin_{\eta, \zeta} \  \left\{\frac{1}{2n_t} \|\bW_t \eta - \zeta\|_2^2- \eta^\top a + \check \lambda_{n_t} \|\eta\|_1 + b_n \|\zeta\|_1 \right\}
\end{equation*}
Therefore, given $\hat \eta_n$, $\hat \zeta_n$ can be written as: 
\begin{equation*}
\textstyle
\hat \zeta_{n, i} = \sign(W_{t, i}^\top \hat \eta_n)(|W_{t, i}^\top \hat \eta_n| - n_t^{1-\alpha})_+, \qquad 1 \le i \le n_t \,.
\end{equation*}
However, we have argue previously that with probabiity going to $1$, $|W_{t, i}^\top \hat \eta_n| \le C\sqrt{\log{(p+q)} + \log{n}} = o(n_t^{1-\alpha})$ and consequently on this even $\hat \zeta_{n, i} =  0$ for all $1 \le i \le n_t$. As $\zeta^{\rm KKT} = \hat \zeta_n/n_t$, we have $\hat \zeta^{\rm KKT} = 0$. Therefore, 
\begin{equation*}
\textstyle
\hat \tau_1^{(2)} = \frac{1}{n_t} \bW_{t,2} \hat \eta_n \implies n\|\hat \tau_1^{(2)} - \tau_1^{(2), *}\|_2^2 = \frac{n}{n_t}\frac{1}{n_t} \|\bW_{t, 2}(\hat \eta_n - \eta^*)\|_2^2= o_p(1) \,,
\end{equation*}
as $n \asymp n_t$. This completes the proof. 

Now coming back to the proof of asymptotic normality, we can write the relevant part as: 
\begin{align*}
    T & := \frac{\sqrt{n}(V_{n, 11}^{\rm cf} - V_n^{\rm cf})}{\sigma_{n, \rm NIE}^{\rm cf, het}} \\
    & = \frac{\sqrt{n}}{2\sigma_{n, \rm NIE}^{\rm cf, het}}\sum_{j = 1}^2 \left[(\hat \omega^{(j)})^\top \bxi_1^{(j)} - (\hat \tau_1^{(j)})^\top \beps^{(j)} - (\hat \tau_2^{(j)})^\top\bU\gamma_1\right] \\
    & = \frac{\sqrt{n}}{2\sigma_{n, \rm NIE}^{\rm cf, het}}\sum_{j = 1}^2 \left[(\hat \omega^{(j)})^\top \bU'\gamma_1 + (\hat \omega^{(j)})^\top \beps^{(j)}  - (\hat \tau_1^{(j)})^\top \beps^{(j)} - (\hat \tau_2^{(j)})^\top\bU\gamma_1\right]\\
    & = \frac{\sqrt{n}}{2\sigma_{n, \rm NIE}^{\rm cf, het}}\sum_{j = 1}^2 \left[(\hat \omega^{(j)} - \hat \tau_1^{(j)})^\top \beps^{(j)} +  (\hat \omega^{(j)})^\top \bU'\gamma_1 - (\hat \tau_2^{(j)})^\top\bU\gamma_1\right] := \frac{L}{\sigma_{n, \rm NIE}^{\rm cf, het}} \,.
\end{align*}
Now, define an oracle version $T^0$ with replacing $\hat \tau_1^{(j)}$ by $\tau^{(j), *}$: 
\begin{align*}
    T^0 & = \frac{\sqrt{n}}{2\sigma_{n, \rm NIE}^{\rm cf, het, 0}}\sum_{j = 1}^2 \left[(\hat \omega^{(j)} - \tau_1^{(j), *})^\top \beps^{(j)} +  (\hat \omega^{(j)})^\top \bU'\gamma_1 - (\hat \tau_2^{(j)})^\top\bU\gamma_1\right] := \frac{L^0}{\sigma_{n, \rm NIE}^{\rm cf, het, 0}} \,.
\end{align*}
where 
\begin{align*}
(\sigma_{n, \rm NIE}^{\rm cf, het, 0})^2 & =  \frac n4\sum_{j = 1}^2 \left[\sum_{i \in \cD_j,A_i = 1}\bbE\left[(\hat\omega_i^{(j)}-\tau_{1,i}^{(j), *})^2\nu_1(X_i, M_i) \mid \cX_n\right] \right. \\
& \qquad \qquad \qquad \left. + v_{U,1,1}^2  \|\hat \omega^{(j)}\|_2^2 + v_{U,0,1}^2 \|\hat \tau_2^{(j)}\|_2^2 
\right] \,.
\end{align*}
Here $\cX_n = \sigma(A_{1:n}, X_{1:n})$. We will also use $\cH_n = \sigma(A_{1:n}, X_{1:n}, \{M_i: A_i = 1\})$ and $\cG_n = \sigma(A_{1:n}, X_{1:n}, M_{1:n})$. Clearly $\cX_n \subseteq \cH_n \subseteq \cG_n$. 
We first argue $T^0 \xrightarrow{d} \cN(0,1)$. Towards that end, define 
\begin{align*}
\textstyle
\zeta_{n, i}^{\rm het, NIE} & = \frac{\sqrt{n}}{2\sigma_{n, \rm NIE}^{\rm cf, het, 0}}\sum_{j = 1}^2 \left[\left\{(\hat \omega^{(j)}_i - \tau_{1,i}^{(j), *})\eps_i + \hat \omega^{(j)}_i(U'_i)^\top \gamma_1 \right\}\mathds{1}_{\substack{i \in \cD_j \\ A_i = 1}} \right. \\
& \hspace{20em}\left. - \hat \tau_{2, i}^{(j)} U_i^\top \gamma_1\mathds{1}_{\substack{i \in \cD_j \\ A_i = 0}}\right]
\end{align*}
It is immediate that conditional on $\cX_n$, we have $\bbE[\zeta_{n, i}^{\rm het, NIE} \mid \cX_n] = 0$. This follows from the fact that conditional on $\cX_n$, $\hat \omega^{(j)}$ and $\hat \tau_2^{(j)}$ are fixed. Furthermore: 
\begin{align*}
\bbE[\tau_{1,i}^{(j), *} \eps_i \mid \cX_n] & = \bbE\left[\bbE[\tau_{1,i}^{(j), *} \eps_i \mid \cH_n] \mid \cX_n\right] = \bbE\left[\tau_{1,i}^{(j), *}\bbE[\eps_i \mid \cH_n] \mid \cX_n\right] = 0\,.
\end{align*}
Here we have used the fact that $\tau_1^{(j),*}$ is measurable with respect to $\cH_n$. Furthermore, for $i \in \cD_j$ with $A_i = 1$
\begin{align*}
    & \var(\zeta_{n, i}^{\rm het, NIE} \mid \cX_n) \\
    & = \frac{n}{4(\sigma_{n, \rm NIE}^{\rm cf, het, 0})^2}\left\{\bbE[(\hat \omega^{(j)}_i - \tau_{1,i}^{(j), *})^2\eps^2_i \mid \cX_n] + \bbE[(\hat \omega^{(j)}_i)^2 ((U'_i)^\top \gamma_1)^2 \mid \cX_n] \right. \\
    & \hspace{15em} + \left. 2 \bbE[(\hat \omega^{(j)}_i - \tau_{1,i}^{(j), *})(\omega^{(j)}_i)\eps_i (U'_i)^\top \gamma_1 \mid \cX_n] \right\}\\
    & = \frac{n}{4(\sigma_{n, \rm NIE}^{\rm cf, het, 0})^2}\left\{\bbE[(\hat \omega^{(j)}_i - \tau_{1,i}^{(j), *})^2 \nu_1(X_i, M_i) \mid \cX_n] + (\hat \omega^{(j)}_i)^2  \gamma_1^\top \Sigma_{U, 1} \gamma_1\right\} 
\end{align*}
The expectation of product/cross term is $0$ as $\bbE[\eps_i \mid \cH_n] = 0$ and $\hat \omega_i^{(j)}$ and $(U_i')^\top \gamma_1$ is measurable with respect to $\cH_n$. 
On the other hand, for $i \in\cD_j$ with $A_i = 0$, we have: 
\begin{equation*}
\textstyle
 \var(\zeta_{n, i}^{\rm het, NIE} \mid \cX_n) = \frac{n}{4(\sigma_{n, \rm NIE}^{\rm cf, het, 0})^2}\bbE[(\hat \tau_{2, i}^{(j)} U_i^\top \gamma_1)^2 \mid \cX_n] = \frac{n}{4(\sigma_{n, \rm NIE}^{\rm cf, het, 0})^2}(\hat \tau_{2, i}^{(j)})^2 \gamma_1^\top \Sigma_{U, 0} \gamma_1 \,.
\end{equation*}
Furthermore, the terms $\{\zeta_{n, i}^{\rm het, NIE}\}$ are independent across $i$ conditionally on $\cX_n$. As a consequence, 
\begin{align*}
& \var\left(\sum_{i= 1}^n \zeta_{n, i}^{\rm het, NIE} \mid \cX_n\right) \\
& = \frac{n}{4(\sigma_{n, \rm NIE}^{\rm cf, het, 0})^2}\sum_{j = 1}^2 \left[\sum_{i \in \cD_j,A_i = 1}\bbE\left[(\hat\omega_i^{(j)}-\tau_{1,i}^{(j), *})^2\nu_1(X_i, M_i) \mid \cX_n\right] \right. \\
& \hspace{20em} \left. + v_{U,1,1}^2  \|\hat \omega^{(j)}\|_2^2 + v_{U,0,1}^2 \|\hat \tau_2^{(j)}\|_2^2 
\right] \\
& = 1\,.
\end{align*}
Therefore, all we need to show is 
\begin{equation*}
\textstyle
\sum_{i = 1}^n \bbE\left[\left|\zeta_{n, i}^{\rm het, NIE}\right|^3 \mid \cX_n\right] \xrightarrow{p} 0 \,.
\end{equation*}
To that end, 
\begin{align*}
    & \sum_{i = 1}^n \bbE\left[\left|\zeta_{n, i}^{\rm het, NIE}\right|^3 \mid \cX_n\right]  \\
    & = \frac{n^{3/2}}{8(\sigma_{n, \rm NIE}^{\rm cf, het,0})^3}\sum_i \sum_{j = 1}^2 \left\{\bbE\left[|(\hat \omega^{(j)}_i - \tau_{1,i}^{(j), *})\eps_i + \hat \omega^{(j)}_i(U'_i)^\top \gamma_1|^3 \mid \cX_n\right]\mathds{1}_{\substack{i \in \cD_j \\ A_i = 1}} \right. \\
    & \hspace{20em} \left. + |\hat \tau_{2, i}^{(j)}|^3\bbE\left[|U_i^\top \gamma_1|^3 \mid \cX_n\right]\mathds{1}_{\substack{i \in \cD_j \\ A_i = 0}}\right\} \\
    & \le \frac{n^{3/2}}{8(\sigma_{n, \rm NIE}^{\rm cf, het,0})^3}\sum_i \sum_{j = 1}^2 \left\{4\bbE\left[|(\hat \omega^{(j)}_i - \tau_{1,i}^{(j), *})\eps_i|^3 \mid \cX_n\right]\mathds{1}_{\substack{i \in \cD_j \\ A_i = 1}} \right. \\
    & \hspace{5em} \left. + 4|\hat \omega^{(j)}_i|^3\bbE\left[|(U'_i)^\top \gamma_1|^3 \mid \cX_n\right]\mathds{1}_{\substack{i \in \cD_j \\ A_i = 1}} + |\hat \tau_{2, i}^{(j)}|^3\bbE\left[|U_i^\top \gamma_1|^3 \mid \cX_n\right]\mathds{1}_{\substack{i \in \cD_j \\ A_i = 0}}\right\} 
\end{align*}
Now from the fact $\|\gamma_1\|_2 \le C_\gamma$ (Assumption \ref{assm:params}) and $U_i$ are sub-Gaussian (Assumption \ref{assm:cov_err}), $\bbE[|U_i^\top \gamma|^3] \vee \bbE[|(U'_i)^\top \gamma_1|^3]\le C$. Furthermore, from uniform sub-Gaussianity of $\eps$ (Assumption \ref{assm:cov_err}), we have; $\bbE[|\eps_i|^3 \mid \cH_n] \le C$. Therefore, 
\begin{align*}
   & \sum_{i = 1}^n \bbE\left[\left|\zeta_{n, i}^{\rm het, NIE}\right|^3 \mid \cX_n\right]  \\
   & \le \frac{Cn^{3/2}}{8(\sigma_{n, \rm NIE}^{\rm cf, het})^3} \sum_i \sum_{j = 1}^2 \left\{4\bbE\left[|\hat \omega^{(j)}_i - \tau_{1,i}^{(j), *}|^3 \mid \cX_n\right]\mathds{1}_{\substack{i \in \cD_j \\ A_i = 1}} \right. \\
   & \qquad \qquad \qquad \qquad \qquad \qquad \qquad \qquad \left. + 4|\hat \omega^{(j)}_i|^3\mathds{1}_{\substack{i \in \cD_j \\ A_i = 1}} + |\hat \tau_{2, i}^{(j)}|^3\mathds{1}_{\substack{i \in \cD_j \\ A_i = 0}}\right\} \\
   & \le \frac{C_1 n^{3/2}}{8(\sigma_{n, \rm NIE}^{\rm cf, het})^3}  \sum_i \sum_{j = 1}^2 \left\{ |\hat \omega^{(j)}_i|^3\mathds{1}_{\substack{i \in \cD_j \\ A_i = 1}} + \bbE\left[|\tau_{1,i}^{(j), *}|^3 \mid \cX_n\right]\mathds{1}_{\substack{i \in \cD_j \\ A_i = 1}} + |\hat \tau_{2, i}^{(j)}|^3\mathds{1}_{\substack{i \in \cD_j \\ A_i = 0}}\right\} \\
   & \le \frac{C_1 n^{3/2}}{8(\sigma_{n, \rm NIE}^{\rm cf, het})^3} \sum_{j = 1}^2\left\{\|\hat \omega^{(j)}\|_2^2\|\hat \omega^{(j)}\|_\infty + \|\hat \tau_2^{(j)}\|_2^2 \|\hat \tau_2^{(j)}\|_\infty \right. \\
   & \hspace{20em} \left. + \sum_i \bbE\left[|\tau_{1,i}^{(j), *}|^3 \mid \cX_n\right]\mathds{1}_{\substack{i \in \cD_j \\ A_i = 1}}\right\}
\end{align*}
In Lemma \ref{lem:feasibility} we have established that $n\|\hat \tau_2^{(j)}\|_2^2 = O_p(1)$. By the exact same argument one can prove that $n\|\hat \omega^{(j)}\|_2^2 = O_p(1)$ (as they are obtained from similar optimization problem with $\bX_{c, j}$ replaced by $\bX_{t, j}$). Furthermore, by the constraint in Equation \eqref{eq:balancing_opt_1} (resp. \eqref{eq:balancing_opt_3}) we have $\|\hat \tau_2^{(j)}\|_\infty \le n_{c, j}^{-\alpha} \le C n^{-\alpha}$ (resp. $\|\hat \omega^{(j)}\|_\infty \le n_{t, j}^{-\alpha} \le Cn^{-\alpha}$ where we have used the fact that $n_{c, j} \asymp n_{t, j} \asymp n$. Therefore, as $\alpha > 1/2$, we have: 
\begin{equation*}
\textstyle
n^{3/2} \|\hat \omega^{(j)}\|_2^2\|\hat \omega^{(j)}\|_\infty + n^{3/2} \|\hat \tau_2^{(j)}\|_2^2 \|\hat \tau_2^{(j)}\|_\infty = O_p(n^{(-1-\alpha + 3/2)}) = o_p(1)
\end{equation*}
Next we argue that $n^{3/2}\sum_j \sum_i \bbE[|\tau_{1,i}^{(j), *}|^3 \mid \cX_n]\mathds{1}_{\substack{i \in \cD_j, A_i = 1}} = o_p(1)$. Define $Q_{n, j}$ as: 
\begin{equation*}
\textstyle
Q_{n, j} = \frac{1}{n_j} \sum_{i \in \cD_j}  \bbE\left[|W_i^\top \eta^*|^3 \mathds{1}_{A_i= 1}\mid \cX\right] 
\end{equation*}
Therefore, 
\begin{equation*}
\textstyle
\bbE[Q_{n, j}] = \bbE\left[|W^\top \eta^*|^3 \mathds{1}_{A = 1}\right] \le C \,.
\end{equation*}
Here we have used the fact that $W \mid A = 1$ is sub-Gaussian and $\|\eta^*\|_2 \le \|\eta^*\|_1 \le C_\eta$. Therefore, $Q_{n, j}= O_p(1)$. Now, 
\begin{align*}
    n^{3/2}\sum_j \sum_{i \in \cD_j} \bbE\left[|\tau_{1,i}^{(j), *}|^3 \mid \cX_n \mathds{1}_{A_i = 1}\right] & = n^{3/2}\sum_j \frac{1}{n_{t, j}^{3}} \sum_{i \in \cD_j}\bbE\left[|W_i^\top \eta^*|^3 \mathds{1}_{A_i = 1}\right]  \\
    & = n^{3/2}\sum_j \frac{n_j}{n_{t, j}^{3}} \frac{1}{n_j}\sum_{i \in \cD_j}\bbE\left[|W_i^\top \eta^*|^3 \mathds{1}_{A_i = 1}\right] \\
    & = n^{3/2}\sum_j \frac{n_j}{n_{t, j}^{3}} Q_{n, j} = O_p(n^{-1/2}) = o_p(1)\,.
\end{align*}
Here the last line follows from the fact that $Q_{n, j}= O_p(1)$, $n_j \asymp n_{t, j} \asymp n$. Now, we argue that $\sigma_{n, \rm NIE}^{\rm cf, het, 0}$ is bounded away from $0$ with probability going to $1$. This is very similar to the argument we used to prove $\sigma_{n, \rm NIE}^{\rm cf, het}$ is bounded away from $0$ with probability going to $1$ by observing that: 
\begin{equation*}
\textstyle
(\sigma_{n, \rm NIE}^{\rm cf, het, 0})^2 \ge (n/4) \|\hat \tau_2^{(1)}\|_2^2 v^2_{U, 0, 1} \ge (\lambda_{
\min}(\Sigma_{U, 0})c_\gamma^2/4) \ n\|\hat \tau_2^{(1)}\|_2^2 
\end{equation*}
and the fact that $n\|\hat \tau_2^{(1)}\|_2^2$ is bounded away from $0$ with probability going to $1$. Therefore, by conditional Lindeberg-Feller CLT, we conclude that $T^0 \xrightarrow{d} \cN(0,1)$. Finally, to conclude the proof, we show that $T - T^0 = o_p(1)$. Using the same expansion as in Equation \ref{eq:t_t0_diff}, we have: 
\begin{equation*}
\textstyle
T - T^0 = \frac{L}{\sigma_{n, \rm NIE}^{\rm cf, het}} - \frac{L^0}{\sigma_{n, \rm NIE}^{\rm cf, het, 0}} = \frac{L - L^0}{\sigma_{n, \rm NIE}^{\rm cf, het}} + \frac{L^0}{\sigma_{n, \rm NIE}^{\rm cf, het, 0}}\left(\frac{\sigma_{n, \rm NIE}^{\rm cf, het, 0}}{\sigma_{n, \rm NIE}^{\rm cf, het}} - 1\right) 
\end{equation*}
So to conclude $T - T^0 = o_p(1)$, all we need to prove is $L - L^0 = o_p(1)$ and $|\sigma_{n, \rm NIE}^{\rm cf, het} - \sigma_{n, \rm NIE}^{\rm cf, het, 0}| = o_p(1)$ as these, along with the fact that $\sigma_{n, \rm NIE}^{\rm cf, het}$ is bounded away from $0$ with probability going to $1$ concludes the proof. For the former, 
\begin{equation*}
\textstyle
L - L^0 = \frac{\sqrt{n}}{2} \sum_{j = 1}^2 (\tau_1^{(j), *} - \hat \tau_1^{(j)})^\top \beps^{(j)}
\end{equation*}
Therefore, 
\begin{align*}
\bbE[(L - L^0)^2 \mid \cG_n] & = \frac{n}{4}\sum_{j = 1}^2 \sum_{i \in \cD_j, A_i = 1}(\tau_{1, i}^{(j), *} - \hat \tau_{1, i}^{(j)})^2 \nu_1(X_i, M_i) \\
& \le \frac{c_+}{4}\sum_j n \|\tau_1^{(j), *} - \hat \tau_1^{(j)}\|_2^2 = o_p(1)
\end{align*}
where the last line follows from Equation \eqref{eq:closeness_tau_tau_star}. For the latter, 
\begin{align*}
    & \left|(\sigma_{n, \rm NIE}^{\rm cf, het})^2 - (\sigma_{n, \rm NIE}^{\rm cf, het, 0})^2\right| \\
    & = \frac{n}{4}\sum_{j = 1}^2 \left|\sum_{\substack{i \in \cD_j \\ A_i = 1}}\bbE[(\hat \omega_i^{(j)} - \tau_{1, i}^{(j), *})^2 \nu_1(X_i, M_i) \mid \cX_n] - (\hat \omega_i^{(j)} - \hat \tau_{1, i}^{(j)})^2 \nu_1(X_i, M_i)\right| \\
    & \le \frac{n}{4}\sum_{j = 1}^2 \left|\sum_{\substack{i \in \cD_j \\ A_i = 1}}\bbE[(\hat \omega_i^{(j)} - \tau_{1, i}^{(j), *})^2 \nu_1(X_i, M_i) \mid \cX_n] - (\hat \omega_i^{(j)} - \tau_{1, i}^{(j),*})^2 \nu_1(X_i, M_i)\right| \\
    & \qquad + \frac{n}{4}\sum_{j = 1}^2 \left|\sum_{\substack{i \in \cD_j \\ A_i = 1}}(\hat \omega_i^{(j)} - \tau_{1, i}^{(j),*})^2 \nu_1(X_i, M_i) - (\hat \omega_i^{(j)} - \hat \tau_{1, i}^{(j)})^2 \nu_1(X_i, M_i)\right| \\
    & := S_1 + S_2 \,.
\end{align*}
To bound $S_2$, 
\begin{align*}
    S_2 & \le \frac{nc_+}{4} \sum_{j = 1}^2 \sum_{i \in \cD_j, A_i = 1} \left|(\hat \omega_i^{(j)} - \tau_{1, i}^{(j),*})^2 - (\hat \omega_i^{(j)} - \hat \tau_{1, i}^{(j)})^2\right| \\
    & \le \frac{nc_+}{4}\sum_{j = 1}^2 \|2\hat \omega^{(j)} - \tau_1^{(j), *} - \hat \tau_1^{(j)}\|_2\|\tau_1^{(j), *}- \hat \tau_1^{(j)}\|_2
\end{align*}
We have already established that  $\sqrt{n}\|\tau_1^{(j), *}- \hat \tau_1^{(j)}\|_2 = o_p(1)$ (Equation \eqref{eq:closeness_tau_tau_star}). Furthermore, we know that $n\|\hat \tau_1^{(j)}\|_2^2 = O_p(1)$ from Lemma \ref{lem:feasibility}. By exact same argument, we also have $n\|\hat \omega^{(j)}\|_2^2 = O_p(1)$. Now, from the definition of $\tau_1^{(j), *}$ we have: 
\begin{equation*}
\textstyle
    n \|\tau_1^{(j), *}\|_2^2 = \frac{n}{n_{t, j}^2}\sum_{i \in \cD_j}|W_i^\top \eta^*|^2\mathds{1}_{A_i = 1}
\end{equation*}
Now, from the sub-Gaussianity of $W \mid A = 1$, we have $\bbE[|W^\top \eta^*|^2 \mathds{1}(A = 1)] \le C < \infty$. As a consequence, 
\begin{equation*}
\textstyle
\frac{1}{n_j}\sum_{i \in \cD_j}|W_i^\top \eta^*|^2 \mathds{1}_{A_i = 1} = O_p(1) \,.
\end{equation*}
Therefore, 
\begin{equation*}
\textstyle
n \|\tau_1^{(j), *}\|_2^2 = \frac{nn_j}{n_{j,t}^2} \times \frac{1}{n_j} \sum_{i \in \cD_j}|W_i^\top \eta^*|^2 \mathds{1}_{A_i = 1} = O_p(1) \,.
\end{equation*}
Here, the last conclusion follows from $n_{j, t} \asymp n_j \asymp n$. Therefore, $S_2 = o_p(1)$. Now we argue that $S_1 = o_p(1)$. Set 
\begin{equation*}
\textstyle
\upsilon_{i}^{(j)} = (\hat \omega_i^{(j)} - \tau_{1, i}^{(j), *})^2 \nu_1(X_i, M_i)\mathds{1}_{A_i = 1} - \bbE\left[(\hat \omega_i^{(j)} - \tau_{1, i}^{(j), *})^2 \nu_1(X_i, M_i)\mathds{1}_{A_i = 1} \mid \cX_n\right]
\end{equation*}
Then, 
\begin{equation*}
\textstyle
S_1 = \frac{n}{4}\sum_j n_j \left|\frac{1}{n_j}\sum_{i \in \cD_j} \upsilon^{(j)}_i \right|\,.
\end{equation*}
Now, for any $i \in \cD_j$, 
\begin{align*}
    \var\left(\upsilon_i^{(j)} \mid \cX_n\right) & \le \bbE\left[(\hat \omega_i^{(j)} - \tau_{1, i}^{(j), *})^4 \nu_1^2(X_i, M_i)\mathds{1}_{A_i = 1} \mid \cX_n\right] \\
    & \le 8c_+^2 \left\{(\hat \omega_i^{(j)})^4\mathds{1}(A_i = 1) + \bbE[(\tau_{1, i}^{(j), *})^4\mathds{1}(A_i = 1) \mid \cX_n]\right\}
\end{align*}
Hence, 
\begin{align*}
\var\left(\frac{1}{n_j}\sum_{i \in \cD_j} \upsilon_i \mid \cX_n\right) & = \frac{1}{n_j^2}\sum_{i \in \cD_j} \var(\upsilon_i \mid \cX_n) \\
& \le \frac{8c_+^2}{n_j^2} \|\hat \omega^{(j)}\|_2^2\|\hat \omega^{(j)}\|_\infty^2 + \frac{8c_+^2}{n_j^2} \sum_{i \in \cD_j} \bbE[(\tau_{1, i}^{(j), *})^4\mathds{1}(A_i = 1) \mid \cX_n]
\end{align*}
Now, we have already argued that $\|\hat \omega^{(j)}\|_2^2 = O_p(n^{-1})$ and by feasibility of $\|\hat \omega^{(j)}\|_\infty^2 \le O_p(n^{-2\alpha})$. Consequently $ \|\hat \omega^{(j)}\|_2^2\|\hat \omega^{(j)}\|_\infty^2/n_j^2 = O_p(n^{-3-2\alpha})$. On other other using the fact $\tau^{(j)}_{1, i} = (W_i^\top \eta^*)/n_{t, j}$ and $W \mid A = 1$ is sub-Gaussian along with $\|\eta^*\|_2 \le \|\eta^*\|_1 \le C_\eta$, we have: 
\begin{align*}
    \frac{8c_+^2}{n_j^2} \sum_{i \in \cD_j} \bbE[(\tau_{1, i}^{(j), *})^4 \mid \cX_n] & = \frac{8c_+^2}{n_j n_{t, j}^4} \frac{1}{n_j}\sum_{i \in \cD_j} \bbE[|W_i^\top \eta^*|^4 \mathds{1}(A_i = 1)] = O_p(n_j^{-5}) \,.
\end{align*}
Here the last conclusion follows from the fact that $n_{t, j} \asymp n_j \asymp n$. Therefore, 
\begin{equation*}
\textstyle
\var\left(\frac{1}{n_j}\sum_{i \in \cD_j} \upsilon_i \mid \cX_n\right) = O_p(n^{-3-2\alpha}) \implies \frac{1}{n_j}\sum_{i \in \cD_j} \upsilon_i = O_p(n^{-3/2 - \alpha}) \,.
\end{equation*}
Therefore, 
\begin{equation*}
\textstyle
S_1 = \frac{n}{4}\sum_j n_j \frac{1}{n_j}\sum_{i \in \cD_j} \upsilon^{(j)}_i = O_p(n^{2-3/2-\alpha}) = O_p(n^{1/2 - \alpha}) = o_P(1) \,,
\end{equation*}
as $\alpha > 1/2$. This completes the proof. 

Finally, we establish the asymptotic normality of $\hat \theta_{\rm NIE}^{\rm cf}$ around the population parameter $\theta_{\rm NIE}^{\rm pop}$. The proof follows the same argument as the population extensions for $\theta_{10}$ and $\theta_{\rm NDE}$; therefore, we only highlight the changes specific to the NIE. To keep notational similarity, define $A_n = \sqrt{n}(\hat\theta_{\rm NIE}^{\rm cf} - \theta_{\rm NIE})$ and $B_n = \sqrt{n}(\theta_{\rm NIE} - \theta_{\rm NIE}^{\rm pop})$. Since $\theta_{\rm NIE} = \bar X^\top\Lambda_{\rm NIE}$ and $\theta_{\rm NIE}^{\rm pop} = \mu_X^\top\Lambda_{\rm NIE}$, we can write 
\begin{equation*}
\textstyle
B_n = \frac1{\sqrt{n}}\sum_{i=1}^n(X_i-\mu_X)^\top\Lambda_{\rm NIE} \,.
\end{equation*}
Furthermore, it is immediate from the definition that $B_n$ is measurable with respect to $\cX_n$. A similar argument, used to prove Lemma \ref{lem:conditional-clt} can be used to establish that for any $u \in \reals$, 
\begin{equation*}
\textstyle
\bbE\left[\exp\left\{iu\frac{A_n}{\sigma_{n,\rm NIE}^{\rm cf,het}}\right\} \mid \cX_n\right] \xrightarrow{p} e^{-u^2/2}.
\end{equation*}
Consequently, by the same conditional characteristic-function calculation as in the proof of Corollary~\ref{cor:pop_extension}, we have, for any $u,v\in\reals$,
\begin{align}
\label{eq:joint_cf_population_NIE}
& \left|\bbE\left[\exp\left\{iu\frac{A_n}{\sigma_{n,\rm NIE}^{\rm cf,het}} + ivB_n\right\}\right] - e^{-u^2/2}\bbE[e^{ivB_n}]\right| \longrightarrow 0 \,.
\end{align}
Now, as $\omega_{n,\rm NIE}^{\rm het}$ admits an asymptotically deterministic equivalent, there exists a deterministic sequence $\{\bar \omega_{n, \rm NIE}^{\rm het}\}_{n \in \bbN}$ with $\bar \omega_{n, \rm NIE}^{\rm het} \in [0, 1]$ such that $\omega_{n,\rm NIE}^{\rm het} - \bar \omega_{n,\rm NIE}^{\rm het} \to 0$ in probability. 
Fix an arbitrary subsequence. Since
$\bar\omega_{n,\rm NIE}^{\rm het}\in[0,1]$ and
$\sigma_{X,\rm NIE}^2\le C$, there exists a further subsequence, which we continue to index by $n$, such that
\begin{equation*}
\textstyle
\bar\omega_{n,\rm NIE}^{\rm het} \to \omega\in[0,1], \qquad \sigma_{X,\rm NIE}^2 \to \zeta^2
\end{equation*}
for some $\zeta\ge0$. Consequently, along this subsequence, $\omega_{n,\rm NIE}^{\rm het} \to \omega$ in probability. 
First, assume $\zeta>0$. By Lindeberg-Feller central limit theorem (as $X_i$'s are sub-Gaussian and $\|\Lambda_{\rm NIE}\|_2 \le C_{\rm NIE}$ is bounded (Assumption \ref{assm:params}), 
\begin{equation*}
\textstyle
\frac{B_n}{\sigma_{X,\rm NIE}} \xrightarrow{d}\cN(0,1).
\end{equation*}
Combining this with Equation~\eqref{eq:joint_cf_population_NIE} yields: 
\begin{equation*}
\textstyle
\left(\frac{A_n}{\sigma_{n,\rm NIE}^{\rm cf,het}}, \frac{B_n}{\sigma_{X,\rm NIE}}\right) \xrightarrow{d} (Z_1,Z_2), \quad Z_1, Z_2 \overset{\iid}{\sim} \cN(0, 1) \,. 
\end{equation*}
Define $D_{n,\rm NIE}^2 = (\sigma_{n,\rm NIE}^{\rm cf,het})^2 + \sigma_{X,\rm NIE}^2$. Then, by Slutsky's theorem, 
\begin{equation*}
\textstyle
\frac{A_n+B_n}{D_{n,\rm NIE}} = \sqrt{\omega_{n,\rm NIE}^{\rm het}}\frac{A_n}{\sigma_{n,\rm NIE}^{\rm cf,het}} + \sqrt{1-\omega_{n,\rm NIE}^{\rm het}} \frac{B_n}{\sigma_{X,\rm NIE}} \xrightarrow{d}\sqrt{\omega}\,Z_1 + \sqrt{1-\omega}\,Z_2 \sim\cN(0,1).
\end{equation*}
Now, if $\zeta=0$, then it is immediate that $\bbE[B_n^2] = \sigma_{X,\rm NIE}^2 \to 0$ and consequently $B_n=o_p(1)$. Moreover, we have already established that $(\sigma_{n,\rm NIE}^{\rm cf,het})^2$ is bounded away from zero with probability going to one. Consequently, $D_{n,\rm NIE}/\sigma_{n,\rm NIE}^{\rm cf,het} \to 1$ in probability and $B_n/D_{n,\rm NIE} = o_p(1)$. Hence, 
\begin{equation*}
\textstyle
\frac{A_n+B_n}{D_{n,\rm NIE}} = \frac{A_n}{\sigma_{n,\rm NIE}^{\rm cf,het}} + o_p(1) \xrightarrow{d} \cN(0,1) \,.
\end{equation*}
Thus every subsequence contains a further subsequence along which
\begin{equation*}
\textstyle
\frac{A_n+B_n}{\sqrt{(\sigma_{n,\rm NIE}^{\rm cf,het})^2 + \sigma_{X,\rm NIE}^2}} \xrightarrow{d} \cN(0,1) \,,
\end{equation*}
and the limit is same for any subsequence. Therefore, the entire sequence converges to $\cN(0, 1)$. This completes the proof.
\end{proof}

\section{{\color{black}Proof of Additional Lemmas}
\label{sec:additional_lemmas}}

\subsection{A Lemma on Sample Restricted Eigenvalue Conditions}
{\color{black}
\begin{lemma}
\label{lem:sample_RE}
{\color{black}Let $s_{X,n}=(1+s)k, \ \ s_{W,n}=k$. 
Under Assumptions~\ref{assm:cov_err} and \ref{assm:params}, suppose
$$
s_{X,n}\log\left(\frac{ep}{r_{X,n}}\right)=o(n), \qquad s_{W,n}\log\left(\frac{e(p+q)}{r_{W,n}}\right)=o(n).
$$
Then, for any fixed cone constant $L<\infty$, there exist constants $c_{X,\rm RE},c_{W,\rm RE}>0$, independent of $(n,p,q)$, such that, with probability going to one, simultaneously over both arms and both data folds,
$$
\inf_{\substack{|S|\le s_{X,n}\\
\delta \neq 0, \ \|\delta_{S^c}\|_1\le L\|\delta_S\|_1}}
\frac{\delta^\top\left(\bX_{a,j}^\top\bX_{a,j}/n_{a,j}\right)\delta}{\|\delta\|_2^2} \ge c_{X,\rm RE},
$$
and
$$
\inf_{\substack{|S|\le s_{W,n}\\ \delta\neq0,\, \|\delta_{S^c}\|_1\le L\|\delta_S\|_1}}\frac{\delta^\top\left(\bW_{a,j}^\top\bW_{a,j}/n_{a,j}\right)\delta}{\|\delta\|_2^2} \ge c_{W,\rm RE}.
$$
Consequently, the corresponding compatibility constants are bounded away from zero with probability tending to one.}
\end{lemma}
}

\begin{proof}
    As the proof is the same for the first and the second conclusion of the lemma, we present the proof in a unified way. Let $V$ denote either $X$ or $W$. We will use the notation $d = \dim(V)$; when $V= X$ then $d = p$ and when $V = W$ then $d = p+q$. Fix $a \in \{0, 1\}$ and let $\mu_{V, a} = \bbE[V \mid A = a]$ and $Q_{V, a} = \bbE[VV^\top \mid A = a]$ be the conditional mean and second moment of $V$ given $A = a$. Now writing $\Sigma_{V, a} = \var(V \mid A = a)$, we have $Q_{V, a} = \Sigma_{V, a} +  \mu_{V, a}\mu_{V, a}^\top$. By Assumptions \ref{assm:cov_err} and \ref{assm:params}, along with Remark \ref{rem:sub-G_W}, there exist constants $0<c_V<C_V<\infty$ (independent of $(n,p,q)$) such that: 
\begin{equation*}
\textstyle
\lambda_{\min}(Q_{V,a})\geq c_V,
\qquad
\max_{1\leq \ell\leq d}(Q_{V,a})_{\ell\ell} \le C_V \,.
\end{equation*}
Therefore, if we define $Z_{V, a} = Q_{V, a}^{-1/2}V$, then we have $\bbE[Z_{V, a}Z_{V, a}^\top \mid A = a] = \bI_d$. Furthermore, conditional on $A = a$, $Z_{V, a}$ is sub-Gaussian random vector as
\begin{equation*}
\textstyle
\|u^\top Z_{V,a}\|_{\psi_2\mid A=a} \le \|u^\top Q_{V,a}^{-1/2}(V-\mu_{V,a})\|_{\psi_2\mid A=a} + C|u^\top Q_{V,a}^{-1/2}\mu_{V,a}| \le C\|u\|_2 \,.
\end{equation*}
Here, we have used the fact that $\lambda_{\min}(Q_{V,a})\geq c_V$ and $\|Q_{V,a}^{-1/2}\mu_{V,a}\|_2^2\leq 1$ which follows from the fact that $\mu_{V,a}\mu_{V,a}^\top\preceq Q_{V,a}$. Let $I_{a,j}:=\{i:i\in\mathcal D_j,\ A_i=a\}$ denotes that set of all observations with $A_i = a$ in the $j^{th}$ fold and let $\cI_n = \sigma(A_{1:n})$, the sigma-field generated by the treatment indicators. It is immediate that $n_{a, j} = |I_{a, j}|$ and conditionally on $\cI_n$, $n_{a, j}$ is fixed and the rows of $\bV_{a, j}$ are i.i.d. with the common distribution $V \mid A = a$. Hence, we have: 
\begin{equation*}
\textstyle
\bbE\left[\frac{\bV_{a,j}^{\top}\bV_{a,j}}{n_{a,j}} \mid \cI_n\right] = \bbE[VV^\top\mid A=a] = Q_{V,a}.
\end{equation*}
Therefore, if we write $\bV_{a,j} = \Psi_{V,a,j}Q_{V,a}^{1/2}$, then, conditionally on $\cI_n$, where the rows of $\Psi_{V,a,j}$ are independent along with second moment (conditionally on $A = a$) being $\bI_d$. Therefore, an application of Theorem 15 of \cite{rudelson2012reconstruction} with $A = Q_{V, a}^{1/2}$, sparsity level $s_{V, n}$ implies: 
\begin{equation*}
\textstyle
n_{a,j} \gtrsim s_{V,n}\log\left(\frac{ed}{s_{V,n}}\right) \implies \frac{1}{n_{a,j}}\|\bV_{a,j}\delta\|_2^2 \geq c_{V,\mathrm{RE}}\|\delta\|_2^2
\end{equation*}
simultaneously for all sets $S$ with $|S|\leq s_{V,n}$ and all $\delta \neq 0$ satisfying $\|\delta_{S^c}\|_1\leq L\|\delta_S\|_1$. 
However, to apply this theorem, we need to show $Q_{V,a}^{1/2}$ satisfies the RE condition. This immediately follows from our assumptions; since $Q_{V,a} = \Sigma_{V,a}+\mu_{V,a}\mu_{V,a}^\top \succeq \Sigma_{V,a} \succeq c_V \bI_d$. we have, for every $\delta \in \reals^{d}$,
\begin{equation*}
\textstyle
\|Q_{V,a}^{1/2}\delta\|_2 \ge \sqrt{c_V}\|\delta\|_2.
\end{equation*}
Therefore, for every $S$ with $|S|\leq s_{V,n}$ and every $\delta$ satisfying $\|\delta_{S^c}\|_1\leq 3L\|\delta_S\|_1$, we have $\|Q_{V,a}^{1/2}\delta\|_2 \ge \sqrt{c_V}\|\delta_S\|_2$. Hence $Q_{V,a}^{1/2}$ satisfies $\mathrm{RE}(s_{V,n},3L)$, with $K(s_{V,n},3L,Q_{V,a}^{1/2}) \le c_V^{-1/2}$.
Finally, for any $\delta$ in the corresponding cone, we have $\|\delta_S\|_1^2 \le |S|\|\delta_S\|_2^2 \le s_{V,n}\|\delta\|_2^2$. Therefore, 
\begin{equation*}
\textstyle
\frac{
s_{V,n}\delta^\top(\bV_{a,j}^\top\bV_{a,j}/n_{a,j})\delta}{\|\delta_S\|_1^2} \ge c_{V,\mathrm{RE}},
\end{equation*}
so the corresponding compatibility constants are also bounded away from zero with probability going to one.
\end{proof}

\subsection{Proof of Lemma \ref{lem:tau_concentration_revised}}
\label{sec:proof_lem_tau_conc}
\begin{proof}
    We only prove the case for $j = 2$ (as it is exactly stated in Algorithm \ref{algomain_modified}), since the case for $j = 1$ is analogous. 
    For notational simplicity, we prove the result for \(j=2\) and suppress the fold indices throughout the proof. Define
\begin{equation*}
\textstyle
a:=a^{(1)}
=
\begin{pmatrix}
\bar X\\
\hat m_c
\end{pmatrix},
\qquad
a^0
=
\begin{pmatrix}
\bar X\\
B_0\bar X
\end{pmatrix},
\qquad
a^*:=a_{10}^*
=
\begin{pmatrix}
\mu_X\\
B_0\mu_X
\end{pmatrix},
\end{equation*}
where \(\hat m_c=\hat m_{c,1}\). Thus, \(a\) is the actual random contrast used in Algorithm \ref{algomain_modified}, \(a^0\) is its oracle sample counterpart obtained by replacing \(\hat m_c\) with \(B_0\bar X\), and \(a^*\) is the corresponding population contrast. 
We also write $\eta^*:=\eta_{10}^*=\Omega_{W,1}^{-1}a^*$. 
The optimization problem in Equation \eqref{eq:balancing_opt_2} can be cleanly written as: 
\begin{equation}
\label{eq:generic_balancing_opt}
\textstyle
\begin{aligned}
\min_{\tau \in \reals^{n_t}} \quad & \frac{1}{2} \|\tau\|_2^2 \\
\text{subject to} \quad & \bigl\| a - \mathbf{W}_t^\top \tau \bigr\|_\infty 
    \le \check{\lambda}_n := K_1 \sqrt{\frac{\log{(p+q)}}{n_t}} \,, \\
& \|\tau\|_\infty \le b_n := n_t^{-\alpha} \,,
\end{aligned}
\end{equation}
and $\hat \tau(a)$ is the unique optimizer of the above optimization problem (it exists and is unique as the objective is strongly convex over a convex compact set).
For any \(v\in\mathbb R^{p+q}\), let \(\hat\tau(v)\) denote the unique optimizer of Equation \eqref{eq:generic_balancing_opt} with \(a\) replaced by \(v\). Therefore, it is immediate that $\hat\tau(a)=\hat\tau_2^{(1)}$ and $\hat\tau(a^0)=\hat\tau_2^{(1),0}$. The Lagrangian of the above optimization problem can be written as, given non-negative vectors (co-ordinatewise) $\{\eta_{i, +}\}_{i = 1}^{p+q}, \{\eta_{i, -}\}_{i = 1}^{p+q}$ and $\{\zeta_{i, +}\}_{i = 1}^{n_t}, \{\zeta_{i, -}\}_{i = 1}^{n_t}$: 
\allowdisplaybreaks
\begin{align*}
    & \bbL\left(\tau, \{\eta_{i, +}\}, \{\eta_{i, -}\}, \{\zeta_{i, +}\}, \{\zeta_{i, -}\}\right) \\
    & = \frac{1}{2} \|\tau\|_2^2 + \sum_{i = 1}^{p+q} \eta_{i,+}\left((a - \bW_t^\top \tau)_i - \check \lambda_{n_t}\right) +   \sum_{i = 1}^{p+q} \eta_{i,-}\left((\bW_t^\top \tau - a)_i - \check \lambda_{n_t}\right) \\
    & \hspace{10em} +  \sum_{i = 1}^{n_t} \zeta_{i,+}\left(\tau_i - b_n\right) + \sum_{i = 1}^{n_t} \zeta_{i,-}\left(-\tau_i - b_n\right) \\
    & = \frac{1}{2} \|\tau\|_2^2 + \sum_{i = 1}^{p+q} (\eta_{i,+} - \eta_{i, -})(a - \bW_t^\top \tau)_i - \check \lambda_{n_t} \sum_{i = 1}^{p+q}(\eta_{i,+} + \eta_{i, -}) \\
    & \hspace{10em}  + \sum_{i = 1}^{n_t} (\zeta_{i,+} - \zeta_{i, -})\tau_i - b_n \sum_{i = 1}^{n_t} (\zeta_{i,+} + \zeta_{i, -})
\end{align*}
Define the signed multipliers $\eta:=\eta_{+}-\eta_{-}$, and $\zeta:=\zeta_{+}-\zeta_{-}$. For any fixed signed vector $\eta$, every nonnegative decomposition
$\eta=\eta_{+}-\eta_{-}$ satisfies, coordinate-wise, $\eta_{j,+}+\eta_{j,-}\ge |\eta_j|,$ with equality for the canonical decomposition $\eta_{j,+}=(\eta_j)_+$ and $\eta_{j,-}=(-\eta_j)_+$. Since the dual objective contains $-\check\lambda_{n_t}\mathbf 1^\top(\eta_{+}+\eta_{-})$, while all its remaining terms depend on $(\eta_{+},\eta_{-})$ only through $\eta_{+}-\eta_{-}$, any overlap between $\eta_{+}$ and $\eta_{-}$ can be removed without changing the signed multiplier and while weakly
increasing the dual objective. Hence, without loss of generality, $\mathbf 1^\top(\eta_{+}+\eta_{-})=\|\eta\|_1.$ The same argument gives $\mathbf 1^\top(\zeta_{+}+\zeta_{-})=\|\zeta\|_1.$ Consequently, the Lagrangian may be written in terms of the signed multipliers as
\begin{equation}
    \label{eq:lagrangian_compact}
    \bbL(\tau, \{\eta_i\}, \{\zeta_i\}) = \frac12 \|\tau\|_2^2 + \eta^\top(a - \bW_t^\top \tau) + \zeta^\top \tau - \check \lambda_{n_t} \|\eta\|_1 - b_n \|\zeta\|_1 \,.
\end{equation}
We will also use this reduced form in our analysis. Henceforth, we will use the notation $\hat \eta^{\rm KKT}(a), \hat \zeta^{\rm KKT}(a)$ to denote the dual optimal parameters of this optimization problem. From the complementary slackness of the KKT condition, we have: 
\begin{align*}
\hat\eta^{\mathrm{KKT}}_{j,+}(a)\left\{(a-\bW_t^\top\hat\tau(a))_j -\check\lambda_{n_t}\right\} &=0, \\
\hat\eta^{\mathrm{KKT}}_{j,-}(a)\left\{(\bW_t^\top\hat\tau(a)-a)_j -\check\lambda_{n_t}\right\} &=0.
\end{align*}
Similarly, for every \(i=1,\ldots,n_t\),
\begin{align*}
\hat\zeta^{\mathrm{KKT}}_{i,+}(a)\left\{\hat\tau_i(a)-b_n\right\} =0,\qquad
\hat\zeta^{\mathrm{KKT}}_{i,-}(a)\left\{-\hat\tau_i(a)-b_n\right\} =0.
\end{align*}
Let us first focus on $\hat \eta^{\rm KKT}$. From the above condition, it is immediate that when $|(a - \bW_t^\top \hat \tau(a))_j| \neq \check \lambda_{n_t}$, then $\hat \eta^{\rm KKT}_{j, +}(a) = \hat \eta^{\rm KKT}_{j, -}(a) = 0$ and hence $\hat \eta^{\rm KKT}_j(a)  = 0$. On the other hand when $(a - \bW_t^\top \hat \tau(a))_j = \check \lambda_{n_t}$, then $\hat \eta^{\rm KKT}_{j, -}(a) = 0$ and consequently $\eta^{\rm KKT}_j \ge 0$. Similarly, when $(a - \bW_t^\top \hat \tau(a))_j = -\check \lambda_{n_t}$, then $\hat \eta^{\rm KKT}_{j, +}(a) = 0$ and consequently $\eta^{\rm KKT}_j \le 0$. Therefore, $\hat \eta^{\rm KKT}(a) \in \cN_{\cB_\infty(\check \lambda_{n_t})}(a - \bW_t^\top \hat \tau(a))$, where $\cN_\cA(x)$ is the normal cone of $\cA$ at $x$ defined as:
\begin{equation}
\textstyle
\label{def:normal_cone}
\cN_\cA(x) = \left\{g: \langle g, y - x \rangle \le 0, \ \forall \ y \in \cA\right\} \,.
\end{equation}
By a similar logic, we also conclude that $\hat\zeta^{\rm KKT}(a)_i = 0$ for $|\hat \tau(a)_i| \neq b_n$, $\hat\zeta^{\rm KKT}(a)_i \ge 0$ for $\hat \tau(a)_i  = b_n$ and $\hat \zeta^{\rm KKT}_i(a) \le 0$ for $\hat \tau(a)_i = -b_n$. Therefore, $\hat \zeta^{\rm KKT}(a) \in \cN_{\cB_\infty(b_n)}(\hat \tau(a))$. 
The following lemma relates the difference $\hat \tau(a) - \hat \tau(a^0)$ to that of $(a - a^0)$ and the dual variable $(\hat \eta^{\rm KKT}(a) - \hat \eta^{\rm KKT}(a^0))$: 
{\color{black}
\begin{lemma}
    \label{lem:rel_tau_a_eta}
   {\color{black} The primal optimizer $(\hat \tau(a), \hat \tau(a^0))$ and the dual optimizer $(\hat \eta^{\rm KKT}(a) - \hat \eta^{\rm KKT}(a^0))$ satisfies: 
    \begin{equation*}
    \textstyle
    \|\hat \tau(a) - \hat \tau(a^0)\|_2^2 \le \|a - a^0\|_\infty \|\hat \eta^{\rm KKT}(a) - \hat \eta^{\rm KKT}(a^0)\|_1 \,.
    \end{equation*}}
\end{lemma}
}

\begin{proof}
    The key ingredient of the proof of this lemma is a simple fact about the normal cone: let $g \in \cN_\cA(x)$ and $g' \in \cN_{\cA}(x')$. Then $\langle g - g', x - x'\rangle \ge 0$. The proof is simple: taking \(y=x'\) in the definition of \(g\in N_A(x)\) and \(y=x\) in the definition of \(g'\in N_A(x')\) gives
\(\langle g,x'-x\rangle\leq 0\) and \(\langle g',x-x'\rangle\leq 0\), which together imply
\(\langle g-g',x-x'\rangle\geq 0\).
    From the first KKT condition, we have, from Equation \eqref{eq:lagrangian_compact}: 
    \begin{equation}
    \label{eq:KKT_cond_tau}
    \textstyle
    \hat \tau(a) - \bW_t \hat \eta^{\rm KKT}(a) + \hat \zeta^{\rm KKT}(a) = 0 \implies \hat \tau(a) = \bW_t \hat \eta^{\rm KKT}(a) - \hat \zeta^{\rm KKT}(a) \,.
    \end{equation}
    For notational simplicity, define $\Delta \hat \tau = \hat \tau(a) - \hat \tau(a^0)$ (and analogously define $\Delta \hat \eta$ and $\Delta \hat \zeta$). Then, from the above equation, we have: 
    \begin{equation*}
    \textstyle
    \Delta \hat \tau = \bW_t \Delta \hat \eta - \Delta \hat \zeta \implies \|\Delta \hat \tau\|_2^2 = (\bW_t^\top \Delta \hat \tau)^\top \Delta \hat \eta - \Delta \hat \tau^\top \Delta \hat \zeta \,.
    \end{equation*}
    Now, $\hat \zeta(a) \in \cN_{\cB_\infty(b_n)}(\hat \tau(a))$ and $\hat \zeta(a^0) \in \cN_{\cB_\infty(b_n)}(\hat \tau(a^0))$. Therefore, from the property of the normal cone described above, we have $\Delta \hat \tau^\top \Delta \hat \zeta \ge 0$, and consequently $\|\Delta \hat \tau\|_2^2 \le (\bW_t^\top \Delta \hat \tau)^\top \Delta \hat \eta$. Now for the next part, we rewrite: 
    \begin{equation*}
    \textstyle
    \bW_t^\top \Delta \hat \tau = (\bW_t^\top \hat \tau(a) - a) - (\bW_t^\top \hat \tau(a^0) - a^0) + (a - a^0) \,.
    \end{equation*}
    Hence, 
    \begin{equation*}
    \textstyle
    (\bW_t^\top \Delta \hat \tau)^\top \Delta \hat \eta = \left((\bW_t^\top \hat \tau(a) - a) - (\bW_t^\top \hat \tau(a^0) - a^0)\right)^\top \Delta\hat \eta + (a - a^0)^\top \Delta \hat \eta
    \end{equation*}
    The first summand on the right-hand side is again negative from the property of the normal cone, and consequently: 
    \begin{equation*}
    \textstyle
    \|\Delta \hat \tau\|_2^2 \le (\bW_t^\top \Delta \hat \tau)^\top \Delta \hat \eta \le (a - a^0)^\top \Delta \hat \eta \le \|a - a^0\|_\infty \|\hat \eta^{\rm KKT}(a) - \hat \eta^{\rm KKT}(a^0)\|_1
    \end{equation*}
    where the last inequality follows from the H\"older inequality. This completes the proof. 
\end{proof}
In the rest of the analysis, we will establish that $n\|\hat \eta^{\rm KKT}(a)\|_1 \vee n\|\hat \eta^{\rm KKT} (a^0)\|_1 = O_p(1)$. This, along with the fact that $\|a - a^0\|_\infty = o_p(1)$ establishes the claim. Towards that end, we will start with the reduced form Lagrangian in Equation \eqref{eq:lagrangian_compact}. Now for any fixed $(\eta, \zeta)$, $\bbL(\tau, \{\eta_i\}, \{\zeta_i\})$ is minimized (with respect to $\tau$) at $\tau^*(\eta, \zeta) = \bW_t \eta - \zeta$. As a consequence: 
\begin{align*}
    \min_{\tau} \bbL(\tau, \{\eta_i\}, \{\zeta_i\}) & = \frac12 \|\bW_t \eta - \zeta\|_2^2 + \eta^\top a - \|\bW_t \eta - \zeta\|_2^2 - \check \lambda_{n_t} \|\eta\|_1 - b_n \|\zeta\|_1 \\
    & = -\left\{\frac12 \|\bW_t \eta - \zeta\|_2^2 - \eta^\top a + \check \lambda_{n_t} \|\eta\|_1 + b_n \|\zeta\|_1 \right\} \,.
\end{align*}
Therefore, the dual optimization problem is:
\begin{align*}
    & \max_{\eta, \zeta} \  -\left\{\frac12 \|\bW_t \eta - \zeta\|_2^2 - \eta^\top a + \check \lambda_{n_t} \|\eta\|_1 + b_n \|\zeta\|_1 \right\}\\
    = &- \min_{\eta, \zeta} \  \left\{\frac12 \|\bW_t \eta - \zeta\|_2^2 - \eta^\top a + \check \lambda_{n_t} \|\eta\|_1 + b_n \|\zeta\|_1 \right\}
\end{align*}
Now let us re-parametrize $\eta_n = n_t \eta$ and $\zeta_n = n_t \zeta$. 
With this reparametrization, all we need to show is $\|\hat \eta(a)\|_1 \vee \|\hat \eta(a^0)\|_1 = O_p(1)$. 
Then, we can rewrite the dual problem with this new definition of $(\eta, \zeta)$ as: 
\begin{equation}
\textstyle
\label{eq:rescaled_dual_KKT}
-\min_{\eta, \zeta} \  \left\{\frac{1}{2n_t} \|\bW_t \eta - \zeta\|_2^2- \eta^\top a + \check \lambda_{n_t} \|\eta\|_1 + b_n \|\zeta\|_1 \right\}
\end{equation}
Now, for a fixed $\eta$, the optimization over $\zeta$ is simple, as the objective function is separable over coordinates: 
\begin{equation*}
\textstyle
\min_\zeta \ \left\{\frac{1}{2n_t} \|\bW_t \eta - \zeta\|_2^2 + b_n \|\zeta\|_1\right\} = \min_{\zeta} \ \left\{\sum_{j = 1}^{n_t} \left(\frac{1}{2n_t}((\bW_t \eta)_i - \zeta_i)^2 + b_n |\zeta_i|\right)\right\}
\end{equation*}
It is well-known that for each coordinate the objective is the proximal operator of $\ell_1$ norm and hence the minimizer $\zeta_i^* = \sign((\bW_t \eta)_i)(|(\bW_t \eta)_i| - n_tb_n)_+$ and the minimum value is the Huber function: 
$$
\min_{\zeta_i} \frac{1}{2n_t}((\bW_t \eta)_i - \zeta_i)^2 + b_n |\zeta_i| = 
\begin{cases}
    \frac{(\bW_t \eta)_i^2}{2n_t}, & \text{ if } |(\bW_t \eta)_i| \le n_tb_n \\
        b_n|(\bW_t \eta)_i| - \frac{n_tb_n^2}{2}, & \text{ if } |(\bW_t \eta)_i| > n_tb_n
\end{cases}
$$
Therefore, if we define Huber function $H_\lambda(x)$ as $x^2/2$ when $|x| \le \lambda$ and $\lambda |x| - \lambda^2/2$ for $|x| > \lambda$, then we have: 
\begin{align*}
& \min_{\eta, \zeta} \  \left\{\frac{1}{2n_t} \|\bW_t \eta - \zeta\|_2^2- \eta^\top a + \check \lambda_{n_t} \|\eta\|_1 + b_n \|\zeta\|_1 \right\} \\
& \qquad \qquad \qquad \qquad = \min_{\eta} \left\{\frac{1}{n_t}
\sum_{i=1}^{n_t}
H_{n_tb_n}\left((\bW_t\eta)_i\right) - \eta^\top a + \check \lambda_{n_t} \|\eta\|_1\right\}
\end{align*}
Therefore, we have: 
\begin{align}
\label{eq:KKT_dual_opt_prob}
\hat \eta_n(a) & = \argmin_\eta \left\{\frac{1}{n_t}
\sum_{i=1}^{n_t}
H_{n_tb_n}\left((\bW_t\eta)_i\right) - \eta^\top a + \check \lambda_{n_t} \|\eta\|_1\right\} \notag \\
& := \argmin_\eta \ \left\{\cL_{n, a}(\eta) + \check \lambda_{n_t} \|\eta\|_1\right\} \,.
\end{align}
Furthemore, we have $\hat \eta_n(a) =  n_t \hat \eta^{\rm KKT}(a)$. A similar relation holds for $a^0$ as well. By Assumption~\ref{assm:params}, we have $\|\eta^*\|_1\le C_\eta$. We next bound \(\|a^0-a^*\|_\infty\) and
\(\|a-a^*\|_\infty\): 
\begin{align}
\label{eq:a0_astar_close}
\|a^0-a^*\|_\infty &\le \max\left\{\|\bar X-\mu_X\|_\infty,\,\|\bB_0(\bar X-\mu_X)\|_\infty\right\} \notag\\
&\le (1+\|\bB_0\|_{1,\infty}) \|\bar X-\mu_X\|_\infty = O_p\left(\sqrt{\frac{\log d}{n}}\right) \\
\label{eq:ahat_astar_close}
\|a-a^*\|_\infty  & \le (1+\|\bB_0\|_{1,\infty}) \|\bar X-\mu_X\|_\infty + \|\hat m_c-\bB_0\bar X\|_\infty \notag\\
&= O_p\left(\sqrt{\frac{\log d}{n}}\right).
\end{align}
Let $\hat \Omega_{W, 1} = (\bW_t^\top \bW_t)/n_t$.  Since the rows of $\bW_t$ are uniformly sub-Gaussian with bounded coordinate means, we have $\|\hat\Omega_{W,1}-\Omega_{W,1}\|_{\max} = O_p(\sqrt{\log{(p+q)}/n_t})$. Consequently, simultaneously for $v\in\{a,a^0\}$,
\begin{equation}
\textstyle
\|\hat\Omega_{W,1}\eta^*-v\|_\infty \le \|\hat\Omega_{W,1}-\Omega_{W,1}\|_{\max} \|\eta^*\|_1 + \|a^*-v\|_\infty \le \frac{\check \lambda_{n_t}}{2}
\label{eq:fixed_comparator_gradient}
\end{equation}
with probability tending to one, after choosing the fixed constant
$K_1$ is sufficiently large. Furthermore, from the sub-Gaussianity of $W_i$ (Assumption \ref{assm:cov_err} and Remark \ref{rem:sub-G_W}), we also have
\begin{equation*}
\textstyle
\max_{1\le i\le n_t}|W_i^\top\eta^*|
\le
\|\eta^*\|_1
\max_{1\le i\le n_t}\|W_i\|_\infty
=
O_p\left(\sqrt{\log(nd)}\right).
\end{equation*}
Since $n_tb_n=n_t^{1-\alpha} \gg \sqrt{\log(nd)}$, it follows that, with probability going to one, $\max_{1\le i\le n_t}|W_i^\top\eta^*| < n_tb_n$ and consequently the Huber loss is quadratic at $\eta^*$ and, for $v \in \{a,a^0\}$, $\nabla\cL_{n_t,v}(\eta^*) = \hat \Omega_{W,1}\eta^*-v$. 
The same calculations also show that following the oracle balancing weight $\tau_i^\circ= (W_i^\top \eta^*)/n_t$ for $1 \le i \le n_t$, 
satisfies: 
\begin{equation*}
\textstyle
\|v-\bW_t^\top\tau^\circ\|_\infty \le \check \lambda_{n_t}/2 < \check \lambda_{n_t}, \qquad \|\tau^\circ\|_\infty = O_p\left(\frac{\sqrt{\log(nd)}}{n}\right) =o_p(b_n),
\end{equation*}
Now, for $v\in\{a,a^0\}$, let $\hat\eta_n(v)$ denote the rescaled Huber-dual minimizer:
\begin{equation*}
\textstyle
\hat\eta_n(v) \in \argmin_{\eta\in\reals^d} \left\{\cL_{n,v}(\eta)+ \check \lambda_{n_t}\|\eta\|_1\right\},
\end{equation*}

Optimality of $\hat \eta_n(v)$ and convexity of
$\cL_{n,v}$ imply
\begin{align*}
\check \lambda_{n_t}\|\hat \eta_n(v)\|_1 &\le \left\|\nabla\cL_{n,v}(\eta^*)\right\|_\infty\|\hat\eta_n(v)-\eta^*\|_1 + \check \lambda_{n_t}\|\eta^*\|_1\\ &\le \frac{\check \lambda_{n_t}}{2}\left(\|\hat\eta_n(v)\|_1+\|\eta^*\|_1 \right) + \check \lambda_{n_t}\|\eta^*\|_1.
\end{align*}
Therefore,
\begin{equation}
\textstyle
\label{eq:bound_eta_hat_l1}
\frac{\check \lambda_{n_t}}{2} \|\hat\eta_n(v)\|_1 \le \frac{3\check \lambda_{n_t}}{2}\|\eta^*\|_1 \implies \|\hat\eta_n(v)\|_1 \le 3\|\eta^*\|_1 \le 3C_\eta.
\end{equation}
Applying this argument to both \(\hat\eta_n(a)\) and
\(\hat\eta_n(a^0)\), we obtain
\begin{equation*}
\textstyle
\|\hat\eta_n(a)-\hat\eta_n(a^0)\|_1 \le 6C_\eta \implies \|\hat\eta^{\mathrm{KKT}}(a) - \hat\eta^{\mathrm{KKT}}(a^0)\|_1 \le \frac{6C_\eta}{n_t} \,.
\end{equation*}
Therefore, by Lemma \ref{lem:rel_tau_a_eta},
\begin{align*}
\|\hat\tau(a)-\hat\tau(a^0)\|_2^2 \le \|a-a^0\|_\infty\left\|\hat\eta^{\mathrm{KKT}}(a) - \hat\eta^{\mathrm{KKT}}(a^0)\right\|_1 & \le \frac{6C_\eta}{n_t}\|a-a^0\|_\infty\\
& = \frac{6C_\eta}{n_t}\|\hat m_c-B_0\bar X\|_\infty.
\end{align*}
Consequently, since \(n\asymp n_t\),
\begin{equation*}
\textstyle
n\|\hat\tau(a)-\hat\tau(a^0)\|_2^2 \le \frac{6C_\eta n}{n_t}
\|\hat m_c-B_0\bar X\|_\infty = O_p\left(\sqrt{\frac{\log p}{n}} + \sqrt{\frac{\log q}{n}}\right) = o_p(1).
\end{equation*}
Hence, $\sqrt{n}\|\hat\tau(a)-\hat\tau(a^0)\|_2 = o_p(1)$, which completes the proof.
\end{proof}

\subsection{{\color{black}Proof of Lemma \ref{lem:conditional-clt}}}
\label{sec:proof_lem_cond_clt}
\begin{proof}
Recall from Proposition~\ref{prop:bias_structure} that we have $\hat\theta_{10}^{\rm cf}-\theta_{10} = \Delta_n^{\rm cf}+V_n^{\rm cf}$, where
\begin{equation*}
\textstyle
V_n^{\rm cf} = \frac12 \left[\sum_{\substack{i\in\mathcal D_2\\A_i=1}}
\hat\tau_{1,i}^{(2)}\epsilon_i + \sum_{\substack{i\in\mathcal D_1\\A_i=0}} \hat\tau_{2,i}^{(1)}U_i^\top\gamma_1 + \sum_{\substack{i\in\mathcal D_1\\A_i=1}} \hat\tau_{1,i}^{(1)}\epsilon_i + \sum_{\substack{i\in\mathcal D_2\\A_i=0}} \hat\tau_{2,i}^{(2)}U_i^\top\gamma_1\right].
\end{equation*}
For $j\in\{1,2\}$, let $\hat\tau_1^{(j),0}$ denote the oracle first balancing weight obtained by replacing the random mediator target $\hat m_{c,j}$ by $\bB_0\bar X$ (same as in the proof of Theorem \ref{thm:main_revised}), and correspondingly define an oracle version of $V_n^{\rm cf}$ as: 
\begin{equation*}
\textstyle
V_n^{\rm cf, 0} = \frac12 \left[\sum_{\substack{i\in\mathcal D_2\\A_i=1}}
\hat\tau_{1,i}^{(2), 0}\epsilon_i + \sum_{\substack{i\in\mathcal D_1\\A_i=0}} \hat\tau_{2,i}^{(1)}U_i^\top\gamma_1 + \sum_{\substack{i\in\mathcal D_1\\A_i=1}} \hat\tau_{1,i}^{(1), 0}\epsilon_i + \sum_{\substack{i\in\mathcal D_2\\A_i=0}} \hat\tau_{2,i}^{(2)}U_i^\top\gamma_1\right].
\end{equation*}
along with its corresponding variance scale as $(\sigma_n^{\rm cf,0})^2$ defined in Equation \eqref{eq:oracle_asym_var}. Furthermore, for notational simplicity, let us define $L_n^{\rm cf,0} = \sqrt{n}V_n^{\rm cf,0}/\sigma_n^{\rm cf,0}$ and $T_n^{\rm cf} = \sqrt{n}(\hat\theta_{10}^{\rm cf}-\theta_{10})/\sigma_n^{\rm cf}$. We first establish that
\begin{equation}
\label{eq:conditional_clt_oracle_replacement}
T_n^{\rm cf}-L_n^{\rm cf,0}=o_p(1).
\end{equation}
Towards that end, define $\cG_n = \sigma(A_{1:n},X_{1:n},M_{1:n})$. Conditional on $\cG_n$, both the actual and oracle first
balancing weights are fixed, whereas the treated outcome errors remain
centered and independent. Therefore,
\begin{equation*}
\textstyle
\bbE\left[\left(\sqrt{n}\left(V_n^{\rm cf}-V_n^{\rm cf,0}\right)\right)^2 \,\middle|\, \cG_n\right]  = \frac{n\sigma_1^2}{4}\sum_{j=1}^2\left\|\hat\tau_1^{(j)}-\hat\tau_1^{(j),0}\right\|_2^2=o_p(1),
\end{equation*}
where the last conclusion follows from
Lemma~\ref{lem:tau_concentration_revised}. Consequently, by a conditional version of Markov's inequality, $\sqrt{n}\left(V_n^{\rm cf}-V_n^{\rm cf,0}\right) =o_p(1)$. Furthermore, we have already established in Equation \eqref{eq:diff_oracle_actual_sigma} that $|(\sigma_n^{\rm cf})^2 - (\sigma_n^{\rm cf,0})^2| = o_p(1)$. Since $\sigma_n^{\rm cf}$ is bounded away from zero with probability going to one, the same conclusion holds for $\sigma_n^{\rm cf,0}$, and using the same logic as in for the proof of lower bounding $(\sigma_n^{\rm cf})^2$ in the proof of Theorem \ref{thm:main_revised}, we conclude: 
$$
\frac{\sigma_n^{\rm cf,0}}{\sigma_n^{\rm cf}}-1=o_p(1).
$$
Now, going back to $L_n^{\rm cf,0}$ and $T_n^{\rm cf}$, we have the following decomposition: 
\begin{equation*}
\textstyle
T_n^{\rm cf}-L_n^{\rm cf,0} = \frac{\sqrt n\,\Delta_n^{\rm cf}}{\sigma_n^{\rm cf}} + \frac{\sqrt n\left(V_n^{\rm cf}-V_n^{\rm cf,0}\right)}{\sigma_n^{\rm cf}} + L_n^{\rm cf,0}\left(\frac{\sigma_n^{\rm cf,0}}{\sigma_n^{\rm cf}}-1\right).
\end{equation*}
The first term is $o_p(1)$ by the bias part of the proof of Theorem~\ref{thm:main_revised}; the second term is $o_p(1)$ by the preceding calculation; and the third term is $o_p(1)$ because $L_n^{\rm cf,0}=O_p(1)$ and $\sigma_n^{\rm cf,0}/\sigma_n^{\rm cf}-1=o_p(1)$. This proves \eqref{eq:conditional_clt_oracle_replacement}.
It remains to establish the conditional asymptotic normality of $L_n^{\rm cf,0}$. Define $\cH_n = \sigma\left(A_{1:n}, X_{1:n}, \{M_i:A_i=1\}\right).$
Then $\cX_n\subseteq\cH_n$. Conditional on $\cH_n$, the oracle first balancing weights $\{\hat\tau_1^{(j),0}\}_{j=1}^2$, the control balancing weights $\{\hat\tau_2^{(j)}\}_{j=1}^2$, and $\sigma_n^{\rm cf,0}$ are fixed. Moreover, the treated outcome errors and the control mediator errors in $V_n^{\rm cf,0}$ are conditionally centered and independent across $i$. Now, define the triangular-array summands
$$
\xi_{n,i}^0
=
\frac{\sqrt n}{2\sigma_n^{\rm cf,0}}
\begin{cases}
\hat\tau_{1,i}^{(2),0}\epsilon_i,
&
i\in\mathcal D_2,\ A_i=1,\\[0.3em]
\hat\tau_{2,i}^{(1)}U_i^\top\gamma_1,
&
i\in\mathcal D_1,\ A_i=0,\\[0.3em]
\hat\tau_{1,i}^{(1),0}\epsilon_i,
&
i\in\mathcal D_1,\ A_i=1,\\[0.3em]
\hat\tau_{2,i}^{(2)}U_i^\top\gamma_1,
&
i\in\mathcal D_2,\ A_i=0.
\end{cases}
$$
Then, it is immediate that $L_n^{\rm cf,0}=\sum_{i=1}^n\xi_{n,i}^0$ and we also have $\bbE[\xi_{n,i}^0\mid\cH_n]=0$, along with
\begin{equation*}
\textstyle
\sum_{i=1}^n\bbE\left[(\xi_{n,i}^0)^2\,\middle|\,\cH_n\right] =\frac{n}{4(\sigma_n^{\rm cf,0})^2}\sum_{j=1}^2\left\{\sigma_1^2\|\hat\tau_1^{(j),0}\|_2^2+\gamma_1^\top\Sigma_{U,0}\gamma_1\|\hat\tau_2^{(j)}\|_2^2\right\} =1.
\end{equation*}
We next verify the conditional Lyapunov condition. By the uniform conditional sub-Gaussian assumptions on $\epsilon_i$ and $U_i$, and the boundedness of $\|\gamma_1\|_2$, there exists a constant $C>0$ such that
\begin{equation*}
\textstyle
\sum_{i=1}^n\bbE\left[|\xi_{n,i}^0|^3\,\middle|\,\cH_n\right] \le (Cn^{3/2}/(\sigma_n^{\rm cf,0})^3)\sum_{j=1}^2
\{\|\hat\tau_1^{(j),0}\|_\infty\|\hat\tau_1^{(j),0}\|_2^2 +\|\hat\tau_2^{(j)}\|_\infty\|\hat\tau_2^{(j)}\|_2^2\}.
\end{equation*}
From Lemma~\ref{lem:feasibility},
Lemma~\ref{lem:tau_concentration_revised}, and the constraints in optimization problems Equation \eqref{eq:balancing_opt_1} and \eqref{eq:balancing_opt_2}, we have $n\|\hat\tau_1^{(j),0}\|_2^2 + n\|\hat\tau_2^{(j)}\|_2^2 = O_p(1)$ and $\|\hat\tau_1^{(j),0}\|_\infty \vee \|\hat\tau_2^{(j)}\|_\infty = O_p(n^{-\alpha})$. 
Since $\sigma_n^{\rm cf,0}$ is bounded away from zero with probability going to one, it follows that
\begin{equation*}
\textstyle
\sum_{i=1}^n\bbE\left[|\xi_{n,i}^0|^3\,\middle|\,\cH_n\right]=O_p(n^{1/2-\alpha})=o_p(1),
\end{equation*}
where the last conclusion follows from $\alpha>1/2$. Therefore, by the conditional Lyapunov central limit theorem, for every $s\in\reals$,
\begin{equation}
\textstyle
\label{eq:oracle_conditional_cf}
\bbE\left[e^{isL_n^{\rm cf,0}}\,\middle|\,\cH_n\right] \longrightarrow e^{-s^2/2}, \qquad\text{in probability}.
\end{equation}
To go towards $\cX_n$ from $\cH_n$, by law of total expectation, 
\begin{equation*}
\textstyle
\bbE\left[e^{isL_n^{\rm cf,0}} \,\middle|\, \cX_n \right] = \bbE\left[\bbE\left(e^{isL_n^{\rm cf,0}}\,\middle|\,\cH_n\right)\,\middle|\,\cX_n\right].
\end{equation*}
Now, define
\begin{equation*}
\textstyle
Z_n(s) = \left|\bbE\left(e^{isL_n^{\rm cf,0}}\,\middle|\,\cH_n\right) - e^{-s^2/2}\right|.
\end{equation*}
By \eqref{eq:oracle_conditional_cf}, $Z_n(s)\to0$ in probability, and $0\le Z_n(s)\le2$. Consequently,
$Z_n(s)\to0$ in $L_1$. Hence,
\begin{equation*}
\textstyle
\left|\bbE\left[e^{isL_n^{\rm cf,0}} \,\middle|\,\cX_n\right] - e^{-s^2/2} \right| \le \bbE\left[Z_n(s)\,\middle|\,\cX_n\right] \longrightarrow 0, \quad\text{in probability},
\end{equation*}
where the last conclusion follows from Markov's inequality and
$\bbE[Z_n(s)]\to0$. Finally, define $D_n = T_n^{\rm cf}-L_n^{\rm cf,0}$. We have already established that $D_n=o_p(1)$. Therefore, for any  $s \in \reals$,
\begin{align*}
\left|\bbE\left[e^{isT_n^{\rm cf}}\,\middle|\, \cX_n\right] - \bbE\left[e^{isL_n^{\rm cf,0}}\,\middle|\, \cX_n\right]\right|
& \le \bbE\left[\left|e^{isD_n}-1\right| \,\middle|\, \cX_n\right] \\
& \le \bbE\left[\min\{2,|s||D_n|\} \,\middle|\, \cX_n\right] \longrightarrow 0, \quad\text{in probability}.
\end{align*}
Combining the last two displays, we conclude that, for every
$s\in\reals$,
\begin{equation*}
\textstyle
\bbE\left[e^{isT_n^{\rm cf}}\,\middle|\,\cX_n\right]\longrightarrow e^{-s^2/2}, \qquad\text{in probability}.
\end{equation*}
Since the limiting characteristic function is continuous at the origin, the conditional laws of $T_n^{\rm cf}$ given $\cX_n$ converge weakly in probability to the standard normal distribution. Since $\Phi$ is continuous, this implies that, for every
$t\in\reals$,
\begin{equation*}
\textstyle
\bbP\left(T_n^{\rm cf}\le t\,\middle|\,\cX_n\right)\longrightarrow \Phi(t), \qquad\text{in probability}.
\end{equation*}
This completes the proof.
\end{proof}
{\color{black}
\begin{lemma}
    \label{lem:rate_b_hat}
   {\color{black} Under Assumption \ref{assm:cov_err} and \ref{assm:params}, the estimate $\hat b$, as obtained in Step 4 of Algorithm \ref{algomain_modified} satisfies: 
    \begin{equation*}
        \textstyle
    \|\hat b- b\|_1 = O_p\left(sk_\gamma \sqrt{\frac{\log{p}}{n}} + k\sqrt{\frac{\log{p+q}}{n}}\right) \,.
    \end{equation*}}
\end{lemma}
}
\begin{proof}
For simplicity of presentation, we will omit the fold number here. We define the response $\check \by_c = \bM_c \hat \gamma_1$ (the responses used in Step 4 of Algorithm \ref{algomain_modified}). We start from the basic inequality. From the definition of $\hat b$, we have:  
    \begin{equation*}
        \textstyle
        \hat b = \argmin_{\tilde b} \left\{\frac{1}{2n_c} \|\check \by_c - \bX_c \tilde b\|_2^2 + \lambda \|\tilde b\|_1\right\}
    \end{equation*}
Hence, we have: 
\begin{equation*}
\textstyle
\frac{1}{2n_c}\|\check \by_c - \bX_c \hat b\|_2^2 + \lambda \|\hat b\|_1 \le \frac{1}{2n_c}\|\check \by_c - \bX_c  b\|_2^2 + \lambda \| b\|_1
\end{equation*}
Defining $\Delta = \hat b - b$ and using the fact that $\check \by_c = \bX_c b + \bU_c \hat \gamma_1$, we have by some simple algebra: 
\begin{equation}
\label{eq:lasso_b_bound_1}
\textstyle
\frac{1}{2n_c} \|\bX_c \Delta\|_2^2 + \lambda \|\hat b\|_1 \le \left\|\frac{1}{n_c} \bX_c^\top \bU_c \hat \gamma_1\right\|_\infty \|\Delta\|_1 + \lambda \|b\|_1 \,.
\end{equation}
Now, suppose we choose $\lambda$ such that 
\begin{equation}
\textstyle
\label{eq:bounding_l_inf_error}
\bbP\left(\left\|\frac{1}{n_c} \bX_c^\top \bU_c \hat \gamma_1\right\|_\infty \le \frac{\lambda}{2}\right) \longrightarrow 1, \quad \text{as } n \uparrow \infty \,.
\end{equation}
We will later show that one can take $\lambda = K_1 \sqrt{\log{p}/n_c}$ for some suitably chosen large constant $K_1$. Let us call the above event $\Omega_{n, 1}$. Then on $\Omega_{n, 1}$, we have, from Equation \eqref{eq:lasso_b_bound_1},
\begin{equation}
\label{eq:lasso_b_bound_2}
\textstyle
\frac{1}{2n_c} \|\bX_c \Delta\|_2^2  \le \frac{\lambda}{2}\|\Delta\|_1 + \lambda (\|b\|_1 - \|b + \Delta\|_1) \,.
\end{equation}
Now, let $S$ denotes that active set of $b_0 = \bB_0^\top \gamma_1$. Then we have $|S| \le sk_\gamma$. Furthermore, 
\allowdisplaybreaks
\begin{align*}
    \|b\|_1 - \|b + \Delta\|_1 & = \|b_S\|_1 - \|b_S + \Delta_S\|_1 + \|b_{S^c}\|_1 - \|b_{S^c} + \Delta_{S^c}\|_1 \\
    & \le \|\Delta_S\|_1 + \|b_{S^c}\|_1 - (\|\Delta_{S^c}\|_1 - \|b_{S^c}\|_1) 
\end{align*}
Hence, using this in Equation \eqref{eq:lasso_b_bound_2}, along with the fact that $(\lambda/2) \|\Delta\|_1 = (\lambda/2)\|\Delta_S\|_1 + (\lambda/2)\|\Delta_{S^c}\|_1)$
we have: 
\begin{equation}
\label{eq:lasso_b_bound_3}
\textstyle
\frac{1}{2n_c} \|\bX_c \Delta\|_2^2 + \frac{\lambda}{2}\|\Delta_{S^c}\|_1 \le \frac{3\lambda}{2}\|\Delta_S\|_1 + 2\lambda\|b_{S^c}\|_1 \,.
\end{equation}
As $\|\bX_c \Delta\|_2^2\ge 0$, it is immediate from Equation \eqref{eq:lasso_b_bound_3}
\begin{equation}
\label{eq:lasso_b_bound_4}
\textstyle
\|\Delta_{S^c}\|_1 \le 3\|\Delta_S\|_1 + 4\|b_{S^c}\|_1 \,.
\vspace{-1em}
\end{equation}
We divide the rest of the analysis into two cases, depending on whether $\|\Delta_S\|_1$ is larger than $\|b_{S^c}\|_1$ or not: 

\noindent
\underline{\bf Case 1: $4\|b_{S^c}\|_1 \le 3\|\Delta_S\|_1$.} In this case, we have from Equation \eqref{eq:lasso_b_bound_4}, $\|\Delta_{S^c}\|_1 \le 6\|\Delta_S\|_1$. Also from Equation \eqref{eq:lasso_b_bound_3}, we have: 
\begin{equation}
\label{eq:lasso_b_bound_5}
    \textstyle
\frac{1}{2n_c}\|\bX_c \Delta\|_2^2 + \frac{\lambda}{2}\|\Delta_{S^c}\|_1 \le 3\lambda \|\Delta_S\|_1
\end{equation}
Furthermore, the compatibility condition holds, and define the event $\Omega_{n, 2}$, on which $\bX_c$ satisfies the RE condition. Then by Assumption \ref{assm:cov_err} and Lemma \ref{lem:sample_RE}, $\bbP(\Omega_{n, 2}) \to 1$ as $n \uparrow \infty$ and consequently, on $\Omega_{n, 2}$, we have for some constant $\phi_0$: 
\begin{equation}
\label{eq:delta_S_1_bound}
    \textstyle
    \frac{\|\bX_c \Delta\|_2^2}{n_c} \ge \phi_0 \|\Delta\|_2^2 \implies \|\Delta_S\|_1 \le \sqrt{sk_\gamma}\|\Delta_S\|_2 \le \sqrt{sk_\gamma}\|\Delta\|_2 \le \sqrt{sk_\gamma}\frac{\|\bX_c \Delta\|_2}{\sqrt{n_c}\sqrt{\phi_0}} \,.
\end{equation}
Hence, from Equation \eqref{eq:lasso_b_bound_5}, we have: 
\begin{equation}
\label{eq:lasso_b_bound_6}
    \textstyle
\frac{1}{2n_c}\|\bX_c \Delta\|_2^2 + \frac{\lambda}{2}\|\Delta_{S^c}\|_1 \le 3\lambda \sqrt{sk_\gamma}\frac{\|\bX_c \Delta\|_2}{\sqrt{n_c}\sqrt{\phi_0}} \le \frac{1}{4n_c}\|\bX_c \Delta\|_2^2 + \frac{9}{\phi_0} sk_\gamma \lambda^2 
\end{equation}
Hence, changing sides, yields: 
\begin{equation*}
    \textstyle
\frac{1}{n_c}\|\bX_c \Delta\|_2^2 \le \frac{36}{\phi_0} sk_\gamma \lambda^2 \implies \frac{1}{\sqrt{n_c}}\|\bX_c \Delta\|_2 \le \frac{6}{\sqrt{\phi_0}} \lambda \sqrt{sk_\gamma}
\end{equation*}
This, along with Equation \eqref{eq:delta_S_1_bound} implies $\|\Delta_S\|_1 \le (6\lambda sk_\gamma)/\phi_0$ and consequently, using the fact under this case 1 $\|\Delta_{S^c}\|_1 \le 6\|\Delta_S\|_1$, we conclude that:
\begin{equation}
\label{eq:lasso_b_bound_7}
    \textstyle
\|\Delta_1\|_1 = \|\Delta_S\|_1 + \|\Delta_{S^c}\|_1 \le 7\|\Delta_S\|_1 \le \frac{42}{\phi_0}\lambda sk_\gamma  = O_p\left(sk_\gamma \sqrt{\frac{\log{p}}{n}}\right) \,.
\end{equation}
Here in the last line we have used the fact $\lambda = K_1 \sqrt{\log{p}/n_c}$, a fact that we will prove at the end, and $n_t \sim n_c \sim n$. 

\noindent
\underline{\bf Case 2: $4\|b_{S^c}\|_1 > 3\|\Delta_S\|_1$:} In this case, we have from Equation \eqref{eq:lasso_b_bound_3}, 
\begin{equation}
    \textstyle
\frac{1}{2n_c}\|\bX_c \Delta\|_2^2 + \frac{\lambda}{2}\|\Delta_{S^c}\|_1 \le 4\lambda \|b_{S^c}\|_1 \implies \|\Delta_{S^c}\|_1 \le 8 \|b_{S^c}\|_1 \implies \|\Delta\|_1 \le \frac{28}{3}\|b_{S^c}\|_1 \,.
\end{equation}
Now, $\|b_{S^c}\|_1 = \|(b - b_0)_{S^c}\|_1 \le \|b - b_0\|_1$, where in the first equality, we have used the fact that $(b_0)_{S^c} \equiv 0$ by definition of $S$. On the other have, we have: 
\begin{equation*}
\textstyle
\|b - b_0\|_1 = \|\bB_0^\top(\hat \gamma_1 - \gamma_1)\|_1 \le \|\bB_0\|_{1, \infty}\|\hat \gamma_1 - \gamma_1\|_1 = O_p\left(k \sqrt{\frac{\log{(p+q)}}{n}}\right)
\end{equation*}
where the last conclusion follows from the fact that $\|\bB_0\|_{1, \infty} \le C_\bB < \infty$ and $\|\hat \gamma_1 - \gamma_1\|_1 \le \|\hat \phi_1 - \phi_1\|_1 = O_p(k\sqrt{\log{(p+q)/n}})$. This completes the proof. Now, all we need to show is that $\lambda = K_1 \sqrt{\log{p}/n_c}$ is a valid choice for Equation \eqref{eq:bounding_l_inf_error} to hold. As $\hat \gamma_1$ is obtained from separate data (treatment data), and $U_i$'s are sub-Gaussian conditional on $X$, it is immediate from the sub-Gaussian concentration inequality that for a large enough constant $C$, 
\begin{equation*}
\textstyle
\bbP\left(\frac{1}{n_c\|\hat \gamma_1\|_2}\|\bX_c^\top \bU_c\hat \gamma_1\|_\infty > C \sqrt{\frac{\log{p}}{n_c}} \mid \hat \gamma_1\right) \to 0, \quad \text{ as } n \uparrow \infty\,.
\end{equation*}
Consequently $\|\bX_c^\top \bU_c\hat \gamma_1\|_\infty/n_c \le C\|\hat \gamma_1\|_2 \sqrt{\log{p}/n_c}$ with probability going to $1$. Now $\|\hat \gamma_1\|_2 \le \|\gamma_1\|_2 + \|\hat \gamma_1 - \gamma_1\|_2 \le 1 + C_\gamma$ with probability going to $1$ as $\|\hat \gamma_1 - \gamma_1\|_2 \le \|\hat \phi_1 - \phi_1\|_2 = o_p(1)$ which follows from the consistency of the lasso estimator. Therefore, $\|\bX_c^\top \bU_c\hat \gamma_1\|_\infty/n_c \le C(1+C_\gamma) \sqrt{\log{p}/n_c}$. Taking $K_1 = 2C(1 + C_\gamma)$ would conclude the proof.  
\end{proof}

\subsection{Calculation of Effects}
\label{effects}

 By taking expectations of the outcome and mediator regressions conditional on $(X,M,A)$ and $(X,A)$, respectively, we obtain
\begin{align*}
    \bbE[Y(1,M(1))]
    &=
    \alpha_1+\bbE[X]^\top\beta_1
    +
    \bbE\!\left[\bbE[M\mid X,A=1]\right]^\top\gamma_1  \\
    &=
    (\alpha_1+\delta_1^\top\gamma_1)
    +
    \bbE[X]^\top(\beta_1+\mathbf B_1^\top\gamma_1),
    \\
    \bbE[Y(0,M(0))]
    &=
    \alpha_0+\bbE[X]^\top\beta_0
    +
    \bbE\!\left[\bbE[M\mid X,A=0]\right]^\top\gamma_0  \\
    &=
    (\alpha_0+\delta_0^\top\gamma_0)
    +
    \bbE[X]^\top(\beta_0+\mathbf B_0^\top\gamma_0),
    \\
    \bbE[Y(1,M(0))]
    &=
    \alpha_1+\bbE[X]^\top\beta_1
    +
    \bbE\!\left[\bbE[M\mid X,A=0]\right]^\top\gamma_1  \\
    &=
    (\alpha_1+\delta_0^\top\gamma_1)
    +
    \bbE[X]^\top(\beta_1+\mathbf B_0^\top\gamma_1),
    \\
    \bbE[Y(0,M(1))]
    &=
    \alpha_0+\bbE[X]^\top\beta_0
    +
    \bbE\!\left[\bbE[M\mid X,A=1]\right]^\top\gamma_0  \\
    &=
    (\alpha_0+\delta_1^\top\gamma_0)
    +
    \bbE[X]^\top(\beta_0+\mathbf B_1^\top\gamma_0).
\end{align*}
Therefore, the natural indirect effect simplifies as
\begin{align*}
   \mathrm{NIE}
   &=
   \bbE[Y(1,M(1))]-\bbE[Y(1,M(0))] \\
   &=
   (\alpha_1+\delta_1^\top\gamma_1)
   +
   \bbE[X]^\top(\beta_1+\mathbf B_1^\top\gamma_1)
   -
   (\alpha_1+\delta_0^\top\gamma_1)
   -
   \bbE[X]^\top(\beta_1+\mathbf B_0^\top\gamma_1) \\
   &=
   (\delta_1-\delta_0)^\top\gamma_1
   +
   \bbE[X]^\top
   (\beta_1-\beta_1+\mathbf B_1^\top\gamma_1-\mathbf B_0^\top\gamma_1) \\
   &=
   (\delta_1-\delta_0)^\top\gamma_1
   +
   \bbE[X]^\top(\mathbf B_1-\mathbf B_0)^\top\gamma_1.
\end{align*}
Similarly, the natural direct effect simplifies as
\begin{align*}
    \mathrm{NDE}
    &=
    \bbE[Y(1,M(0))]-\bbE[Y(0,M(0))] \\
    &=
    (\alpha_1+\delta_0^\top\gamma_1)
    +
    \bbE[X]^\top(\beta_1+\mathbf B_0^\top\gamma_1)
    -
    (\alpha_0+\delta_0^\top\gamma_0)
    -
    \bbE[X]^\top(\beta_0+\mathbf B_0^\top\gamma_0) \\
    &=
    \alpha_1-\alpha_0+\delta_0^\top(\gamma_1-\gamma_0)
    +
    \bbE[X]^\top
    (\beta_1+\mathbf B_0^\top\gamma_1-\beta_0-\mathbf B_0^\top\gamma_0) \\
    &=
    \alpha_1-\alpha_0+\delta_0^\top(\gamma_1-\gamma_0)
    +
    \bbE[X]^\top(\beta_1-\beta_0)
    +
    \bbE[X]^\top\mathbf B_0^\top(\gamma_1-\gamma_0).
\end{align*}

\addtocontents{toc}{\protect\setcounter{tocdepth}{1}}

\section{{\color{black}Connection with the Existing Literatures}}
\label{appendix:existing-high-dimensional-mediation-comparison}

\subsection{{\color{black}Comparison with \cite{guo2022high, guo2023statistical, guo2024estimations} and the role of treatment interactions}}
\label{appendix:comparison-guo-trilogy}

In this section, we elaborate on the relationship between our method and existing high-dimensional mediation inference methods, with particular emphasis on the trilogy of \citep{guo2022high, guo2023statistical, guo2024estimations}. These works make significant contributions to high-dimensional mediation analysis, and our inferential objective is aligned with theirs at a high level: namely, to conduct valid statistical inference for mediated treatment effect, e.g., the natural indirect effect (NIE) and the natural direct effect (NDE) in high-dimensional settings. However, the framework studied in our paper extends this line of work primarily in three directions:
\begin{enumerate}
    \item We allow ultra-high-dimensional (UHD) covariates \(X\) in addition to UHD mediators \(M\).
    \item More importantly, we allow treatment interactions in both stages of the data-generating mechanism; in particular, we allow \(A\)--\(X\) and \(A\)--\(M\) interactions in the generation of the response, and \(A\)--\(X\) interactions in the generation of the mediators.
    \item Our assumptions required for high-dimensional inference are weaker than those of \citep{guo2022high, guo2023statistical, guo2024estimations}
\end{enumerate}
It is the combination of UHD covariates and treatment interactions that fundamentally changes the inferential structure. Note that there exist several works in the causal inference literature discussing the challenges added by interaction effects (e.g., the textbook \cite{vanderweele2015explanation}), yet such investigations are, to the best of our knowledge, scarce in the high-dimensional mediation analysis literature. To clarify the source of the difficulty, it is useful to compare the basic mediation functional $\theta_0:=\bbE[Y^{(1,M^{(0)})}]$ under a no-interaction model and under our interaction-rich model: 

\setcounter{algorithm}{0}
\renewcommand{\thealgorithm}{A\arabic{algorithm}}
\begin{algorithm}[tb]
\caption{{\color{black}Debiased estimation of the natural indirect effect in the absence of interactions}}
\label{alg:ours_modified}
\begingroup
\small
\setlength{\abovedisplayskip}{0.45em}
\setlength{\belowdisplayskip}{0.45em}

\begin{algorithmic}[1]

\State \textbf{Step 1.}
Regress $Y_i$ on $(A_i,\mathbf X_i)$ with $\ell_1$ penalty on $\beta$:
\[
\hat \phi_1 = (\hat\alpha_1^{(1)},\hat\beta^{(1)})
=
\arg\min_{\alpha_1,\beta}
\sum_{i}
\left(
Y_i - A_i \alpha_1 - \mathbf X_i^\top \beta
\right)^2
+ \lambda_1 \|\beta\|_1.
\]
Let $\mathbf W_i^{(1)} = (A_i,\mathbf X_i^\top)^\top$ and
$a=(1,\mathbf 0^\top)^\top$. The vector $a$ extracts the treatment coefficient in the reduced outcome regression. Compute $\tau_a$ via
\[
\tau_a
=
\arg\min_{\tilde\tau}
\|\tilde\tau\|_2^2
\ \text{s.t.}\ 
\left\|
a - (\mathbf W^{(1)})^\top \tilde\tau
\right\|_\infty
\le K_a \sqrt{\frac{\log(1+p)}{n}},
\ 
\|\tilde\tau\|_\infty \le n^{-2/3}.
\]
Set
\[
\hat{\mathrm{TE}}
=
a^\top \hat \phi_1
+
\sum_{i}
\tau_{a,i}
\left(
Y_i - A_i \hat\alpha_1^{(1)} - \mathbf X_i^\top \hat\beta^{(1)}
\right).
\]

\Statex
\State \textbf{Step 2.}
Regress $Y_i$ on $(A_i,\mathbf X_i,\mathbf M_i)$ with $\ell_1$ penalty on $(\beta,\gamma)$:
\[
\hat \phi_2
=
(\hat\alpha_1^{(2)},\hat\beta^{(2)},\hat\gamma)
=
\arg\min_{\alpha_1,\beta,\gamma}
\sum_{i}
\left(
Y_i - A_i \alpha_1 - \mathbf X_i^\top \beta - \mathbf M_i^\top \gamma
\right)^2
+ \lambda_2(\|\beta\|_1+\|\gamma\|_1).
\]
Let $\mathbf W_i^{(2)}=(A_i,\mathbf X_i^\top,\mathbf M_i^\top)^\top$
and $b=(1,\mathbf 0^\top,\mathbf 0^\top)^\top$. The vector $b$ extracts the treatment coefficient after adjusting for both covariates and mediators. Compute $\tau_b$ analogously and set
\[
\hat{\mathrm{NDE}}
=
b^\top \hat \phi_2
+
\sum_{i}
\tau_{b,i}
\left(
Y_i - A_i \hat\alpha_1^{(2)}
- \mathbf X_i^\top \hat\beta^{(2)}
- \mathbf M_i^\top \hat\gamma
\right).
\]

\Statex
\State \textbf{Step 3.}
Compute the natural indirect effect as the difference between the total effect and the natural direct effect:
\[
\hat{\mathrm{NIE}}
=
\hat{\mathrm{TE}} - \hat{\mathrm{NDE}}.
\]

\Statex
\State \textbf{Return} $\hat{\mathrm{NIE}}$.

\end{algorithmic}
\endgroup
\end{algorithm}

\medskip
\noindent\textbf{1. The role of the simultaneous presence of interaction and UHD $X$.}
To isolate the role of interaction, suppose we naturally extend \citep{guo2022high, guo2023statistical, guo2024estimations} by allowing UHD covariates \(X\), but without treatment interactions:
\begin{equation}
\label{eq:our_no_interaction}
\begin{aligned}
Y &= \alpha_0 + A\alpha_1 + X^\top \beta + M^\top \gamma + \varepsilon, \\
M &= \delta_0 + A\delta_1 + B X + U.
\end{aligned}
\end{equation}
Under this model, the sample average mediation functional is
$$
\theta_0 = \bbE[Y^{(1,M^{(0)})}] = \alpha_0+\alpha_1+\delta_0^\top\gamma + \bar X^\top(\beta+B^\top\gamma) \,.
$$
Although this expression contains the term \(B^\top\gamma\), in the no-interaction model, this term can be absorbed into a single reduced-form outcome regression. Indeed, substituting the mediator equation into the outcome equation gives
$$
Y = (\alpha_0+\delta_0^\top\gamma) + A(\alpha_1+\delta_1^\top\gamma) + X^\top(\beta+B^\top\gamma) + U^\top\gamma+\varepsilon.
$$
Thus, the high-dimensional component $\bar X^\top(\beta+B^\top\gamma)$ is now a standard linear functional of a single reduced-form regression coefficient $\beta+B^\top\gamma$ of $X$. Consequently, even if \(X\) is UHD, the problem remains similar to standard high-dimensional linear functional inference: one may first estimate this regression coefficient using penalized regression of $Y$ on $(A, X)$ and then debias it in the direction \(\bar X\) using the techniques of \citep{zhang2014confidence,van2014asymptotically,javanmard2014confidence}.

\begin{algorithm}[tb]
\caption{{\color{black}\cite{guo2022high} estimator for the natural indirect effect}}
\label{alg:guo_original}
\begingroup
\small
\setlength{\abovedisplayskip}{0.45em}
\setlength{\belowdisplayskip}{0.45em}

\begin{algorithmic}[1]

\State \textbf{Step 1.}
Estimate
\[
(\hat\gamma,\hat\alpha_1,\hat\beta)
=
\arg\min_{\gamma,\alpha_1,\beta}
\frac1n
\sum_{i=1}^n
\left(
Y_i
-
A_i\alpha_1
-
X_i^\top\beta
-
M_i^\top\gamma
\right)^2
+
\sum_{j=1}^q
p_\lambda(|\gamma_j|),
\]
where $p_\lambda(\cdot)$ denotes the SCAD penalty. The tuning parameter $\lambda$ is selected by HBIC.

\Statex
\State \textbf{Step 2.}
Define the selected mediator set
$
\hat{\mathcal A}
=
\{j:\hat\gamma_j\neq0\}
$
and construct the restricted mediator matrix
$
M_{\hat{\mathcal A}}.
$

\Statex
\State \textbf{Step 3.}
Refit the outcome model on the selected mediators:
\[
(\hat\gamma_{\hat{\mathcal A}}^{\mathrm{rf}},
\hat\alpha_1^{\mathrm{rf}},
\hat\beta^{\mathrm{rf}})
=
\arg\min_{\gamma_{\hat{\mathcal A}},\alpha_1,\beta}
\sum_{i=1}^n
\left(
Y_i
-
A_i\alpha_1
-
X_i^\top\beta
-
M_{i,\hat{\mathcal A}}^\top
\gamma_{\hat{\mathcal A}}
\right)^2.
\]
Set
$
\hat{\mathrm{NDE}}
=
\hat\alpha_1^{\mathrm{rf}}.
$

\Statex
\State \textbf{Step 4.}
Using the reduced-form representation
\[
Y
=
\alpha_0
+
\delta_0^\top\gamma
+
A(\alpha_1+\delta_1^\top\gamma)
+
X^\top(\beta+\mathbf B^\top\gamma)
+
U^\top\gamma
+
\epsilon,
\]
estimate the total effect by regressing $Y_i$ on $(A_i,X_i)$:
\[
(\hat\theta_A,\hat\theta_X)
=
\arg\min_{\theta_A,\theta_X}
\sum_{i=1}^n
\left(
Y_i
-
A_i\theta_A
-
X_i^\top\theta_X
\right)^2.
\]
Set
$
\hat{\mathrm{TE}}
=
\hat\theta_A.
$

\Statex
\State \textbf{Step 5.}
Compute
\[
\hat{\mathrm{NIE}}
=
\hat{\mathrm{TE}}
-
\hat{\mathrm{NDE}}
=
\hat\theta_A
-
\hat\alpha_1^{\mathrm{rf}}.
\]

\Statex
\State \textbf{Return} $\hat{\mathrm{NIE}}$.

\end{algorithmic}
\endgroup
\end{algorithm}

The situation changes fundamentally once treatment interactions are introduced. Under our model,
\begin{align*}
Y & = (1-A)(\alpha_0+X^\top\beta_0+M^\top\gamma_0+\varepsilon') + A(\alpha_1+X^\top\beta_1+M^\top\gamma_1+\varepsilon), \\
M &= (1-A)(\delta_0+B_0X+U)+A(\delta_1+B_1X+U'),
\end{align*}
the same mediation functional becomes
$$
\theta_0 = \bbE[Y^{(1,M^{(0)})}] = \alpha_1+\delta_0^\top\gamma_1 + \bar X^\top(\beta_1+B_0^\top\gamma_1),
$$
Suppressing intercept terms, which are not the main difficulty, the challenging part is $\bar X^\top B_0^\top\gamma_1$. Here \(B_0\) is identified from the mediator model in the control arm, whereas \(\gamma_1\) and \(\beta_1\) are identified from the outcome model in the treated arm. Therefore, $\bar X^\top B_0^\top\gamma_1$ is not a fixed linear functional of a single high-dimensional regression coefficient. It is a cross-arm functional involving a UHD matrix $B_0$ (whose rows and columns can be much larger than the sample size) and a UHD vector $\gamma_1$.
This is precisely the structural feature that prevents a direct reduction to a standard one-step debiased Lasso problem. To see this point from another angle, if the \(A\)--\(X\) interaction in the mediator model were absent, then \(B_0=B_1\). In that case, the high-dimensional coefficient \(\beta_1+B_0^\top\gamma_1\) would coincide with the coefficient of \(X\) in the treated-arm reduced-form regression. However, when \(B_0\neq B_1\), the coefficient obtained from the treated-arm reduced-form regression is \(\beta_1+B_1^\top\gamma_1\), whereas the counterfactual mediation functional \(\theta_0\) requires $\beta_1+B_0^\top\gamma_1$. This cross-arm structure is absent in the no-interaction model and is also absent in standard single-model high-dimensional linear functional inference.
\begin{algorithm}[tb]
\caption{{\color{black}\cite{guo2022high} estimator for the natural indirect effect with modification for high-dimensional covariates}}
\label{alg:guo_modified}
\begingroup
\small
\setlength{\abovedisplayskip}{0.45em}
\setlength{\belowdisplayskip}{0.45em}

\begin{algorithmic}[1]

\State \textbf{Step 1.}
Estimate
\[
(\hat\gamma,\hat\alpha_1,\hat\beta)
=
\arg\min_{\gamma,\alpha_1,\beta}
\frac1n
\sum_{i=1}^n
\left(
Y_i
-
A_i\alpha_1
-
X_i^\top\beta
-
M_i^\top\gamma
\right)^2
+
\sum_{j=1}^q
p_\lambda(|\gamma_j|)
+
\sum_{k=1}^p
p_\lambda(|\beta_k|),
\]
where the tuning parameter $\lambda$ is selected by HBIC.

\Statex
\State \textbf{Step 2.}
Define
$
\hat{\mathcal A}_M
=
\{j:\hat\gamma_j\neq0\}
$
and
$
\hat{\mathcal A}_X
=
\{k:\hat\beta_k\neq0\}.
$
Construct
$
M_{\hat{\mathcal A}_M}
$
and
$
X_{\hat{\mathcal A}_X}.
$

\Statex
\State \textbf{Step 3.}
Refit the outcome model on the selected mediators and covariates:
\[
(\hat\gamma_{\hat{\mathcal A}_M}^{\mathrm{rf}},
\hat\alpha_1^{\mathrm{rf}},
\hat\beta_{\hat{\mathcal A}_X}^{\mathrm{rf}})
=
\arg\min_{\gamma_{\hat{\mathcal A}_M},\alpha_1,\beta_{\hat{\mathcal A}_X}}
\sum_{i=1}^n
\left(
Y_i
-
A_i\alpha_1
-
X_{i,\hat{\mathcal A}_X}^\top
\beta_{\hat{\mathcal A}_X}
-
M_{i,\hat{\mathcal A}_M}^\top
\gamma_{\hat{\mathcal A}_M}
\right)^2.
\]
Set
$
\hat{\mathrm{NDE}}
=
\hat\alpha_1^{\mathrm{rf}}.
$

\Statex
\State \textbf{Step 4.}
Using the reduced-form representation
\[
Y
=
\alpha_0
+
\delta_0^\top\gamma
+
A(\alpha_1+\delta_1^\top\gamma)
+
X^\top(\beta+\mathbf B^\top\gamma)
+
U^\top\gamma
+
\epsilon,
\]
estimate the total effect by regressing $Y_i$ on $(A_i,X_{i,\hat{\mathcal A}_X})$:
\[
(\hat\theta_A,\hat\theta_X)
=
\arg\min_{\theta_A,\theta_X}
\sum_{i=1}^n
\left(
Y_i
-
A_i\theta_A
-
X_{i,\hat{\mathcal A}_X}^\top
\theta_X
\right)^2.
\]
Set
$
\hat{\mathrm{TE}}
=
\hat\theta_A.
$

\Statex
\State \textbf{Step 5.}
Compute
\[
\hat{\mathrm{NIE}}
=
\hat{\mathrm{TE}}
-
\hat{\mathrm{NDE}}
=
\hat\theta_A
-
\hat\alpha_1^{\mathrm{rf}}.
\]

\Statex
\State \textbf{Return} $\hat{\mathrm{NIE}}$.

\end{algorithmic}
\endgroup
\end{algorithm}

\medskip
\noindent
\textbf{2. Comparison of assumptions.}
Another important difference between our approach and that of Guo et al. is the set of conditions under which high-dimensional inference is performed. In \citep{guo2022high, guo2023statistical, guo2024estimations}, the authors do not use a debiasing step. They do not penalize the low-dimensional parameter of interest and penalize only the high-dimensional nuisance component. To illustrate the idea, let us consider the simple high-dimensional linear model
$$
Y_i = X_i^\top \beta_0 + \varepsilon_i = X_{i1}\beta_{0,1} + X_{i,-1}^\top \beta_{0,-1} + \varepsilon_i,
$$
where the goal is to infer on the first coordinate \(\beta_{0,1}\). An analogue of the partial-penalization approach of \cite{guo2023statistical} would estimate \(\beta_0=(\beta_{0,1},\beta_{0,-1})\) by
$$
(\hat\beta_{1},\hat\beta_{-1}) =  \arg\min_{b_1,b_{-1}} \left\{\frac{1}{2n}\sum_{i=1}^n \left(Y_i-X_{i1}b_1-X_{i,-1}^\top b_{-1}\right)^2 + \sum_{j=2}^p \rho_{\lambda_n}(|b_j|)\right\},
$$
where $\rho_{\lambda_n}$ is typically a nonconvex penalty such as SCAD or MCP. Importantly, the first coordinate is not penalized. Under their assumptions, \cite{guo2023statistical} then show that the resulting estimator of the low-dimensional target parameter (here $\beta_{0, 1}$) is $\sqrt{n}$-consistent and asymptotically normal, i.e., asymptotic normality is achieved without an additional debiasing step.

However, the achievement of asymptotic normality without debiasing comes at the cost of stronger assumptions. In particular, their Condition (A1) includes a support-recovery-oriented irrepresentability type of condition, namely an \(L_{2,\infty}\)-type bound such as $\|M_{\mathcal A^c}^{\top}(X,M_{\mathcal A})\|_{2,\infty}=O_p(n)$ (where $\mathcal A$ is the active set and in our linear regression example above it roughly translates to $\|X_{-1, \mathcal A^c}^{\top}(X_1,X_{-1, \mathcal A})\|_{2,\infty}=O_p(n)$), 
and, more importantly, their Condition (A2) imposes a 
$\beta$-min condition on the nonzero entries of the high-dimensional regression vector, i.e., 
\begin{equation}
\label{eq:beta_min_guo}
d_n=\frac12\min_{j\in A}|\alpha^*_{0j}| = \frac12\min_{j\in \mathcal A}|\beta_{0,j}| \gg \lambda_n \gg  \max\left\{\sqrt{s/n},\sqrt{\log p/n}\right\} \,.
\end{equation}
Hence, under their condition, all active coefficients must be sufficiently separated from zero. This is precisely the type of condition that debiased-Lasso methods are designed to avoid. For example, \cite{zhang2014confidence} emphasize that the main distinction between their low-dimensional projection(LDP)/debiasing approach and variable-selection-based approaches is the latter's reliance on a uniform signal strength condition and to quote: ``However, a main complaint about the variable selection approach is the practicality of the uniform signal strength condition, and the crucial difference between the two approaches is precisely in the case where the condition fails to hold. Without the condition, the LDP approach is still able to select correctly all coefficients above a threshold of order $C\sqrt{(2/n)\log{p}}$ and all zero coefficients" (kindly see the Introduction section of \cite{zhang2014confidence} for more details). In other words, debiased-Lasso inference of \cite{zhang2014confidence} (or \cite{van2014asymptotically,javanmard2014confidence}) does not require a $\beta$-min condition on the nonzero nuisance coefficients; weak but nonzero coefficients are also allowed, provided the usual sparsity and other regularity conditions hold. 

Condition (A2) (Equation \eqref{eq:beta_min_guo}) of \cite{guo2023statistical} also has further implications. This condition requires \(d_n\gg \sqrt{s/n}\). Now, in the standard debiased Lasso approach, we require $s\log{p}/\sqrt{n} \to 0$, i.e., we can allow sparsity up to order $\sqrt{n}/\log{p}$. Therefore, if we set $s \sim \sqrt{n}/(\log{p})^2$, then the minimal signal strength condition of \cite{guo2023statistical} requires $d_n \gg \sqrt{s/n} \gg n^{-1/4}(\log{p})^{-1}$, which is fairly strong condition on the minimal active signal strength and rules out relatively weak nonzero coefficients. Furthermore, the asymptotic normality result, Theorem 1 and Corollary 1 of \cite{guo2023statistical}, requires \(s=o(n^{1/3})\), whereas, as mentioned before, debiased-Lasso analyses typically allow sparsity of order $(s\log{p})/\sqrt{n} \to 0$. Since our approach is also motivated by the debiased-Lasso idea, we \emph{do not impose a minimal signal strength condition on the regression coefficients}, and our analysis also \emph{allows the relevant sparsity levels \((s_\beta,s_\gamma,s_k)\) to be of order $o(\sqrt{n})$, up to the logarithmic factors}. In this sense, our inferential theory requires weaker signal-strength and sparsity conditions than \citep{guo2022high, guo2023statistical, guo2024estimations}. This distinction is also reflected in our simulations: when the sparsity level is high (Table \ref{tab:simu_comparison_sparsity}) or the minimum active coefficient is small (Table \ref{tab:simu_comparison_signal}), our method outperforms theirs.

In summary, when treatment interactions are absent, the mediation functional reduces to a standard linear functional of a high-dimensional reduced-form regression coefficient, making conventional Lasso-plus-debiasing arguments more directly applicable. By contrast, in the interaction-rich model considered in our work, even the basic mediation functional $\bbE[Y^{(1,M^{(0)})}]$ involves the composite cross-arm term $\bar X^\top B_0^\top\gamma_1$, which is the primary source of the additional technical difficulty and motivates the proposed multi-step debiasing procedure. Furthermore, our assumptions on sparsity are weaker than \citep{guo2022high, guo2023statistical, guo2024estimations}, and we do not require any assumption on the minimal signal strength. 
Thus, although our inferential objective is aligned with \citep{guo2022high, guo2023statistical, guo2024estimations}, the structural model studied in this paper is fundamentally more challenging. The simultaneous presence of UHD covariates, UHD mediators, and treatment interactions changes the structure of the mediation functional in a way that is not covered by settings where either UHD covariates or interaction effects are absent. This distinction necessitates both the multi-step debiasing strategy developed in our paper and a separate theoretical analysis.

\medskip
\noindent
{\bf Comparison via numerical experiments:} We next provide an additional numerical comparison with \cite{guo2022high}. The purpose of this comparison is to evaluate the inferential behavior of the proposed debiased estimator relative to an existing inference-oriented method under a common data-generating framework where both methods can be implemented. Since existing high-dimensional mediation methods, such as \citet{guo2022high, guo2023statistical, guo2024estimations}, do not directly accommodate the full interaction-rich UHD covariate setting, we conduct this comparison under the reduced no-interaction setting in Equation \eqref{eq:our_no_interaction}. This ensures that both the proposed method and the method of \cite{guo2022high} can be applied.

We compare the specialization of our proposed debiased estimator to the no-interaction data-generating process in Equation \eqref{eq:our_no_interaction}, summarized in Algorithm \ref{alg:ours_modified}, with a version of the estimator proposed by \cite{guo2022high}.
Under this reduced model, the target admits a reduced-form representation; hence, this special-case implementation does not require estimation of the full mediator-on-covariate coefficient matrix at all.
Since the original procedure in \cite{guo2022high} does not account for UHD covariates, we consider a natural modification of their method that incorporates penalization of both the mediator and covariate coefficients. For completeness, Algorithm \ref{alg:guo_original} records the baseline estimator following \cite{guo2022high}, while Algorithm \ref{alg:guo_modified} presents the modified version used for high-dimensional covariates. 
We next describe our simulation setup. The generative model for $(X, A, Y, M)$ remains the same; the two different regimes we consider differ in terms of the strength of the nonzero coefficients, as specified below.
\begin{enumerate}
    \item \textbf{Covariates.}
    For each simulation run, we generate \(X_i \overset{iid}{\sim} \mathcal N(\mu \mathbf 1, \Sigma_X)\), where \(\mu = 0.5\) and \(\Sigma_X = R_X \Lambda_X R_X^\top\) with \(R_X\) orthonormal and \(\Lambda_X = \mathrm{diag}(\lambda_1,\ldots,\lambda_p)\), where \(\lambda_j \sim \mathcal U(0.1,0.5)\) (generated once and kept fixed throughout the Monte Carlo iterations).

    \item \textbf{Treatment assignment.}
    The binary treatment \(A\) is generated from
    \[
    \bbP(A=1\mid X)
    =
    \mathrm{logit}^{-1}(X^\top \alpha_A),
    \]
    where \(\|\alpha_A\|_0 = s\) is with nonzero entries sampled from \(\mathcal U(0,2)\).

    \item \textbf{Mediators.}
    Given \((A,X)\), we generate mediators from \(M = A \delta_1 + B X + U\),
    where \(\|\delta_1\|_0 = s\) and \(B\) is row-sparse with $\|B_{i,*}\|_0 = s_B$, with non-zero entries sampled from \(\mathcal U(0.05,0.2)\). The noise \(U \sim \mathcal N(0,\Sigma_U)\), where \(\Sigma_U = R_U \Lambda_U R_U^\top\) with \(R_U\) orthonormal and \(\lambda_j \sim \mathcal U(0.1,0.5)\).

    \item \textbf{Outcome.}
    The outcomes are generated from the model \(Y = \alpha_1 A + X^\top \beta + M^\top \gamma + \epsilon\), where \(\alpha_1=0.15\); \(\|\beta\|_0  = \|\gamma_0\|_0 = s\) and \(\epsilon \sim \mathcal N(0,1)\).
\end{enumerate}
We set \(p=q \in \{50,400,750,1250\}\) and
\(n \in \{250,500,750,1000,1250\}\). Across all regimes the sparsity levels grow
with the ambient dimension according to
\[
(s,s_B)=
\begin{cases}
(5,5), & p=50,\\
(10,10), & p=400 \text{ or } 750,\\
(20,20), & p=1250.
\end{cases}
\]
The two regimes differ only in how the nonzero entries of the mediation and
outcome coefficients \((\delta_1,\beta,\gamma)\) are generated:
\begin{itemize}
    \item \textbf{Increasing sparsity with moderate minimum active signal strength:} Every nonzero coefficient is a moderate
    signal sampled from \(\mathcal U(0.05,0.2)\). 
    \item \textbf{Increasing sparsity with vanishing minimum active signal strength:} Half of the nonzero coefficients (chosen at
    random) are moderate signals sampled from \(\mathcal U(0.05,0.2)\), while the
    remaining half are weak signals of order \(n^{-1/2}\), sampled from
~$\mathcal U(0.8,1.2)/\sqrt{n_{\max}}$ with \(n_{\max}=1250\). 
\end{itemize}

\begin{figure}[!ht]
    \centering
    \includegraphics[width=0.8\linewidth]{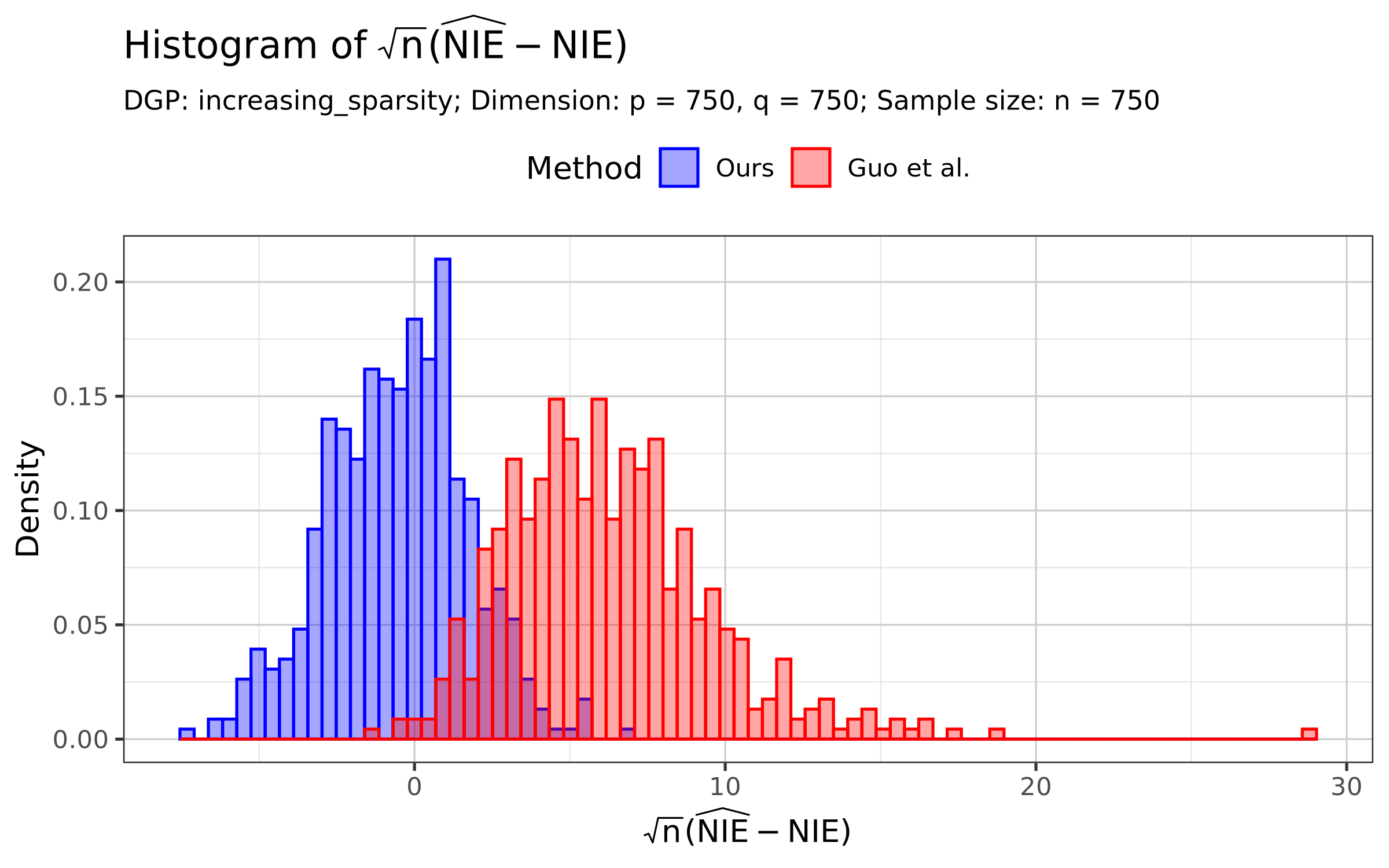}
    \caption{{\color{black}Comparison with \cite{guo2022high}}}
    \label{fig:simu_comparison_t}
\end{figure}

We report RMSE, empirical standard deviation, and confidence interval length, and empirical coverage to compare the performance. The results are summarized in Tables \ref{tab:simu_comparison_sparsity} and \ref{tab:simu_comparison_signal}, with the corresponding visualizations provided in Figure \ref{fig:simu_comparison_t}. Across both regimes, the comparison reveals that the modified procedure based on \cite{guo2022high} can perform reasonably in some lower-dimensional or larger-sample settings, but it becomes unstable in the ultra-high-dimensional regimes, especially when the sample size is small. In contrast, the proposed debiased estimator demonstrates more stable inferential performance across the regimes considered, with empirical coverage generally closer to the nominal level relative to the competing estimator.

\begin{table}[ht!]
\centering
\scriptsize
\renewcommand{\arraystretch}{1.3}
\resizebox{1\textwidth}{!}{
\begin{tabular}{cc|cccc|ccc|cccc|ccc} \hline
\multicolumn{2}{c|}{increasing sparsity} & \multicolumn{14}{c}{$\sigma^2 = 1$}\\ \hline
\multicolumn{2}{c|}{Estimators} & \multicolumn{7}{c|}{Debiasing (ours)} & \multicolumn{7}{c}{\citep{guo2022high}}\\ \hline
 & & \multicolumn{4}{c|}{Monte Carlo CI} & \multicolumn{3}{c|}{Analytical CI} & \multicolumn{4}{c|}{Monte Carlo CI} & \multicolumn{3}{c}{Analytical CI}\\
$p+q$ & \multicolumn{1}{c|}{$n$} & RMSE & SD & CI Len & Cov & SE & CI Len & Cov & RMSE & SD & CI Len & Cov & SE & CI Len & Cov\\ \hline
\multirow{5}{*}{$100$} & \multicolumn{1}{c|}{$250$} & 0.101 & 0.095 & 0.372 & 0.936 & 0.099 & 0.387 & 0.870& 0.065 & 0.065 & 0.253 & 0.966 & 0.026 & 0.102 & 0.568\\
 & \multicolumn{1}{c|}{$500$} & 0.087 & 0.079 & 0.311 & 0.916 & 0.084 & 0.330 & 0.936 & 0.050 & 0.049 & 0.193 & 0.974 & 0.017 & 0.067 & 0.534\\
 & \multicolumn{1}{c|}{$750$} & 0.078 & 0.074 & 0.289 & 0.940 & 0.077 & 0.303 & 0.948 & 0.044 & 0.044 & 0.171 & 0.958 & 0.018 & 0.069 & 0.626\\
 & \multicolumn{1}{c|}{$1000$} & 0.069 & 0.065 & 0.253 & 0.928 & 0.071 & 0.279 & 0.952 & 0.043 & 0.041 & 0.159 & 0.954 & 0.017 & 0.068 & 0.626\\
 & \multicolumn{1}{c|}{$1250$} & 0.066 & 0.062 & 0.241 & 0.928 & 0.052 & 0.204 & 0.948 & 0.049 & 0.036 & 0.141 & 0.854 & 0.018 & 0.072 & 0.504\\ \hline
\multirow{5}{*}{$800$} & \multicolumn{1}{c|}{$250$} & 0.126 & 0.115 & 0.451 & 0.926 & 0.099 & 0.390 & 0.880 & 0.171 & 0.097 & 0.378 & 0.732 & 0.063 & 0.247 & 0.452\\
 & \multicolumn{1}{c|}{$500$} & 0.096 & 0.091 & 0.357 & 0.948 & 0.083 & 0.325 & 0.914 & 0.216 & 0.105 & 0.413 & 0.610 & 0.048 & 0.187 & 0.098\\
 & \multicolumn{1}{c|}{$750$} & 0.079 & 0.076 & 0.300 & 0.950 & 0.081 & 0.316 & 0.906& 0.268 & 0.134 & 0.525 & 0.688 & 0.042 & 0.165 & 0.068\\
 & \multicolumn{1}{c|}{$1000$} & 0.074 & 0.075 & 0.292 & 0.952 & 0.070 & 0.276 & 0.936 & 0.333 & 0.163 & 0.640 & 0.622 & 0.039 & 0.154 & 0.056\\
 & \multicolumn{1}{c|}{$1250$} & 0.071 & 0.070 & 0.276 & 0.946 & 0.061 & 0.241 & 0.962 & 0.426 & 0.239 & 0.936 & 0.718 & 0.037 & 0.146 & 0.072\\ \hline
\multirow{5}{*}{$1500$} & \multicolumn{1}{c|}{$250$} & 0.141 & 0.128 & 0.503 & 0.916 & 0.099 & 0.387 & 0.838 & 0.161 & 0.102 & 0.398 & 0.796 & 0.064 & 0.250 & 0.562\\
 & \multicolumn{1}{c|}{$500$} & 0.132 & 0.099 & 0.388 & 0.876 & 0.084 & 0.330 & 0.792 & 0.196 & 0.108 & 0.425 & 0.714 & 0.050 & 0.196 & 0.252\\
 & \multicolumn{1}{c|}{$750$} & 0.085 & 0.083 & 0.324 & 0.936 & 0.076 & 0.300 & 0.922 & 0.259 & 0.122 & 0.478 & 0.576 & 0.044 & 0.174 & 0.058\\
 & \multicolumn{1}{c|}{$1000$} & 0.079 & 0.076 & 0.297 & 0.936 & 0.072 & 0.281 & 0.920 & 0.311 & 0.137 & 0.539 & 0.510 & 0.041 & 0.161 & 0.052\\
 & \multicolumn{1}{c|}{$1250$} & 0.070 & 0.069 & 0.272 & 0.942 & 0.061 & 0.239 & 0.922& 0.398 & 0.199 & 0.779 & 0.660 & 0.038 & 0.151 & 0.042\\ \hline
\multirow{5}{*}{$2500$} & \multicolumn{1}{c|}{$250$} & 0.186 & 0.146 & 0.574 & 0.870 & 0.102 & 0.399 & 0.666 & 3.169 & 1.177 & 4.612 & 0.312 & 0.258 & 1.012 & 0.002\\
 & \multicolumn{1}{c|}{$500$} & 0.162 & 0.128 & 0.502 & 0.894 & 0.089 & 0.349 & 0.702 & 3.950 & 0.890 & 3.487 & 0.034 & 0.190 & 0.743 & 0.002\\
 & \multicolumn{1}{c|}{$750$} & 0.131 & 0.097 & 0.382 & 0.830 & 0.081 & 0.317 & 0.734 & 3.911 & 1.130 & 4.431 & 0.128 & 0.165 & 0.648 & 0.000\\
 & \multicolumn{1}{c|}{$1000$} & 0.097 & 0.080 & 0.313 & 0.898 & 0.072 & 0.283 & 0.860 & 3.956 & 1.189 & 4.661 & 0.144 & 0.142 & 0.556 & 0.000\\
 & \multicolumn{1}{c|}{$1250$} & 0.087 & 0.084 & 0.330 & 0.942 & 0.065 & 0.254 & 0.934& 3.860 & 1.385 & 5.427 & 0.214 & 0.125 & 0.488 & 0.004\\ \hline
\end{tabular}%
}
\caption{{\color{black}No-interaction NIE comparison, increasing sparsity with moderate minimum active signal strength DGP, $\sigma^2=1$: our debiased estimator vs.\ \citet{guo2022high}. $s=5,10,15,20$ for $p+q=100,800,1500,2500$. Averaged over 500 replications, $K_1=K_2=2.75$.}}
\label{tab:simu_comparison_sparsity}
\end{table}

\begin{table}[ht!]
\centering
\scriptsize
\renewcommand{\arraystretch}{1.3}
\resizebox{1\textwidth}{!}{
\begin{tabular}{cc|cccc|ccc|cccc|ccc} \hline
\multicolumn{2}{c|}{minimal signal} & \multicolumn{14}{c}{$\sigma^2 = 1$}\\ \hline
\multicolumn{2}{c|}{Estimators} & \multicolumn{7}{c|}{Debiasing (ours)} & \multicolumn{7}{c}{\citep{guo2022high}}\\ \hline
 & & \multicolumn{4}{c|}{Monte Carlo CI} & \multicolumn{3}{c|}{Analytical CI} & \multicolumn{4}{c|}{Monte Carlo CI} & \multicolumn{3}{c}{Analytical CI}\\
$p+q$ & \multicolumn{1}{c|}{$n$} & RMSE & SD & CI Len & Cov & SE & CI Len & Cov & RMSE & SD & CI Len & Cov & SE & CI Len & Cov\\ \hline
\multirow{5}{*}{$100$} & \multicolumn{1}{c|}{$250$} & 0.087 & 0.076 & 0.298 & 0.916 & 0.098 & 0.385 & 0.972 & 0.078 & 0.078 & 0.306 & 0.946 & 0.022 & 0.087 & 0.400\\
 & \multicolumn{1}{c|}{$500$} & 0.084 & 0.077 & 0.301 & 0.922 & 0.085 & 0.331 & 0.942 & 0.067 & 0.067 & 0.263 & 0.958 & 0.015 & 0.058 & 0.320\\
 & \multicolumn{1}{c|}{$750$} & 0.074 & 0.069 & 0.269 & 0.934 & 0.077 & 0.304 & 0.958 & 0.063 & 0.063 & 0.246 & 0.948 & 0.014 & 0.055 & 0.356\\
 & \multicolumn{1}{c|}{$1000$} & 0.068 & 0.062 & 0.245 & 0.922 & 0.071 & 0.279 & 0.954 & 0.054 & 0.053 & 0.206 & 0.958 & 0.015 & 0.060 & 0.472\\
 & \multicolumn{1}{c|}{$1250$} & 0.065 & 0.061 & 0.237 & 0.944 & 0.052 & 0.204 & 0.974 & 0.062 & 0.051 & 0.202 & 0.926 & 0.017 & 0.068 & 0.362\\ \hline
\multirow{5}{*}{$800$} & \multicolumn{1}{c|}{$250$} & 0.113 & 0.111 & 0.436 & 0.956 & 0.100 & 0.392 & 0.922 & 0.167 & 0.091 & 0.355 & 0.724 & 0.053 & 0.207 & 0.332\\
 & \multicolumn{1}{c|}{$500$} & 0.087 & 0.086 & 0.339 & 0.944 & 0.082 & 0.321 & 0.930 & 0.219 & 0.102 & 0.399 & 0.576 & 0.042 & 0.166 & 0.046\\
 & \multicolumn{1}{c|}{$750$} & 0.075 & 0.075 & 0.295 & 0.958 & 0.080 & 0.312 & 0.970 & 0.305 & 0.160 & 0.628 & 0.736 & 0.038 & 0.150 & 0.014\\
 & \multicolumn{1}{c|}{$1000$} & 0.074 & 0.074 & 0.291 & 0.958 & 0.070 & 0.273 & 0.938 & 0.372 & 0.172 & 0.674 & 0.586 & 0.036 & 0.143 & 0.004\\
 & \multicolumn{1}{c|}{$1250$} & 0.067 & 0.066 & 0.259 & 0.948 & 0.061 & 0.238 & 0.922 & 0.470 & 0.219 & 0.858 & 0.602 & 0.035 & 0.139 & 0.010\\ \hline
\multirow{5}{*}{$1500$} & \multicolumn{1}{c|}{$250$} & 0.133 & 0.123 & 0.484 & 0.936 & 0.100 & 0.393 & 0.872 & 0.148 & 0.102 & 0.399 & 0.842 & 0.056 & 0.219 & 0.606\\
 & \multicolumn{1}{c|}{$500$} & 0.126 & 0.092 & 0.362 & 0.848 & 0.084 & 0.329 & 0.802 & 0.226 & 0.140 & 0.548 & 0.804 & 0.046 & 0.182 & 0.254\\
 & \multicolumn{1}{c|}{$750$} & 0.085 & 0.079 & 0.311 & 0.938 & 0.076 & 0.298 & 0.924 & 0.338 & 0.177 & 0.694 & 0.704 & 0.044 & 0.173 & 0.048\\
 & \multicolumn{1}{c|}{$1000$} & 0.081 & 0.074 & 0.292 & 0.916 & 0.071 & 0.279 & 0.910 & 0.406 & 0.189 & 0.741 & 0.594 & 0.041 & 0.161 & 0.014\\
 & \multicolumn{1}{c|}{$1250$} & 0.068 & 0.067 & 0.264 & 0.950 & 0.061 & 0.237 & 0.926 & 0.512 & 0.217 & 0.849 & 0.528 & 0.040 & 0.156 & 0.010\\ \hline
\multirow{5}{*}{$2500$} & \multicolumn{1}{c|}{$250$} & 0.157 & 0.139 & 0.544 & 0.928 & 0.104 & 0.410 & 0.816 & 2.056 & 0.722 & 2.829 & 0.284 & 0.173 & 0.680 & 0.000\\
 & \multicolumn{1}{c|}{$500$} & 0.125 & 0.116 & 0.455 & 0.934 & 0.089 & 0.347 & 0.838 & 2.548 & 0.484 & 1.897 & 0.012 & 0.130 & 0.510 & 0.000\\
 & \multicolumn{1}{c|}{$750$} & 0.101 & 0.090 & 0.351 & 0.914 & 0.080 & 0.313 & 0.878 & 2.593 & 0.497 & 1.947 & 0.034 & 0.114 & 0.446 & 0.000\\
 & \multicolumn{1}{c|}{$1000$} & 0.079 & 0.077 & 0.303 & 0.954 & 0.071 & 0.280 & 0.880 & 2.589 & 0.526 & 2.061 & 0.042 & 0.098 & 0.385 & 0.000\\
 & \multicolumn{1}{c|}{$1250$} & 0.084 & 0.084 & 0.328 & 0.950 & 0.064 & 0.250 & 0.920 & 2.528 & 0.652 & 2.557 & 0.086 & 0.087 & 0.340 & 0.000\\ \hline
\end{tabular}%
}
\caption{{\color{black}No-interaction NIE comparison, increasing sparsity with vanishing minimum active signal strength DGP, $\sigma^2=1$: our debiased estimator vs.\ \citet{guo2022high}. Sparsity fixed at $s=3$. Averaged over 500 replications, $K_1=K_2=2.75$.}}

\label{tab:simu_comparison_signal}
\end{table}

\subsection{{\color{black}Relation to \cite{rakshit2024statistical}}}
\label{appendix:comparison-rakshit-guo-poisson}

We next elaborate on the similarity and distinction between our work and \cite{rakshit2024statistical}. The Poisson-link mediation model studied by \cite{rakshit2024statistical} is more general than the linear model in terms of the outcome link function. Thus, our contribution should not be interpreted as a generalization in the sense of moving from a linear model to a nonlinear exponential-link model. Rather, our work extends the high-dimensional mediation literature along a different modeling dimension: we allow ultra-high-dimensional covariates \(X\) and ultra-high-dimensional mediators \(M\), along with \(A\)--\(X\) and \(A\)--\(M\) interactions in the outcome model for \(Y\), as well as \(A\)--\(X\) interactions in the mediator model.
\cite{rakshit2024statistical} makes an important contribution by developing inference for high-dimensional sparse Poisson regression, with an application to mediation analysis for count outcomes and, in particular, testing treatment--mediator interactions. In their mediation formulation, the outcome model can be written as
\[
g\big(\bbE[Y \mid A, M]\big)
=
\alpha_0 + \alpha_1 A + M^\top \beta_0 + A M^\top \beta_1
=
\alpha_0 + \alpha_1 A + (\beta_0^\top, \beta_1^\top) 
\begin{pmatrix}
M \\
AM
\end{pmatrix},
\]
where \(g\) is the Poisson GLM link function. Thus, the treatment--mediator interaction is encoded by the coefficient \(\beta_1\). However, they do not include UHD covariates, and consequently their model does not allow for \(A\)--\(X\) interactions in either the mediator-generation or outcome-generation model. This is precisely why the structural problem considered in our paper is fundamentally different from that of \cite{rakshit2024statistical}. In particular, our work extends this line of work by allowing UHD covariates and UHD mediators simultaneously, together with their interactions with the treatment variable, although we focus on a linear link rather than a Poisson link. We have also elaborated in Section \ref{appendix:comparison-guo-trilogy} that the simultaneous presence of UHD mediators and covariates, together with their interactions with the treatment, makes estimation of the mediation functional substantially more involved, since the target depends on cross-arm nuisance parameters involving both a UHD matrix and a UHD vector.

\subsection{{\color{black}Relation to other methods on high-dimensional mediation analysis}}
\label{appendix:comparison-other-hd-mediation}

We also discuss several other recent works in the high-dimensional mediation analysis literature, including \citep{jones2025causal,chakrabortty2018inference,shi2022testing,loh2022nonlinear}. These papers make important contributions to the broader literature, but their settings, parameters of interest, or inferential goals are different from those considered in this paper.

First, \citep{jones2025causal} studies a similar mediation analysis problem from the perspective of post-selection inference. Their target parameter is related to the mediation functional considered here, and their modeling strategy is more general in some respects. However, their paper does not consider interactions with treatment and, more importantly, focuses on \emph{fixed-dimensional mediators}. We therefore view \citep{jones2025causal} as addressing an interesting complementary setting rather than being directly comparable to the interaction-rich UHD mediation problem studied in this paper. 

Next, \cite{chakrabortty2018inference} and \cite{shi2022testing} study important high-dimensional mediation analysis problems under a linear structural equation model (LSEM) over a causal graph. In a simplified notation, their model can be written as
\begin{equation*}
\textstyle
V-\mu = \mathbf G^\top(V-\mu)+\varepsilon,
\end{equation*}
where \(V\) collects the treatment, covariates, mediators, and outcome, and \(G_{ij}\neq 0\) encodes a directed edge from \(V_i\) to \(V_j\). Equivalently, each node is generated as a linear combination of its parent nodes plus an additive error:
\begin{equation*}
\textstyle
V_j-\mu_j=\sum_{k\in \mathrm{pa}(j)}G_{kj}(V_k-\mu_k)+\varepsilon_j.
\end{equation*}
Mapping this framework to our notation, the corresponding mediator and outcome models are of the form
\begin{align*}
Y &= \alpha_0+\alpha_A A+ X^\top\beta + M^\top\gamma+\varepsilon, \\
M &= \delta_0+\delta_A A+ B_X X + H M+U,
\end{align*}
where \(H\) may encode directed relationships among mediators. Thus, after absorbing constants and coefficients, this falls into the no-treatment-interaction structure discussed in Section \ref{appendix:comparison-guo-trilogy}. In particular, there are no \(A\)--\(X\) interactions in the mediator model and no \(A\)--\(X\) or \(A\)--\(M\) interactions in the outcome model. As discussed extensively in Section \ref{appendix:comparison-guo-trilogy}, in such a no-interaction model, the mediation functional, and consequently the NDE and NIE, have a substantially simpler structure and can be reduced to a standard high-dimensional regression/debiasing problem. In contrast, our model requires a multi-step debiasing strategy due to the presence of composite cross-arm nuisance terms such as $\bar X^\top B_0^\top\gamma_1$. Furthermore, the primary goals of \cite{chakrabortty2018inference} and \cite{shi2022testing} are \emph{causal discovery} and the estimation of \emph{individual mediation effects}. For mediation analysis, they focus on controlled direct and indirect effects, which are fundamentally different from the natural direct and indirect effects considered in this paper. Finally, \cite{chakrabortty2018inference} allows \(p+q\) to grow only polynomially with \(n\), whereas our methodology permits \(p+q\) to be exponentially large in \(n\).

\cite{loh2022nonlinear} proposes an outcome-regression-based averaging approach for estimating interventional direct and indirect effects in the presence of high-dimensional mediators. Our work differs from theirs in several important respects. First and foremost, \cite{loh2022nonlinear} does not provide any theoretical guarantee for their estimation procedure. Second, the causal targets are different: interventional direct and indirect effects are distinct from the natural direct and indirect effects considered in this paper; see \cite{vanderweele2014effect} for a detailed discussion of this distinction. Third, their procedure fits outcome-regression models after high-dimensional variable selection and averages the resulting imputed potential outcomes over estimated counterfactual mediator distributions using Monte Carlo integration. In ultra-high-dimensional regimes, where \(\dim(M)\gg n\), such regression-based plug-in estimators can suffer from substantial regularization bias unless an additional debiasing step is used; this is precisely one of the main issues addressed by our method. Thus, while \cite{loh2022nonlinear} makes an important contribution to estimating interventional direct and indirect effects with high-dimensional mediators, their parameter of interest, methodology, and focus differ from ours.

\subsection{{\color{black}Relation to Existing Debiased Inference for High-Dimensional Functionals}}
\label{appendix:existing-debiasing-functional}
In this section, we elaborate on the relationship between our proposed debiasing strategy and the existing literature on debiased inference for high-dimensional functionals. The estimation of debiased inference for linear functionals of high-dimensional regression parameters has a long history, dating back to the seminal works of \citep{javanmard2014confidence,van2014asymptotically,zhang2014confidence}. More recently \cite{cai2021optimal} develop debiasing techniques for linear functionals \(x^\top(\beta_1 - \beta_2)\) where $\beta_1$ and $\beta_2$ comes from two regression model, and \cite{guo2019optimal} developed debiasing techniques for quadratic/bilinear forms \(\beta^\top\gamma\) in high-dimensional linear models. Although these results are highly relevant to the broader literature on high-dimensional inference, the mediation functional considered in our paper is structurally different and cannot be reduced directly to either of these canonical settings.
To see the difficulty, consider the mediation functional under our model, suppressing intercepts for simplicity:
\begin{equation*}
\textstyle
\theta_0 = \bar X^\top(\beta_1+\bB_0^\top\gamma_1).
\end{equation*}
The key nonstandard component is the cross-arm nuisance component $\bar X^\top \bB_0^\top\gamma_1$, where \(\bB_0\) is identified from the mediator model in the control arm and \(\gamma_1\) is identified from the outcome model in the treated arm. There are two natural ways to relate this term to existing debiasing results:
\begin{equation*}
\textstyle
\bar X^\top \bB_0^\top\gamma_1 = (\bB_0\bar X)^\top\gamma_1 \qquad\text{or}\qquad \bar X^\top \bB_0^\top\gamma_1 = \bar X^\top(\bB_0^\top\gamma_1).
\end{equation*}
We explain why neither route gives an immediate application of existing linear- or bilinear-functional debiasing methods.

First, consider the representation $\bar X^\top \bB_0^\top\gamma_1 = (\bB_0\bar X)^\top\gamma_1$. This representation views the target component as an inner product between two high-dimensional vectors, $\bB_0 \bar X$ and $\gamma_1$, and therefore appears related to the inner-product debiasing method of \cite{guo2019optimal}. However, their framework requires both vectors in the inner product to be sparse regression coefficient vectors arising from two high-dimensional linear models, which can be estimated consistently with respect to $\ell_1/\ell_2$ norm at a standard Lasso rate. In our case, \(\gamma_1\) is sparse, but \(\bB_0\bar X\) need not be sparse at all. Indeed, \(\bB_0\bar X\) is not a regression coefficient; it is a derived \(q\)-dimensional quantity, with $q \gg n$ in our setup, obtained by multiplying the unknown mediator coefficient matrix \(\bB_0\) by the sample mean \(\bar X\). Row-sparsity of \(\bB_0\) does not imply sparsity of \(\bB_0\bar X\). For example, suppose each row of \(\bB_0\) has only one nonzero entry, say
\begin{equation*}
\textstyle
\bB_{0,j1}=1,\qquad \bB_{0,j\ell}=0\ \text{for all }\ell\neq 1,
\qquad j=1,\ldots,q.
\end{equation*}
Then $\bB_0\bar X = \bar X_1 \mathbf 1_q$, which is dense whenever \(\bar X_1\neq 0\), even though each row of \(\bB_0\) is only \(1\)-sparse. More generally, since \(\bar X\) is typically dense, the inner products $(\bB_0\bar X)_j=\langle \bB_{0,j\cdot},\bar X\rangle$ will generally be nonzero for many, if not all, \(j\)'s. 
Although each coordinate $(\bB_0\bar X)_j=\langle \bB_{0,j\cdot},\bar X\rangle$ could in principle be estimated by debiasing the row-wise regression of \(M_j\) on \(X\) in the direction of $\bar X$, this would only provide coordinate-wise control. The entire debiased vector $\bB_0 \bar X$ can not be consistently estimated with respect to $\ell_1/\ell_2$ norm as $q \gg n$. Thus, the inner-product debiasing theory for two estimable high-dimensional regression vectors does not apply directly.

Second, one may instead write $\bar X^\top \bB_0^\top\gamma_1 = \bar X^\top b_0$, where $b_0 = \bB_0^\top\gamma_1$. If \(b_0\) were the coefficient vector of a single observed high-dimensional linear regression, then standard debiasing methods for linear functionals \citep{zhang2014confidence,van2014asymptotically,javanmard2014confidence} would be directly applicable. The key difficulty is that $b_0$ is not an observed regression coefficient from a single model. It is a composite parameter: the mediator relationship represented by $\bB_0$ is associated with the control arm, whereas $\gamma_1$ is learned from the treated-arm outcome model. If \(\gamma_1\) were known, one could regress \(M^\top\gamma_1\) on \(X\) in the control arm to directly estimate $b_0$ and then debias in the direction \(\bar X\). In reality, however, \(\gamma_1\) is unknown and must be estimated from the treated sample. Replacing \(\gamma_1\) by \(\hat\gamma_1\) introduces an additional first-stage estimation error into the control-arm regression. The resulting bias is not present in the standard single-regression linear functional setting. The product-form bias decomposition in Proposition \ref{prop:bias_structure} shows how the estimation errors from these different nuisance components interact and why additional debiasing is needed. Our multi-step debiasing procedure is designed precisely to address this cross-arm bias propagation.

In summary, our parameter of inference is neither a linear functional of a single sparse regression coefficient nor a bilinear functional of two separately estimable sparse regression coefficients. Rather, it contains a cross-arm and cross-equation component whose relevant parts must be learned and carefully coupled to remove the effect of estimation/regularization in the first order.
This structural feature is absent from standard high-dimensional functional inference frameworks and necessitates the multi-step debiasing strategy developed in this paper.

\subsection{{\color{black}Relation to Riesz-representer and balancing-based debiasing methods}}
Our construction is also related to the literature on Riesz-representer and balancing-based approaches to debiasing \citep{chernozhukov2022automatic,hirshberg2021augmented,ben2021balancing} and to \cite{athey2018approximate} for the estimation of ATE in the presence of UHD covariates. In particular, the optimization problems used to compute the weights $\tau_1$ and $\tau_2$ in Steps 1 and 3 of Algorithm \ref{algomain_modified} are closely related to this line of work; they can be viewed as approximate balancing for estimating the relevant Riesz representers. 
However, the connection with Riesz regression appears only at the level of the individual linear-functional debiasing subproblems. 
The full mediation-functional estimation problem, however, is not a standard linear-functional debiasing problem. The key difficulty is the composite term $\bar X^\top \bB_0^\top\gamma_1$ where the relevant information of $B_0$ must be learned from the control-arm mediator-covariate relationship, whereas $\gamma_1$ must be learned from the treated-arm outcome
model, and $\bar X$ is computed using observations from both treatment arms. If $\bB_0\bar X$ were known, one could debias the treated-arm outcome regression in the fixed direction $(\bar X^\top,(\bB_0\bar X)^\top)^\top$. 
If $\gamma_1$ were known, one could regress $M^\top\gamma_1$ on $X$ in the control arm and debias the resulting linear functional in the direction $\bar X$. However, neither object is known in our setting. Moreover, $\bB_0\bar X$ can be completely dense and is not generally consistently estimable with respect to the standard $\ell_1/\ell_2$ norm. Consequently, the debiasing directions in the subsequent regressions are themselves estimated from different treatment arms, which introduces additional regularization and plug-in biases that must be carefully removed to obtain a $\sqrt{n}$-CAN estimate of $\theta_{10}$. 

The proposed procedure (Algorithm \ref{algomain_modified}) does not estimate the full matrix $\bB_0$. Instead, the control-side balancing program directly constructs $\hat m_c=\mathbf M_c^\top\hat\tau_2$ as an approximation of $\bB_0\bar X$. This estimated quantity enters the direction used in the treated-arm debiasing problem. The procedure subsequently regresses the pseudo-outcome $M^\top\hat\gamma_1$ on $X$ in the control arm and reuses the same weight $\hat\tau_2$ in the control-side correction. Thus, although the individual optimization problems defining $\hat\tau_1$ and $\hat\tau_2$ are closely related to standard Riesz-representer or approximate-balancing constructions, these weights are embedded in a coupled
multi-step procedure in which both a debiasing direction and a pseudo-outcome contain estimated high-dimensional quantities. Controlling the resulting cross-arm bias propagation and establishing the product-form remainder in
Proposition \ref{prop:bias_structure} are the main roles of our theoretical analysis. Standard Riesz regression or linear/bilinear functional debiasing tools do not directly yield this result; see Section \ref{appendix:existing-debiasing-functional} for further discussion.

The relationship with \cite{liu2024bias} is complementary. Their approach also starts from the mediation influence-function representation and uses balancing ideas to stabilize estimation of the nuisance components that enter the leading bias terms. By contrast, we study a sparse UHD linear structural model in which both $X$ and $M$ may substantially exceed the sample
size and treatment interactions are present in both the mediator and outcome models. In this setting, the main difficulty is not only instability associated with inverse-probability-type weighting, but also the regularization and cross-arm plug-in biases arising from the estimated direction $\hat m_c$ and the estimated pseudo-outcome $M^\top\hat\gamma_1$. Thus, although the balancing steps in our method are analogous in spirit to those in \cite{liu2024bias}, they play a different technical role. Rather than estimating nuisance functions for a fixed influence-function estimator and plugging them into the influence-function expression, our balancing weights are embedded in a multi-step debiasing procedure for penalized high-dimensional regressions. Establishing that the resulting product-form remainder is $o_p(n^{-1/2})$ under sparse UHD assumptions is a principal component of our analysis.

Regarding the connection with \cite{athey2018approximate}, our work extends the high-dimensional debiasing and balancing approach from average treatment
effect estimation to causal mediation analysis. \cite{athey2018approximate} focus on estimating the total effect of a treatment in the presence of high-dimensional covariates. Estimation of the NDE and NIE is more involved because it requires estimating counterfactual mediation functionals rather than only the total effect. In our interaction-rich UHD setting, these functionals contain composite cross-arm nuisance terms such as $\bar X^\top \bB_0^\top\gamma_1$, which require an estimated debiasing direction, an estimated pseudo-outcome, and coupled treatment- and control-side corrections. Consequently, the proposed construction goes beyond applying a
single balancing-weight correction separately to two standard linear
functionals.

A nonlinear extension based on Auto-DML \citep{chernozhukov2022automatic} or RieszBoost \citep{lee2025rieszboost} is an interesting but nontrivial direction for future work, rather than a direct consequence of existing theory. In a fully nonparametric mediation model, the target involves the conditional law of the UHD mediator vector $M$ given UHD covariates $X$, and estimating either this law or the corresponding Riesz representer is generally infeasible without strong structural assumptions. Even if the target can be reduced to a non-linear functional of just the conditional mean $\bbE[M\mid X,A=0]$, the problem is still not an immediate application of existing Riesz regression theory. The mediator regression is vector-valued:
\begin{align*}
\nu_0(X) & = \bbE[M\mid X, A=0] = (\nu_{0,1}(X),\ldots,\nu_{0,q}(X))^\top, \\
\nu_{0, j}(x) & = \bbE[M_j \mid X = x, A = 0] \,.
\end{align*}
The linear construction developed in this paper avoids estimating these $q$ conditional-mean components separately by exploiting the exact identity $\mathbf M_c=\mathbf X_cB_0^\top+\mathbf U_c$ and constructing $\mathbf M_c^\top\hat\tau_2$ directly. There is no
immediate analog of this reduction for a generic nonlinear mediator model.
This leads to a growing collection of nuisance regressions, with the number of components of order \(q\), and in our UHD setting \(q\gg n\) (can even be exponentially large in $n$). Existing Auto-DML theory is not directly designed to handle a growing number of nuisance components, especially when that number far exceeds the sample size; the number of nuisance components is always fixed in \cite{chernozhukov2022automatic}. As a matter of fact, in Example 5 of \cite{chernozhukov2022automatic}, the authors assumed the mediators are categorical random variables with $K$ different levels, where $K$ is a fixed constant that does not change/grow with $n$, whereas our model allows the mediators to be discrete or continuous and the number of mediators can be much larger than the sample size.  
Consequently, a non-linear extension in the UHD regime would require a careful analysis of the interplay between the complexity of UHD nonparametric function classes, an appropriate notion of sparsity (e.g., sparse multi-index or sparse additive structure), and the fact that the ambient dimensions themselves grow with the sample size. The relevant asymptotic framework would no longer be the classical fixed-dimensional setting in which the population model is held fixed while \(n \uparrow \infty\). Rather, it would be a triangular-array-type regime, where the statistical model, the nuisance parameter spaces, and the ambient dimensions \(p\) and \(q\) all vary with \(n\). Therefore, new tools would be needed to handle nonparametric estimation (and debiasing) of mediation functionals in this regime, which we view as an interesting direction for future research.
\addtocontents{toc}{\protect\setcounter{tocdepth}{2}}

\section{Phenotype Features \label{appendix:pheno}}

\newcommand{\feature}[1]{\texttt{\detokenize{#1}}}

\begingroup
\small
\setlength{\LTleft}{0pt}
\setlength{\LTright}{0pt}
\renewcommand{\arraystretch}{1.18}
\rowcolors{3}{gray!4}{white}

\begin{longtable}{
@{}
>{\raggedright\arraybackslash}p{0.34\textwidth}
>{\raggedright\arraybackslash}p{0.62\textwidth}
@{}
}
\caption{Description of the phenotype variables used in the lung cancer data analysis.}
\label{tab:pheno}
\\

\toprule
\rowcolor{gray!15}
\textbf{Feature Name} & \textbf{Explanation} \\
\midrule
\endfirsthead

\toprule
\rowcolor{gray!15}
\textbf{Feature Name} & \textbf{Explanation} \\
\midrule
\endhead

\midrule
\multicolumn{2}{r}{\footnotesize Continued on next page} \\
\endfoot

\bottomrule
\endlastfoot

\feature{gender}
& Patient's gender, such as male or female. \\

\feature{age}
& Age of the patient at diagnosis. \\

\feature{ABSOLUTE_Purity}
& Tumor purity score estimated through molecular analysis. \\

\feature{number_pack_years_smoked}
& Number of cigarette pack-years smoked by the patient. \\

\feature{initial_weight}
& Patient's initial body weight at diagnosis. \\

\feature{radiation_therapy}
& Indicator of whether the patient underwent radiation therapy. \\

\feature{dlco_predictive_percent}
& Predicted percentage of the diffusing capacity of the lungs for carbon monoxide. \\

\feature{intermediate_dimension}
& Intermediate dimension of the tumor, corresponding to a size-related tumor measurement. \\

\feature{longerest_dimension}
& Longest dimension of the tumor. \\

\feature{karnofsky_performance_score}
& Performance status score quantifying the patient's ability to perform daily activities. \\

\feature{anatomic_neoplasm_subdivision}
& Anatomic location of the tumor in the lungs. \\

\feature{histological_type}
& Histological classification of the tumor, such as adenocarcinoma or squamous cell carcinoma. \\

\feature{days_to_collection}
& Number of days from diagnosis to sample collection. \\

\makecell[l]{\feature{new_tumor_event}\\\feature{after_initial_treatment}}
& Indicator of whether a new tumor event occurred after initial treatment. \\

\makecell[l]{\feature{post_bronchodilator}\\\feature{fev1_percent}}
& Predicted percentage of forced expiratory volume in one second after bronchodilator use. \\

\feature{shortest_dimension}
& Shortest dimension of the tumor. \\

\feature{pathologic_stage}
& Pathologic stage of the cancer, such as Stage I or Stage II. \\

\feature{sample_type}
& Type of the sample, such as primary tumor or normal tissue. \\

\feature{targeted_molecular_therapy}
& Indicator of whether the patient received molecularly targeted therapy. \\

\feature{status}
& Survival status of the patient, such as alive or deceased. \\

\end{longtable}
\endgroup

\section{{\color{black}More on Simulation Study and Sensitivity Analysis \label{Appendix:more_simu_sensi}}} 

This section provides additional simulation results and sensitivity analyses that complement the main simulation study in Section \ref{sec:simulation}. We use the same data-generating mechanisms, estimators, and evaluation metrics described there, and report results for the simulation regimes omitted from the main text for brevity. In particular, the additional experiments examine the performance of the proposed estimator under lower noise levels, increasing sparsity, weak nonzero signals, and conditional heteroskedasticity, as well as its sensitivity to the balancing constants $K_1$ and $K_2$.

Table~\ref{tab:simu_combined_base_sigma0p1} reports the results for the base DGP with a smaller noise variance, $\sigma^2=0.1$. Table~\ref{tab:simu_sparsity_sigma1} considers the increasing-sparsity regime, where the number of active coefficients grows with the ambient dimension. Table~\ref{tab:simu_signal_sigma1} considers the minimal-signal regime, in which a subset of the nonzero coefficients is generated near the detection boundary. Tables~\ref{tab:simu_hetero_sigma1} and~\ref{tab:simu_hetero_sigma0p5} evaluate the estimator under conditional heteroskedasticity at two different overall noise levels. Finally, Tables~\ref{tab:sens_diag_n750} and~\ref{tab:sens_diag_n1250} examine the sensitivity of the RMSE for $K_1, K_2$ at two sample sizes, while Table~\ref{tab:sens_grid} varies $K_1$ and $K_2$ separately to assess the robustness of the estimator to their joint choice.

\begin{table}[!ht]
\centering
\small
\renewcommand{\arraystretch}{1.4}
\resizebox{1\textwidth}{!}{
\begin{tabular}{cc|cccc|ccc|cccc} \hline
\multicolumn{2}{c|}{$s=3,\ k_1=k_2=3$} & \multicolumn{11}{c}{$\sigma^2 = 0.1$}\\ \hline
\multicolumn{2}{c|}{Estimators} & \multicolumn{7}{c|}{$\hat\theta_0^{\mathrm{cf}}$ (Debiasing)} & \multicolumn{4}{c}{$\hat \theta^{\rm plug-in}$}\\ \hline
 & & \multicolumn{4}{c|}{Monte Carlo CI} & \multicolumn{3}{c|}{Analytical CI} & \multicolumn{4}{c}{Monte Carlo CI}\\
$p+q$ & \multicolumn{1}{c|}{$n$} & RMSE & SD & CI Len & Cov & SE & CI Len & Cov & RMSE & SD & CI Len & Cov\\ \hline
\multirow{5}{*}{$100$} & \multicolumn{1}{c|}{$250$}  & 0.720 & 0.321 & 1.260 & 0.500 & 0.296 & 1.159 & 0.416 & 0.921 & 0.229 & 0.899 & 0.032\\
 & \multicolumn{1}{c|}{$500$}  & 0.182 & 0.168 & 0.657 & 0.786 & 0.176 & 0.691 & 0.804& 0.661 & 0.140 & 0.548 & 0.008\\
 & \multicolumn{1}{c|}{$750$}  & 0.203 & 0.130 & 0.510 & 0.926& 0.139 & 0.546 & 0.938 & 0.575 & 0.106 & 0.416 & 0.000\\
 & \multicolumn{1}{c|}{$1000$} & 0.117 & 0.115 & 0.450 & 0.932& 0.121 & 0.475 & 0.922& 0.519 & 0.090 & 0.354 & 0.000\\
 & \multicolumn{1}{c|}{$1250$} & 0.125 & 0.119 & 0.465 & 0.942 & 0.114 & 0.446 & 0.958 & 0.490 & 0.083 & 0.324 & 0.000\\ \hline
\multirow{5}{*}{$800$} & \multicolumn{1}{c|}{$250$}  & 0.231 & 0.230 & 0.901 & 0.888& 0.392 & 1.538 & 0.996 & 0.366 & 0.213 & 0.837 & 0.694\\
 & \multicolumn{1}{c|}{$500$}  & 0.176 & 0.142 & 0.557 & 0.892 & 0.199 & 0.781 & 0.978 & 0.356 & 0.045 & 0.175 & 0.000\\
 & \multicolumn{1}{c|}{$750$}  & 0.150 & 0.125 & 0.492 & 0.942& 0.144 & 0.564 & 0.946 & 0.352 & 0.042 & 0.166 & 0.000\\
 & \multicolumn{1}{c|}{$1000$} & 0.115 & 0.107 & 0.420 & 0.942 & 0.126 & 0.493 & 0.966 & 0.302 & 0.034 & 0.132 & 0.000\\
 & \multicolumn{1}{c|}{$1250$} & 0.090 & 0.089 & 0.350 & 0.944 & 0.106 & 0.416 & 0.984 & 0.274 & 0.030 & 0.118 & 0.000\\ \hline
\multirow{5}{*}{$1500$} & \multicolumn{1}{c|}{$250$}  & 0.337 & 0.216 & 0.848 & 0.772 & 0.345 & 1.352 & 0.988 & 0.190 & 0.181 & 0.709 & 0.956\\
 & \multicolumn{1}{c|}{$500$}  & 0.159 & 0.144 & 0.563 & 0.890 & 0.197 & 0.773 & 0.994 & 0.286 & 0.036 & 0.140 & 0.000\\
 & \multicolumn{1}{c|}{$750$}  & 0.148 & 0.118 & 0.463 & 0.934& 0.142 & 0.556 & 0.944 & 0.243 & 0.028 & 0.108 & 0.000\\
 & \multicolumn{1}{c|}{$1000$} & 0.095 & 0.093 & 0.363 & 0.934 & 0.115 & 0.451 & 0.976 & 0.216 & 0.023 & 0.090 & 0.000\\
 & \multicolumn{1}{c|}{$1250$} & 0.096 & 0.089 & 0.351 & 0.942 & 0.102 & 0.402 & 0.966 & 0.199 & 0.020 & 0.079 & 0.000\\ \hline
\multirow{5}{*}{$2500$} & \multicolumn{1}{c|}{$250$}  & 0.635 & 0.294 & 1.153 & 0.551 & 0.436 & 1.710 & 0.831 & 0.263 & 0.190 & 0.747 & 0.843\\
 & \multicolumn{1}{c|}{$500$}  & 0.184 & 0.151 & 0.592 & 0.828 & 0.226 & 0.885 & 0.990 & 0.211 & 0.024 & 0.095 & 0.000\\
 & \multicolumn{1}{c|}{$750$}  & 0.183 & 0.128 & 0.502 & 0.890& 0.167 & 0.655 & 0.940 & 0.254 & 0.025 & 0.098 & 0.000\\
 & \multicolumn{1}{c|}{$1000$} & 0.110 & 0.108 & 0.424 & 0.930 & 0.136 & 0.532 & 0.988 & 0.198 & 0.020 & 0.078 & 0.000\\
 & \multicolumn{1}{c|}{$1250$} & 0.106 & 0.103 & 0.404 & 0.954& 0.118 & 0.461 & 0.960 & 0.137 & 0.014 & 0.054 & 0.000\\ \hline
\end{tabular}%
}
\caption{{\color{black}Simulation results for the debiased ($\hat\theta_0^{\mathrm{cf}}$) and naive ($\hat \theta^{\rm plug-in}$) estimators under the base DGP with $\sigma^2=0.1$. Numbers are averaged over 500 replications with $K_1=K_2=2.75$.}}
\label{tab:simu_combined_base_sigma0p1}
\end{table}

\begin{table}[!ht]
\centering
\small
\renewcommand{\arraystretch}{1.4}
\resizebox{1\textwidth}{!}{
\begin{tabular}{cc|cccc|ccc|cccc} \hline
\multicolumn{2}{c|}{increasing sparsity} & \multicolumn{11}{c}{$\sigma^2 = 1$}\\ \hline
\multicolumn{2}{c|}{Estimators} & \multicolumn{7}{c|}{$\hat\theta_0^{\mathrm{cf}}$ (Debiasing)} & \multicolumn{4}{c}{$\hat \theta^{\rm plug-in}$}\\ \hline
 & & \multicolumn{4}{c|}{Monte Carlo CI} & \multicolumn{3}{c|}{Analytical CI} & \multicolumn{4}{c}{Monte Carlo CI}\\
$p+q$ & \multicolumn{1}{c|}{$n$} & RMSE & SD & CI Len & Cov & SE & CI Len & Cov & RMSE & SD & CI Len & Cov\\ \hline
\multirow{5}{*}{$100$} & \multicolumn{1}{c|}{$250$}  & 0.456 & 0.233 & 0.914 & 0.605 & 0.278 & 1.088 & 0.741 & 1.789 & 0.303 & 1.189 & 0.000\\
 & \multicolumn{1}{c|}{$500$}  & 0.264 & 0.173 & 0.678 & 0.790 & 0.191 & 0.747 & 0.830 & 1.162 & 0.218 & 0.854 & 0.000\\
 & \multicolumn{1}{c|}{$750$}  & 0.177 & 0.152 & 0.595 & 0.903 & 0.159 & 0.624 & 0.917 & 0.939 & 0.170 & 0.666 & 0.000\\
 & \multicolumn{1}{c|}{$1000$} & 0.144 & 0.129 & 0.505 & 0.916 & 0.136 & 0.534 & 0.938 & 0.815 & 0.136 & 0.532 & 0.000\\
 & \multicolumn{1}{c|}{$1250$} & 0.136 & 0.119 & 0.466 & 0.915 & 0.126 & 0.494 & 0.940& 0.792 & 0.134 & 0.527 & 0.000\\ \hline
\multirow{5}{*}{$800$} & \multicolumn{1}{c|}{$250$}  & 0.839 & 0.304 & 1.192 & 0.254 & 0.462 & 1.809 & 0.631 & 1.361 & 0.508 & 1.992 & 0.326\\
 & \multicolumn{1}{c|}{$500$}  & 0.341 & 0.190 & 0.747 & 0.683 & 0.222 & 0.869 & 0.787 & 1.365 & 0.181 & 0.709 & 0.000\\
 & \multicolumn{1}{c|}{$750$}  & 0.211 & 0.151 & 0.590 & 0.834 & 0.182 & 0.713 & 0.913 & 1.570 & 0.048 & 0.187 & 0.000\\
 & \multicolumn{1}{c|}{$1000$} & 0.138 & 0.131 & 0.513 & 0.933 & 0.159 & 0.625 & 0.978 & 1.606 & 0.056 & 0.219 & 0.000\\
 & \multicolumn{1}{c|}{$1250$} & 0.123 & 0.118 & 0.461 & 0.943 & 0.143 & 0.560 & 0.973 & 1.495 & 0.060 & 0.236 & 0.000\\ \hline
\multirow{5}{*}{$1500$} & \multicolumn{1}{c|}{$250$}  & 0.601 & 0.600 & 2.351 & 0.933 & 0.724 & 2.838 & 0.967 & 2.413 & 0.709 & 2.778 & 0.056\\
 & \multicolumn{1}{c|}{$500$}  & 0.677 & 0.336 & 1.317 & 0.580 & 0.458 & 1.796 & 0.774 & 0.796 & 0.377 & 1.478 & 0.540\\
 & \multicolumn{1}{c|}{$750$}  & 0.606 & 0.246 & 0.963 & 0.390 & 0.288 & 1.129 & 0.529 & 2.187 & 0.048 & 0.189 & 0.000\\
 & \multicolumn{1}{c|}{$1000$} & 0.382 & 0.223 & 0.873 & 0.710 & 0.241 & 0.945 & 0.749 & 2.286 & 0.035 & 0.136 & 0.000\\
 & \multicolumn{1}{c|}{$1250$} & 0.199 & 0.188 & 0.738 & 0.940 & 0.223 & 0.874 & 0.976 & 2.243 & 0.031 & 0.122 & 0.000\\ \hline
\multirow{5}{*}{$2500$} & \multicolumn{1}{c|}{$250$}  & 0.806 & 0.669 & 2.623 & 0.375 & 0.790 & 3.098 & 0.625 & 3.734 & 0.537 & 2.105 & 0.000\\
 & \multicolumn{1}{c|}{$500$}  & 0.670 & 0.404 & 1.583 & 0.761& 0.553 & 2.167 & 0.876 & 2.061 & 0.560 & 2.197 & 0.052\\
 & \multicolumn{1}{c|}{$750$}  & 0.699 & 0.286 & 1.122 & 0.875& 0.380 & 1.490 & 0.875& 1.162 & 0.296 & 1.162 & 0.038\\
 & \multicolumn{1}{c|}{$1000$} & 0.483 & 0.205 & 0.802 & 0.810& 0.256 & 1.004 & 0.819& 2.094 & 0.042 & 0.164 & 0.000\\
 & \multicolumn{1}{c|}{$1250$} & 0.318 & 0.218 & 0.853 & 0.810 & 0.223 & 0.872 & 0.917& 2.089 & 0.031 & 0.123 & 0.000\\ \hline
\end{tabular}%
}
\caption{{\color{black}Simulation results under the increasing-sparsity DGP with $\sigma^2=1$. $s=s_B=5,10,15,20$ for $p+q=100,800,1500,2500$; Averaged over 500 replications, $K_1=K_2=2.75$.}}
\label{tab:simu_sparsity_sigma1}
\end{table}

\begin{table}[!ht]
\centering
\small
\renewcommand{\arraystretch}{1.4}
\resizebox{1\textwidth}{!}{
\begin{tabular}{cc|cccc|ccc|cccc} \hline
\multicolumn{2}{c|}{minimal signal} & \multicolumn{11}{c}{$\sigma^2 = 1$}\\ \hline
\multicolumn{2}{c|}{Estimators} & \multicolumn{7}{c|}{$\hat\theta_0^{\mathrm{cf}}$ (Debiasing)} & \multicolumn{4}{c}{$\hat \theta^{\rm plug-in}$}\\ \hline
 & & \multicolumn{4}{c|}{Monte Carlo CI} & \multicolumn{3}{c|}{Analytical CI} & \multicolumn{4}{c}{Monte Carlo CI}\\
$p+q$ & \multicolumn{1}{c|}{$n$} & RMSE & SD & CI Len & Cov & SE & CI Len & Cov & RMSE & SD & CI Len & Cov\\ \hline
\multirow{5}{*}{$100$} & \multicolumn{1}{c|}{$250$}  & 0.513 & 0.304 & 1.193 & 0.724 & 0.300 & 1.176 & 0.720 & 0.832 & 0.236 & 0.926 & 0.066\\
 & \multicolumn{1}{c|}{$500$}  & 0.165 & 0.165 & 0.648 & 0.844 & 0.182 & 0.714 & 0.860 & 0.614 & 0.147 & 0.576 & 0.026\\
 & \multicolumn{1}{c|}{$750$}  & 0.197 & 0.139 & 0.545 & 0.944& 0.147 & 0.575 & 0.972& 0.555 & 0.123 & 0.481 & 0.010\\
 & \multicolumn{1}{c|}{$1000$} & 0.118 & 0.117 & 0.459 & 0.946 & 0.127 & 0.497 & 0.954& 0.484 & 0.099 & 0.389 & 0.000\\
 & \multicolumn{1}{c|}{$1250$} & 0.121 & 0.118 & 0.462 & 0.954 & 0.117 & 0.457 & 0.960& 0.434 & 0.101 & 0.396 & 0.020\\ \hline
\multirow{5}{*}{$800$} & \multicolumn{1}{c|}{$250$}  & 0.208 & 0.206 & 0.806 & 0.954 & 0.336 & 1.316 & 1.000 & 0.421 & 0.194 & 0.762 & 0.490\\
 & \multicolumn{1}{c|}{$500$}  & 0.154 & 0.142 & 0.557 & 0.890 & 0.185 & 0.725 & 0.974 & 0.345 & 0.048 & 0.189 & 0.000\\
 & \multicolumn{1}{c|}{$750$}  & 0.149 & 0.124 & 0.485 & 0.936& 0.140 & 0.549 & 0.938 & 0.361 & 0.050 & 0.195 & 0.000\\
 & \multicolumn{1}{c|}{$1000$} & 0.113 & 0.113 & 0.442 & 0.940& 0.122 & 0.480 & 0.972 & 0.303 & 0.039 & 0.152 & 0.000\\
 & \multicolumn{1}{c|}{$1250$} & 0.097 & 0.092 & 0.361 & 0.956& 0.106 & 0.414 & 0.976 & 0.268 & 0.034 & 0.135 & 0.000\\ \hline
\multirow{5}{*}{$1500$} & \multicolumn{1}{c|}{$250$}  & 0.336 & 0.188 & 0.735 & 0.681 & 0.270 & 1.057 & 0.934 & 0.215 & 0.139 & 0.543 & 0.768\\
 & \multicolumn{1}{c|}{$500$}  & 0.145 & 0.125 & 0.491 & 0.918 & 0.151 & 0.592 & 0.956 & 0.279 & 0.035 & 0.139 & 0.000\\
 & \multicolumn{1}{c|}{$750$}  & 0.109 & 0.100 & 0.393 & 0.932 & 0.118 & 0.461 & 0.958 & 0.228 & 0.027 & 0.104 & 0.000\\
 & \multicolumn{1}{c|}{$1000$} & 0.083 & 0.083 & 0.325 & 0.952 & 0.099 & 0.387 & 0.982 & 0.200 & 0.022 & 0.087 & 0.000\\
 & \multicolumn{1}{c|}{$1250$} & 0.089 & 0.080 & 0.314 & 0.920 & 0.087 & 0.342 & 0.942 & 0.181 & 0.021 & 0.083 & 0.000\\ \hline
\multirow{5}{*}{$2500$} & \multicolumn{1}{c|}{$250$}  & 0.267 & 0.223 & 0.872 & 0.904 & 0.313 & 1.227 & 0.980 & 0.276 & 0.124 & 0.486 & 0.492\\
 & \multicolumn{1}{c|}{$500$}  & 0.135 & 0.132 & 0.519 & 0.934 & 0.171 & 0.670 & 0.992 & 0.194 & 0.023 & 0.091 & 0.000\\
 & \multicolumn{1}{c|}{$750$}  & 0.137 & 0.113 & 0.445 & 0.890 & 0.130 & 0.510 & 0.944 & 0.226 & 0.025 & 0.099 & 0.000\\
 & \multicolumn{1}{c|}{$1000$} & 0.104 & 0.095 & 0.370 & 0.922 & 0.109 & 0.427 & 0.962 & 0.176 & 0.018 & 0.072 & 0.000\\
 & \multicolumn{1}{c|}{$1250$} & 0.089 & 0.088 & 0.344 & 0.946 & 0.096 & 0.376 & 0.960 & 0.112 & 0.012 & 0.048 & 0.000\\ \hline
\end{tabular}%
}
\caption{{\color{black}Interaction setting, minimal-signal DGP, $\sigma^2=1$: debiased ($\hat\theta_0^{\mathrm{cf}}$) vs.\ naive ($\hat \theta^{\rm plug-in}$). $s=k_1=k_2=3$ fixed; half of nonzero coefficients are weak ($\sim n_{\max}^{-1/2}$). Averaged over 500 replications, $K_1=K_2=2.75$.}}
\label{tab:simu_signal_sigma1}
\end{table}

\begin{table}[!ht]
\centering
\small
\renewcommand{\arraystretch}{1.4}
\resizebox{1\textwidth}{!}{
\begin{tabular}{cc|cccc|ccc|cccc} \hline
\multicolumn{2}{c|}{heteroscedastic noise} & \multicolumn{11}{c}{$\sigma^2 = 1$}\\ \hline
\multicolumn{2}{c|}{Estimators} & \multicolumn{7}{c|}{$\hat\theta_0^{\mathrm{cf}}$ (Debiasing)} & \multicolumn{4}{c}{$\hat \theta^{\rm plug-in}$}\\ \hline
 & & \multicolumn{4}{c|}{Monte Carlo CI} & \multicolumn{3}{c|}{Analytical CI} & \multicolumn{4}{c}{Monte Carlo CI}\\
$p+q$ & \multicolumn{1}{c|}{$n$} & RMSE & SD & CI Len & Cov & SE & CI Len & Cov & RMSE & SD & CI Len & Cov\\ \hline
\multirow{5}{*}{$100$} & \multicolumn{1}{c|}{$250$}  & 0.691 & 0.337 & 1.320 & 0.598 & 0.305 & 1.196 & 0.500 & 0.972 & 0.248 & 0.970 & 0.024\\
 & \multicolumn{1}{c|}{$500$}  & 0.198 & 0.178 & 0.698 & 0.804 & 0.189 & 0.740 & 0.830 & 0.717 & 0.160 & 0.628 & 0.010\\
 & \multicolumn{1}{c|}{$750$}  & 0.214 & 0.143 & 0.561 & 0.912& 0.152 & 0.596 & 0.924& 0.620 & 0.126 & 0.495 & 0.002\\
 & \multicolumn{1}{c|}{$1000$} & 0.129 & 0.122 & 0.479 & 0.934 & 0.132 & 0.518 & 0.926& 0.558 & 0.109 & 0.426 & 0.000\\
 & \multicolumn{1}{c|}{$1250$} & 0.133 & 0.127 & 0.496 & 0.932 & 0.123 & 0.481 & 0.946 & 0.512 & 0.107 & 0.421 & 0.004\\ \hline
\multirow{5}{*}{$800$} & \multicolumn{1}{c|}{$250$}  & 0.238 & 0.224 & 0.879 & 0.942 & 0.385 & 1.508 & 0.996 & 0.490 & 0.216 & 0.845 & 0.440\\
 & \multicolumn{1}{c|}{$500$}  & 0.169 & 0.147 & 0.575 & 0.898 & 0.205 & 0.804 & 0.978 & 0.442 & 0.059 & 0.232 & 0.000\\
 & \multicolumn{1}{c|}{$750$}  & 0.158 & 0.130 & 0.510 & 0.910& 0.153 & 0.600 & 0.946 & 0.437 & 0.056 & 0.220 & 0.000\\
 & \multicolumn{1}{c|}{$1000$} & 0.119 & 0.119 & 0.466 & 0.938 & 0.133 & 0.521 & 0.964 & 0.358 & 0.042 & 0.167 & 0.000\\
 & \multicolumn{1}{c|}{$1250$} & 0.098 & 0.098 & 0.384 & 0.952 & 0.114 & 0.447 & 0.976 & 0.329 & 0.040 & 0.157 & 0.000\\ \hline
\multirow{5}{*}{$1500$} & \multicolumn{1}{c|}{$250$}  & 0.446 & 0.258 & 1.011 & 0.696 & 0.375 & 1.469 & 0.961 & 0.247 & 0.175 & 0.684 & 0.834\\
 & \multicolumn{1}{c|}{$500$}  & 0.174 & 0.150 & 0.586 & 0.910 & 0.200 & 0.783 & 0.980 & 0.377 & 0.046 & 0.180 & 0.000\\
 & \multicolumn{1}{c|}{$750$}  & 0.168 & 0.125 & 0.491 & 0.850 & 0.148 & 0.579 & 0.914 & 0.326 & 0.037 & 0.145 & 0.000\\
 & \multicolumn{1}{c|}{$1000$} & 0.098 & 0.098 & 0.383 & 0.954 & 0.121 & 0.475 & 0.986 & 0.285 & 0.030 & 0.119 & 0.000\\
 & \multicolumn{1}{c|}{$1250$} & 0.104 & 0.096 & 0.374 & 0.928 & 0.108 & 0.423 & 0.960 & 0.252 & 0.027 & 0.106 & 0.000\\ \hline
\multirow{5}{*}{$2500$} & \multicolumn{1}{c|}{$250$}  & 0.719 & 0.312 & 1.221 & 0.465 & 0.464 & 1.820 & 0.781 & 0.446 & 0.189 & 0.743 & 0.405\\
 & \multicolumn{1}{c|}{$500$}  & 0.170 & 0.159 & 0.623 & 0.772 & 0.226 & 0.884 & 0.992 & 0.271 & 0.030 & 0.118 & 0.000\\
 & \multicolumn{1}{c|}{$750$}  & 0.207 & 0.131 & 0.513 & 0.936& 0.170 & 0.665 & 0.902 & 0.328 & 0.032 & 0.127 & 0.000\\
 & \multicolumn{1}{c|}{$1000$} & 0.118 & 0.114 & 0.446 & 0.942 & 0.138 & 0.542 & 0.984 & 0.260 & 0.026 & 0.102 & 0.000\\
 & \multicolumn{1}{c|}{$1250$} & 0.110 & 0.108 & 0.422 & 0.936 & 0.122 & 0.479 & 0.970 & 0.178 & 0.019 & 0.075 & 0.000\\ \hline
\end{tabular}%
}
\caption{{\color{black}Interaction setting, heteroscedastic-noise DGP, $\sigma^2=1$: debiased ($\hat\theta_0^{\mathrm{cf}}$) vs.\ naive ($\hat \theta^{\rm plug-in}$). The noise variance is $\sigma^2 v(X_i,M_i)$ with $\bbE[v]=1$; sparsity fixed at $s=k_1=k_2=3$. Averaged over 500 replications, $K_1=K_2=2.75$.}}
\label{tab:simu_hetero_sigma1}
\end{table}

\begin{table}[!ht]
\centering
\small
\renewcommand{\arraystretch}{1.4}
\resizebox{1\textwidth}{!}{
\begin{tabular}{cc|cccc|ccc|cccc} \hline
\multicolumn{2}{c|}{heteroscedastic noise} & \multicolumn{11}{c}{$\sigma^2 = 0.5$}\\ \hline
\multicolumn{2}{c|}{Estimators} & \multicolumn{7}{c|}{$\hat\theta_0^{\mathrm{cf}}$ (Debiasing)} & \multicolumn{4}{c}{$\hat \theta^{\rm plug-in}$}\\ \hline
 & & \multicolumn{4}{c|}{Monte Carlo CI} & \multicolumn{3}{c|}{Analytical CI} & \multicolumn{4}{c}{Monte Carlo CI}\\
$p+q$ & \multicolumn{1}{c|}{$n$} & RMSE & SD & CI Len & Cov & SE & CI Len & Cov & RMSE & SD & CI Len & Cov\\ \hline
\multirow{5}{*}{$100$} & \multicolumn{1}{c|}{$250$}  & 0.716 & 0.327 & 1.280 & 0.526 & 0.297 & 1.164 & 0.432 & 0.872 & 0.235 & 0.919 & 0.046\\
 & \multicolumn{1}{c|}{$500$}  & 0.186 & 0.169 & 0.664 & 0.922 & 0.179 & 0.700 & 0.936 & 0.596 & 0.144 & 0.566 & 0.032\\
 & \multicolumn{1}{c|}{$750$}  & 0.211 & 0.133 & 0.521 & 0.772 & 0.142 & 0.558 & 0.808 & 0.519 & 0.112 & 0.439 & 0.008\\
 & \multicolumn{1}{c|}{$1000$} & 0.120 & 0.116 & 0.455 & 0.936 & 0.124 & 0.485 & 0.954 & 0.456 & 0.095 & 0.371 & 0.002\\
 & \multicolumn{1}{c|}{$1250$} & 0.129 & 0.121 & 0.474 & 0.934 & 0.116 & 0.455 & 0.924 & 0.420 & 0.091 & 0.356 & 0.010\\ \hline
\multirow{5}{*}{$800$} & \multicolumn{1}{c|}{$250$}  & 0.229 & 0.224 & 0.876 & 0.940& 0.384 & 1.504 & 0.998 & 0.396 & 0.214 & 0.840 & 0.626\\
 & \multicolumn{1}{c|}{$500$}  & 0.174 & 0.142 & 0.556 & 0.888 & 0.199 & 0.781 & 0.974 & 0.368 & 0.048 & 0.188 & 0.000\\
 & \multicolumn{1}{c|}{$750$}  & 0.153 & 0.126 & 0.494 & 0.894 & 0.146 & 0.571 & 0.946 & 0.356 & 0.046 & 0.182 & 0.000\\
 & \multicolumn{1}{c|}{$1000$} & 0.117 & 0.111 & 0.436 & 0.938 & 0.127 & 0.500 & 0.966 & 0.297 & 0.036 & 0.141 & 0.000\\
 & \multicolumn{1}{c|}{$1250$} & 0.093 & 0.092 & 0.359 & 0.950 & 0.108 & 0.424 & 0.986 & 0.268 & 0.033 & 0.129 & 0.000\\ \hline
\multirow{5}{*}{$1500$} & \multicolumn{1}{c|}{$250$}  & 0.365 & 0.226 & 0.887 & 0.755 & 0.348 & 1.364 & 0.976 & 0.202 & 0.178 & 0.697 & 0.913\\
 & \multicolumn{1}{c|}{$500$}  & 0.162 & 0.144 & 0.563 & 0.866 & 0.195 & 0.763 & 0.990 & 0.314 & 0.039 & 0.151 & 0.000\\
 & \multicolumn{1}{c|}{$750$}  & 0.154 & 0.119 & 0.468 & 0.924& 0.142 & 0.557 & 0.928 & 0.263 & 0.030 & 0.118 & 0.000\\
 & \multicolumn{1}{c|}{$1000$} & 0.095 & 0.093 & 0.364 & 0.942 & 0.116 & 0.454 & 0.978 & 0.231 & 0.025 & 0.097 & 0.000\\
 & \multicolumn{1}{c|}{$1250$} & 0.098 & 0.091 & 0.357 & 0.943& 0.103 & 0.405 & 0.962 & 0.207 & 0.022 & 0.088 & 0.000\\ \hline
\multirow{5}{*}{$2500$} & \multicolumn{1}{c|}{$250$}  & 0.656 & 0.296 & 1.159 & 0.509 & 0.443 & 1.735 & 0.823 & 0.322 & 0.188 & 0.738 & 0.694\\
 & \multicolumn{1}{c|}{$500$}  & 0.178 & 0.152 & 0.597 & 0.808 & 0.223 & 0.874 & 0.992 & 0.226 & 0.025 & 0.099 & 0.000\\
 & \multicolumn{1}{c|}{$750$}  & 0.190 & 0.128 & 0.500 & 0.900& 0.167 & 0.653 & 0.932 & 0.270 & 0.027 & 0.106 & 0.000\\
 & \multicolumn{1}{c|}{$1000$} & 0.112 & 0.109 & 0.429 & 0.948 & 0.136 & 0.532 & 0.990 & 0.210 & 0.022 & 0.084 & 0.000\\
 & \multicolumn{1}{c|}{$1250$} & 0.107 & 0.104 & 0.407 & 0.940 & 0.119 & 0.465 & 0.968 & 0.141 & 0.015 & 0.060 & 0.000\\ \hline
\end{tabular}%
}
\caption{{\color{black}Interaction setting, heteroscedastic-noise DGP, $\sigma^2=0.5$: debiased ($\hat\theta_0^{\mathrm{cf}}$) vs.\ naive ($\hat \theta^{\rm plug-in}$). The noise variance is $\sigma^2 v(X_i,M_i)$ with $\bbE[v]=1$; sparsity fixed at $s=k_1=k_2=3$. Averaged over 500 replications, $K_1=K_2=2.75$.}}
\label{tab:simu_hetero_sigma0p5}
\end{table}

\begin{table}[!ht]
\centering
\small
\renewcommand{\arraystretch}{1.4}
\resizebox{1\textwidth}{!}{
\begin{tabular}{c|ccccccccc} \hline
\multicolumn{1}{c|}{$n = 750$} & \multicolumn{9}{c}{$K_1 = K_2 = K$}\\ \hline
$p+q$ & $1$ & $1.25$ & $1.5$ & $1.75$ & $2$ & $2.25$ & $2.5$ & $2.75$ & $3$ \\ \hline
$100$ & 0.137 & 0.134 & 0.131 & 0.129 & 0.127 & \textbf{0.126} & \textbf{0.126} & \textbf{0.126} & 0.126\\
$800$ & \textbf{0.102} & \textbf{0.102} & 0.102 & \textbf{0.101} & 0.102 & 0.102 & 0.103 & 0.104 & 0.104\\
$1500$ & 0.139 & 0.137 & \textbf{0.136} & 0.136 & 0.136 & 0.136 & 0.136 & \textbf{0.135} & \textbf{0.135}\\
$2500$ & 0.130 & 0.129 & 0.129 & 0.129 & 0.129 & 0.128 & \textbf{0.127} & \textbf{0.125}& \textbf{0.125}\\
\hline
\end{tabular}%
}
\caption{{\color{black}Sensitivity of the RMSE of $\hat\theta_0^{\mathrm{cf}}$ to the common balancing constant $K_1=K_2=K$ under the base DGP ($s=k_1=k_2=3$, $\sigma^2=0.5$), $n=750$. Averaged over 500 replications; the three smallest values in each row are boldfaced.}}
\label{tab:sens_diag_n750}
\end{table}

\begin{table}[!ht]
\centering
\small
\renewcommand{\arraystretch}{1.4}
\resizebox{1\textwidth}{!}{
\begin{tabular}{c|ccccccccc} \hline
\multicolumn{1}{c|}{$n = 1250$} & \multicolumn{9}{c}{$K_1 = K_2 = K$}\\ \hline
$p+q$ & $1$ & $1.25$ & $1.5$ & $1.75$ & $2$ & $2.25$ & $2.5$ & $2.75$ & $3$ \\ \hline
$100$ & 0.124 & 0.122 & 0.121 & 0.121 & 0.121 & \textbf{0.120} & 0.120 & \textbf{0.119} & \textbf{0.119}\\
$800$ & 0.081 & 0.079 & 0.077 & 0.076 & 0.075 & \textbf{0.074} & \textbf{0.074} & \textbf{0.074} & 0.074\\
$1500$ & 0.100 & 0.099 & 0.098 & 0.097 & \textbf{0.096} & \textbf{0.096} & \textbf{0.096} & 0.096 & 0.096\\
$2500$ & 0.093 & 0.093 & 0.093 & 0.092 & \textbf{0.091} & 0.091 & 0.091 & \textbf{0.090} & \textbf{0.090}\\
\hline
\end{tabular}%
}
\caption{{\color{black}Sensitivity of the RMSE of $\hat\theta_0^{\mathrm{cf}}$ to the common balancing constant $K_1=K_2=K$ under the base DGP ($s=k_1=k_2=3$, $\sigma^2=0.5$), $n=1250$. Averaged over 500 replications; the three smallest values in each row are boldfaced.}}
\label{tab:sens_diag_n1250}
\end{table}

\begin{table}[!ht]
\centering
\small
\renewcommand{\arraystretch}{1.4}
\resizebox{1\textwidth}{!}{
\begin{tabular}{c|ccccccccc} \hline
$K_1 \backslash K_2$ & $1$ & $1.25$ & $1.5$ & $1.75$ & $2$ & $2.25$ & $2.5$ & $2.75$ & $3$ \\ \hline
$1$ & 0.075 & 0.074 & 0.074 & 0.074 & 0.073 & 0.073 & 0.073 & 0.073 & 0.073\\
$1.25$ & 0.075 & 0.075 & 0.074 & 0.074 & 0.074 & 0.074 & 0.074 & 0.074 & 0.074\\
$1.5$ & 0.075 & 0.075 & 0.074 & 0.074 & 0.074 & 0.074 & 0.074 & 0.074 & 0.074\\
$1.75$ & 0.075 & 0.075 & 0.074 & 0.074 & 0.074 & 0.073 & 0.073 & 0.073 & 0.073\\
$2$ & 0.075 & 0.074 & 0.074 & 0.074 & 0.073 & 0.073 & 0.073 & 0.073 & 0.073\\
$2.25$ & 0.075 & 0.074 & 0.074 & 0.073 & 0.073 & \textbf{0.072} & \textbf{0.072} & \textbf{0.072} & \textbf{0.072}\\
$2.5$ & 0.075 & 0.074 & 0.074 & 0.073 & 0.073 & \textbf{0.072} & \textbf{0.072} & \textbf{0.072} & \textbf{0.072}\\
$2.75$ & 0.075 & 0.074 & 0.073 & 0.073 & 0.073 & \textbf{0.072} & \textbf{0.072} & \textbf{0.072} & \textbf{0.072}\\
$3$ & 0.075 & 0.074 & 0.073 & 0.073 & \textbf{0.072} & \textbf{0.072} & \textbf{0.072} & \textbf{0.072} & \textbf{0.072}\\
\hline
\end{tabular}%
}
\caption{{\color{black}Sensitivity of the RMSE of $\hat\theta_0^{\mathrm{cf}}$ to $(K_1, K_2)$ under the base DGP ($s=k_1=k_2=3$, $\sigma^2=0.5$), $p+q=1500$, $n=1250$. Averaged over 500 replications; the five smallest values are boldfaced.}}
\label{tab:sens_grid}
\end{table}

\end{document}